\documentclass[fleqn,usenatbib]{mnras}

\usepackage{newtxtext,newtxmath}

\usepackage[T1]{fontenc}

\DeclareRobustCommand{\VAN}[3]{#2}
\let\VANthebibliography\thebibliography
\def\thebibliography{\DeclareRobustCommand{\VAN}[3]{##3}\VANthebibliography}

\usepackage{graphicx}	
\usepackage{amsmath}	

\usepackage{pdflscape}   
\usepackage{subfig}   

\newcommand{\editone}[1]{#1}
\newcommand{\edittwo}[1]{#1}
\newcommand{\editthree}[1]{#1}

\newcommand{\editfive}[1]{#1}
\newcommand{\editsix}[1]{#1}
\newcommand{\editseven}[1]{#1}
\newcommand{\editeight}[1]{#1}
\newcommand{\editnine}[1]{#1}
\newcommand{\editten}[1]{#1}
\newcommand{\editelev}[1]{#1}
\newcommand{\edittwelve}[1]{#1}

\newcommand{\editfourteen}[1]{#1}
\newcommand{\editfifteen}[1]{#1}
\newcommand{\refone}[1]{#1}
\newcommand{\reftwo}[1]{#1}
\newcommand{\refthree}[1]{#1}
\newcommand{\reffour}[1]{#1}
\newcommand{\reffive}[1]{#1}

\title[Starspots on WTTS]{Starspots on weak-lined T Tauri stars in the $\sigma$ Orionis cluster}

\author[A. J. Wilson et al.]{
	Andrew J. Wilson,$^{1}$\thanks{E-mail: a.j.wilson@exeter.ac.uk, andyjwilson\_uk@hotmail.com}
	Darryl J. Sergison,$^{1}$
	Tim Naylor$^{1}$
	and Robin D. Jeffries$^{2}$
	\\
	$^{1}$Department of Physics and Astronomy, University of Exeter, Stocker Road, Exeter EX4 4QL, UK\\
	$^{2}$Astrophysics Group, Keele University, Keele ST5 5BG, UK
}

\date{Accepted XXX. Received YYY; in original form ZZZ}

\pubyear{2026}

\begin{document}
\label{firstpage}
\pagerange{\pageref{firstpage}--\pageref{lastpage}}
\maketitle

\begin{abstract}
The light curves and spectra of T Tauri stars indicate the presence of cool starspots covering large areas of their photospheres. Heavily spotted stars should have a larger mid-infrared flux than the equivalent single-temperature photospheres. 
\refone{To test this prediction we first fit single and two-temperature models to optical spectra of young stars in the 4\,Myr-old $\sigma$ Orionis cluster and a control set of older field stars. }
\refone{The fits to the young stars favour two temperatures} while the field stars show no preference. 
\refone{Independent tests of our models for nine probable single K-type weak-lined T Tauri stars using} infrared photometry confirms the presence of large cool components. We consider and rule out as explanations of the two temperature photospheres: binary stars \refone{for all but one WTTS}; accretion, as the \refthree{observed} H$\alpha$ equivalent width is too small; gravity darkening as the \refthree{predicted} temperature differences are too small. 
This leaves cool \reftwo{areas akin to starspots} covering 30—95 per cent of the surface as the plausible explanation. 
Our results show no significant difference in \reftwo{the cool temperature} \refthree{for} surface fractions \refthree{between} \refone{0.30--0.75, implying} a common mechanism is responsible. 
\reftwo{These cool surface fractions} would lead to significant overestimates in age and mass when fitting unspotted isochrones in colour-magnitude diagrams. \reftwo{The dispersion in cool surface fraction could also} explain the dispersion in lithium equivalent width seen in the spectra of coeval zero-age main sequence stars.

\end{abstract}




\section{\editten{Introduction}}
\label{sec:introduction}

Starspots are cool regions of the stellar photosphere created where a strong magnetic field breaks through the stellar surface \citep{1955ApJ...121..491P}. The magnetic field inhibits convection, creating a cool inner umbra. This is surrounded by a warmer penumbra, which is composed of horizontal filaments that are created by the Evershed flow \citep{1909MNRAS..69..454E}. It has long been suspected that young stars have cool spots covering large areas of their photospheres. By using \ion{Ca}{ii} \editfourteen{H and K} emission as indicators of stellar activity \citep{1968ApJ...153..221W}, it was found by \citet{1972ApJ...171..565S} that surface magnetic field strength decreases with age, Skumanich's law. \editfourteen{This trend was} confirmed for low mass young stars 20--250\,Myr by \citet{2016MNRAS.457..580F}, with magnetic fields in T Tauri stars (TTS) around ten times stronger than Zero-Age Main Sequence (ZAMS) stars. Using Zeeman-Doppler imaging and Zeeman line broadening to derive magnetic field strengths and their variation, \citet{2025ApJ...985...46Y} inferred the same mechanism is responsible for surface magnetic fields on TTS as main sequence (MS) stars.

The stellar dynamo responsible for magnetic activity is driven by stellar rotation \citep{1955ApJ...122..293P} coupled with the convective zone \citep{2011SoPh..269....3R}. The Rossby number \citep{1984ApJ...279..763N} relates stellar rotation to the convective turnover time, and is inversely correlated with the magnetic field strength. At short rotation periods, corresponding to \edittwelve{small} Rossby numbers, the magnetic field strength saturates with the surface covered by active regions \citep{1984A&A...133..117V,1996IAUS..176..237S,2016MNRAS.457..580F}.

Young low mass stars spend the first few million years after the dispersal of their prenatal clouds as TTS. They start as accreting classical T Tauri stars (CTTS) exhibiting emission lines in their spectra and a strong infrared (IR) excess from their accretion disc. During this evolutionary phase, the rotation of the star and inner disc \edittwelve{may be} locked together, with the disc acting as a \edittwelve{brake} on the stellar rotation \citep{1991ApJ...370L..39K,1993A&A...274..309C}. When accretion stops, they become weak-lined T Tauri stars (WTTS), and may retain a residual disc with little or no IR excess. The star and disc become decoupled and the young star can spin up, increasing the strength of the stellar dynamo. 
\edittwelve{Evidence for the decoupling of the star and disc rotation is provided by the bimodal distribution in the rotation periods of TTS \citep{1992ApJ...398L..61A,1993AJ....106..372E,2001ApJ...554L.197H,Herbst2002,2021MNRAS.506.5989F}, where the discless WTTS tend to be the \editfourteen{faster} rotators with periods of approximately 1--5 days.}

Inflated radii of 3--15 per cent have been observed in low mass pre-main sequence (PMS) stars, \edittwelve{ZAMS stars and in the tidally-locked MS components of close binaries} \citep{2002ApJ...567.1140T,2005ApJ...631.1120L,Jackson2018,2018AJ....155..225K,2019MNRAS.483.1125J,2019MNRAS.489.2615M}. This may be explained if the stars are magnetically active, as convection would be inhibited and cool starspots would reduce the flux emitted from the surface, leading to an inflated stellar radius. Near-IR spectra of a young binary star with an age <10\,Myr were modelled by \citet{2024AJ....168..134K}. Their results indicated the lower mass component had both a larger radius and a larger fraction of its surface covered by cool spots. \citet{2025ApJ...988L...1C} reviewed published data on radii, rotation and starspot coverage for 261 stars from 10--670\,Myr in age. They found radius inflation depends on spot coverage, and bolometric luminosity is preserved under radius inflation. They also showed a relationship between \edittwelve{radius inflation and rotation,} represented by the Rossby number.

The fractional temperature difference between the unblemished photosphere and the cool spots follows a decreasing trend, such that cooler photospheres have cooler and lower contrast spots. For a sample of giants, subgiants, dwarfs and TTS, \citet{2005LRSP....2....8B} found spot temperatures 2000\,K cooler than the unblemished photosphere for G0 stars to 200\,K cooler for M4 stars. In 3D radiative magnetohydrodynamic (MHD) simulations of starspots, \citet{2020ApJ...893..113P} found spots to be 1400\,K cooler for G2V stars and 350\,K cooler for M0V stars.

\edittwelve{As cool spots are darker than the unblemished photosphere, they can reveal their presence through a modulation of the light curve at the rotation period of the star. Periodic variations from the optical to near-IR have uncovered cool spots covering 3--36 per cent of the surfaces of WTTS and CTTS \citep{Rydgren1983,Bouvier1986,Bouvier1989}. The accompanying variation in colour indicates the modulation is driven by hot plages as well as cool spots  \citep{1985AJ.....90..326V,1986ApJ...306..199V,1993A&A...272..176B}.}

Where starspots are distributed across the surface of a star \editfourteen{or located at high latitudes}, the light curve technique will have a tendency to underestimate the surface coverage. As one spot moves off the visible surface, so another may be rotating into view. It is possible to obtain a better estimate of the surface coverage by analysing the spectra of spotted stars, an approach proposed by \citet{1980ApJ...239L.121R}. As the spots will be at a cooler temperature than the immaculate photosphere, spots may be detected where a stellar spectrum is a blend of two or more components at different temperatures. The surface coverage may be estimated by finding the best fitting model by combining either synthetic spectra or template spectra of quiet stars of known spectral type. As the cool component will emit less flux per unit area than the hot component, successful use of the technique requires the cool component covers a sufficiently large area to be detected above the noise. Binary stars can also give rise to blended temperature spectra, so they must be considered when reviewing results.

The technique was first used on older stars \citep{1979PASP...91..616V,1981ApJ...247..975V,1987ApJ...319..392H,Neff1995,ONeal1996} and the secondary stars of cataclysmic variables \citep{Webb2002}. Molecular absorption bands such as TiO usually seen in cool M-type stars were observed in the spectra of G and K-type stars where the surface temperature is too hot for molecules.

In the past decade the technique has been extended to TTS. Using near-IR spectroscopy of the WTTS LkCa 4, \citet{Gully-Santiago2017} found a spot coverage of 80 per cent with cool component at  2700--3000\,K and the hot component at 4100\,K. A spot filling factor of 65 per cent was found for this star by \citet{2022MNRAS.517.2165C} using high-resolution $H$-band spectra. \citet{2023ApJ...946...10P} observed LkCa 4 over 5 nights, finding spot coverages 77-94 per cent, with the cool component at 3060\,K and the hot at 4100--4400\,K. This indicates spot coverage is not uniform across the stellar surface, and may change over longer time periods.
\citet{2024ApJ...967...45P} fitted near-IR spectra of 10 WTTS in the Taurus-Auriga star forming region with models composed of two temperature photospheres, and a warm disc modelled by a blackbody at 1500\,K. Their best fitting models found spot coverages of 42--84 per cent of the photosphere.
\citet{2025ApJ...978...32P} re-examined NIR spectra of 16 CTTS originally analysed by \citet{2011ApJ...730...73F}. Their updated technique fits a model composed of a two-temperature photosphere, with a 1500\,K blackbody for a warm disc, and a 4000--12\,500\,K blackbody to model accretion. This gave spot coverages from 45--87 per cent, explaining the mysterious $IYJ$ excess found by \citet{2011ApJ...730...73F}. 
\citet{2025ApJ...990..205P} fit optical to NIR spectra of 56 WTTS with two-temperature model spectra, finding spot coverages 0--87 per cent, with coverages above 50 per cent common. By including spots, their models increased stellar mass estimates by up to 80 per cent and ages by 0.3--0.5\,dex. 
\refone{They suggest the surface gravity \reftwo{of the umblemished photosphere and cool spots should be be fitted separately}, as the effective surface gravity may be 1--2\,dex higher for starspots. This is because surface gravity inferences from spectra are measuring the pressure where the spectrum is formed, and the magnetic pressure can be significant or the dominant factor within starspots.} 
\citet{2025ApJ...988L...1C} analysed APOGEE NIR spectra for a sample of star clusters with ages 10--670\,Myr, finding spot coverage decreases with age. For 106 members of the Upper Scorpius cluster with an age around 10\,Myr, they found spot coverages of 8--84 percent with a median of 42 per cent.

A rudimentary comparison between the results of the light curve and spectroscopic techniques, confirms that spectroscopy is finding larger surface coverages of cool spots. As we expect the light curve technique to underestimate spot cover\editfifteen{age}, so spectroscopy is likely to be giving a more accurate picture of spot coverage and magnetism \editfourteen{on} the PMS.

\refone{A large fractional coverage by cool \reftwo{areas} will have a significant impact on the structure of a star. \citet{Somers2015b} found the radii of heavily spotted PMS stars could be inflated by up to 10 per cent, decreasing their luminosities and effective temperatures. Using unspotted models on spotted PMS stars can understate their mass by a factor of two, and age by two to ten times. The SPOTS models \citep{2020ApJ...891...29S} address this by providing stellar evolutionary tracks and isochrones for stars with 0--85 per cent spot surface coverage from the PMS to the red giant branch.}

\refone{As the surface coverage by cool spots approaches 50 per cent, the term cool \reftwo{areas} may be a more appropriate \refthree{topological} description than spots. However, to convey our meaning in the most simple terms, we shall \reftwo{sometimes} use the term spots to describe cool areas caused by magnetic inhibition of surface convection regardless of the area covered. More importantly, the unblemished photosphere may have little or no meaning when a large fraction of the surface is composed of cool areas. In such an active photosphere the hotter areas may be dominated by plages and faculae.}

We will investigate whether starspots, and by inference strong magnetic fields, are present in WTTS by trialling a variation of the spectroscopic technique developed by \citet{Webb2002} based on the TiO band head at 7050\,\r{A}. The approach observes K-type stars, as their unspotted photospheres are too hot for molecules of TiO. While the temperature of starspots will be around \edittwelve{1000--1500\,K cooler}, allowing molecules of TiO to form. If the spectra of K-type stars show TiO absorption, this will indicate the presence of a cooler photosphere component. This will be formalised by comparing the best fitting single-temperature and two-temperature model photospheres. The aim is to test whether this technique can detect evidence for two temperatures in the spectra of WTTS. This will not by itself confirm the presence of starspots. So we shall investigate other causes for two-temperature photospheres, including binarity and gravity darkening. \reffour{If starspots are present, then the photosphere may be at multiple temperatures due to (for example) the spot umbra and penumbra, and magnetic activity may cause plages and faculae. In this case the two-temperature model would be a simplification, approximating the two components providing the strongest contribution to the spectrum.}

We shall verify our results by comparing synthetic photometry from our models \refone{(flux calibrated
using \textit{Gaia} DR3 $G_{\rm RP}$)} to IR observations from the Two Micron All Sky Survey \citep[2MASS;][]{Skrutskie2006} and \textit{Spitzer} \citep{2004ApJS..154....1W}. 
Cool spots emit a significant proportion of their flux in the IR, creating an IR excess compared to \refone{an} unspotted star \refone{with the same $G_{\rm RP}$}. 
This will provide an independent test of whether the two-temperature \editfourteen{spotted} models fitted in the optical offer a significant improvement over the single-temperature \editfourteen{unspotted} models \editfourteen{of the same total luminosity}.

To simplify the analysis, discless K-type WTTS were observed. WTTS should exhibit little or no accretion that would otherwise complicate the analysis. Selecting K-type stars with a narrow range of effective temperatures will allow us to investigate whether the cool spot temperature varies with surface coverage.

In Section~\ref{sec:vlt_observations} we provide a summary of the observations, then in Section~\ref{sec:membership_determination} refine the membership of the $\sigma$ Orionis cluster, and in Section~\ref{sec:vlt_data_reduction} we explain how the spectra were reduced. Our technique for assessing whether the stellar spectra are best fit by a single or two-temperature photosphere is described in Section~\ref{sec:fitting_model_spectra}. To enable a straightforward comparison between WTTS and older field stars, we refine our sample selection in Section~\ref{sec:sample_selection}. We then review the results in Section~\ref{sec:review}. We test our results using independent IR photometry in Section~\ref{sec:photometry}. Then, in Section~\ref{sec:blended_spectra_causes} we explore the possible reasons why the WTTS spectra may be best fitted by a blend of two effective temperatures. Finally, we consider the implications of our findings in Section~\ref{sec:discussion}.

\section{Data}
\label{sec:data}

\subsection{Observations}
\label{sec:vlt_observations}

As part of ESO period 92, spectra were obtained using the Very Large Telescope (VLT) for a \editten{sample of K-type TTS in the $\sigma$ Orionis Cluster (SOC) selected from \citet{2013MNRAS.434..806B}. In addition, a set of non-members also selected from \citet{2013MNRAS.434..806B}} were observed in the same field to enable a comparison between YSOs and more evolved stars. \editten{The spectra were taken with \editfifteen{the} FLAMES (Fibre Large Array Element Spectrograph) GIRAFFE spectrograph \citep{2002Msngr.110....1P} fed by the MEDUSA optical fibres \citep{1998ASPC..152...32B}. Each fibre} \editfive{yielded} an aperture of 1.2$\arcsec$ on the sky \citep[table 1]{ESO2014}. The GIRAFFE spectrograph was configured in low resolution mode with the 600\,lines/mm grating and filter 6. This provided a wavelength coverage of 6438--7184\,\r{A}, a \editfourteen{resolving power} of R=8600 \citep[table 12]{ESO2014}, and a dispersion of 0.2\,\r{A} per pixel.

The spectra were \editfive{taken} on 1st November 2013, with two exposures of 1200 seconds of 38 stars. \editten{To enable telluric line removal and instrument response correction, spectra were obtained of the B6 D Herbig Ae/Be star EM* StHA 50 \citep{2019AA...629A.114C}. Herbig Ae/Be are young stars which may retain circumstellar discs \citep{1998ARA&A..36..233W}. EM* StHA 50 has a \textit{WISE} \citep{Wright2010} mid-IR excess from (W3-W4) of 0.8, which may indicate the presence of a debris disc \citep{2012ApJ...751..114S}. The impact of the disc is considered during the response correction, see Section~\ref{sec:vlt_data_reduction}.} There were 12 fibres pointed away from any stars to enable sky background subtraction. The \editfive{flat fields} were acquired separately during the daytime, with the fibres in a different configuration to the observations, \editfive{which potentially changes their throughput.}
 
\editten{To facilitate concise discussion, within the text \edittwelve{the targetted stars are referred to by their allocated fibre numbers, and are} listed in Table~\ref{tab:observed_stars}.}

 \subsection{$\sigma$ Orionis cluster membership determination}
 \label{sec:membership_determination}
 
\begin{figure}
	\centering
	\includegraphics[width=0.45\textwidth]{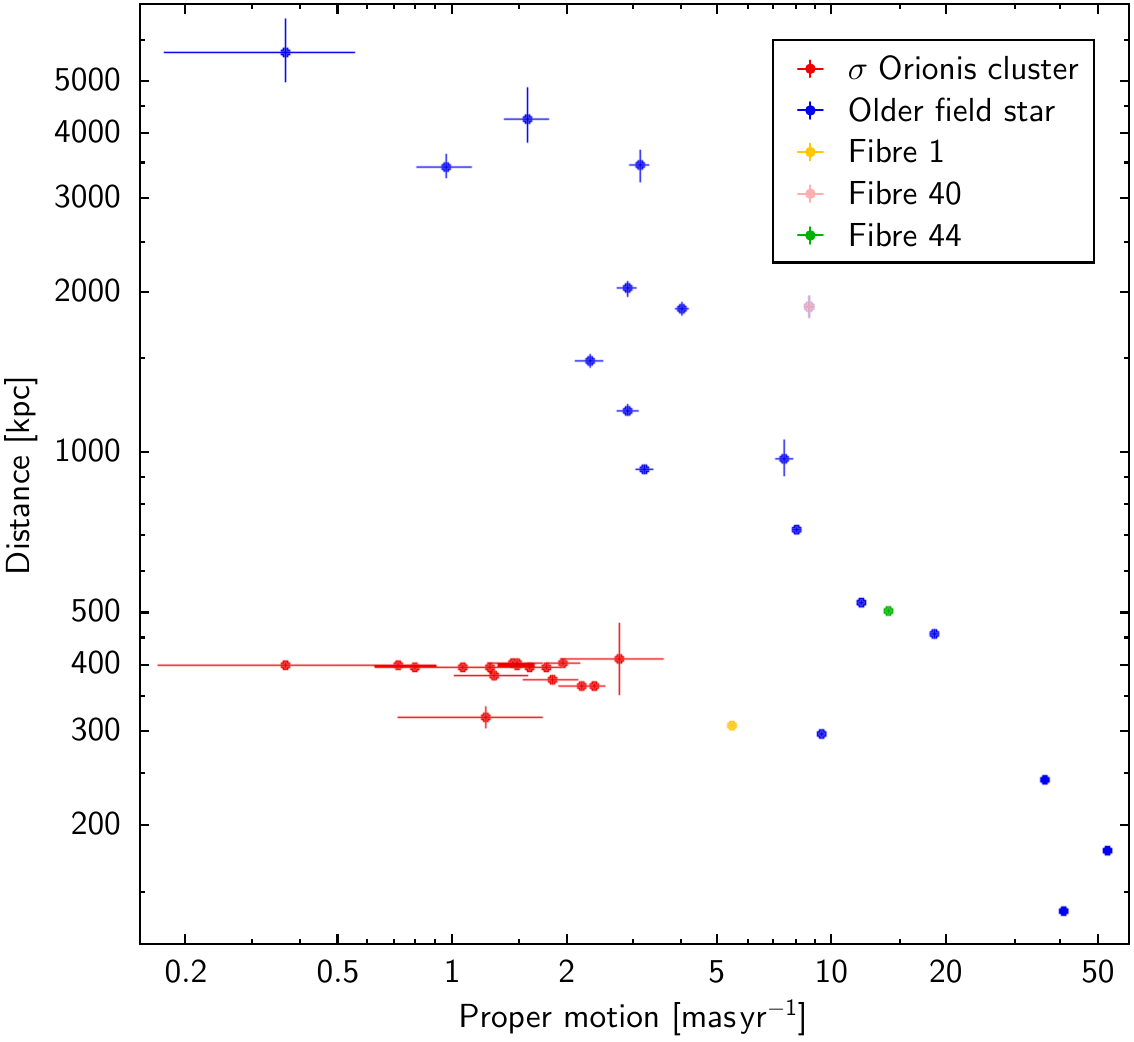}
	\caption{\editelev{The geometric distances \citep{2021AJ....161..147B} and \textit{Gaia} DR3 total proper motions for the sample with logarithmic scales on both axes. The uncertainties are displayed but where their RUWE is large they are likely to be underestimates. The cluster members are red and the older field stars blue. The young field stars are orange for fibre 1, \refone{pink for fibre 40} and green for fibre 44.}}
	\label{fig:Membership}
\end{figure}

\editten{A revised determination of SOC membership was made possible by \textit{Gaia} DR3 astrometry \citep{2016A&A...595A...1G,2023A&A...674A...1G}, \editelev{see Fig.\ref{fig:Membership}}. The target stars were matched to the \textit{Gaia} archive using a radius of 1.1$\arcsec$. The \textit{Gaia} DR3 processing pipeline assumes all sources are single stars. The renormalised unit weight error (RUWE) \citep{GAIA-C3-TN-LU-LL-124-01} indicates how well the data fit the single star model, with RUWE $\leq 1.4$ considered good fits \citep{GAIA-C3-TN-LU-LL-124-01,Fabricius2021}. However, \citet{2022RNAAS...6...18F} found the RUWE of single young stars can be inflated, suggesting RUWE cut-offs of 1.6 for disc-free stars and 2.5 for disc-bearing stars. Further, \citet{2020MNRAS.496.1922B} found that variability can increase the RUWE, and young stars are known to be highly variable. To avoid removing stars from our analysis just because they had high RUWE values, we did not impose a criterion on the RUWE. We instead allowed generous cuts in distance and proper motion to allow for poorly constrained values when selecting SOC members, and performed a follow-up check on our selection by using the Li 6708\,\r{A} line as an indicator of stellar youth. \citet{2024A&A...686A..42H} found a distance of $396.1\pm0.4$\,pc and proper motions of $\mu_{\alpha}\approx 1.48\pm0.36 \rm \,mas\,yr^{-1}$ and $\mu_{\delta} \approx -0.61\pm0.34 \rm \, mas\,yr^{-1}$ for the SOC using \textit{Gaia} DR3. We required the Bayesian estimate of the geometric distance from \textit{Gaia} EDR3 by \citet{2021AJ....161..147B} to be within 300--500\,pc. The proper motions were required to be within 2.50\,$\rm mas\,yr^{-1}$ of the central value for the cluster derived by \citet{2024A&A...686A..42H}. This divided the sample into 17 members and 20 field stars, listed in Table~\ref{tab:observed_stars}.} \editelev{It can be seen from Fig.\ref{fig:Membership} that one of the SOC members is an outlier in distance at 319\,pc. This is fibre 29 and as it has a RUWE of 6.3 its distance will not be as well constrained.}

\begin{table*}
	\centering
	\caption{\editten{The stars observed by the VLT. The top of the table are the $\sigma$ Orionis cluster members, the middle are the young field stars, the lower part are the older field stars and the B6 spectral calibration star is at the bottom. The sky coordinates are those used when making the observations. The distances are geometric distances from \citet{2021AJ....161..147B} based on \textit{Gaia} EDR3. The proper motions and RUWE are from \textit{Gaia} DR3. \editfifteen{The monochromatic 550\,nm extinction ($A_{55}$) were obtained from the \citet{2022A&A...664A.174V} dust map, see Section~\ref{sec:vlt_data_reduction}. }The Li 6708\,\r{A} EW were calculated from the VLT spectra using the method given in Appendix~\ref{app:li_ew}. Membership of the cluster is determined in Section~\ref{sec:membership_determination}.}}
	\label{tab:observed_stars}
	\begin{tabular}{llccccccccc}
		\hline
		Fibre & Designation & RA & Dec & Distance & $\mu_{\mathrm{\alpha}}$ & $\mu_{\mathrm{\delta}}$ & RUWE & $A_{55}$ & Li EW & Set \\
		 &  &  &  & [pc] & [mas yr$^{-1}$] & [mas yr$^{-1}$] & & & [\r{A}] & \\
		\hline
		15 & [BNM2013] 92.01 11 & 05:38:50.38 & -02:26:47.7 & 411 & 2.09 & 1.80 & 8.7 & 0.39 & \refone{0.6} & SOC \\
		18 & [BNM2013] 90.02 63 & 05:39:18.07 & -02:29:28.4 & 365 & 1.96 & -1.30 & 1.0 & 0.39 & 0.4 & SOC \\
		23 & [BNM2013] 90.02 77 & 05:39:07.60 & -02:32:39.1 & 375 & 1.59 & -0.92 & 3.3 & 0.39 & 0.5 & SOC \\
		24 & [BNM2013] 92.01 28 & 05:38:40.27 & -02:30:18.5 & 395 & 0.87 & -0.61 & 1.1 & 0.39 & 0.6 & SOC \\
		25 & [BNM2013] 90.02 74 & 05:38:43.33 & -02:32:00.8 & 384 & 1.21 & -0.45 & 1.1 & 0.39 & 0.6 & SOC \\
		26 & [BNM2013] 90.02 76 & 05:39:05.40 & -02:32:30.3 & 397 & 1.71 & -0.47 & 1.0 & 0.39 & 0.6 & SOC \\
		27 & [BNM2013] 90.02 78 & 05:38:53.37 & -02:33:22.9 & 399 & -0.23 & -0.28 & 1.6 & 0.39 & 0.5 & SOC \\
		29 & [BNM2013] 90.02 79 & 05:38:43.55 & -02:33:25.3 & 319 & 0.60 & -1.06 & 6.3 & 0.39 & \refone{0.7} & SOC \\
		31 & [BNM2013] 92.01 43 & 05:38:35.46 & -02:31:51.6 & 398 & 1.49 & -0.59 & 1.0 & 0.39 & 0.6 & SOC \\
		32 & [BNM2013] 92.01 1006 & 05:38:32.84 & -02:35:39.2 & 400 & 0.62 & -0.37 & 1.2 & 0.39 & 0.6 & SOC \\
		33 & [BNM2013] 92.01 690 & 05:38:34.31 & -02:35:00.0 & 401 & 1.45 & -0.28 & 1.0 & 0.39 & 0.6 & SOC \\
		34 & [BNM2013] 92.01 34 & 05:38:38.22 & -02:36:38.4 & 402 & 1.74 & -0.89 & 1.6 & 0.39 & 0.6 & SOC \\
		35 & [BNM2013] 92.01 32 & 05:38:38.48 & -02:34:55.0 & 397 & 0.80 & 0.00 & 1.3 & 0.39 & 0.6 & SOC \\
		37 & [BNM2013] 93.03 43 & 05:38:29.11 & -02:36:02.6 & 404 & 1.18 & -0.90 & 1.0 & 0.39 & 0.7 & SOC \\
		45 & [BNM2013] 92.01 122 & 05:38:06.74 & -02:30:22.7 & 402 & 1.33 & -0.53 & 1.1 & 0.39 & 0.5 & SOC \\
		46 & [BNM2013] 92.01 686 & 05:38:35.87 & -02:30:43.3 & 366 & 1.38 & -1.70 & 2.5 & 0.39 & 0.5 & SOC \\
		47 & [BNM2013] 92.01 118 & 05:38:07.84 & -02:31:31.3 & 398 & 1.07 & 0.65 & 1.6 & 0.39 & 0.5 & SOC \\
		\hline
		1 & 2MASS J05382993-0223381 & 05:38:29.93 & -02:23:38.2 & 309 & -4.99 & 2.11 & 1.0 & 0.20 & 0.2 & Young field star \\
		40 & 2MASS J05381589-0234412 & 05:38:15.89 & -02:34:41.2 & 1893 & 4.26 & -7.52 & 1.1 & 1.13 & \refone{0.1} & \refone{Young field star} \\
		44 & 2MASS J05380649-0228494 & 05:38:06.49 & -02:28:49.4 & 507 & 6.01 & -12.72 & 1.0 & 0.74 & 0.1 & Young field star \\
		\hline
		2 & 2MASS J05382178-0221494 & 05:38:21.78 & -02:21:49.5 & 2041 & 2.80 & -0.80 & 1.0 & 1.13 & 0.0 & Older field star \\
		3 & 2MASS J05381800-0221103 & 05:38:18.00 & -02:21:10.3 & 3466 & 3.10 & -0.04 & 1.1 & 1.11 & 0.0 & Older field star \\
		5 & [BNM2013] 92.01 684 & 05:38:38.55 & -02:26:44.8 & 1496 & 0.80 & -2.17 & 1.1 & 1.20 & 0.0 & Older field star \\
		6 & 2MASS J05383972-0218376 & 05:38:39.72 & -02:18:37.6 & 1863 & 1.85 & -3.58 & 1.1 & 1.19 & 0.0 & Older field star \\
		7 & 2MASS J05383378-0220417 & 05:38:33.78 & -02:20:41.7 & 1198 & -1.09 & -2.66 & 1.2 & 1.13 & 0.0 & Older field star \\
		8 & 2MASS J05383008-0221198 & 05:38:30.08 & -02:21:19.8 & 5689 & -0.04 & -0.36 & 1.1 & 1.16 & 0.0 & Older field star \\
		10 & UCAC3 176-24681 & 05:38:39.61 & -02:26:49.6 & 180 & 45.59 & -25.77 & 0.9 & 0.07 & 0.0 & Older field star \\
		12 & 2MASS J05390987-0225479 & 05:39:09.87 & -02:25:47.9 & 298 & 0.17 & -9.35 & 1.0 & 0.15 & 0.0 & Older field star \\
		13 & 2MASS J05390301-0221273 & 05:39:03.01 & -02:21:27.3 & 934 & 1.84 & 2.59 & 1.0 & 1.13 & 0.0 & Older field star \\
		14 & 2MASS J05391120-0225490 & 05:39:11.21 & -02:25:49.0 & 523 & -11.57 & 2.74 & 1.0 & 0.91 & 0.0 & Older field star \\
		17 & 2MASS J05391564-0229568 & 05:39:15.64 & -02:29:56.8 & 138 & -0.31 & -40.48 & 0.8 & 0.05 & 0.0 & Older field star \\
		19 & 2MASS J05392307-0228112 & 05:39:23.07 & -02:28:11.2 & 716 & 2.66 & -7.59 & 1.1 & 1.12 & 0.0 & Older field star \\
		38 & UCAC2 30800055 & 05:38:26.39 & -02:34:28.6 & 457 & -11.63 & 14.36 & 1.5 & 0.63 & 0.0 & Older field star \\
		39 & 2MASS J05381914-0235279 & 05:38:19.14 & -02:35:27.9 & 4242 & -1.57 & 0.10 & 1.1 & 1.14 & 0.0 & Older field star \\
		41 & 2MASS J05382949-0235175 & 05:38:29.49 & -02:35:17.5 & 978 & 4.10 & -6.20 & 5.0 & 1.06 & 0.0 & Older field star \\
		42 & 2MASS J05381064-0232574 & 05:38:10.64 & -02:32:57.4 & 3438 & 0.96 & -0.09 & 1.0 & 1.11 & 0.0 & Older field star \\
		43 & [BNM2013] 92.01 95 & 05:38:19.06 & -02:32:01.4 & 243 & -29.18 & 20.82 & 1.0 & 0.12 & 0.0 & Older field star \\
		\hline
		4 & EM* StHA 50 & 05:38:34.44 & -02:28:47.6 & 1471 & 0.67 & -0.41 & 1.0 & 1.18 & 0.0 & Calibration star \\
		\hline
	\end{tabular}
\end{table*}

To check our selection we measured the equivalent width (EW) of the Li 6708\,\r{A} absorption line in the VLT spectra, see Table ~\ref{tab:observed_stars} and Appendix~\ref{app:li_ew} for the method. This did not confirm SOC membership, but it did indicate whether our selection approach had contaminated the SOC sample with older field stars, and whether any of the field stars may be young. \edittwelve{An age of 3.7\,Myr (2.5--4.5\,Myr) for the SOC was found by \citet{2023A&A...673A.114H}. The lithium depletion model of \citet{2023MNRAS.523..802J} gives a Li EW of 0.3--0.6\,\r{A} for a stellar age of 5\,Myr with temperatures 3500--5500\,K, and <0.1\,\r{A} from 300\,Myr to 5\,Gyr.} All of the sources flagged as SOC members had a Li EW $\geq0.4$\,\r{A}. \refone{Of the field stars, those observed by fibres 1, 40 and 44 had a Li EW of 0.1--0.2\,\r{A}}. The \textit{Gaia} DR3 RUWE for these sources are \refone{close to 1}, so their astrometry is likely to be reliable. \refone{Fibres 1 and 44} have similar radial velocities (RVs) to the SOC (the RVs are derived during the model fitting described in Section~\ref{sec:fitting_model_spectra}), see Table~\ref{tab:sigori_results}. However, their \textit{Gaia} DR3 proper motions are inconsistent with SOC membership, see Fig.\ref{fig:Membership}. Their geometric distances from \citet{2021AJ....161..147B} are 308\,pc for fibre 1, \refone{1893\,pc for fibre 40} and 507\,pc for fibre 44. \refone{So fibre 40 is well beyond the SOC,} while fibres 1 and 44 are just within the inner edge and just beyond the outer edge of our generous range for membership. We consider they are all unlikely to be members of the SOC. The two closest may be young field stars that are part of the Orion star forming complex with different ages to the SOC. The remainder of the field stars had a Li EW consistent with zero and hence an older age. The field stars were therefore split into an older field star sample of \refone{17} stars for comparison to the SOC, and a young field star sample of \refone{3} stars, see Table~\ref{tab:observed_stars}.

\subsection{Data reduction}
\label{sec:vlt_data_reduction}

The spectra were downloaded from the ESO archive. The ESO Reflex (Recipe Flexible Execution Workbench) pipeline was used for the initial data reduction of the spectra \citep{Ory2015}, performing bias removal, flat field corrections, and wavelength calibration. Reflex automatically selected the calibration frames using parameter XML files downloaded with the fits data files.

The sky background spectrum was calculated as the median of the normalised sky fibres, and removed from the stellar spectra. The ESO software Molecfit was used to remove the atmospheric telluric lines from the spectra \citep{Smette2015} \editone{by analysing spectrum of} the hot B6 star. The derived parameters were used to remove the telluric lines from all of the stellar spectra, leaving around 1\% residuals relative to the continuum.

\editten{The two exposures for each star were averaged together. Cosmic rays were identified where the flux in one spectrum was greater than the other spectrum by at least five times the uncertainty. The ratio of the normalised fluxes was also required to be above two per cent and the excess above 200 counts. Cosmic rays were removed by replacing the pixel with the value from the unaffected exposure, adjusted to match the intensity of the affected exposure by using the median intensity of the 25 pixels either side of the affected pixel in both exposures.}

\editfive{Monochromatic 550\,nm} extinction values were derived from the \citet{2022A&A...664A.174V} dust map with 25\,pc resolution. \editten{The \citet{2021AJ....161..147B} geometric distances based on $Gaia$ DR3 astrometry were adopted for the field stars.} \editfourteen{The parallax uncertainties of the cluster members would lead to individual distance errors that are larger than the size of the cluster. Instead, a distance of 396\,pc from \citet{2024A&A...686A..42H} based on the combined $Gaia$ DR3 astrometry for the cluster was assumed for the \editten{SOC} members.} The extinction was calculated by summing $A_{55}$ along the line of sight to each star. \editten{Nine of the field stars lie below the lower face of the dust map, which extends to a 400\,pc vertical distance from the Sun's location in the plane of the Galaxy.} The vertical extent of the map captures the majority of the dust as this resides near the plane of the Galactic disc. So the integrated extinction along the sight line within the map is expected to give a reasonable estimate of the total extinction to these nine stars. \editfive{The extinction to the B6 star was used to deredden its spectrum for the instrument response calculation. The extinctions to the \editten{SOC members} and field stars were used to redden their best fit model spectra in the analysis described in Section~\ref{sec:fitting_model_spectra}.}

The instrument response curve was calculated by comparing the \editthree{dereddened observed B6} spectrum with the B6 spectrum from the PyHammer\footnote{\url{https://github.com/BU-hammerTeam/PyHammer}} version 1 spectral typing suite \citep{Kesseli2017}. This is constructed from five co-added BOSS Sloan Digital Sky Survey (SDSS) spectra \citep{2000AJ....120.1579Y} originally from \citet{2007AJ....134.2398C}. \editten{As there were indications of a debris disc around the observed B6 star (see Section~\ref{sec:vlt_observations}), and the template PyHammer spectrum did not include a disc, this will have left a residual trend in the spectra after response correction. This was removed by detrending the spectra during the model fitting, see Appendix~\ref{app:fitting_procedure}.} The B6 spectrum was assumed to be smooth apart from the H$\alpha$ line. So a smoothed representation was created by fitting a power law to the PyHammer B6 template spectrum. The observed B6 spectrum was then divided by the power law function, giving a noisy instrument response curve, which was normalised by its median. A fourth order polynomial summed with a sine wave was fitted to the noisy response curve, using the non-linear least squares fitting \textsc{Python} \textsc{SCIPY} optimisation library. This provided a smooth instrument response curve, accurate to around 1 per cent, that was used to \edittwelve{response-correct} the spectra into relative flux per unit wavelength.

\section{Fitting model spectra to the observations}
\label{sec:fitting_model_spectra}

To assess whether the \editeight{TTS} spectra were consistent with a spotted photosphere, model spectra were fitted to the observed spectra. The BT-Settl CIFIST model atmosphere spectra from \citet{Allard2014} and references therein, with solar abundances from \citet{Caffau2011}, were used for the models. \editfive{These covered an effective temperature range of} 2500--7000\,K in steps of 100\,K, and $\log(g)$ of 2.5--5.5 in steps of 0.5\refone{\,dex}. The wavelengths of the VLT spectra were for air, and the BT-Settl spectra in vacuum. The wavelengths in the BT-Settl spectra were converted to air using the formula from \citet{2000ApJS..130..403M}, originally derived by \citet{1994Metro..31..315B}. To enable faster downstream processing, the BT-Settl spectra were cropped to the wavelength range 6200--7300\,\r{A}. \editelev{The extinction values determined in Section~\ref{sec:vlt_data_reduction} were used to redden the model spectra to match observed spectra, utilising the \citet{2019ApJ...886..108F} extinction curves with $R_{55}=3.02$ for average Galactic dust.}

Two sets of model spectra were fitted. One set composed of a single effective temperature and surface gravity. The other set a blend of two effective temperatures and surface gravities, \refone{with the cool component surface fraction varied from 0.00 to 1.00 in steps of 0.05}. The fits were assessed by calculating the reduced $\chi$ squared ($\chi^2_{\nu}$) between the observed spectra and the models\footnote{$\chi^2_{\nu}$ is $\chi^2$ divided by the number of fitted data points in the spectrum minus the number of free parameters. \refone{The single temperature fits had two parameters where surface gravity was unconstrained and one parameter where the surface gravity was fixed. The two temperature fits had five parameters where the surface gravity was unconstrained (two temperatures, two gravities and the cool surface area fraction), and four parameters where the hot component surface gravity was fixed}.}. \editsix{To enable the calculation of the $\chi^2_{\nu}$, the uncertainties from the spectral extraction were propagated through the analysis}. The best fit from each set was defined as the model with the smallest $\chi^2_{\nu}$. 

\editfifteen{We cannot fit the BT-Settl models directly to }\editfive{our observed spectra. First, the models must be shifted in velocity space to match the RVs of the targets. Next, their lines need to be broadened to allow for the effects of stellar rotation and spectrograph resolution. Finally, the models need to be rebinned to match the pixellation of the observed spectra. Carrying out this series of steps requires measuring both the RV and broadening of the spectra. The process is complicated by the fact that one cannot accurately measure the RV until one has identified the correct spectral type. Conversely the best spectral type cannot be found without knowing the RV. Hence, we used a three-stage process, with the first two stages refining the RVs. In stage 1 we made a crude RV estimate and found the best-fitting BT-Settl model. In stage 2 we used that model to get a good estimate of the RV. Finally, in stage 3 we determined the broadening we needed to apply to the BT-Settl models to match the line widths in the observed data.}

Each stage began by estimating the observed spectrum RV and in later stages the spectral line width. This was done by cross-correlating the observed spectrum with a model spectrum \editten{at its native resolution} in velocity space. A Gaussian was fitted to the cross-correlation results, with its centre giving the RV, and the Gaussian width an approximation of the spectral line width. The RV, and in later stages the line width, were then used to alter the model spectra to match the observed spectrum (separately for each observed spectrum). Finally, the best fit \refone{single and two-temperature models were found for each observed spectrum} by calculating the $\chi^2_{\nu}$ across the \refone{applicable} range of effective temperatures and surface gravities. \editten{The three processing stages are explained in detail in Appendix~\ref{app:fitting_procedure}.}

\refone{The fit of single-temperature models to the field stars searched all of the effective temperatures and surface gravities in the BT-Settl grid. The fit of two-temperature models to the field stars searched for hot components in the range 3000--7000\,K and cool components in the range 2500--5000\,K, limited to combinations where the cool component was cooler than the hot component. All surface gravity combinations were searched for both the hot and cool components. The fit to the SOC members were the same apart from a restriction in surface gravity that will now be discussed.}

\refone{
An initial run for the SOC members without any restrictions on surface gravity found the best fitting single-temperature models had $\log(g)$ of 2.5 and 4.5--5.5, and the hot component (simply assumed to be an unblemished photosphere) of the best fitting two-temperature models had $\log(g)$ of 4.0--5.5. 
As the SOC is around 3--4\,Myr old and the effective temperatures of the K-type stars are around 3900--5300\,K \citep{2013ApJS..208....9P}, the surface gravities of the SOC sample should cover a limited range in $\log(g)$. The PARSEC V2.0 \citep{2019MNRAS.485.4641C,2019A&A...631A.128C,2022A&A...665A.126N} models have $\log(g)\approx4.0.$ for these ages and temperatures. 
If the SOC have large magnetically induced cool \reftwo{areas} on their photospheres, then it is possible they have no unblemished photosphere with plages and faculae dominating the hot regions. \refthree{Hence,} the higher effective surface gravities in the hot regions could be caused by strong magnetic fields affecting the surface pressure. However, we have no independent evidence or a reasoned physical mechanism. So we consider it more likely that non-surface gravity related phenomena in the TTS spectra were giving a fractionally better fit to a variety of $\log(g)$ values. Hence, we restricted the single-temperature models and the hot component of the two-temperature models to $\log(g)=4.0$ for the SOC, while allowing the surface gravity of the cool component to take any value \refthree{as this enabled an exploration of possible effects due to magnetic pressure in the cooler regions}.}
	
\begin{figure*}
	\centering
	\includegraphics[width=0.9\textwidth]{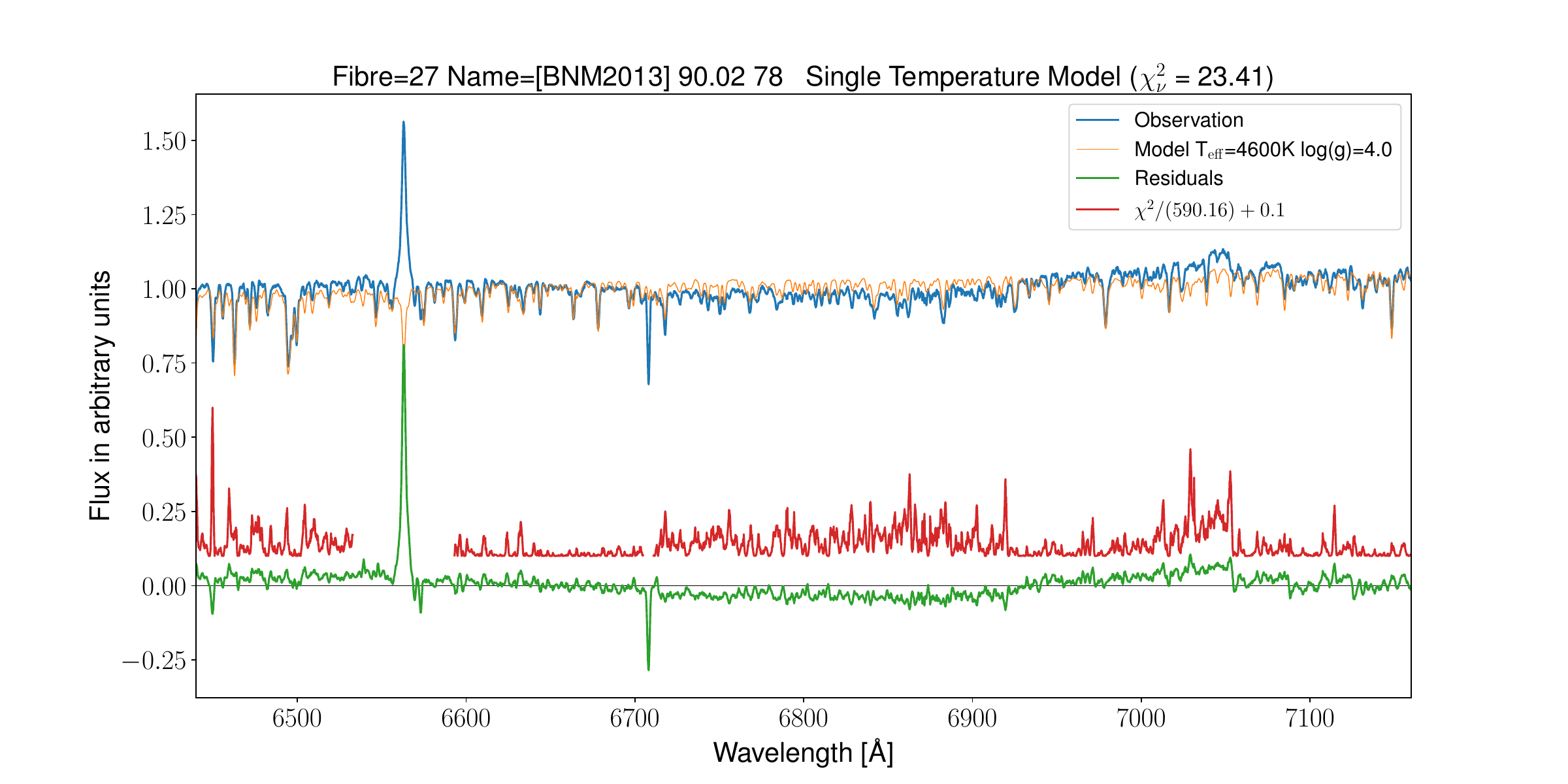}
	\includegraphics[width=0.9\textwidth]{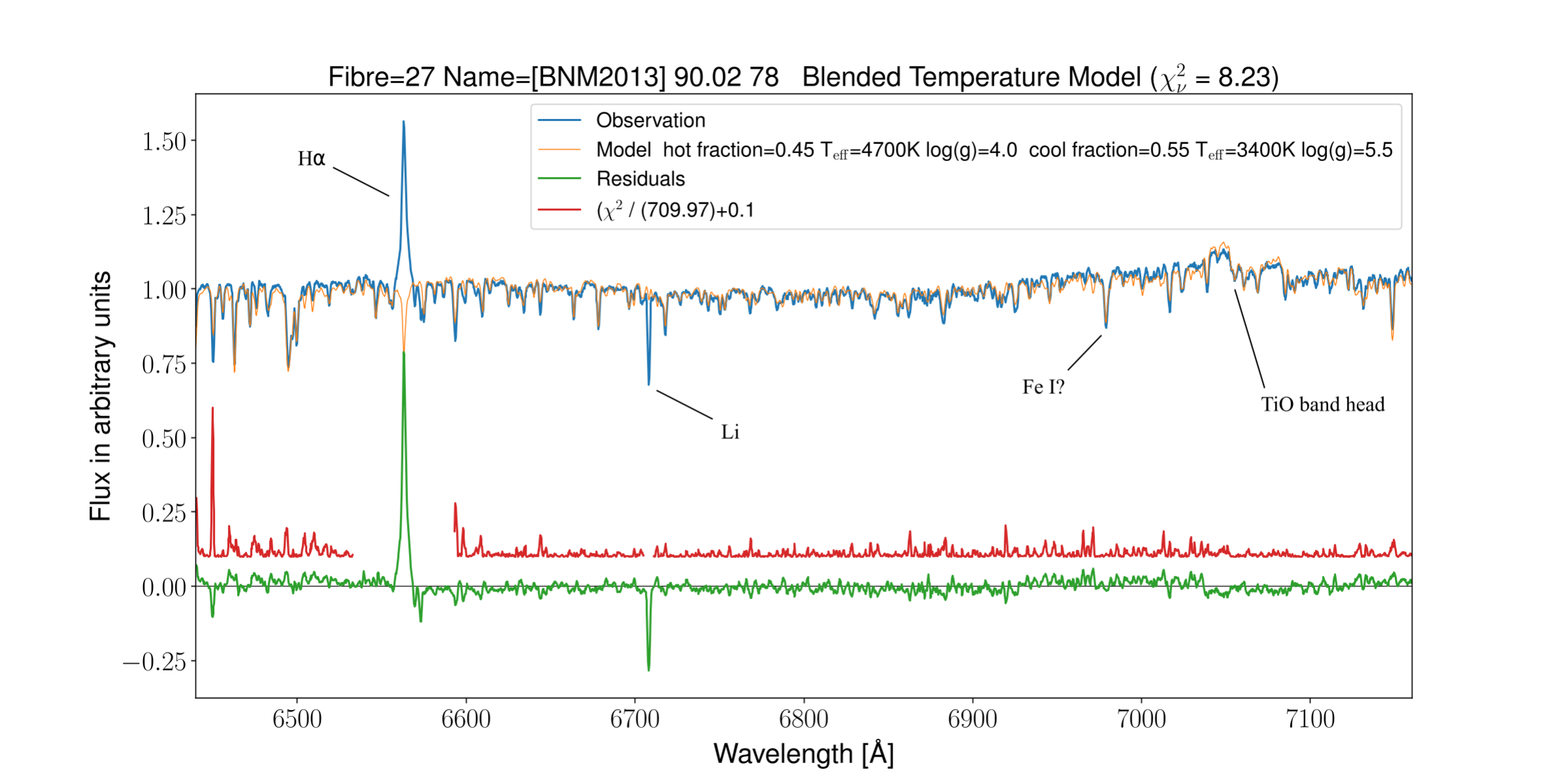}
	\includegraphics[width=0.9\textwidth]{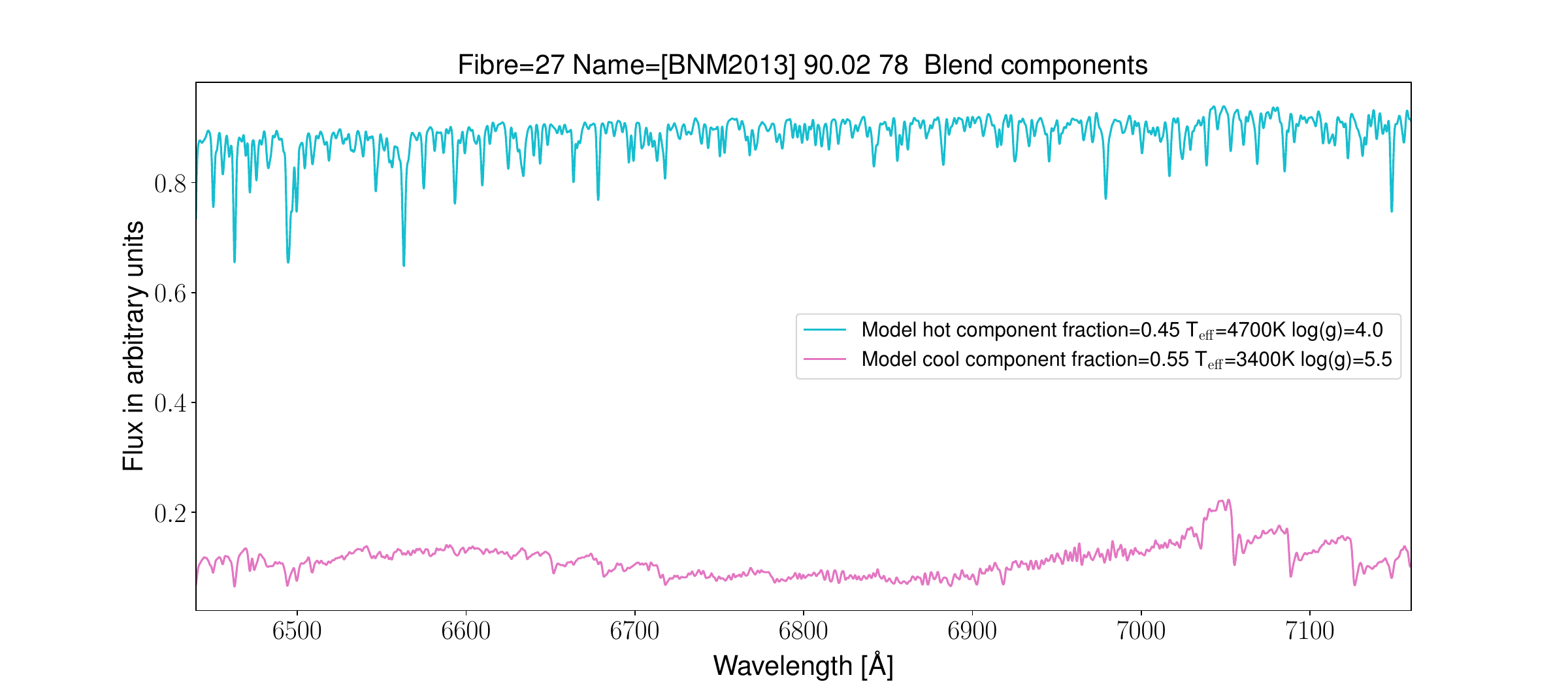}
	\caption{Plots of the spectral fits for fibre 27. Top is the fit to a single photosphere, middle is the fit to the blended photosphere, and bottom shows the component spectra for the blend scaled by their relative contributions. In the top two plots the observed spectrum is blue, the best fit model is orange, the residuals are green, and the $\chi^2$ values are red. The $\chi^2$ values are scaled by the factor in the legend. In the bottom plot the hot and cool components are cyan and pink respectively.}
	\label{fig:fibre_27_spectra}
\end{figure*}

\editelev{The results are summarised in Table~\ref{tab:sigori_results} and example plots for fibre 27 of the best fitting single and two-temperature models with the observed spectrum are presented in Fig.~\ref{fig:fibre_27_spectra}.} In the middle and bottom panels it can be seen that the cool spectrum contributes the molecular absorption bands, most noticeable \editfive{by} the TiO band heads longward of 7550\,\r{A}, and the hot temperature contributes features like the strong absorption line at 6978\,\r{A} (tentatively identified as \ion{Fe}{i} from \citet{2008A&A...489..349S,2021RAA....21..249B}). The presence of both hot and cool features in the spectrum demonstrates how a photosphere composed of a blend of two temperatures is indeed appropriate for \editten{fibre 27. This is affirmed by the $\chi^2_{\nu}$, with a value of 8.2 for the blend of two temperatures compared to 23.4 for the single-temperature model.} \editelev{The spectra of the other WTTS identified in Table~\ref{tab:sigori_review_members} and not presented elsewhere are provided in Appendix~\ref{app:sig_ori_spectra}.}

\begin{landscape}
	\begin{table}
		\vspace{1.0cm}
		\centering
		\caption{\editten{Summary of the best fit model photospheres. The members of the $\sigma$ Orionis cluster in the top of the table, the young field stars are in the middle, and the older field stars are at the bottom of the table. A barycentric correction of 19.0\,kms$^{-1}$ has been applied to the RVs.}}
		\label{tab:sigori_results}
		\begin{tabular}{llccccccccccccc}
			\hline
			& & | & & | & \multicolumn{3}{c}{Fit to single photosphere} & | & \multicolumn{6}{c}{Fit to blended photosphere} \\
			Fibre & Designation & | & RV & | & $T_{\rm{eff}}$ & $\log(g)$ & $\chi^2_{\nu}$ & | & Hot $T_{\rm{eff}}$ & Hot $\log(g)$ & Cool $T_{\rm{eff}}$ & Cool $\log(g)$ & Cool photosphere & $\chi^2_{\nu}$ \\
			& & | & [km\,s$^{-1}$] & | & [K] & [$\log_{10}$(cm\,s$^{-2}$)]& & | & [K] & [$\log_{10}$(cm\,s$^{-2}$)] & [K] & [$\log_{10}$(cm\,s$^{-2}$)] & area & \\
			\hline
			15 & [BNM2013] 92.01 11 & | & \refone{28.9} & | & 3400 & \refone{4.0} & 11.2 & | & \refone{4300} & \refone{4.0} & \refone{3200} & \refone{4.0} & 0.95 & \refone{8.5} \\
			18 & [BNM2013] 90.02 63 & | & \refone{21.7} & | & \refone{5300} & \refone{4.0} & \refone{9.2} & | & \refone{5200} & \refone{4.0} & \refone{3700} & \refone{5.5} & 0.30 & \refone{6.9} \\
			23 & [BNM2013] 90.02 77 & | & \refone{28.7} & | & \refone{4600} & \refone{4.0} & \refone{13.4} & | & \refone{4800} & \refone{4.0} & \refone{3400} & \refone{5.5} & \refone{0.60} & \refone{5.3} \\
			24 & [BNM2013] 92.01 28 & | & \refone{30.2} & | & \refone{5900} & \refone{4.0} & \refone{18.5} & | & 5500 & \refone{4.0} & 3400 & 5.0 & 0.85 & 4.4 \\
			25 & [BNM2013] 90.02 74 & | & \refone{28.5} & | & 3200 & \refone{4.0} & \refone{12.3} & | & \refone{3300} & \refone{4.0} & \refone{3200} & 2.5 & \refone{0.90} & \refone{8.2} \\
			26 & [BNM2013] 90.02 76 & | & \refone{29.4} & | & \refone{4500} & \refone{4.0} & \refone{24.1} & | & \refone{4800} & \refone{4.0} & 3500 & 5.0 & \refone{0.75} & 6.3 \\
			27 & [BNM2013] 90.02 78 & | & \refone{31.6} & | & \refone{4600} & \refone{4.0} & \refone{23.4} & | & \refone{4700} & \refone{4.0} & \refone{3400} & \refone{5.5} & \refone{0.55} & \refone{8.2} \\
			29 & [BNM2013] 90.02 79 & | & \refone{31.0} & | & 3900 & \refone{4.0} & \refone{22.9} & | & \refone{5000} & \refone{4.0} & \refone{3300} & 4.5 & 0.90 & \refone{4.3} \\
			31 & [BNM2013] 92.01 43 & | & 29.6 & | & \refone{4600} & \refone{4.0} & \refone{19.7} & | & \refone{4700} & \refone{4.0} & \refone{3500} & 5.0 & \refone{0.60} & 7.2 \\
			32 & [BNM2013] 92.01 1006 & | & \refone{29.1} & | & \refone{4500} & \refone{4.0} & \refone{21.0} & | & \refone{4600} & \refone{4.0} & 3400 & 5.0 & \refone{0.70} & \refone{6.2} \\
			33 &[BNM2013] 92.01 690 & | & \refone{30.3} & | & \refone{4500} & \refone{4.0} & \refone{17.9} & | & \refone{4700} & \refone{4.0} & 3500 & 5.0 & \refone{0.60} & 5.7 \\
			34 & [BNM2013] 92.01 34 & | & 31.3 & | & \refone{5600} & \refone{4.0} & \refone{32.4} & | & \refone{4800} & \refone{4.0} & 3300 & 5.0 & \refone{0.75} & \refone{8.5} \\
			35 & [BNM2013] 92.01 32 & | & \refone{-6.0} & | & \refone{5700} & \refone{4.0} & \refone{40.7} & | & \refone{5600} & \refone{4.0} & \refone{3500} & \refone{5.5} & 0.60 & \refone{26.9} \\
			37 & [BNM2013] 93.03 43 & | & 28.9 & | & 3700 & \refone{4.0} & \refone{12.3} & | & 5100 & \refone{4.0} & 3300 & 4.5 & 0.95 & \refone{4.1} \\
			45 & [BNM2013] 92.01 122 & | & 28.1 & | & \refone{6200} & \refone{4.0} & \refone{16.7} & | & \refone{5900} & \refone{4.0} & 3400 & 5.0 & 0.90 & 4.1 \\
			46 & [BNM2013] 92.01 686 & | & \refone{29.1} & | & \refone{4500} & \refone{4.0} & \refone{19.7} & | & \refone{4600} & \refone{4.0} & \refone{3300} & 5.0 & \refone{0.65} & \refone{5.3} \\
			47 & [BNM2013] 92.01 118 & | & 29.9 & | & \refone{4700} & \refone{4.0} & \refone{15.9} & | & \refone{4700} & \refone{4.0} & 3400 & \refone{5.0} & \refone{0.45} & \refone{8.8} \\
			\hline
			1 & 2MASS J05382993-0223381 & | & 26.9 & | & 5800 & 5.0 & 3.1 & | & 5800 & 5.0 & \refone{3800} & \refone{5.5} & 0.20 & \refone{2.8} \\
			40 & 2MASS J05381589-0234412 & | & \refone{-29.3} & | & 5000 & 4.0 & 1.6 & | & \refone{5000} & 3.5 & \refone{4000} & \refone{5.5} & \refone{0.15} & 1.6 \\
			44 & 2MASS J05380649-0228494 & | & 25.4 & | & 6900 & 4.5 & 3.1 & | &	 6900 & \refone{4.0} & \refone{4000} & \refone{5.5} & 0.10 & 3.1 \\
			\hline
			2 & 2MASS J05382178-0221494 & | & \refone{62.4} & | & 5300 & 3.0 & 3.6 & | & 5800 & 3.5 & 4400 & 2.5 & 0.55 & 3.4 \\
			3 & 2MASS J05381800-0221103 & | & 92.1 & | & 5100 & 3.0 & 1.9 & | & 5700 & 3.0 & 4500 & 2.5 & 0.75 & 1.8 \\
			5 & [BNM2013] 92.01 684 & | & \refone{91.8} & | & 5300 & 3.5 & 1.5 & | & \refone{5300} & 3.5 & \refone{4000} & \refone{5.5} & \refone{0.15} & 1.5 \\
			6 & 2MASS J05383972-0218376 & | & \refone{80.6} & | & 4400 & 2.5 & 22.0 & | & 4400 & 2.5 & \refone{2700} & \refone{5.5} & \refone{0.50} & 21.0 \\
			7 & 2MASS J05383378-0220417 & | & \refone{10.1} & | & 5400 & 4.0 & 1.8 & | & 5800 & 4.0 & 4600 & 3.5 & 0.60 & 1.7 \\
			8 & 2MASS J05383008-0221198 & | & \refone{86.8} & | & 5300 & 3.0 & 1.5 & | & 5700 & 3.0 & 4500 & 3.0 & 0.60 & 1.4 \\
			10 & UCAC3 176-24681 & | & \refone{96.6} & | & 5600 & 5.5 & 0.5 & | & 5700 & 5.5 & 4000 & 5.5 & 0.30 & 0.5 \\
			12 & 2MASS J05390987-0225479 & | & -21.9 & | & 4900 & 4.5 & 6.4 & | & 4900 & 4.5 & \refone{3700} & \refone{5.5} & 0.15 & \refone{5.9} \\
			13 & 2MASS J05390301-0221273 & | & \refone{39.1} & | & 6900 & 4.5 & 0.9 & |& \refone{7000} & \refone{4.5} & \refone{3600} & \refone{5.5} & 0.40 & 0.8 \\
			14 & 2MASS J05391120-0225490 & | & \refone{12.3} & | & 6700 & 4.5 & 2.3 & | & \refone{6700} & \refone{4.5} & 3700 & \refone{5.5} & 0.30 & \refone{2.0} \\
			17 & 2MASS J05391564-0229568 & | & -49.1 & | & 6100 & 5.0 & 5.1 & | & \refone{6000} & \refone{4.5} & \refone{3700} & \refone{5.5} & 0.20 & \refone{4.7} \\
			19 & 2MASS J05392307-0228112 & | & \refone{-4.2} & | & 6800 & 5.0 & 0.8 & | & 6900 & 5.0 & 3600 & \refone{5.5} & 0.35 & 0.7 \\
			38 & UCAC2 30800055 & | & \refone{11.2} & | & 6300 & 4.5 & 0.9 & | & \refone{6500} & \refone{4.5} & \refone{3700} & 5.5 & \refone{0.40} & 0.7 \\
			39 & 2MASS J05381914-0235279 & | & 122.5 & | & 4500 & 2.5 & 2.1 & | & 4500 & \refone{2.5} & \refone{4100} & \refone{5.5} & \refone{0.05} & 2.1 \\
			41 & 2MASS J05382949-0235175 & | & -15.4 & | & 5400 & 4.0 & 2.1 & | & \refone{5400} & \refone{3.0} & 4500 & \refone{5.5} & \refone{0.20} & 2.0 \\
			42 & 2MASS J05381064-0232574 & | & 77.5 & | & 4800 & 3.0 & 4.0 & | & 6800 & 2.5 & 4500 & 2.5 & 0.95 & 3.8 \\
			43 & [BNM2013] 92.01 95 & | & 29.8 & | & 4900 & 5.0 & 2.6 & | & \refone{5000} & \refone{5.0} & 3800 & 5.5 & \refone{0.35} & \refone{1.9} \\
			\hline
		\end{tabular}
	\end{table}
\end{landscape}

As expected, the \editeight{TTS} spectra showed the H$\alpha$ emission line at 6563\,\r{A} (except for fibre 18 where H$\alpha$ was in absorption but \refone{with a small net emission after subtraction of the two-temperature model}), and the Li absorption \editone{line} at 6708\,\r{A}. \editten{Three of the spectra exhibited \editfourteen{very} strong H$\alpha$ emission, see Table~\ref{tab:sigori_review_members}. They are identified as accreting CTTS in Section~\ref{sec:accretion} and removed from our refined sample in Section~\ref{sec:sample_selection}.} As these spectral features were not present in the model spectra, the ranges 6533--6593\,\r{A} and 6705--6711\,\r{A} were excluded from the $\chi^2_{\nu}$ calculation. \editten{We applied a generous exclusion region around the H$\alpha$ line to treat all the spectra equally and to avoid any low level emission in the line wings from contaminating the fitting process.}

Some of the observed spectra contained one or more absorption and emission spikes or lines, with no counterpart in any of the model spectra. Plots of $\chi^2$ by wavelength showed these spectral features were contributing a large proportion of the $\chi^2_{\nu}$, affecting the selection of the best fitting model. The spikes were due to a mix of residual cosmic ray hits not fully removed during the data reduction, and \editten{sharp} positive and negative features superimposed on the continuum. The narrow width and shape of the spikes meant they could not have originated in the stellar photosphere, as they would have been broadened by stellar physics and the spectrograph optics. \editten{The only genuine spectral lines were a series of emission lines in the spectra from fibres 24 and 45. These are two of the three accreting CTTS mentioned above and discussed in Sections~\ref{sec:sample_selection} and \ref{sec:accretion}.} As they only affected small portions of the spectra, these wavelength sections were excluded from the $\chi^2_{\nu}$ calculation on a spectrum by spectrum basis, allowing the $\chi^2_{\nu}$ calculation to better fit the model spectra.

\section{Refined sample selection}
\label{sec:sample_selection}

Our goal is to assess whether models based on a blend of two temperatures provide a better fit to the spectra of WTTS than single-temperature models. To allow a straightforward analysis, \editten{we divide the SOC members into CTTS, WTTS candidate binary stars and a WTTS clean sample with no evidence of binarity. \refone{As our technique is designed to work on K-type stellar spectra, we further divide the WTTS clean sample into WTTS M and K-type stars.} We start by separating fibres 23, 24 and 45 into a CTTS set as they are identified as accreting in Section~\ref{sec:accretion}.}

If the components of a binary star are separated by less than the 1.2$\arcsec$ aperture of the MEDUSA optical fibres, then the recorded spectrum will be a blend of both stars. As the image of a star at the focal plane of a telescope is represented by a point spread function, a star a little beyond the optical fibre aperture can also contaminate the recorded spectrum. $Gaia$ DR3 is able to resolve all individual stars separated by more than 1.5$\arcsec$, the majority by 0.7$\arcsec$, and a few to smaller separations \citep{2021A&A...649A...1G,Fabricius2021}. The $Gaia$ DR3 source catalogue was searched for additional stars within a radius of $3\arcsec$ of each SOC member. No other stars were found. This gives confidence there are no wide binaries \editfive{(or chance alignments)} with separations of 0.7--3.0$\arcsec$. \editelev{Where the \textit{Gaia} pipeline identifies multiple sources within 0.18$\arcsec$, a single source is retained and flagged as duplicated \citep{Lindegren2021a}. None of the SOC members were flagged as duplicated. This does not guarantee there are no close double stars, simply there were none detected by \textit{Gaia}.}

\begin{figure}
	\centering
	\includegraphics[width=0.5\textwidth]{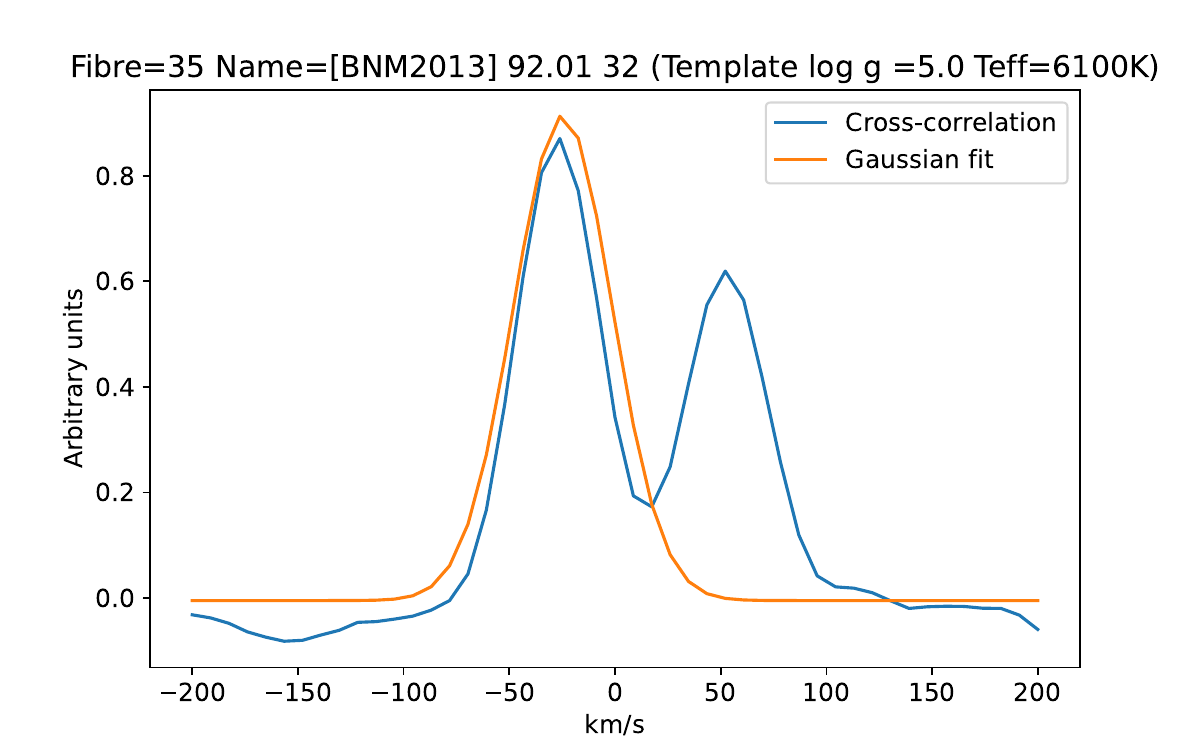}
	\caption{The stage 2 cross-correlation between the observed spectrum and the BT-Settl model atmosphere for fibre 35.}
	\label{fig:fibre35_crosscorr}
\end{figure}

If binary stars are physically close, then during parts of their mutual orbit, their spectra can exhibit distinct spectral lines from the two components. The cross-correlation results are especially adept at identifying spectroscopic binaries, as they will appear as a double bump in the cross-correlation curve. Double bumps were evident for the SOC members corresponding to fibres 35 (see Fig.~\ref{fig:fibre35_crosscorr}) and 46, and field star 38 showed a distinct double and possibly triple bump. We can therefore be confident that SOC members observed by fibres 35 and 46 are binary stars. \editten{This also explains why the RV of fibre 35 is an outlier at -6.0\,kms$^{-1}$, compared to the median of 29.1\,kms$^{-1}$ for the SOC members. It can be seen in Fig.~\ref{fig:fibre35_crosscorr} that the SOC median value lies between the two peaks of the cross-correlation function of fibre 35.}

\editten{Binary stars can potentially be identified from the the \textit{Gaia} RUWE \citep{2022MNRAS.517.2165C,2025ApJ...988L...1C}. As discussed in Section~\ref{sec:membership_determination}, the RUWE can be inflated for young stars. So we adopt the single discless young star threshold from \citet{2022RNAAS...6...18F}, identifying candidate binary stars where RUWE > 1.6. This identifies fibres 15, 23 (CTTS), 29 and 46 (also a spectroscopic binary) as candidate binaries.}

\begin{figure*}
	\centering
	\includegraphics[width=0.9\textwidth]{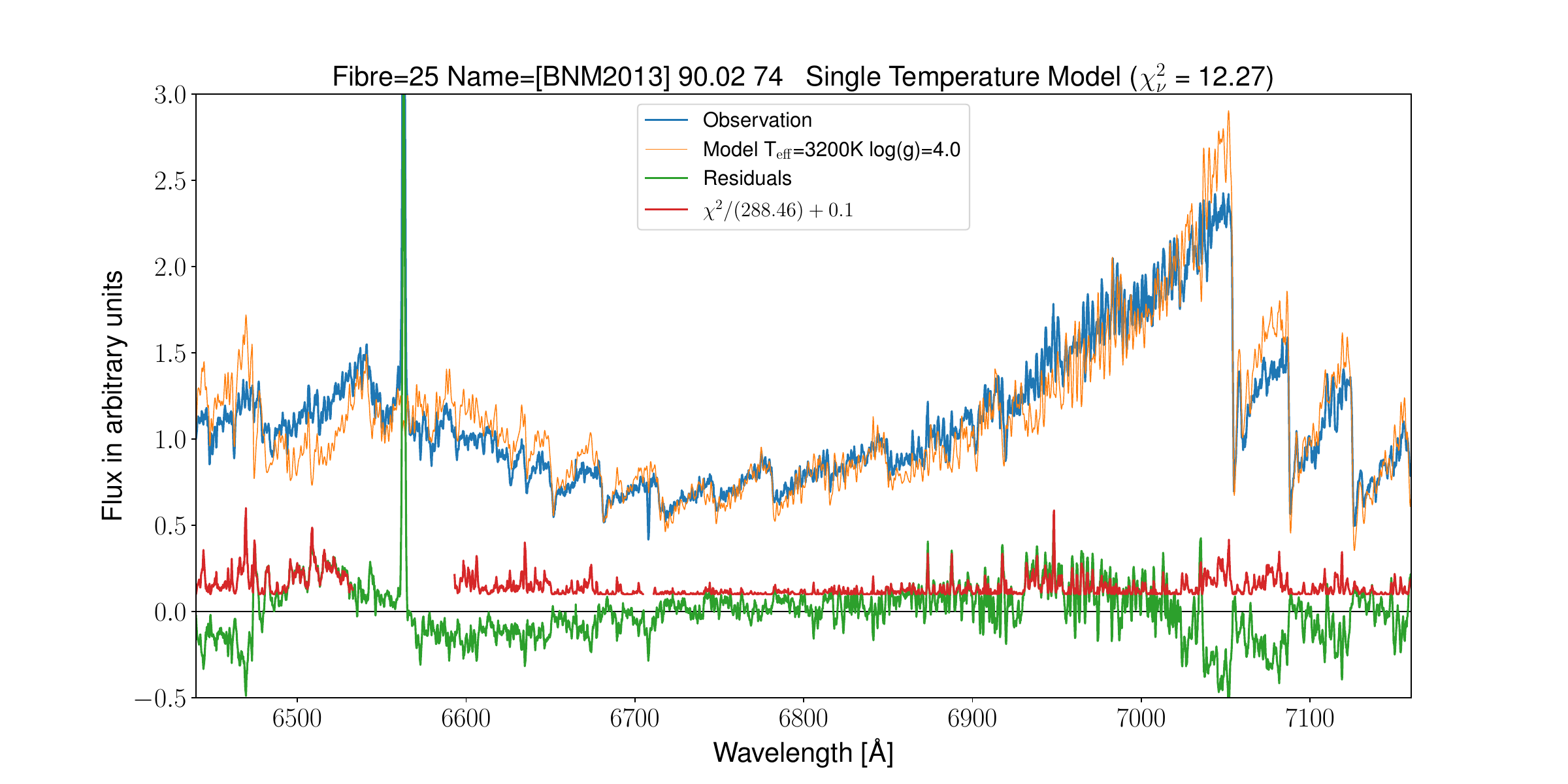}
	\includegraphics[width=0.9\textwidth]{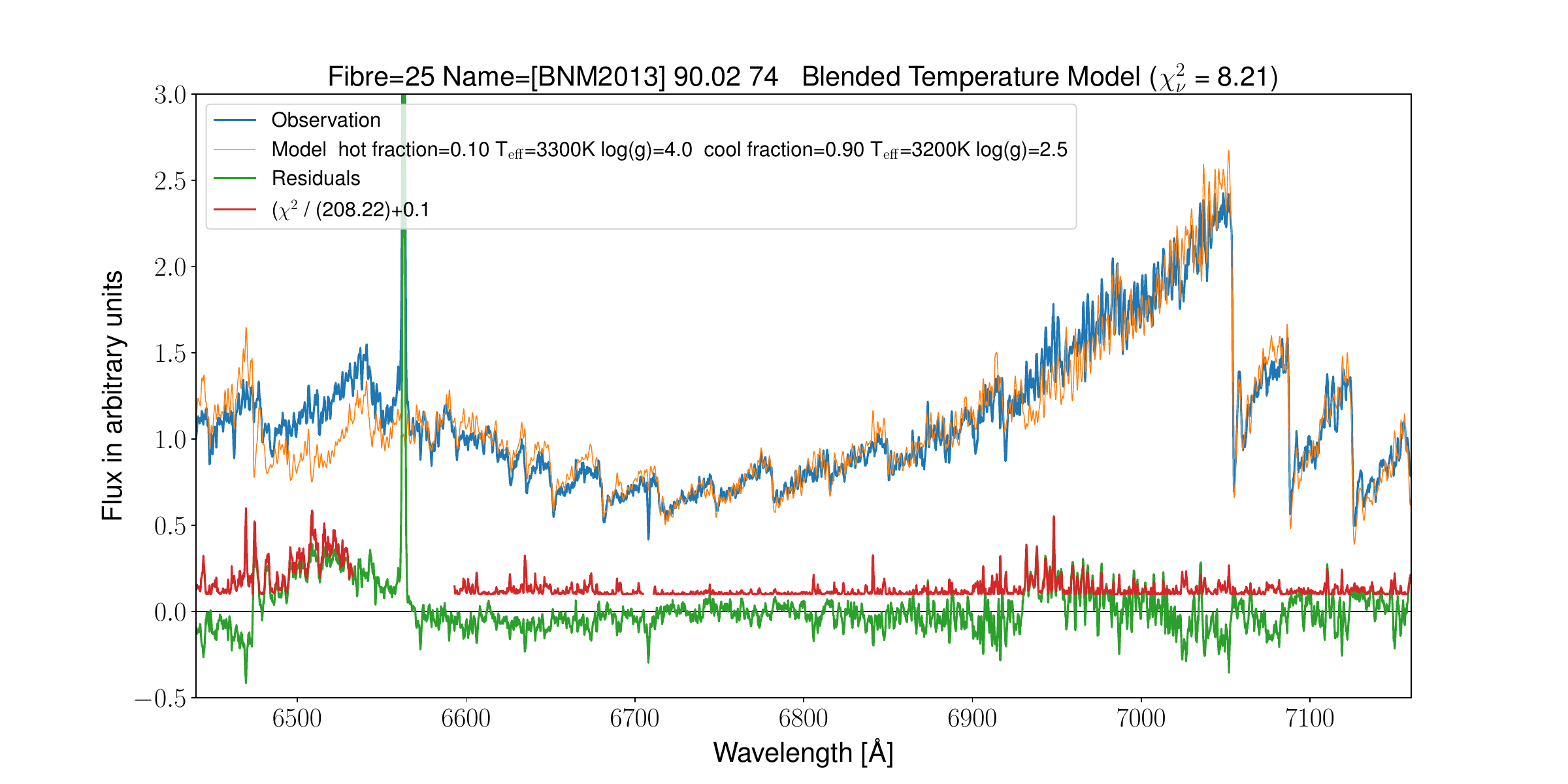}
	\includegraphics[width=0.9\textwidth]{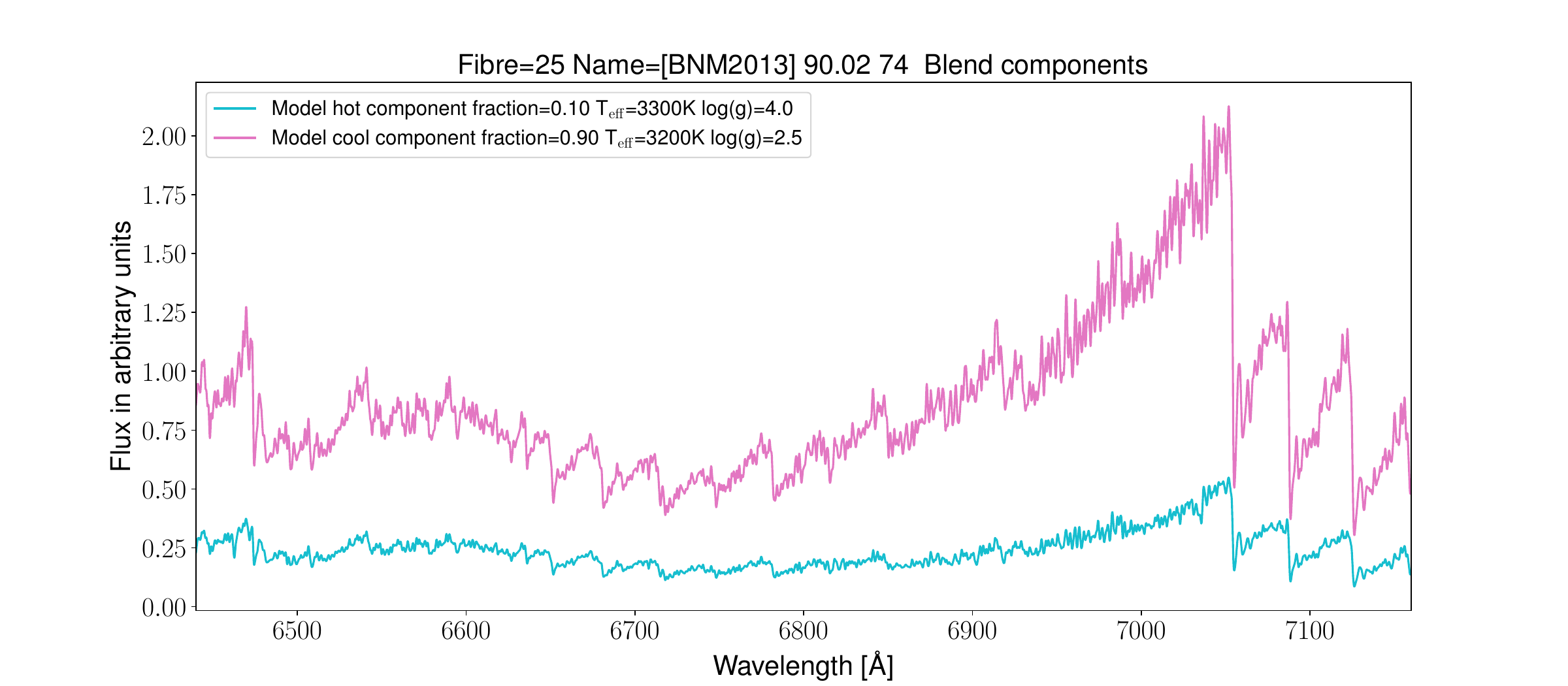}
	\caption{\refone{As Fig.~\ref{fig:fibre_27_spectra} for fibre 25.}}
	\label{fig:fibre_25_spectra}
\end{figure*}

\refone{The spectrum and fit results for the source corresponding to fibre 25 demonstrate it is a cooler M-type star with the two-temperature fit combing two M-type spectra at different temperatures, see Fig.~\ref{fig:fibre_25_spectra}. The spectral range of our method is explicitly designed for detecting starspots on K-type stars, relying on the presence TiO molecular bands in the cool component and on their absence in the hot component. As the TiO molecular bands are present in both the hot and cool components of the M-type star, the technique will be relatively ineffective. This is evidenced by the two-temperature best fit, where the difference between the hot and cool components is only 100\,K and the cool component has $\log(g)=2.5$, see Table~\ref{tab:sigori_results}. This is an anomalously low surface gravity among the SOC results, and a value typically associated with giant stars. The remaining nine stars in the clean sample have temperatures that are broadly consistent with K-type stars \citep{2013ApJS..208....9P}. Hence, \refthree{our decision to} separate our WTTS clean sample into one M-type star and nine probable K-type stars. For shorthand we shall refer to them as the WTTS K and WTTS M samples.}

This splits the SOC members into \refone{WTTS K sample of nine stars, a WTTS M sample containing a single star}, a WTTS binary sample with four stars, and a CTTS sample of three stars, see Table~\ref{tab:sigori_review_members}. We emphasize only two of the WTTS candidate binaries are confirmed binary stars, and the WTTS \refone{K and M samples} are not guaranteed single stars, there is simply no evidence for binarity. Fibre 23 in the CTTS sample is a candidate binary, though we do not subdivide the CTTS sample as our analysis is based on the WTTS. \refone{The subsequent analysis will concentrate on the WTTS K sample.}

\begin{table*}
\begin{minipage}{\textwidth}
	\centering
	\caption{\editten{The $\sigma$ Orionis cluster members with their subsample labels. The H$\mathrm \alpha$ equivalent width is used to identify CTTS, see Section~\ref{sec:accretion}.}}
	\label{tab:sigori_review_members}
	\begin{tabular}{llcccccccc}
		\hline
		& & & | & \multicolumn{3}{c}{H$\mathrm \alpha$ equivalent width} & | & Rotation period & Amplitude\footnote{\citet{2023MNRAS.523..169S} define the amplitude using the 90th--10th percentiles of the relative flux.} \\
		& & & | & \multicolumn{3}{c}{[\r{A}]} & | & [days] & [relative flux] \\
		Fibre & Designation &  Sample & | & This work & \protect\citet{2019AA...629A.114C} & \protect\citet{2021ApJ...923..177S} & | & \multicolumn{2}{c}{\citet{2023MNRAS.523..169S}} \\
		\hline
		15 & [BNM2013] 92.01 11 & WTTS binary & | & \refone{-3.9} &  & -4.1 & | & & \\
		18 & [BNM2013] 90.02 63 & WTTS \refone{K} & | & \refone{-0.6} & 0.4 &  & | & & \\
		23 & [BNM2013] 90.02 77 & CTTS & | & \refone{-6.7} & -20.8 & -13.0 & | & & \\
		24 & [BNM2013] 92.01 28 & CTTS & | & \refone{-9.1} & -11.1 & -11.6 & | & 7.9$\pm$0.2 & 0.25 \\
		25 & [BNM2013] 90.02 74 & WTTS \refone{M} & | & \refone{-6.0} &  & -3.7 & | & & \\
		26 & [BNM2013] 90.02 76 & WTTS \refone{K} & | & \refone{-1.8} & -1.0 & -2.0 & | & & \\
		27 & [BNM2013] 90.02 78 & WTTS \refone{K} & | & \refone{-2.7} & -1.0 &  & | & & \\
		29 & [BNM2013] 90.02 79 & WTTS binary & | & \refone{-3.6} & -4.5 & -4.2 & | & & \\
		31 & [BNM2013] 92.01 43 & WTTS \refone{K} & | & \refone{-1.9} & -3.6 & -2.7 & | & 10.4$\pm$0.3 & 0.17 \\
		32 & [BNM2013] 92.01 1006 & WTTS \refone{K} & | & -8.1 & -2.3 & -2.5 & | & 6.1$\pm$0.1 & 0.04 \\
		33 & [BNM2013] 92.01 690 & WTTS \refone{K} & | & \refone{-1.8} & -1.2 & -1.8 & | & & \\
		34 & [BNM2013] 92.01 34 & WTTS \refone{K} & | & \refone{-2.1} & -1.9 & -2.1 & | & & \\
		35 & [BNM2013] 92.01 32 & WTTS binary & | & \refone{-2.9} & -0.9 & -1.7 & | & & \\
		37 & [BNM2013] 93.03 43 & WTTS \refone{K} & | & \refone{-4.1} &  & -4.9 & | & 5.0$\pm$0.1 & 0.06 \\
		45 & [BNM2013] 92.01 122 & CTTS & | & \refone{-20.2} & -32.0 & -13.0 & | & 7.4$\pm$0.2 & 0.40 \\
		46 & [BNM2013] 92.01 686 & WTTS binary & | & \refone{-3.4} & -3.1 & -3.6 & | & 1.0$\pm$0.0 & 0.16 \\
		47 & [BNM2013] 92.01 118 & WTTS \refone{K} & | & \refone{-1.5} & -0.5 & -1.3 & | & 6.1$\pm$0.1 & 0.08 \\
		\hline
	\end{tabular}
\end{minipage}
\end{table*}

\section{Reviewing the model fits}
\label{sec:review}

\begin{figure}
	\centering
	\includegraphics[width=0.45\textwidth]{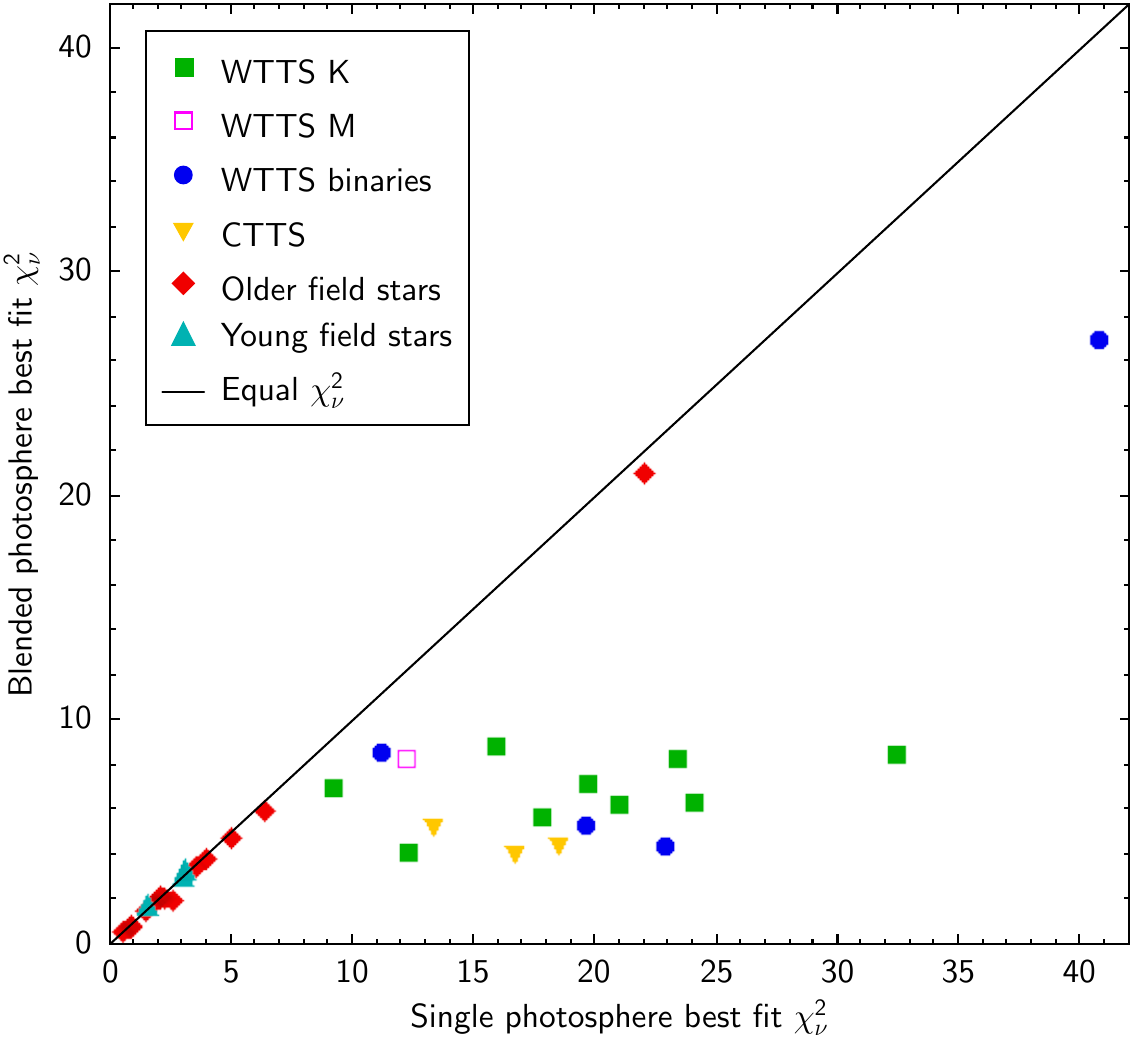}
	\caption{\editseven{Comparison between the two-temperatures and single-temperature model fits to the spectra. \refone{The green filled squares are the WTTS K-type stars from the clean sample, the magneta open square is the WTTS M-type star from the clean sample,} the blue circles are the WTTS binaries, and the orange downward pointing triangles are the CTTS, all members of the $\mathrm \sigma$ Orionis cluster. The red diamonds are the older field stars and the cyan upward pointing triangles are the young field stars. The black line marks equal $\chi^2_{\nu}$ between the two and single-temperature models.}}
	\label{fig:soc_chi_sq}
\end{figure}

To assess whether each star was best fitted by a single temperature photosphere or a blend of two temperatures, the $\chi^2_{\nu}$ for the blend was plotted against that of the single photosphere, see Fig.~\ref{fig:soc_chi_sq}. The older and young field stars exhibit little benefit from the blended photosphere over the single photosphere, staying close to the line of equal $\chi^2_{\nu}$. \refone{The sample of SOC members all exhibit a lower $\chi^2_{\nu}$ for a blend of two photospheric temperatures with the majority showing a significant reduction.}

\editsix{Where the $\chi^2_{\nu}$ of the blend is only slightly smaller than the fit by a single temperature, the result should be treated with caution. This could be due to the two-temperature model being a slightly better match by fitting non-temperature or gravity related phenomena, such as metallicity, line broadening that is not fully modelled by a single Gaussian kernel, and small imperfections in the response correction. The field stars show these small differences. They provide a stark contrast to the large differences seen in the $\chi^2_{\nu}$ of the SOC members. This is evidence that the smaller $\chi^2_{\nu}$ of the \reftwo{two-temperature} photosphere fits to the SOC members are due to modelling of real physics.} \editfourteen{In general, the $\chi^2_{\nu}$ of single-temperature fits to the field stars are smaller than the two-temperature fits to the SOC members. This indicates that greater complexity is needed to adequately model TTS spectra, or that stellar atmospheres threaded by magnetic fields and back-irradiated by chromospheric emission may not be perfectly represented by the BT-Settl model atmospheres.}

\refone{Reviewing the TTS spectra it is clear that the single-temperature models cannot fit all of the spectral lines and molecular bands, while the two-temperature models provide an improved fit to the observations. The WTTS K sample have cool components accounting for 30--95 per cent of their visible surface and hot to cool temperature differences of 1200--1800\,K with a median value of 1300\,K, see Table~\ref{tab:sigori_results}.}

The values of the $\chi^2_{\nu}$ show that none of the models are formally good fits. \reffour{This could be due to unmodelled parameters such as metallicity, issues with the line lists and strengths in the model atmospheres, or because there are more than two temperatures. The surface areas and temperatures should therefore be treated as indicative.} A key assumption of the $\chi^2$ method is the residuals are independent. However, this assumption does not hold for the differences between the model and the observation as they are correlated in wavelength space at a variety of \edittwelve{length scales} corresponding to spectral lines and broader features such as molecular absorption bands. 
\refone{This combination of high $\chi^2_{\nu}$ and correlated residuals makes it impossible to calculate meaningful probabilities \refthree{of a good fit} or confidence intervals on the best fit values.}

\begin{figure*}
	\centering
	\includegraphics[width=0.45\textwidth]{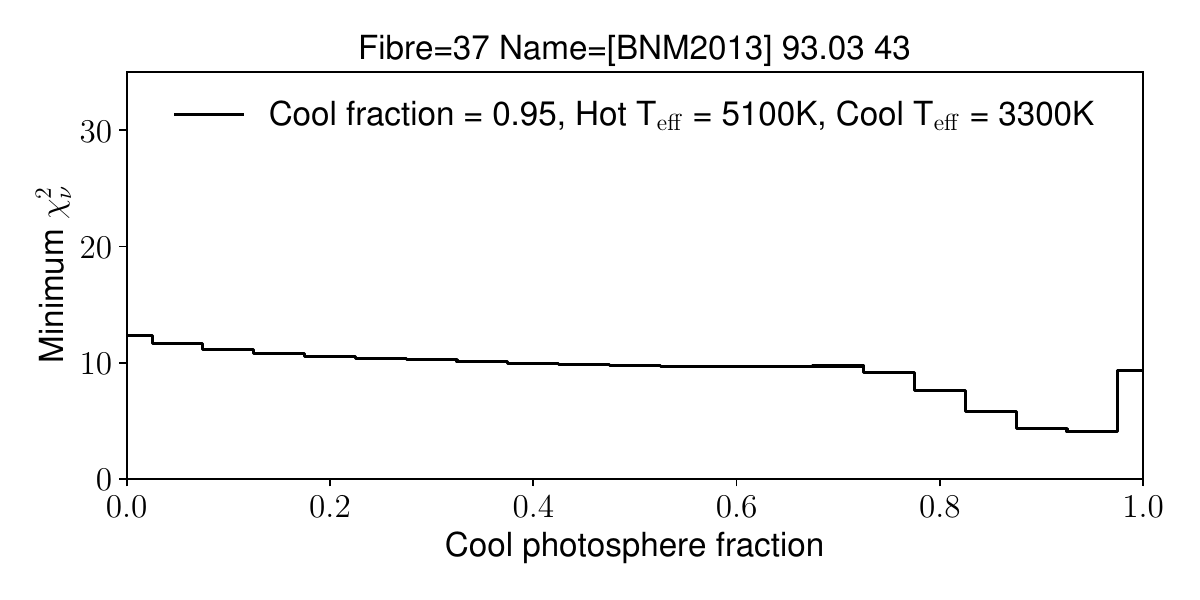}
	\includegraphics[width=0.45\textwidth]{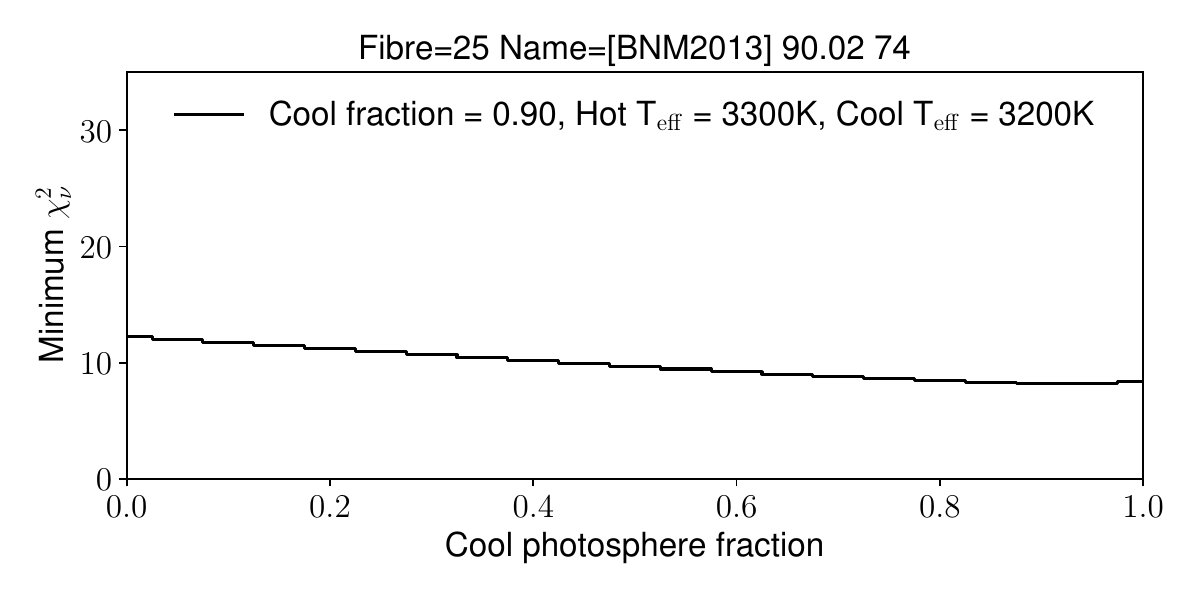}
	\includegraphics[width=0.45\textwidth]{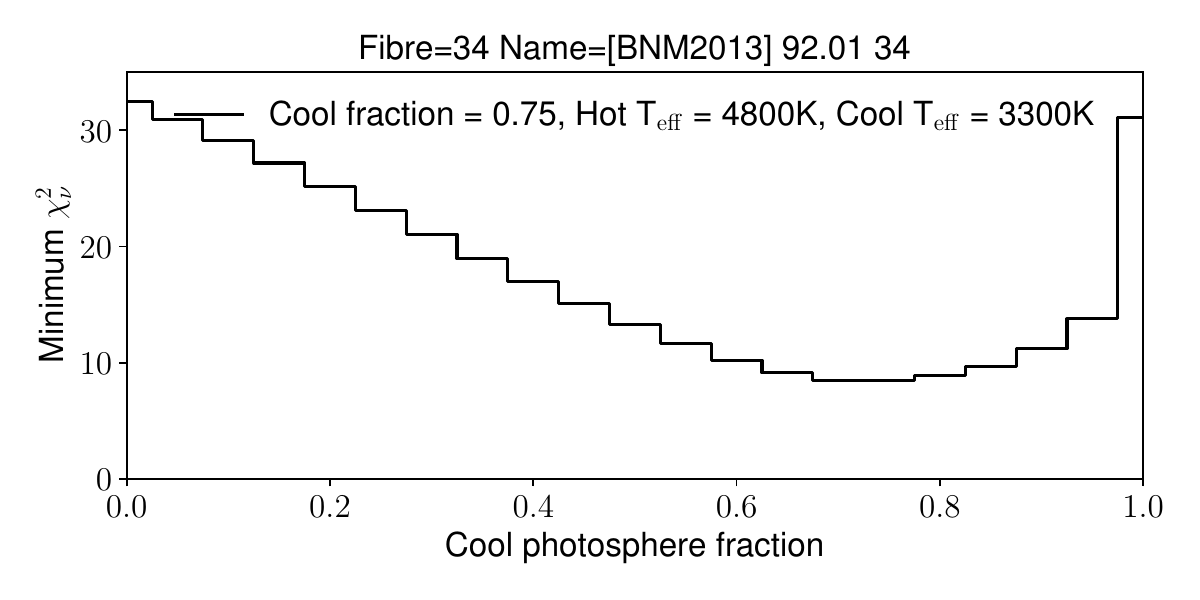}
	\includegraphics[width=0.45\textwidth]{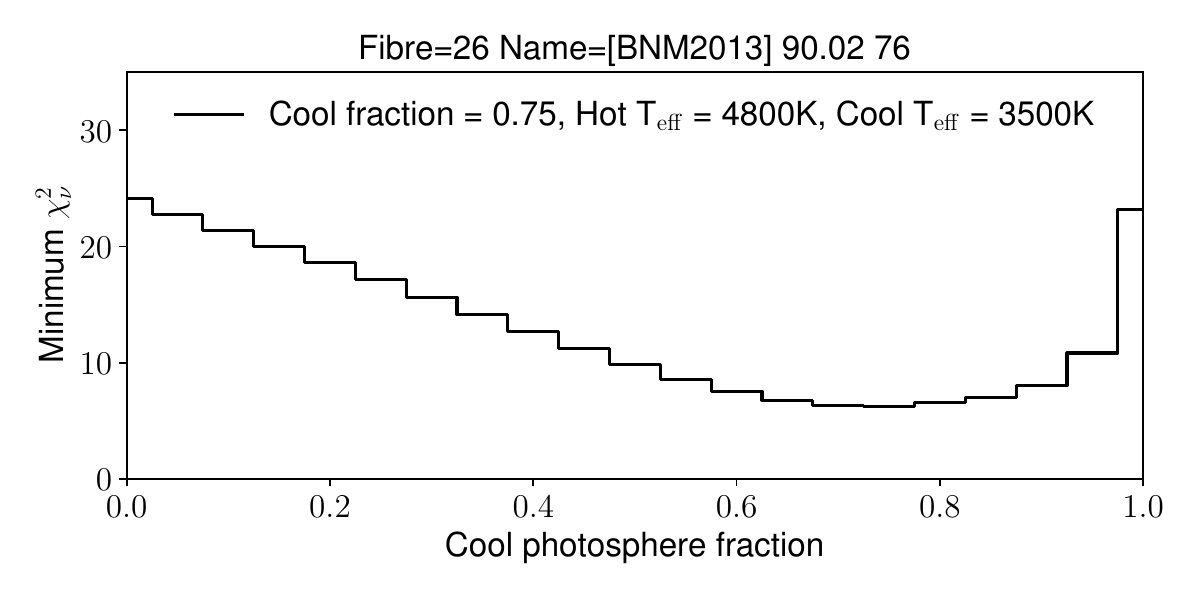}
	\includegraphics[width=0.45\textwidth]{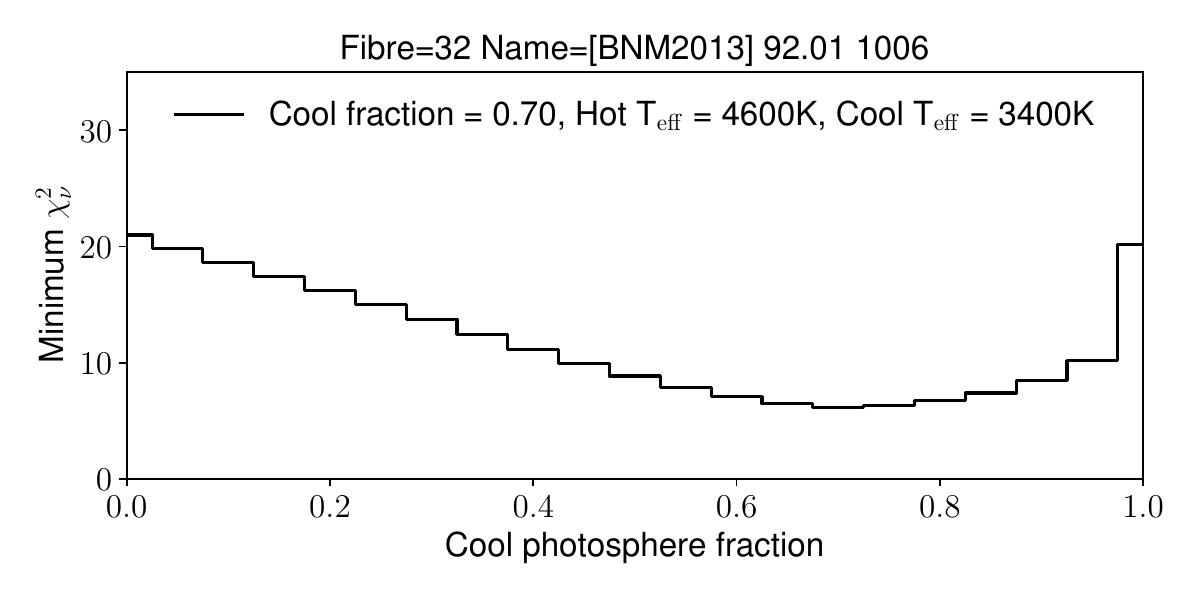}
	\includegraphics[width=0.45\textwidth]{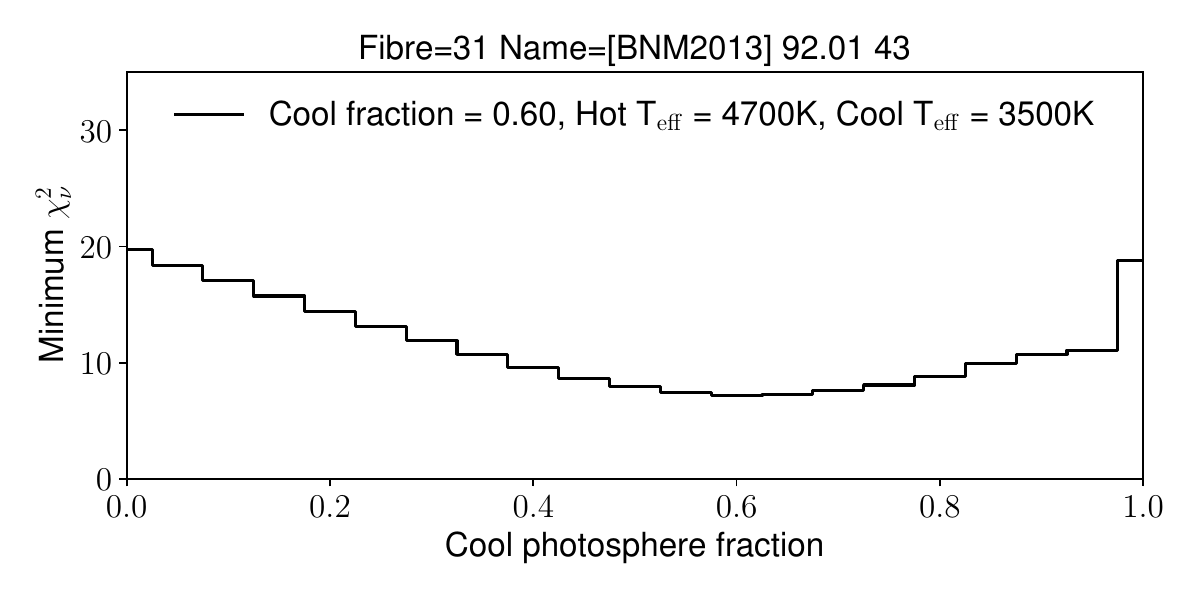}
	\includegraphics[width=0.45\textwidth]{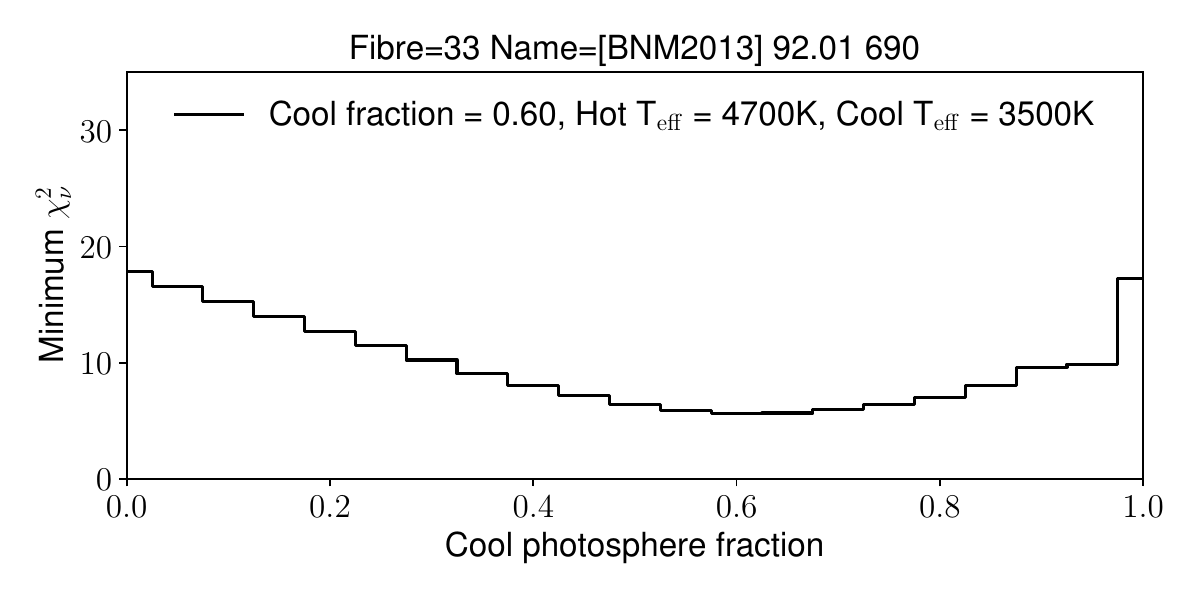}
	\includegraphics[width=0.45\textwidth]{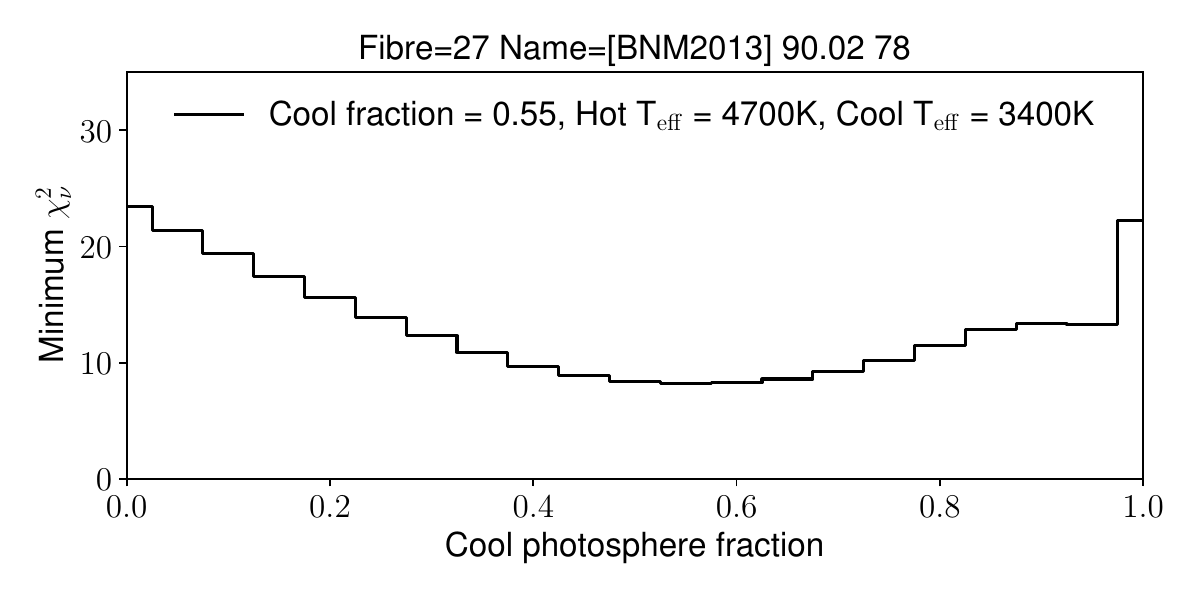}
	\includegraphics[width=0.45\textwidth]{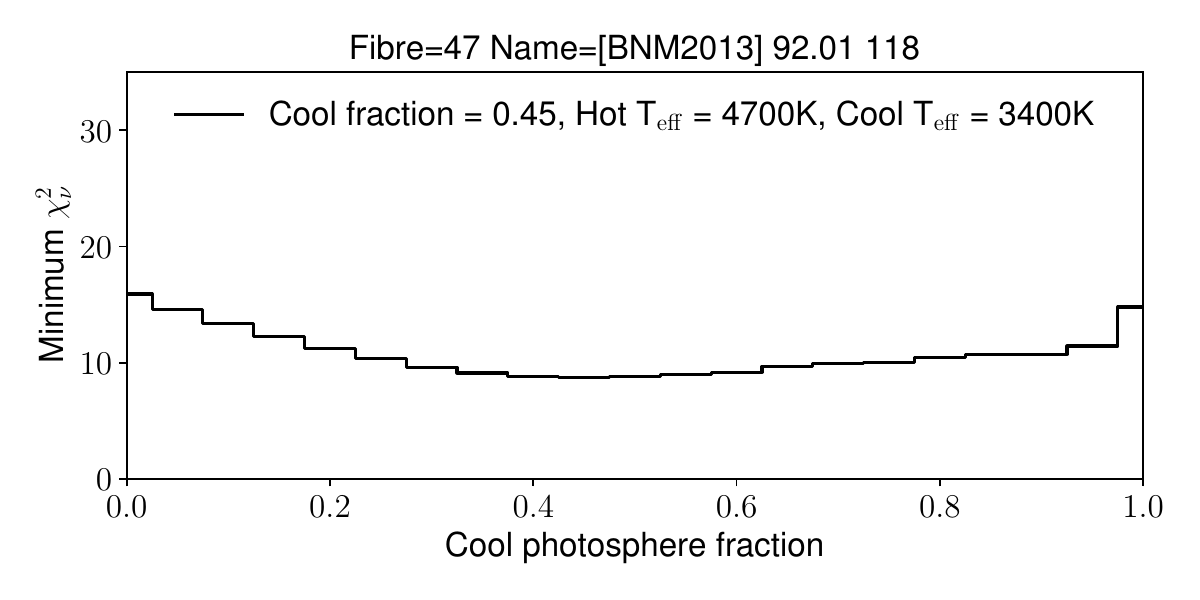}
	\includegraphics[width=0.45\textwidth]{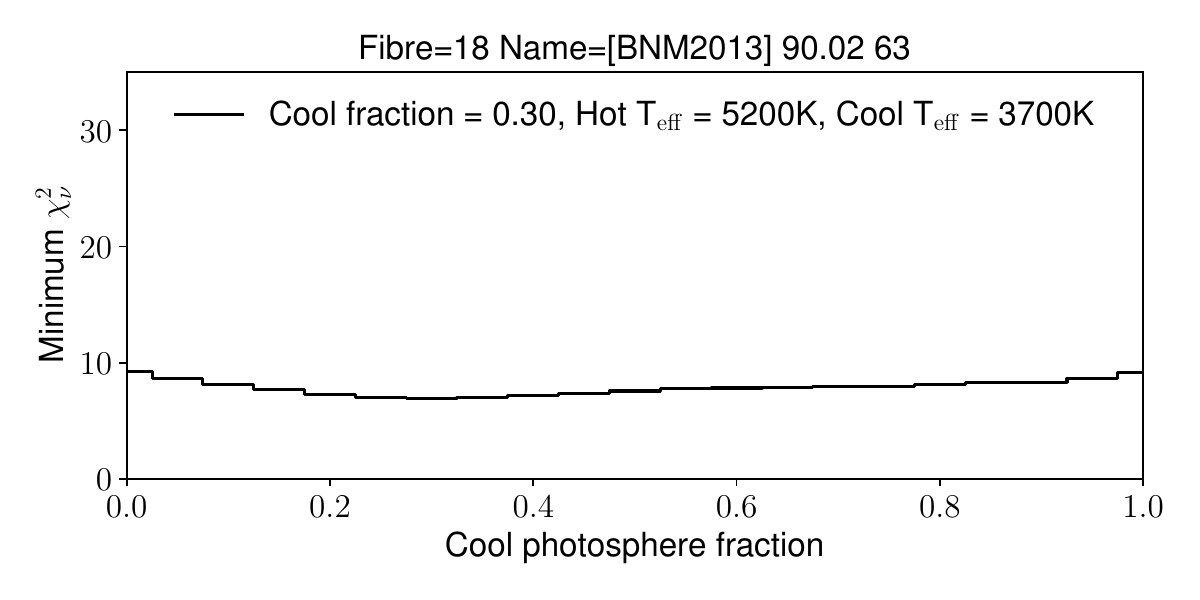}
	\caption{\editten{The minimum $\chi^2_{\nu}$ by cool surface photosphere fraction from stage 3 of the blended model fitting for WTTS clean sample. The change in $\chi^2_{\nu}$ represents a ratio in probability space. The spectra are ordered by decreasing cool photosphere fraction, then by increasing temperature of the hot component and finally by increasing temperature of the cool component.}}
	\label{fig:minredchi_coolfrac}
\end{figure*}

\begin{figure*}
	\centering
	\includegraphics[width=0.9\textwidth]{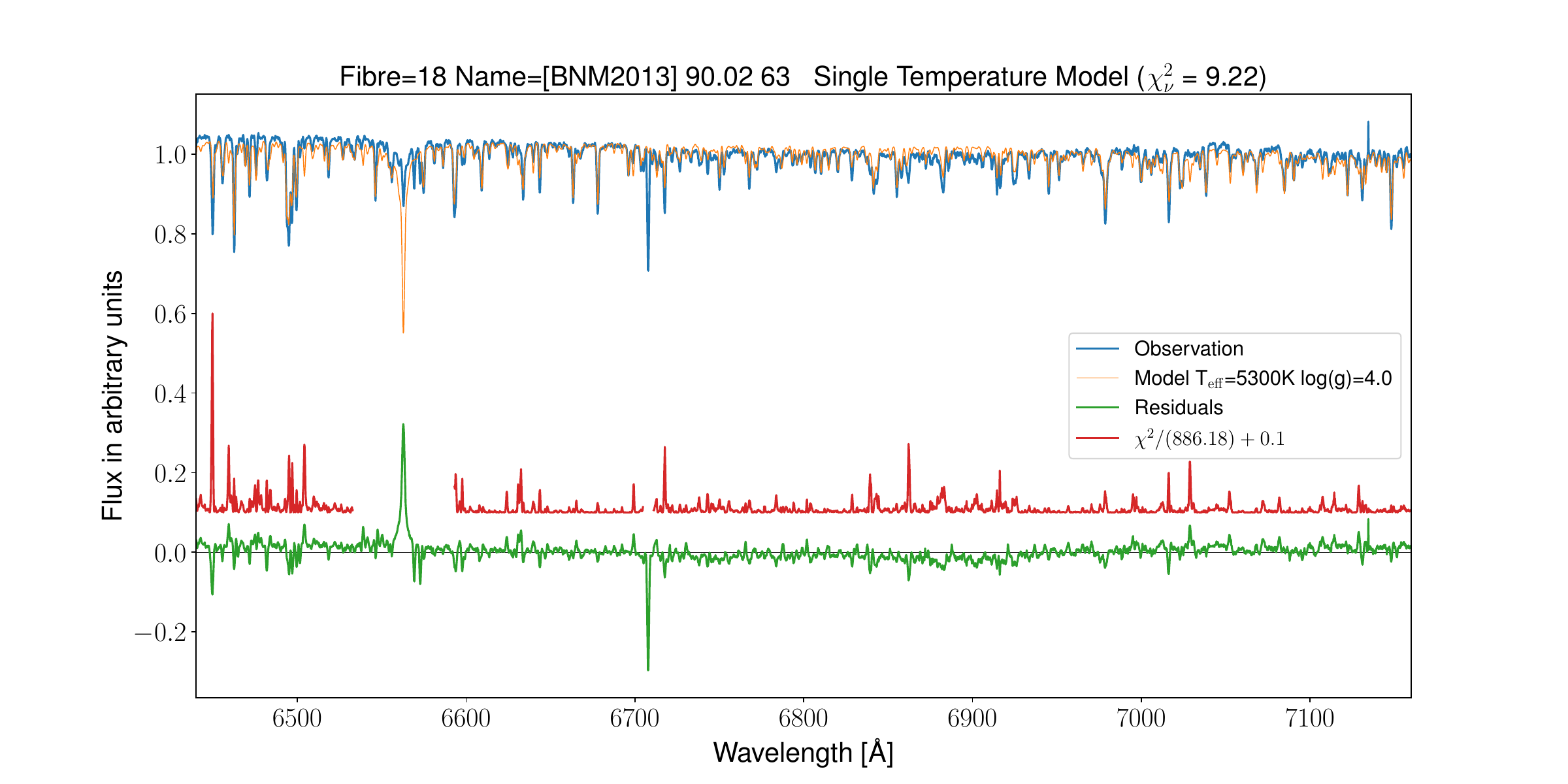}
	\includegraphics[width=0.9\textwidth]{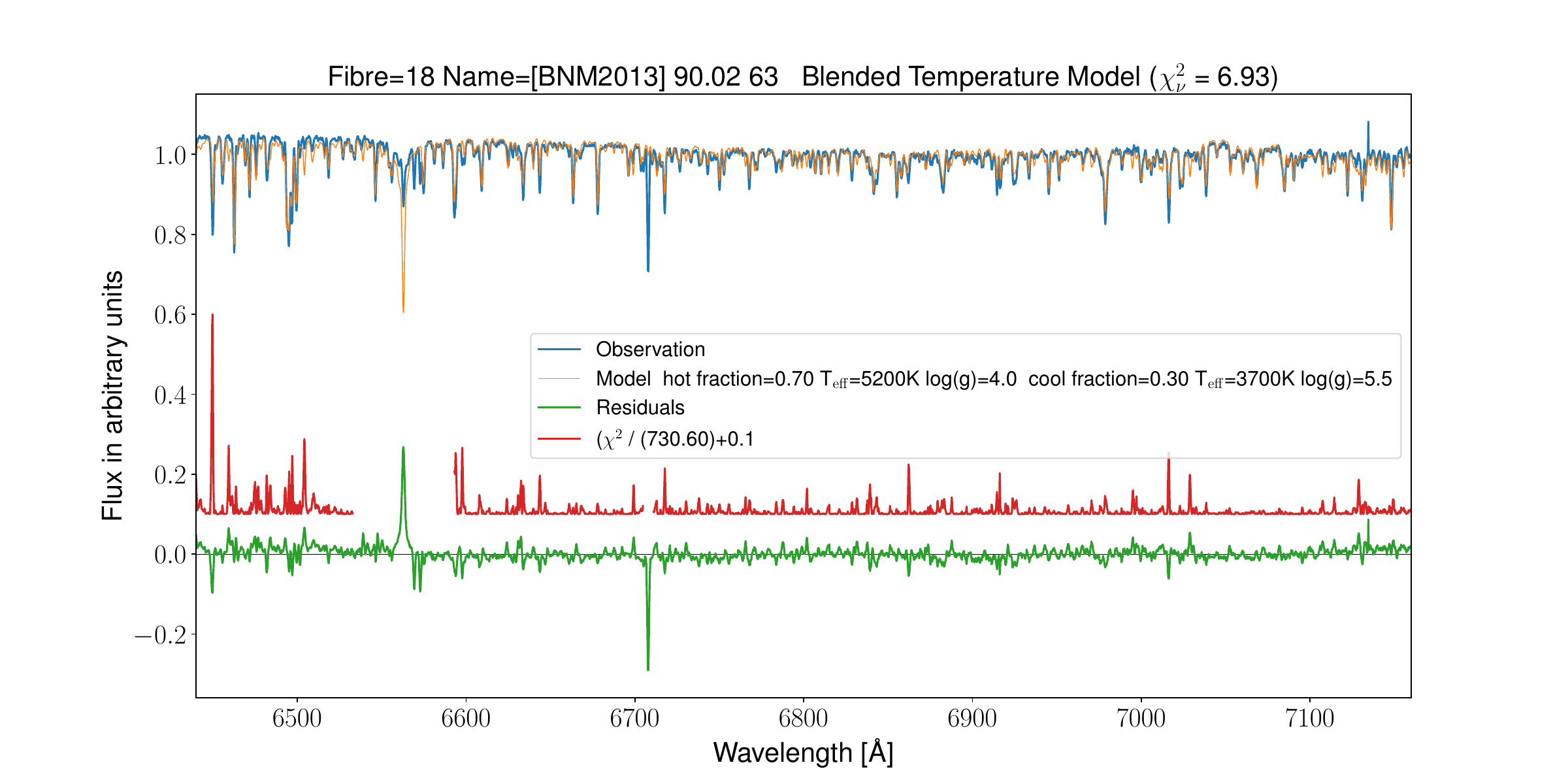}
	\includegraphics[width=0.9\textwidth]{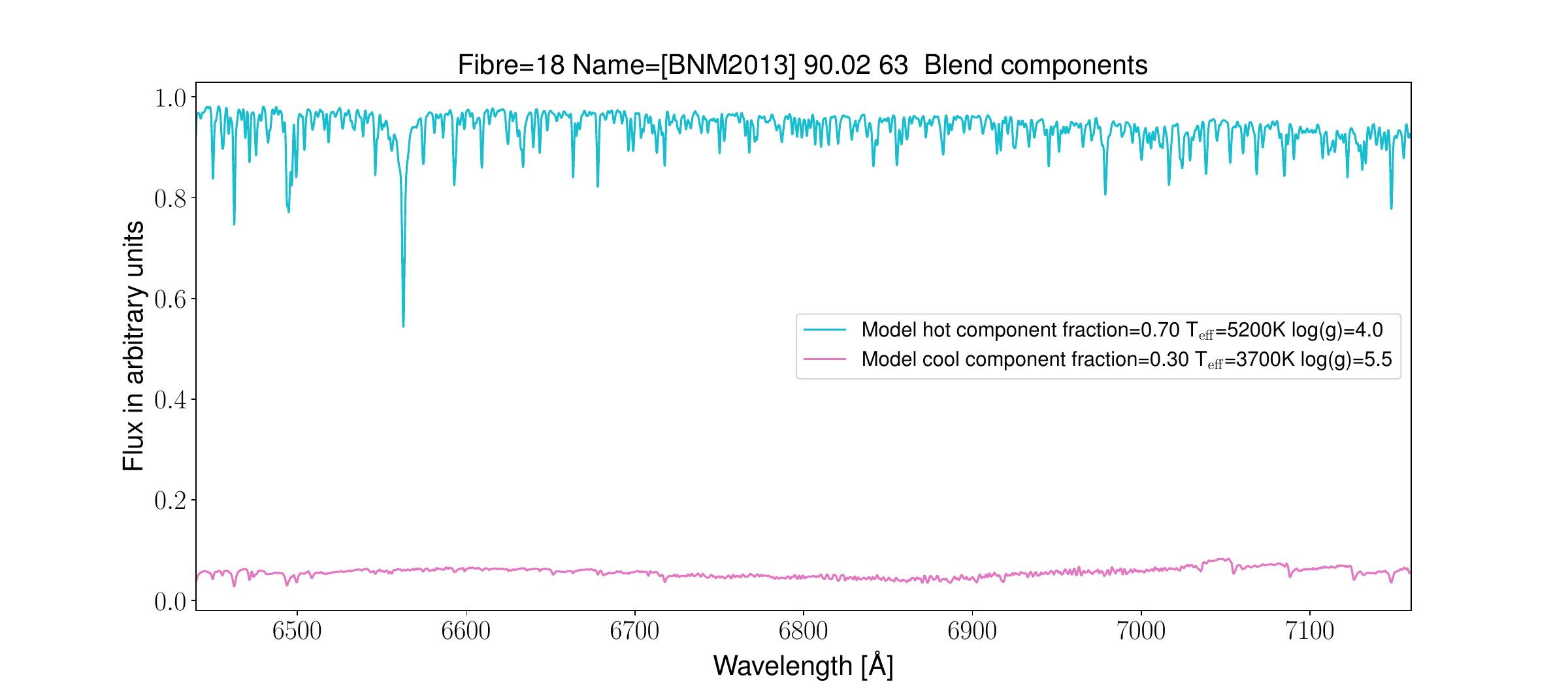}
	\caption{As Fig.~\ref{fig:fibre_27_spectra} for fibre 18.}
	\label{fig:fibre_18_spectra}
\end{figure*}

\begin{figure*}
	\centering
	\includegraphics[width=0.9\textwidth]{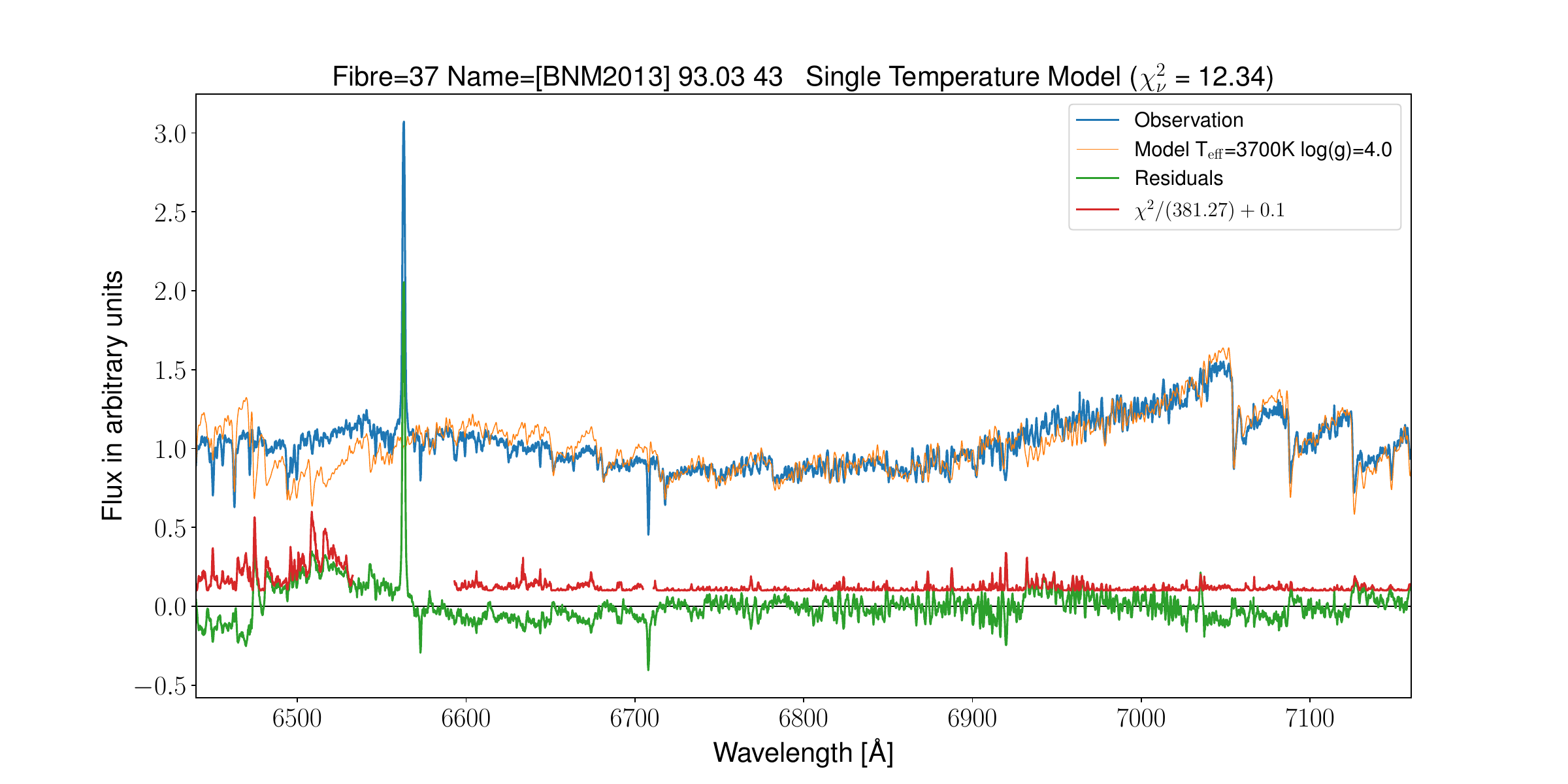}
	\includegraphics[width=0.9\textwidth]{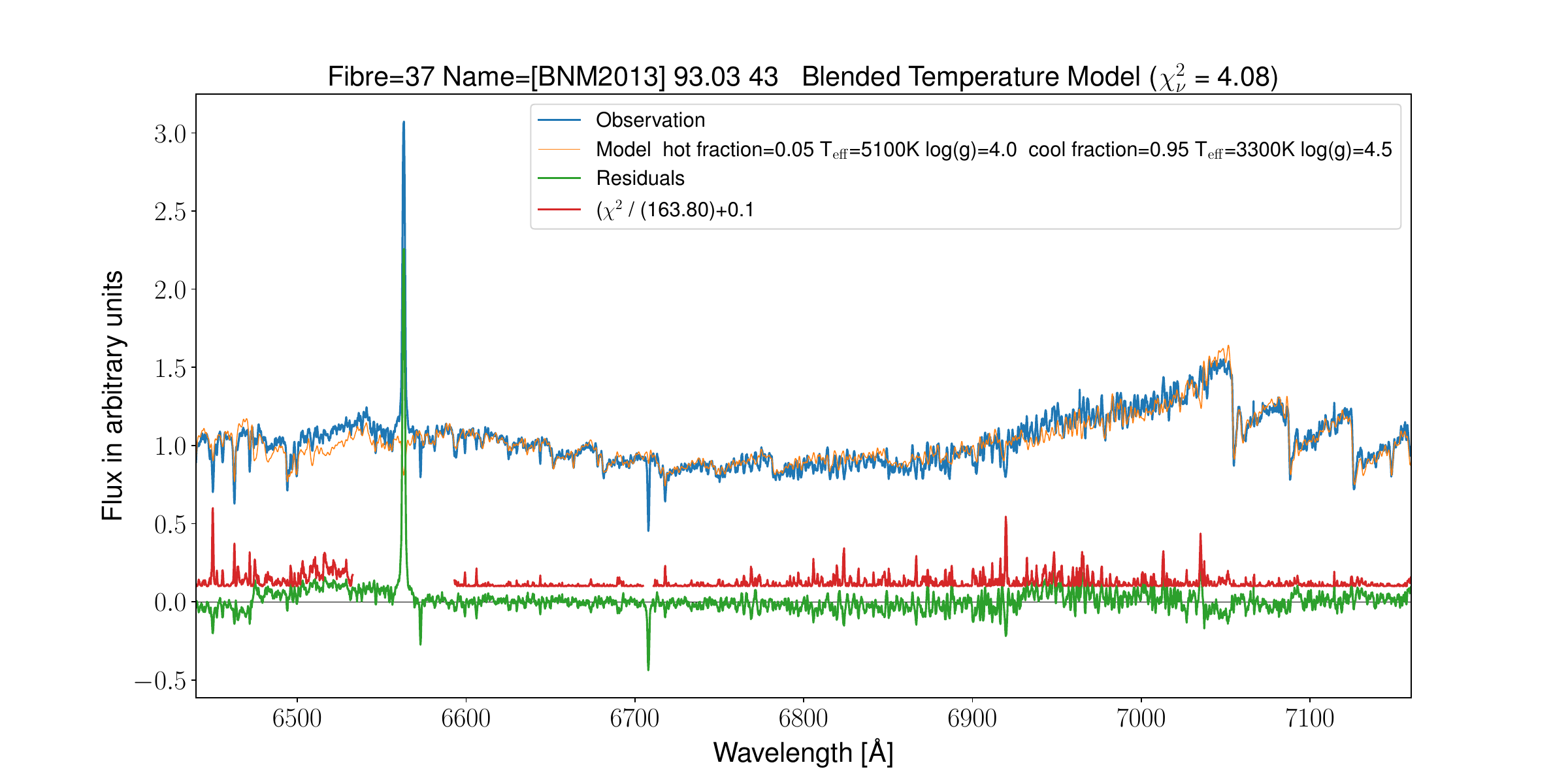}
	\includegraphics[width=0.9\textwidth]{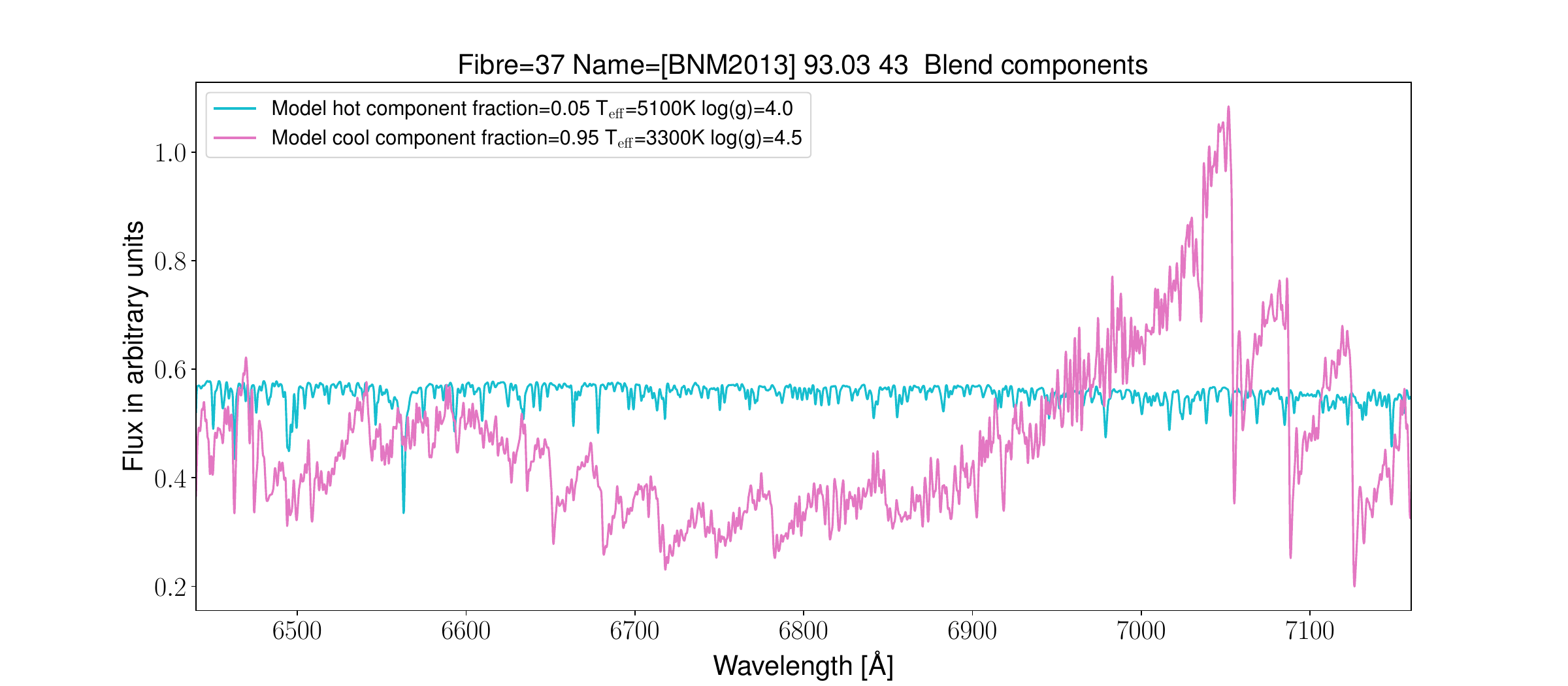}
	\caption{\refone{As Fig.~\ref{fig:fibre_27_spectra} for fibre 37.}}
	\label{fig:fibre_37_spectra}
\end{figure*}

\refone{While a quantitative confidence interval cannot be calculated, a qualitative assessment is possible. Plots of the minimum $\chi^2_{\nu}$ by cool photosphere fraction for the WTTS clean sample are presented in Fig.~\ref{fig:minredchi_coolfrac}, using the same vertical scale on all plots to aid comparison.} 
At the model resolution of 5 per cent in photosphere area fraction, the plots demonstrate the minima of the $\chi^2_{\nu}$ are smooth and single valued. 
\refone{The plot of fibre 25, the M-type star, has a shallow downward slope to high cool surface fraction with a very shallow minimum, further demonstrating that our technique struggles with M-type stars. 
The plot of fibre 18 shows a broad and shallow minimum. This is the star with the lowest cool surface fraction at 0.30 and the weakest result of the SOC members with a $\chi^2_{\nu}=6.9$ for the two-temperature model fit compared to 9.2 for the single-temperature model. \refthree{Despite this,} the spectrum clearly shows contributions from both hot and cool components, confirming the presence of two temperatures. The TiO feature is relative weak, as would be expected from a smaller surface area fraction, see Fig.~\ref{fig:fibre_18_spectra}. 
Fibre 37 has the largest cool surface fraction of the WTTS K sample at 0.95. The plot of minimum $\chi^2_{\nu}$ by cool photosphere fraction shows a clearly defined valley favouring a large cool component. Although its spectrum is dominated by the cool component, the presence of a hot component is clearly evidenced by the spectral feature around 6500\,\r{A}, see Fig.~\ref{fig:fibre_37_spectra}. The remaining plots are consistent with the large cool fractions of their best fit two-temperature models,} \reffour{though the uncertainties in surface fraction are unclear.}

\refone{The surface gravity of the cool component is greater than the hot component for all of the WTTS K sample. 
This may be evidence of higher effective surface gravities in the spectra of cool starspots than the hotter unblemished photosphere, as hypothesised by \citet{2025ApJ...990..205P}, see Section~\ref{sec:introduction}.}

\refone{Returning to fibres 1, 40 and 44, the young field stars with a weak Li 6708\,\r{A} detection in their spectra. Fibres 1 and 44 both have $\chi^2_{\nu}=3.1$, and fibre 40 has $\chi^2_{\nu}=1.6$ for a fit by a single-temperature photosphere. These values} are significantly lower than any of the SOC members and are consistent with the older field stars.

\refone{In summary, the results for the SOC members show a strong preference for two-temperature models over the single-temperature models, while the field stars show little or no preference for two-temperature over a single-temperature models. 
The cool components from the best fitting two-temperature models to the WTTS K sample cover 30--95 per cent of the visible stellar surface. The temperature differences between the hot and cool components are 1200--1800\,K with a median of 1300\,K. The observed spectra of these nine sources show clear contributions from both hot and cool components, reinforcing the results.}

\section{Infrared photometry}
\label{sec:photometry}

\editten{The large fraction of the blended model surface areas taken up by the cool components will make a significant contribution to the IR flux,} and in many cases the cool component will dominate the mid-IR. This allows us to perform an independent test of the credibility of these large cooler areas by comparing the models derived from optical spectroscopy to IR photometry. Our aim is not to demonstrate an exact match between the observations and the models, just to show the expected large mid-IR fluxes are present.

The BT-Settl CIFIST model atmospheres \citep{Allard2014} extend into the infrared, allowing us to construct our blended models at longer wavelengths. Near and mid-IR photometry were obtained for the SOC members from 2MASS \citep{Skrutskie2006} and the \textit{Spitzer} \citep{2004ApJS..154....1W} InfraRed Array Camera (IRAC) \citep{2004ApJS..154...10F} from the \citet{2007ApJ...662.1067H} study of discs in the SOC, see Table~\ref{tab:sigori_photometry}. Since the models were defined by optical spectra, \textit{Gaia} DR3 $G_{\rm BP}$ and $G_{\rm RP}$ mean magnitudes have been included.

\begin{table*}
	\centering
	\caption{Photometry rounded to the nearest tenth of a magnitude from \textit{Gaia}, 2MASS and \textit{Spitzer} for the WTTS clean sample.}
	\label{tab:sigori_photometry}
	\begin{tabular}{llcccccccccccc}
		\hline
		& & | & \multicolumn{2}{c}{\textit{Gaia}} & | & \multicolumn{3}{c}{2MASS} & | &  \multicolumn{4}{c}{\textit{Spitzer}} \\
		Fibre & Designation & | & $G_{\rm BP}$ & $G_{\rm RP}$ & | & $J$ & $H$ & $K$ & | & 3.6\,$\mu$m & 4.5\,$\mu$m & 5.8\,$\mu$m & 8.0\,$\mu$m \\
		\hline
		15 & [BNM2013] 92.01 11 & | & 17.1 & 14.3 & | & 12.5 & 11.8 & 11.5 & | & 11.4 & 11.2 & 11.1 & 10.9 \\
		18 & [BNM2013] 90.02 63 & | & 12.8 & 11.6 & | & 10.7 & 10.3 & 10.1 & | & 10.1 & 10.0 & 10.0 & 10.0 \\
		23 & [BNM2013] 90.02 77 & | & 14.4 & 12.7 & | & 11.3 & 10.6 & 10.3 & | & 9.8 & 9.7 & 9.6 & 9.3 \\
		24 & [BNM2013] 92.01 28 & | & 15.0 & 13.0 & | & 11.5 & 10.8 & 10.4 & | & 9.8 & 9.4 & 8.9 & 8.1 \\
		25 & [BNM2013] 90.02 74 & | & 17.5 & 14.3 & | & 12.2 & 11.6 & 11.3 & | & 11.1 & 11.0 & 11.0 & 11.0 \\
		26 & [BNM2013] 90.02 76 & | & 14.7 & 12.9 & | & 11.5 & 10.9 & 10.7 & | & 10.6 & 10.7 & 10.6 & 10.6 \\
		27 & [BNM2013] 90.02 78 & | & 13.3 & 11.7 & | & 10.6 & 9.9 & 9.7 & | & 9.6 & 9.6 & 9.5 & 9.5 \\
		29 & [BNM2013] 90.02 79 & | & 15.5 & 13.3 & | & 11.7 & 11.0 & 10.8 & | & 10.6 & 10.5 & 10.4 & 10.5 \\
		31 & [BNM2013] 92.01 43 & | & 14.3 & 12.6 & | & 11.3 & 10.6 & 10.5 & | & 10.3 & 10.3 & 10.3 & 10.2 \\
		32 & [BNM2013] 92.01 1006 & | & 14.9 & 13.1 & | & 11.5 & 10.9 & 10.7 & | & 10.6 & 10.5 & 10.6 & 10.5 \\
		33 & [BNM2013] 92.01 690 & | & 14.3 & 12.5 & | & 11.2 & 10.6 & 10.4 & | & 10.3 & 10.4 & 10.2 & 10.2 \\
		34 & [BNM2013] 92.01 34 & | & 14.2 & 12.5 & | & 11.2 & 10.5 & 10.3 & | & 10.2 & 10.2 & 10.2 & 10.1 \\
		35 & [BNM2013] 92.01 32 & | & 12.6 & 11.0 & | & 9.9 & 9.3 & 9.1 & | & 9.0 & 9.1 & 8.9 & 8.8 \\
		37 & [BNM2013] 93.03 43 & | & 16.5 & 14.2 & | & 12.6 & 11.9 & 11.7 & | & 11.5 & 11.5 & 11.5 & 11.5 \\
		45 & [BNM2013] 92.01 122& | & 15.9 & 13.8 & | & 11.8 & 10.9 & 10.5 & | & 9.7 & 9.5 & 9.2 & 8.4 \\
		46 & [BNM2013] 92.01 686 & | & 14.2 & 12.5 & | & 11.2 & 10.6 & 10.4 & | & 10.3 & 10.3 & 10.2 & 10.2 \\
		47 & [BNM2013] 92.01 118 & | & 13.1 & 11.6 & | & 10.6 & 9.9 & 9.8 & | & 9.6 & 9.7 & 9.6 & 9.6 \\
		\hline
	\end{tabular}
\end{table*}

\editsix{To compare our models to the IR photometry we needed to place the \refone{single and two-temperature model spectra onto} an absolute flux scale. As the wavelength range of the VLT spectra lie within a 700\,\r{A} window of the \textit{Gaia} $G_{\rm RP}$ bandpass, we used the \textit{Gaia} DR3 mean $G_{\rm RP}$ \editfourteen{apparent} magnitudes to perform a crude flux calibration \editfourteen{for each star}. Synthetic photometry was generated using \textsc{synphot}\footnote{\url{https://synphot.readthedocs.io/en/latest/}} \citep{2018ascl.soft11001S} by folding the reddened best fitting model spectra through filter bandpasses obtained from the \textsc{SVO Filter Profile Service}\footnote{\url{https://svo2.cab.inta-csic.es/theory/fps/}} \citep{2012ivoa.rept.1015R,2020sea..confE.182R,2024A&A...689A..93R}. We multiplied the model spectrum by the factor required to reproduce the observed \textit{Gaia} DR3 mean $G_{\rm RP}$ magnitude when folded through filter bandpass.}

To plot the observed photometry over the spectra in flux space, the observed magnitudes were converted to fluxes using zero-points and effective wavelengths from the \textsc{SVO Filter Profile Service}. To compare the observed photometry and models in magnitude space, \textsc{synphot} was used to calculate synthetic magnitudes from the model spectrum, which was then subtracted from the observed magnitudes. An example of these comparisons is provided for fibre 34 in Fig.~\ref{fig:fibre_34_photometry} with plots of the other \refone{ WTTS K sample} in Appendix~\ref{app:sig_ori_photometry}.

\refone{It can be seen that the \refone{two-temperature} models follow the IR photometry more closely than the single-temperature models. This demonstrates the IR photometry is consistent with large cool components that are absent from the single-temperature models but captured by the two-temperature models. }
Looking at the \refone{WTTS K sample} in more detail, \refone{the two-temperature models still underpredict the mid-IR photometry, but only by about a tenth of a magnitude. This implies} the cool areas could be \refone{a little} larger than indicated by the models based on optical data, \reftwo{or that a model composed of more than two temperatures is required}. 

By taking into consideration the \textit{Gaia} optical photometry \editelev{it can be seen that many of the \edittwelve{two-temperature (and single-temperature)} models are apparently too warm.} \refone{Seven out of nine of the two-temperature} models overpredict the blue $G_{\rm BP}$ and underpredict the near-IR with a tendency for a larger underprediction in the mid-IR. A trivial explanation is our models are a little too hot, or they are underestimating the contribution from the cool photosphere. It is more likely that our two-temperature models are too simplistic. A three temperature model may be required, \refone{composed from an} unblemished photosphere \refone{with cooler and hotter components}. Another possibility is presented by \citet{2024ApJ...967...45P,2025ApJ...978...32P}, who fit a cool blackbody spectrum to the inner disc and a hot blackbody spectrum for accretion, in addition to a two temperature photosphere. \editten{Though we are confident our sample of WTTS} are not accreting, see Section~\ref{sec:accretion}. The TiO bands in the spectroscopy demonstrate that the cool component in our models is photospheric in origin rather than cool dust. Though it is plausible that the small additional mid-IR flux in the photometry could be due to the warm inner region of a protoplanetary disc.

\refone{A potential issue with this analysis are the differing epochs of the observations. The $G_{\rm RP}$ magnitude is the mean of photometric measurements acquired by \textit{Gaia} over the 34 months between July 2014 and May 2017 \citep{2021A&A...649A...1G}. The 2MASS photometry was acquired between 1997 and 2001 \citep{Skrutskie2006} and the \textit{Spitzer} photometry in 2004 \citep{2007ApJ...662.1067H}. Hence, both the total magnitudes of the stars and their spot coverages may have changed between sets of observations and the VLT spectra acquired in November 2013. Further, H$\alpha$ is in strong emission in these young stars and highly variable, so it will have an undue influence on the $G_{\rm RP}$ magnitude. These problems are partially mitigated as we calibrate both the single and two-temperature models to the same $G_{\rm RP}$ mean magnitude of each star. Any change in $G_{\rm RP}$ will shift the absolute flux calibration but not the relative flux between the single and two-temperature models. If there were significant changes to the total magnitudes or the contribution from the cool components, then we would expect the different epochs to introduce a shift between the photometric bandpasses of the different surveys. However, the photometric comparisons show trends that are maintained across the different surveys. This indicates the photometry is stable to around a tenth of a magnitude across the survey epochs.}

In summary, the two-temperature models of the \refone{WTTS K sample provide a better fit to} the IR photometry \refone{than the single-temperature models}. \refone{The two-temperature models show} a small underprediction of the \refone{IR flux \reftwo{of} around a tenth of a magnitude}. This validates the two-temperature models in a way that is independent of our spectroscopic technique.

\begin{landscape}
	\begin{figure}
		\centering
		\subfloat[]{\includegraphics[width=0.7\textwidth]{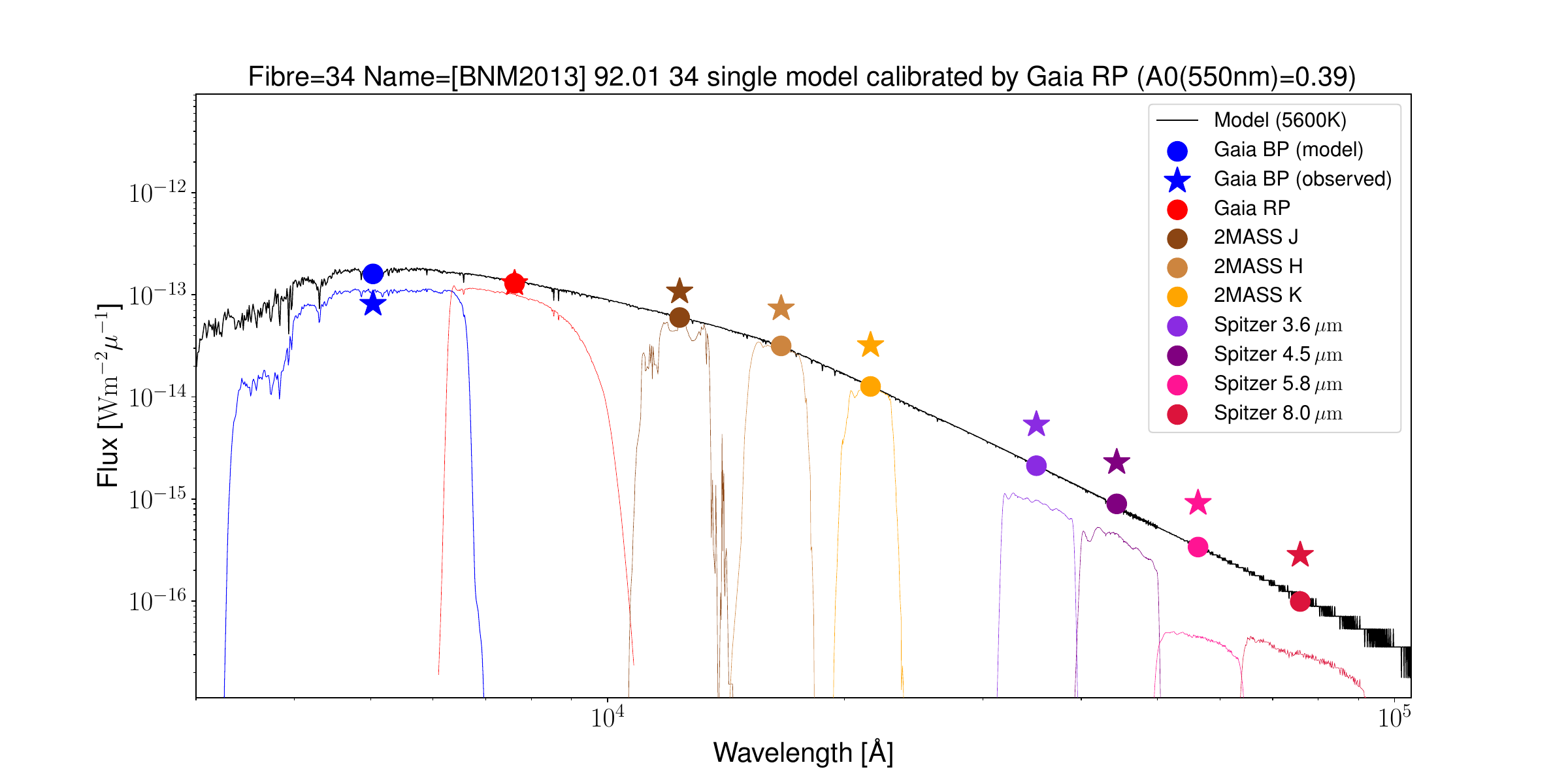}}
		\subfloat[]{\includegraphics[width=0.7\textwidth]{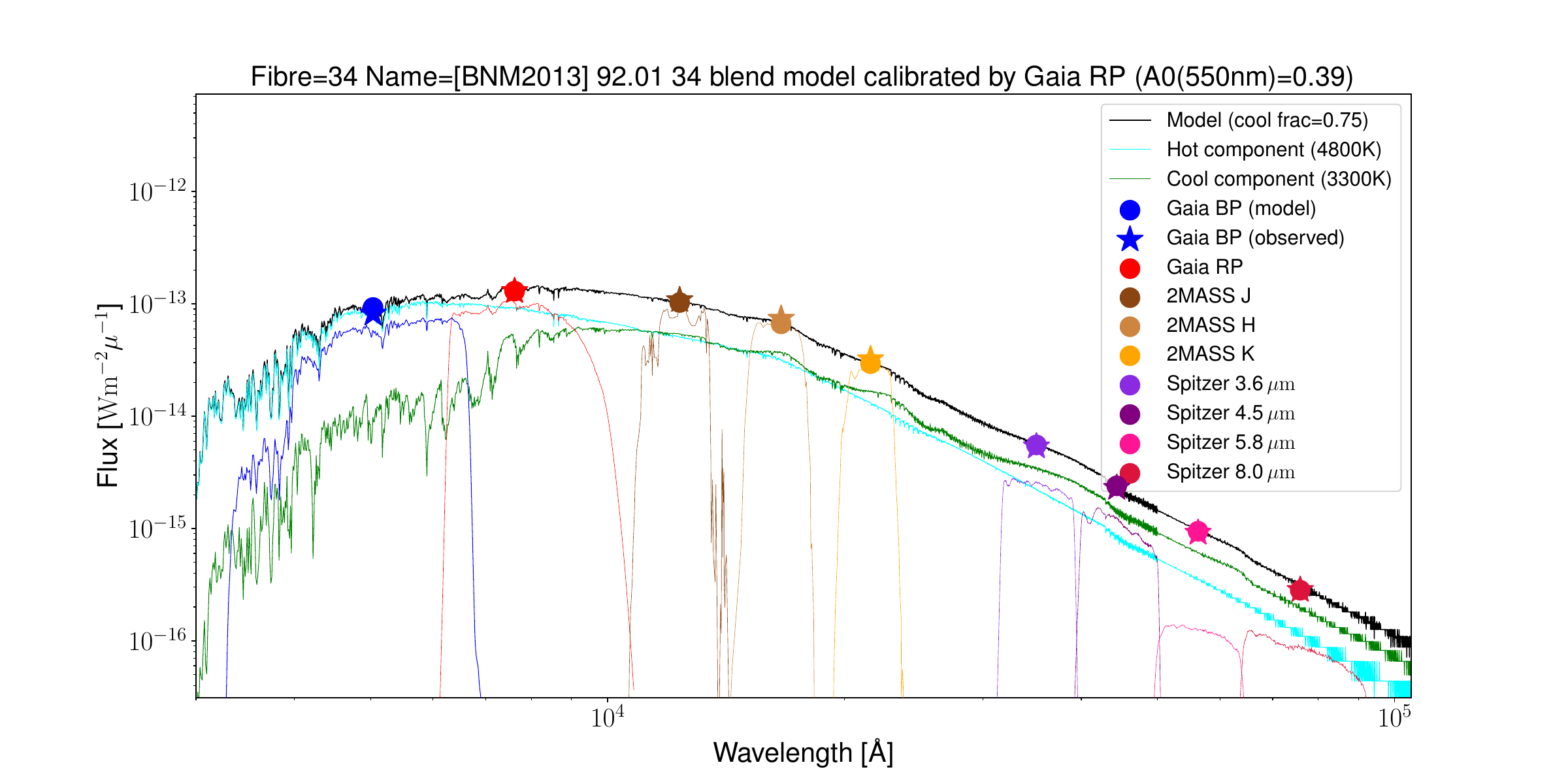}}\\
		\subfloat[]{\includegraphics[width=0.7\textwidth]{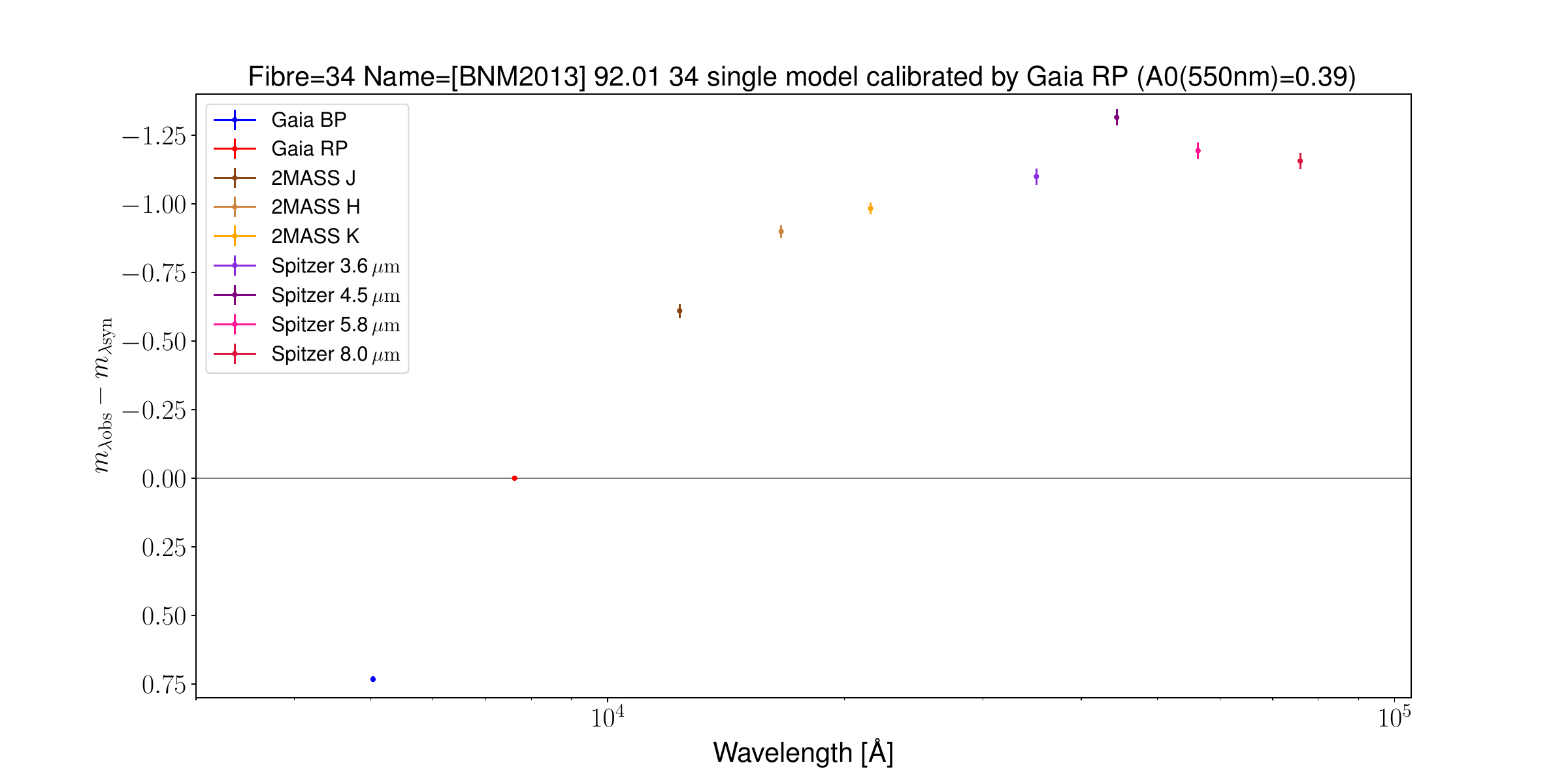}}
		\subfloat[]{\includegraphics[width=0.7\textwidth]{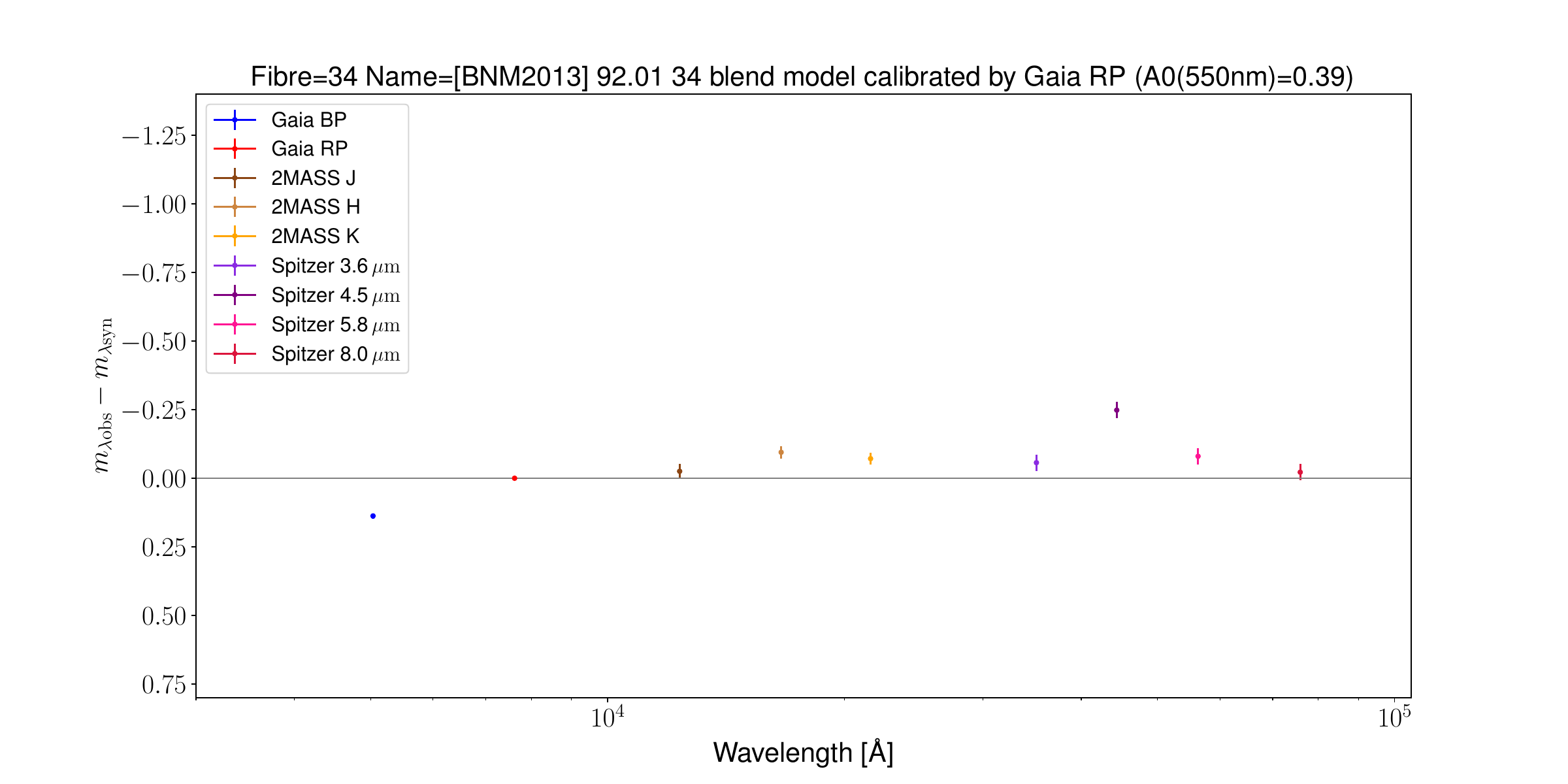}}
		\caption{Plots of the photometry for fibre \refone{34}. The single-temperature model is displayed in the left plots, and the two-temperature model in the right plots. The top plots overlay the photometry on the model spectrum. The symbols for the synthetic photometry are coloured circles, and the observed photometry are stars of the same colour. For clarity, the legend only includes the observed photometry for the $Gaia$ $G_{\rm RP}$ band. The bottom plots give the excess \edittwelve{ratio (in magnitudes)} in the observed \editten{photometry over the models, with the same vertical scale used to allow easy comparison between the single and blended temperature models.}}
		\label{fig:fibre_34_photometry}
	\end{figure}
\end{landscape}

\section{Possible causes of blended spectra}
\label{sec:blended_spectra_causes}

There are many reasons why a blend of two photospheres at different temperatures may produce a better fit to an observed spectrum than a single temperature. Binary and multiple star systems are common, and there can be chance alignments of unconnected stars on the sky. Fast rotating stars can become oblate spheroids, with gravity darkening leading to a cooler effective temperature at the equator than at the poles \citep{1924MNRAS..84..665V}. In young stars, accretion can give rise to a range of temperatures across the surface. It may also be due to cool spots in the stellar photosphere, created by strong magnetic fields.

\subsection{Double and binary stars}
\label{sec:blended_spectra_binary}

\editfive{Double stars can be due to a chance alignment on the sky, or a binary star system where the pair are gravitationally bound. If the blended temperature spectra were due to chance alignments leading to double stars, then the field stars should suffer about the same rate of contamination as the SOC members. As \editten{all 20 of the young and older} field stars show no preference for a blended spectrum \edittwelve{at the same level of significance as the WTTS clean sample}, we can rule out chance alignments as the cause for the majority of the blended SOC spectra.}

It is possible that all stars are born in multiple systems, which quickly break apart due to dynamical interactions \citep{Larson1972,Reipurth2014}. We therefore cannot \edittwelve{completely exclude} the effect of binarity by simply comparing the SOC members with the field stars. \editten{We can however investigate whether the blended temperature models of our WTTS clean and binary samples are compatible with theoretical models of coeval binary stars.}

We simulated binary stars by combining PARSEC V2.0 models (\url{http://stev.oapd.inaf.it/cmd}) \edittwelve{with an age of 3\,Myr as this gave a reasonable match to a \textit{Gaia} CMD of the cluster.} The temperature of the secondary component in the simulated binaries was set to the closest model around \refone{1300\,K} cooler than the primary, \refone{the median temperature difference in the two-temperature models of the WTTS K sample}. The fractional cool surface areas of \editten{the blended temperature fits to the spectra are significantly larger than the cool component areas of the binary star models, except for fibre 18 (WTTS \refone{K} sample) with a cool surface area fraction of 0.3, see Fig.~\ref{fig:Binary_Area}.} This is unsurprising as the lower mass star in a coeval pair will be cooler, smaller and emit less flux. So we should reasonably expect the cooler component of a binary star to contribute less than half of the combined visible area of the two stars. {Thus, \refone{eight of the nine WTTS K sample} are inconsistent with binarity.

\editten{The cool photosphere fractions of the WTTS binary sample are also inconsistent with binarity. This does not mean they are not binary stars, \editelev{simply that binarity cannot explain the 60--95 per cent cool surface area fractions found in our two-temperature models.} If single WTTS have photospheres comprised of two temperatures, then we should expect binary WTTS to have photospheres comprised from four temperatures. The more massive star would contribute the majority of the flux, so it is plausible that fitting two temperatures to a WTTS binary would fit the two temperatures seen in the photosphere of the more massive component.}

\begin{figure}
	\centering
	\includegraphics[width=0.45\textwidth]{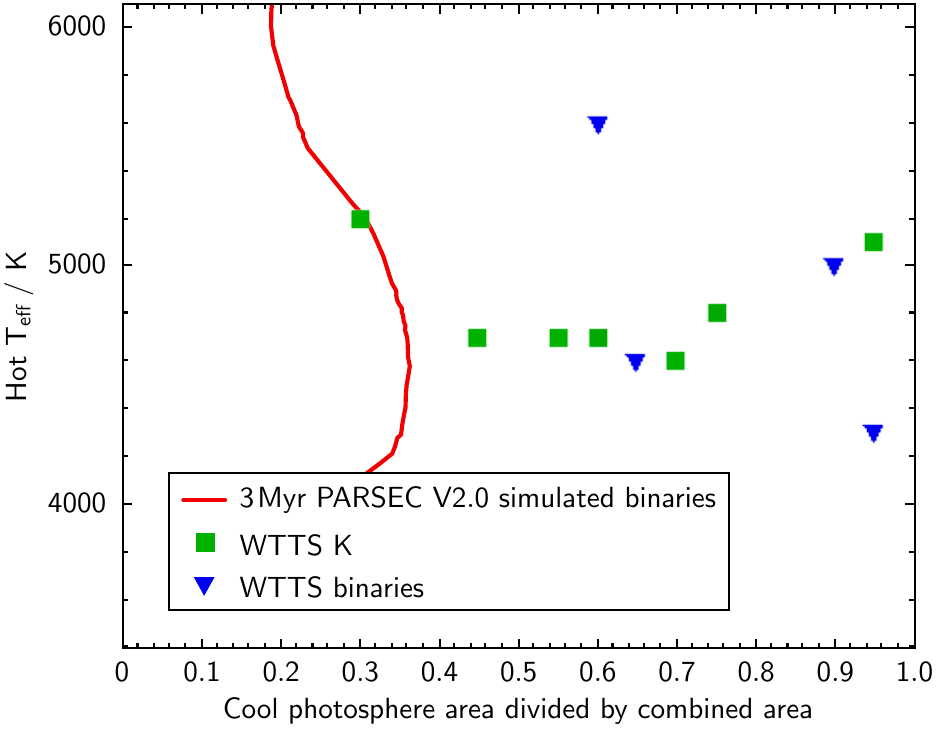}
	\caption{\editseven{A plot comparing our two-temperature fits of the $\sigma$ Orionis cluster members with simulated binaries. \editten{The green squares are the \refone{WTTS K sample} and the blue triangles are the WTTS binaries, both using parameters from the two-temperature fits to their spectra.} The red line is simulated binary stars with a fixed \refone{1300\,K} temperature difference between the components, created from 3\,Myr PARSEC V2.0 models. The lowest effective temperature of the PARSEC models is about 2700\,K at 3\,Myr, limiting the effective temperature of the hot component to 4200\,K.}}
	\label{fig:Binary_Area}
\end{figure}

\subsection{Gravity darkening}
\label{sec:blended_spectra_gravity_darkening}

\editfive{As YSOs tend to rotate quickly, gravity darkening is another possible cause of the blended spectra. Rapidly rotating stars are oblate, with a larger radius at their equator than their poles. This causes the surface gravity to be weaker at the equator than the poles. Also, the apparent centrifugal force gives rise to an outward acceleration that counters gravity. Both of these effects contribute to the fall in gravity between the poles and the equator.}

The pressure, density and temperature are equal along the equipotentials that define the stellar structure including the surface. However, the lower gravity means the equipotential surfaces are further apart at the equator than at the poles. Hence, when looking at the stellar surface, for a given optical depth we see a greater contribution from cooler material at the equator than the poles, an effect analogous to limb darkening. The overall result is a cooler effective temperature at the equator than the poles. This effect was identified by \citet{1924MNRAS..84..665V} and is quantified by von Zeipel's law
\begin{equation}
	T_{\rm eff}^4 \propto g_{\rm eff}^\beta ,
	\label{eq:von_zeipel_law}
\end{equation}
where $g_{\rm eff}$ is the surface gravity at the location on the stellar surface for $T_{\rm eff}$, and $\beta$ is the gravity darkening coefficient.

The original version of the von Zeipel law \citep{1924MNRAS..84..665V} used $\beta=1.00$ for stars with a radiative atmosphere. For stars with a convective atmosphere, \citet{1967ZA.....65...89L} found $\beta=0.32$. The situation is now understood to be a complex interplay of temperature and the observed photometric band, with values of $\beta$ broadly between 1.00 and 0.32 \citep{2021RNAAS...5..256T} for the \editfive{range} of two effective temperatures in \editseven{the blended model fits to our WTTS sample}. 
\editten{To keep the analysis simple, we found an approximate upper limit for gravity darkening by performing calculations with a $\beta$ of 1.00.} From equation~\ref{eq:von_zeipel_law}, we can construct an equation for the effective temperature at the equator
\begin{equation}
	T_{\rm eq} = T_{\rm pol} \left( \frac{g_{\rm eq}}{g_{\rm pol}} \right)^{\beta/4} ,
	\label{eq:von_zeipel_equatorial_temp}
\end{equation}
where $T_{\rm pol}$ is the effective temperature at the pole, $g_{\rm pol}$ is the effective surface gravity at the pole, and $g_{\rm eq}$ is the effective surface gravity at the equator. \editfive{We give our derivation of the ratio of the effective gravities in terms of temperatures and rotation rate in Appendix~\ref{app:gravity_darkening}.}

\editten{A literature search found rotation periods for four of the WTTS \refone{K} sample, giving periods of 5--10\,days \citep{2023MNRAS.523..169S}, see Table~\ref{tab:sigori_review_members}. As we have rotation periods for less than half of the sample, we shall instead use} \edittwelve{the distribution of WTTS rotation periods found in the literature, typically 1 to 5 days, see Section~\ref{sec:introduction}.} Since gravity darkening will have the greatest effect on the fastest rotating YSOs, \editten{we select a short period of 1 day for our analysis.}

\editfive{We used a 3\,Myr PARSEC V2.0 model (see Section~\ref{sec:blended_spectra_binary}) to estimate the mass and radius of a typical SOC member.} A model star with an effective temperature of \refone{4700\,K (median of the hot components in our two-temperature fits to the WTTS K sample)} has an approximate mass of \refone{1.2\,$\rm M_{\odot}$} and a spherical radius of \refone{1.2\,$\rm R_{\odot}$} that we shall take for the polar radius of our oblate star. \editten{From equations~\ref{eq:von_zeipel_equatorial_temp} and~\ref{eq:final_gravity_ratio}, the effective temperature at the equator is \refone{4400\,K} for $\beta=1.00$.}

Thus, oblateness may make a measurable contribution to the blended temperature photospheres. However, for a realistic extremely fast rotation period of 1\,day, the maximum temperature difference due to gravity darkening is around 300\,K. \refone{The temperature differences in the two-temperature models of the WTTS K sample are 4--6 times this value.} Hence, while gravity darkening may contribute to the blended temperature spectra, it cannot be the dominant factor.

\subsection{Accretion}
\label{sec:accretion}

\editsix{Young stars undergoing accretion will have a range of temperatures in their photospheres. Material from the inner edge of the accretion disc will follow magnetic field lines, free falling onto the stellar surface. The optically thin accretion columns cause the photospheric absorption lines to appear shallower than for a normal stellar photosphere, due to a combination of a blue continuum and emission lines  \citep{Hartigan1989,1993AJ....106..372E,Edwards1994,2013ApJ...767..112I}, an effect known as veiling. Where the material impacts the stellar surface, it generates temperatures of around 1\,MK \citep{1998ApJ...509..802C}. The optical depth to the shock region is very high, so the hot spot on the visible photosphere has a temperature around 6000-12000\,K \citep{1972ApJ...175...89W,1998ApJ...509..802C}.}

To avoid the complicating affects of accretion, WTTS members of the SOC were selected for \editten{analysis}. An H$\mathrm \alpha$ EW greater than 10\,\r{A} is the traditional threshold that divides accreting CTTS from non-accreting WTTS \citep{Strom1989}. As our spectra include the H$\mathrm \alpha$ line, this allowed us to check this selection by measuring the EW, which we supplemented with measurements from \citet{2019AA...629A.114C,2021ApJ...923..177S}, see Table~\ref{tab:sigori_review_members}. The three sources observed by fibres 23, 24 and 45 have H$\mathrm \alpha$ EWs greater than 10\,\r{A} in at least two of the three sets of measurements. These three objects also exhibit the largest excess in mid-IR photometry over the blended models, at respectively \refone{2.7, 6.3 and 9.4} times the model flux in the \textit{Spitzer} IRAC 80\,$\mathrm \mu$m bandpass. Their classification as CTTS is confirmed by \citet{2021ApJ...923..177S}. \refone{All of the WTTS} have have H$\mathrm \alpha$ below 10\,\r{A} in all three sets of measurements. \refone{The WTTS K sample have \textit{Spitzer} IRAC 80\,$\mathrm \mu$m flux excesses less than a third above their two-temperature model values.} Hence, they are consistent with a classification of discless WTTS.

\subsection{Starspots}
\label{sec:starspots}

The blended temperature spectra could be due to cool spots on the stellar surface caused by strong magnetic fields. \reffour{Although we have modelled this as two temperatures there could be multiple temperatures in the photosphere. Our two-temperature models would then approximate the range of temperatures in the photosphere, see Section~\ref{sec:introduction}. As} spot coverage decreases with age \citep{2020MNRAS.491.2706N,2020ApJ...893...67M,2025ApJ...988L...1C}, we would expect the SOC members to be more spotted than the \editten{older} field stars.

The cool photosphere fraction by the hot to cool temperature difference from the two-temperature models of the WTTS \refone{K} sample stars are presented in Fig.~\ref{fig:SpotFrac_HotMinCoolTeff}. \refone{Eight out of nine have a tight distribution in their hot to cool temperature difference of 1200--1500\,K. 
These results are consistent with the literature review of spot temperatures by \citet{2005LRSP....2....8B}. They found a 1400\,K temperature difference for a single K-type TTS, 1030--1525\,K for K-dwarfs and 950--1560\,K for K-giants. 
The \reffour{two-temperature model} of the other K-type star in our sample (fibre 37), has a larger hot to cool temperature difference of 1800\,K, and \reffour{the most} extreme cool component surface fraction of 95 per cent. The model has hot component temperature of 5100\,K and a cool component at 3300\,K, which are within the bounds of the other WTTS K sample.}

\begin{figure}
	\centering
	\includegraphics[width=0.45\textwidth]{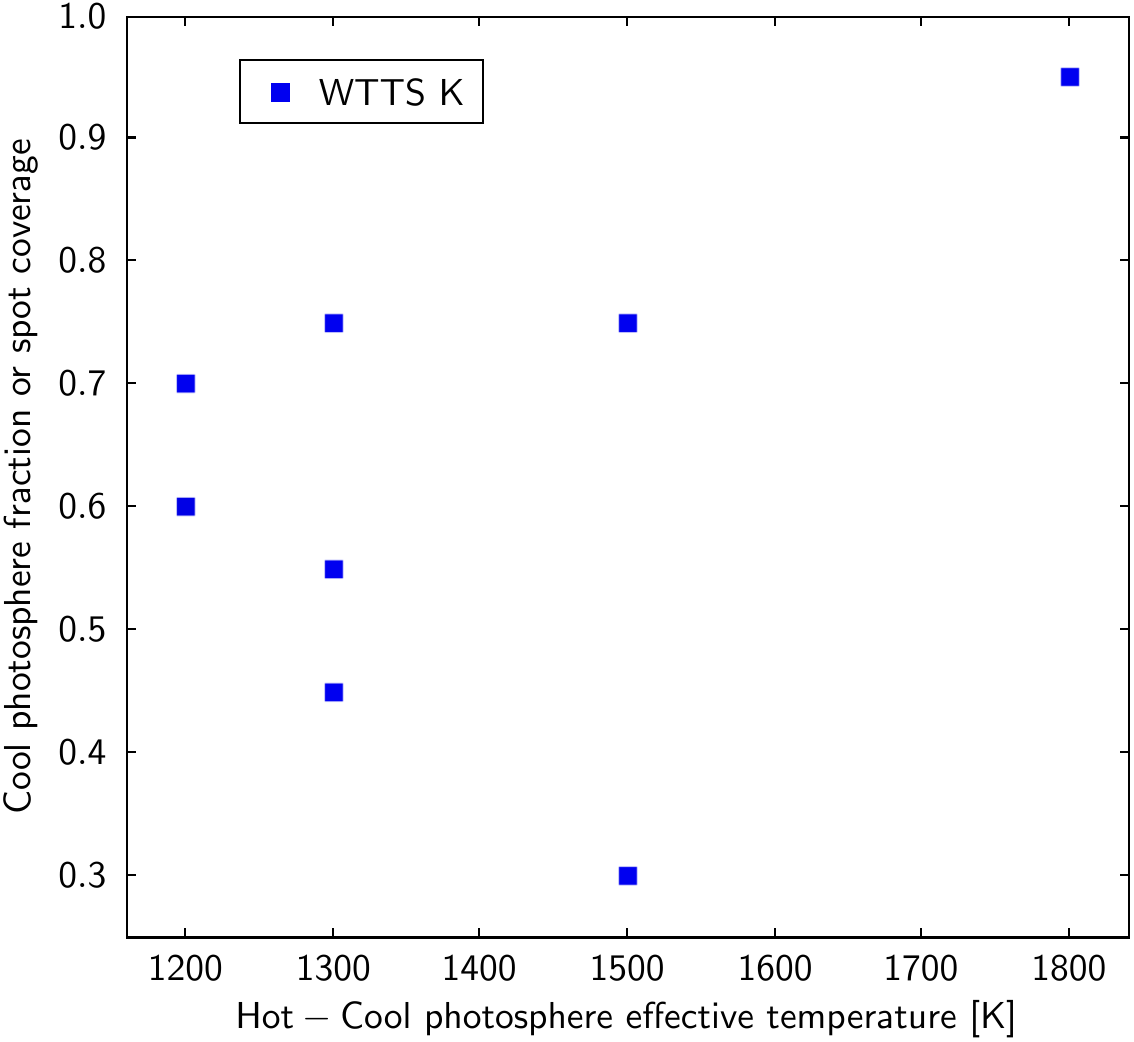}
	\caption{\editten{Results from the blended photosphere fitting for the WTTS \refone{K} sample. The vertical axis is the fractional surface area of the cool component and the horizontal axis the hot-cool component temperature difference.}}
	\label{fig:SpotFrac_HotMinCoolTeff}
\end{figure}

\refone{The WTTS K sample have $\log(g)$ 0.5--1.5\,dex higher in the cool regions than the hot regions. \citet{2025ApJ...990..205P} hypothesised that strong magnetic fields could increase the measured surface gravity at starspots by up to 1--2\,dex, see Section~\ref{sec:introduction}. This provides further evidence that the cool regions could be starspots.}

If the results are interpreted as cool starspots, they indicate surface area coverages around 30--\refone{75} per cent, \refone{and as high as 95 per cent if we include fibre 37}. \reffour{These results are consistent with the cool spot coverages found on young stars by other recent studies, see Section~\ref{sec:introduction}. Due to the limitations of our two-temperature models, the values should be treated as indicative of large cool areas rather than} \reffive{an exact description.}

\reftwo{Terms such as cool areas may be a better topologically description than spots when the cool surface fraction exceeds 50 per cent. These large cool areas could be composed from} many small spots or a few large areas such as polar spots. The amplitude of the photometric variability gives an indication of the \reftwo{cool area} morphology. In their study of stellar rotation in Orion, \citet{2023MNRAS.523..169S} found a flux amplitude of 0.04--0.17 for four of our WTTS clean sample, see Table~\ref{tab:sigori_review_members}. Hence, it is unlikely these stars have purely polar spots. We can estimate an extreme non-polar morphology by assuming one hemisphere of the star is entirely at the cool temperature and the other is at the hot temperature, with these hemispheres rotating in and out of view. Using the hot and cool temperatures from our models gives flux amplitudes of \refone{2.4}--3.6, over ten times the measured amplitudes. 
\reftwo{This \reffour{suggests} the cool areas are relatively evenly distributed in longitude, such that there are relatively small net changes in the visible cool area fraction as the star rotates. This implies multiple cool areas distributed across the stellar surface, or cool areas that are mostly bound in latitude.}

\subsection{Summary of possible causes for blended spectra}
\label{sec:blended_spectra_summary}

In reality, the observed spectra are likely to have contributions from one or more of the causes discussed. Also, these causes are likely to give rise to a blend of more than two temperatures. \editthree{In such cases, our fitting procedure will find the two effective temperatures that best approximate the stellar spectrum.} \editeight{The question we address in this subsection is what is the dominant cause of the blended spectra.}

Our analysis indicates that \refone{the spectra of the WTTS K sample} are a blend of at least two temperatures. In Section~\ref{sec:blended_spectra_binary}, we ruled out chance alignment double stars. We then showed that the fractional contribution of the cool photospheres was inconsistent with binarity for all but one of the \refone{stars} \editten{(fibre 18)}. In Section~\ref{sec:blended_spectra_gravity_darkening} we demonstrated the temperature differences resulting from gravity darkening are too small to explain the results. We showed the WTTS \editten{clean} sample \editten{are} unlikely to be accreting in Section~\ref{sec:accretion}. Finally, we considered cool starspots in Section~\ref{sec:starspots}. \refone{The two temperatures in the blended models for eight of the nine WTTS K sample are consistent with cool starspots. \reftwo{At such high cool surface fractions, cool areas may be a better description than spots.} The exception was the star with the largest cool component of 95 per cent, with a temperature difference between the hot and cool components higher than expected for starspots. Hence, \reftwo{cool areas akin to} starspots are the plausible explanation for eight of the nine stars in the WTTS K sample.}

\section{\editten{Discussion}}
\label{sec:discussion}

Our results for \refone{WTTS K-sample} are consistent with \reftwo{magnetically induced cool areas akin to starspots} covering 30--\refone{75} per cent \refone{and possibly as high as 95 per cent} of the stellar surface, see Section~\ref{sec:blended_spectra_causes}. This agrees with an emerging narrative of strong magnetic fields at the stellar surface causing very high levels of spot coverage during the early stages of stellar evolution. As described by Skumanich's law, this activity reduces with age as stars evolve towards the MS and through the beginning of the MS.

A key finding of our analysis is the \reftwo{cool component} temperature does not vary \refone{significantly with \reftwo{surface} coverages of 30--75 per cent}. 
As the hot effective temperatures are not all a single value, the effect of a warmer cool \reftwo{component} being associated with a \refone{warmer hot component} is removed by comparing the hot to cool temperature difference, see Fig.~\ref{fig:SpotFrac_HotMinCoolTeff}. No trend is evident and there is a tight range of hot to cool temperature difference of \refone{1200-1500\,K}. Hence, we conclude \reftwo{the cool area fraction} does not effect the temperature of the \reftwo{cool areas up to a surface coverage of 75 per cent, implying a common mechanism may be responsible}.

Assuming that magnetism is dominating the stellar surface, then we may be seeing a mix of \editten{\reftwo{cool areas akin to starstops} composed of a cool umbra and warmer penumbra, along with hot faculae and plages. Faculae will not make a significant contribution to the hotter temperatures in our models, as their effective temperatures are only a few 10s\,K higher than the unblemished photosphere \citep{2021MNRAS.504.4751J}. However, the contribution from plages could be important as solar plages are around 200-700\,K hotter than the unblemished photosphere.} This \edittwelve{would be} consistent with the finding of \citet{1986ApJ...306..199V} that hot plages are more significant than cool spots in driving the photometric variability of TTS. \editten{More recently the tomographic imaging results of \citet{2014MNRAS.444.3220D} for the highly spotted WTTS LkCa 4 indicated bright plages and a large polar spot. Regardless of the details, our two-temperature fitting process is only capable of finding the two strongest contributors to the stellar spectrum. As surface magnetism is likely to result in regions at a variety of temperatures, it is unsurprising that our $\chi^2_{\nu}$ values fail to correspond to formally good \edittwelve{statistical} fits between the models and observations.} It may be that at the highest cool area coverages, there is little or no unblemished photosphere as the hot regions may be dominated \edittwelve{by} plages. \refone{This could explain why the hot to cool temperature difference for fibre 37 with the 95 per cent cool surface is a few hundred kelvin higher, see Section~\ref{sec:starspots}. This result would then be consistent with \reftwo{the cool areas found for other members of the WTTS K-sample}, as the two dominant components would be \reftwo{magnetically induced cool areas} and hot plages. The same mechanism may therefore be responsible for cool \reftwo{areas covering} 30--95 per cent \reftwo{of the stellar surface}.}

The strong magnetic field associated with \reftwo{cool areas} will inhibit convection and thus energy transport through the outer layers of stars. Also, the presence of cool \reftwo{areas} in the photosphere reduces the flux emitted by the stellar surface per unit area. The SPOTS models \citep{2020ApJ...891...29S} include these effects, and they indicate that increasing levels of spot coverage result in cooler effective temperatures and inflated radii. The range of \reftwo{cool area fractions} in our results, could be due to a range in magnetic activity, which would explain the spread in radius inflation, see Section~\ref{sec:introduction}. The effect of \reftwo{cool areas} and magnetism is to slow the stellar contraction towards the MS. \edittwelve{We can gain insight into the impact on the stellar age and mass estimates inferred from CMDs by comparing our WTTS \refone{K} sample with the SPOTS models. We take the closest SPOTS model to the median \textit{Gaia} $G$ and $G_{\rm BP}-G_{\rm RP}$ of the WTTS \refone{K} sample, equivalent to a representative K-type star. \editfourteen{The unspotted model has an age of 2.2\,Myr and a mass of 0.75\,$\rm M_{\odot}$. Moving to a SPOTS model with} a 68 per cent cool spot coverage increases the age by a factor of 3.0 and the mass by 1.4.}

As lithium burns at a lower temperature than hydrogen, lithium depletion begins on the PMS and continues until it is burnt or a radiative core develops. Measuring the strength of the Li 6708\,\r{A} line has been widely used to derive the ages of star clusters \citep{1998ApJ...499L.199S,Palla2005,Jeffries2013}. However, there is a dispersion in lithium abundance \citep{1983ApJ...271..663D,1999MNRAS.309..189J}. For the Pleiades K/M dwarfs, \citet{2025MNRAS.539.3364J} found the dispersion in Li EW could not be explained by \editfourteen{the small measured dispersion in the spectroscopic estimates of moderate (20 per cent) spot coverage at the ZAMS (from \citet{2022MNRAS.517.2165C})}. \editfourteen{When winding back the results from the Pleiades to younger ages when the Li was depleted, they found that spotted evolutionary models with the same range of spot coverage as the ZAMS stars underpredict the dispersion in the Li EW by a factor of two.} \editfourteen{Instead,} they postulate that a \edittwelve{larger and rotation dependent} spot coverage at ages \editfifteen{of 10--20\,Myr, during the main phase of PMS Li depletion,} could resolve this discrepancy. Our \reftwo{results indicating cool areas covering} 30--95 per cent \reftwo{of the stellar surface of} WTTS at around \editfourteen{3--4\,Myr} is direct evidence of this at early ages.

Theory predicts that faster rotating stars are more active, see Section~\ref{sec:introduction}. We have rotation periods for four of our WTTS \refone{K} sample with periods of 5--10\, days which show no obvious correlation with \reftwo{cool area} coverage, see Table~\ref{tab:sigori_review_members}. While detailed conclusions cannot be extracted from this small sample, it does indicate that WTTS can be extremely active \edittwelve{in terms of \reftwo{magnetically induced cool area} coverage, even} at moderate rotation periods.

\section{Conclusions}
\label{sec:conclusions}

To assess whether WTTS are best fitted by a model photosphere at a single temperature or a blend of two temperatures, we analysed the spectra of 17 members of the $\sigma$ Orionis cluster and 20 non-member field stars, Section~\ref{sec:fitting_model_spectra}. \editten{Based on the presence of lithium in their spectra, we split the field stars into \refone{17} older field stars and \refone{3} young field stars, Section~\ref{sec:membership_determination}. We divided the $\sigma$ Orionis cluster members into \refone{3 CTTS, 4 WTTS binaries, 1 WTTS M and 9 WTTS K samples (composed from M and K-type stars respectively),} Section~\ref{sec:sample_selection}.} \refone{Our analysis focused on the WTTS K sample as our fitting technique is designed for K-type stellar spectra and the sample contained no known or suspected binary stars.}

In Section~\ref{sec:review} we saw a clear distinction between the WTTS \editten{clean} sample and the \editten{older} field stars. \refone{The field stars exhibited little or no preference for a fit by one or two temperature photospheres, while the SOC members showed a clear preference for a fit by two temperatures. 
The best fit two-temperature models of the WTTS K sample had cool \reftwo{areas} covering 30--95 per cent of their visible stellar surfaces, with hot to cool temperature differences of 1200--1800\,K.}

In Section~\ref{sec:photometry}, we compared the model predictions to independent mid-IR photometry \refone{by flux calibrating the model spectra using \textit{Gaia} DR3 $G_{\rm RP}$. 
The single-temperature models showed an underprediction of the flux in the IR, while the two-temperature models were a significantly closer match with only a small underprediction of around a tenth of a magnitude. 
This confirmed the presence of mid-IR excesses that are modelled by the cool components of the two-temperature models. }
The \refone{small underprediction} could indicate that our models are too simplistic. It may be that three photosphere temperatures are needed to model the WTTS, an unblemished photosphere combined with both cool and hot \refone{components}. Alternatively, there may be weak contributions in the IR from the warm inner regions of protoplanetary discs, not accounted for by our models \citep{2024ApJ...967...45P,2025ApJ...978...32P}.

In Section~\ref{sec:blended_spectra_binary} we investigated double and binary stars as a potential cause of the blended temperature photospheres. As the field stars did not exhibit any preference for blended spectra, we ruled out chance alignment double stars as the cause of the blended spectra. By comparing the \editten{WTTS} to models \refone{of coeval} binaries, we demonstrated that the cool surface area was inconsistent with binarity for \refone{8 of the 9 WTTS K sample and also the WTTS binary sample.}

Gravity darkening was explored as a potential explanation for the blended spectra, see Section~\ref{sec:blended_spectra_gravity_darkening}. The hot to cool temperature differences created by gravity darkening were too small \refone{by a factor of four to six to explain the temperature differences.} We considered whether accretion could be the cause, see Section~\ref{sec:accretion}. However, we demonstrated the WTTS clean sample were not accreting.

In Section~\ref{sec:starspots}, we showed the \refone{two-temperature models for 8 of the 9 K-type stars are consistent with starspots. \reftwo{At such high surface coverages, magnetically induced cool areas may be a more appropriate description.} The other star had a hot to cool temperature difference a few hundred kelvin higher than is found for starspots on K-type stars. This star has an extreme cool surface fraction of 95 per cent. It could be consistent with \reftwo{starspot like cool areas} if the two dominant components are cool starspots and hot plages, see Section~\ref{sec:discussion}. Thus \reftwo{magnetically induced cool areas} are a plausible explanation for the two-temperature photospheres found for the WTTS K sample.} \edittwelve{The rotational photometric variability of 4--17 per cent found by \citet{2023MNRAS.523..169S} for four of the stars, combined with our measured \reftwo{cool area} coverages suggests the cool \reftwo{areas} are relatively evenly distributed \reftwo{in longitude across} the surface of the stars.}

\refone{The inferred cool components cover 30--95 per cent of the visible stellar surface}. While these are high \refone{fractions for starspots}}, they are not unprecedented as \citet{Gully-Santiago2017,2022MNRAS.517.2165C,2024ApJ...967...45P,2025ApJ...988L...1C} find similar spot coverage in TTS. 
\reffour{If starspots are the cause of the blended temperature spectra, then our two-temperature models may be approximations of multi-temperature photospheres. Hence, the temperatures and cool component fractions should not be treated as} \reffive{an exact description} {but indicative of large cool areas.}

\edittwelve{In Section~\ref{sec:discussion}, we used the SPOTS models \citep{2020ApJ...891...29S} to show that a typical K-type star in our WTTS clean sample would have its age underestimated by a factor of 3.0 and its mass by 1.4 when fitting an unspotted model compared to a model with 68 per cent spot coverage.}

\editten{The dispersion in \reftwo{cool area} coverage of 30--95 per cent at an age of \editfourteen{around 3--4\,Myr}, could explain the dispersion in Li EW found \edittwelve{amoung coeval low-mass ZAMS stars} by \citet{2025MNRAS.539.3364J}.} 

\refone{The temperature difference between the hot and cool \reffour{model} \reftwo{components} exhibit a tight range without a trend for the WTTS K sample with 30--75 per cent cool surface fractions. This implies the same physical mechanism could be maintaining \reftwo{the cool areas}. If the larger temperature difference for the \reffour{model} with a 95 per cent cool surface is due to plages, then this implies the same mechanism is responsible across the 30--95 per cent cool surface areas.}

\section*{Acknowledgements}
\label{sec:acknowledgements}

\edittwo{This work was partially supported by the UK Science and Technology Facilities Council (STFC) grant number ST/Y001826/1; and DJS was funded by an STFC studentship. This research was based on observations collected at the European Southern Observatory under ESO programme 092.C-0324(A).}

\refone{We thank the referee for their feedback on surface gravities for members of the sigma Orionis cluster. This significantly improved the analysis and gave a more consistent set of results.}

This work has made use of data from the European Space Agency (ESA) mission {\it Gaia} (\url{https://www.cosmos.esa.int/gaia}), processed by the {\it Gaia} Data Processing and Analysis Consortium (DPAC, \url{https://www.cosmos.esa.int/web/gaia/dpac/consortium}). Funding for the DPAC has been provided by national institutions, in particular the institutions participating in the {\it Gaia} Multilateral Agreement.

\editfifteen{This work is based in part on observations made with the Spitzer Space Telescope, which was operated by the Jet Propulsion Laboratory, California Institute of Technology under a contract with NASA.}

\editfifteen{This publication makes use of data products from the Two Micron All Sky Survey, which is a joint project of the University of Massachusetts and the Infrared Processing and Analysis Center/California Institute of Technology, funded by the National Aeronautics and Space Administration and the National Science Foundation.}

\edittwo{This research made use of \textsc{astropy}\footnote{\url{http://www.astropy.org}}, a community-developed core \textsc{python} package for Astronomy \citep{2013A&A...558A..33A, 2018AJ....156..123A}, \textsc{numpy} \citep{harris2020array}, TOPCAT \citep{Taylor2005} for analysis, and the SIMBAD database \citep{Wenger2000}, operated at CDS, Strasbourg, France.}

\editthree{This research has made use of the SVO Filter Profile Service "Carlos Rodrigo", funded by MCIN/AEI/10.13039/501100011033/ through grant PID2023-146210NB-I00.}

\section*{Data Availability}
\label{sec:data_availability}

\edittwo{The raw data are available for download from the ESO archive\footnote{\url{https://www.eso.org/sso/login}}. They are stored on the La Silla Paranal data (raw) archive service and can be located by searching on program id 092.C-0324(A).} The reduced spectra are openly available for download from the University of Exeter's Open Research Exeter (ORE) at \url{https://doi.org/10.24378/exe.34056594} and Zenodo at \url{https://doi.org/10.5281/zenodo.23104192}.

\edittwo{The BT-Settl CIFIST model atmosphere spectra were downloaded from \url{https://phoenix.ens-lyon.fr/Grids/BT-Settl/CIFIST2011_2015/FITS/BT-Settl_M-0.0a+0.0.tar} when the date stamp was 16 March 2015. As this website is no longer available, the primary author can provide the files by private communication if requested.}



\bibliographystyle{mnras}
\bibliography{library_mnras} 




\appendix

\section{Calculation of the equivalent width of lithium 6708\,\r{A}}
\label{app:li_ew}

\editten{Several of the spectra contain other spectral lines close to the Li 6708\,\r{A} absorption line. To prevent them from contaminating the Li 6708\,\r{A} EW we removed them using the \refone{best fit single-temperature model spectra for the field stars and the best fit two-temperature model spectra for the SOC members}. The model and observed spectra were first normalised and detrended based on their flux within the 6703--6713\,\r{A} wavelength range. Then the model spectrum was subtracted from the observed spectrum, and 1.0 was added to the residuals, returning a normalised spectrum with the majority of spectral lines removed. The EW was then calculated on the flux within 6706--6710\,\r{A}, with the continuum calculated as the upper quartile of the flux in regions 6703--6706\,\r{A} and 6710--6713\,\r{A}. To avoid the residuals from dominating the result where the EW was <0.1\,\r{A}, the EW was recalculated on the narrower 6707--6709\,\r{A} range, with the continuum based on 6704--6707\,\r{A} and 6709--6712\,\r{A}.}

\section{Single and two temperature fitting procedure}
\label{app:fitting_procedure}

\refone{Our model fitting was performed in three stages. The surface gravity was unconstrained for the field stars. The surface gravity of the SOC members was fixed to $\log(g)=4.0$ for the single-temperature models and the hot component of the two-temperature models, see Section~\ref{sec:fitting_model_spectra}. The best fits were found by calculating the $\chi^2_{\nu}$ for all models in temperature and surface gravity, then selecting the model with the smallest $\chi^2_{\nu}$.}

\begin{description}
	
	\item[\textbf{Stage 1 spectrum parameters.}] An initial estimate of the RV was made using \refone{a model spectrum at an effective temperature of 4000\,K with a $\log(g)=4.0$ for the SOC members and $\log(g)=4.5$ for the field stars}. This was sufficiently close to the spectral type of the observed spectra to obtain the initial estimates. \editthree{For each possible RV, the full resolution BT-Settl model spectrum was first shifted by that velocity and then rebinned onto the pixels of the observed spectrum, allowing us to evaluate the value of the cross-correlation function at that velocity.} The RV was estimated from the centre of the Gaussian fit to the cross-correlation results.
	
	\item[\textbf{Stage 1 best fit model.}] To enable fast processing \editten{in the first two stages where an approximate model is sufficient}, we created a smoothed and rebinned version of each BT-Settl spectrum that roughly matched the binning and resolution of the VLT spectra. These low resolution model spectra were wavelength shifted to match the RV of the star in question and interpolated onto the wavelength steps of the observed spectrum using quadratic interpolation. Both the observed and model spectra were normalised to unity, using the median of the flux. To remove small bulk differences due to \editten{an imperfect} response correction and uncertainty in the extinction to the source, the spectrum was detrended by fitting a straight line to the ratio of the model to observed spectrum, and dividing the model spectrum by the straight line fit. The $\chi^2_{\nu}$ was calculated between each single photosphere model and the observed spectrum. The best fit was identified as the model with the smallest $\chi^2_{\nu}$.
	
	\item[\textbf{Stage 2 spectrum parameters.}] The RV was refined by cross-correlating the observed spectrum with the best fit single photosphere model found in stage 1. In all other respects this process was the same as stage 1. The width of the Gaussian fitted to the cross-correlation function was used in stage 3, as the starting point for finding the width of the observed spectral lines.
	
	\item[\textbf{Stage 2 best fit model.}] The stage 2 best fit to a single photosphere was identified in an identical way to stage 1, other than the refined RV was used to wavelength shift the model spectra. This gave an improved estimate of the best fit single photosphere.
	
	\item[\textbf{Stage 3 spectrum parameters.}] The third and final stage began by deriving the line broadening needed to match the line width of the model spectrum to the observed spectrum. The width of the Gaussian fit to the cross-correlation function found in stage 2 only provided a rough estimate of the spectral line width. A robust measure of the broadening was derived by cross-correlating the full resolution model spectrum with a broadened copy of itself. The model was broadened using a Gaussian kernel across a variety of widths. The kernel widths were varied from 0.7 to 1.5 times the width estimate found in stage 2, in steps of 0.05. The Gaussian kernel also enabled the model to be wavelength shifted onto the observed spectrum using the RV from stage 2, and rebinned onto the observed spectrum wavelength steps. This smoothed model was cross-correlated with its full resolution master copy, and the cross-correlation function fitted with a Gaussian as in the earlier stages. The kernel width broadening that gave the closest match to the Gaussian fit to the observed spectrum from stage 2, was taken as the kernel width needed to broaden the model spectrum to match the observed spectrum.
	
	\item[\textbf{Stage 3 best fit models.}] All of the full resolution BT-Settl model spectra were broadened, shifted in wavelength, and rebinned onto the model spectrum wavelength steps, by convolving them with a Gaussian kernel of width found in stage 3 and centre from the RV found in stage 2. This gave a set of models whose line widths and wavelengths matched the observed spectra.
	
	\item[\textbf{Stage 3 best fit single photosphere.}] To find the best fit single photosphere model, the $\chi^2_{\nu}$ was calculated between the observed and model spectra (broadened, RV shifted, rebinned, \editten{and detrended}) as in stages 1 and 2. The best fit single effective temperature and surface gravity model, was the model that gave smallest $\chi^2_{\nu}$. \refone{This searched for 322 models for the field stars and 46 at the single surface gravity of $\log(g)=4.0$ for the SOC members.}
	
	\item[\textbf{Stage 3 best fit blended photosphere.}] To find the best fit blended spectrum, pairs of model spectra were combined. As the BT-Settl spectra were in units of $\rm W m^{-2} \micron^{-1}$, this was performed in fractions of unit surface area. The cool photosphere \editfive{fractional area} was varied from 0.00 to 1.00 in steps of 0.05, mirrored by the hot component from 1.00 to 0.00. To avoid repetitions and unlikely combinations, blends were only derived where the second photosphere was cooler than the first photosphere. The hot photosphere was limited to \editten{3000-7000\,K} and the cool to 2500--5000\,K. \refone{The best fit with the smallest $\chi^2_{\nu}$ was found from 880\,824 models for the field stars and 125\,832 models for the SOC members. Fewer models were searched for the SOC members as the hot component surface gravity was fixed at $\log(g)=4.0$.}
	
\end{description}

\section{Stellar oblateness of fast rotating stars}
\label{app:gravity_darkening}

\editfive{To understand whether gravity darkening contributes to the blended temperature photospheres, an estimate of the ratio between the equatorial to polar effective gravities is required by equation~\ref{eq:von_zeipel_equatorial_temp}. For this estimate we make the simplifying assumption of solid body rotation.} The stellar surface will follow equipotentials
\begin{equation}
	\mathit{\Phi}(\theta) = -\frac{\mathrm{G} M}{r} - \frac{1}{2} \mathit{\Omega}^2 r^2 \rm{sin^2} \theta ,
	\label{eq:rotating_star_equipotential}
\end{equation}
adapted from \citet[Chapter 25]{Lamers2017}, where $\mathrm G$ is the gravitational constant, $r$ is the stellar radius, $M$ is the stellar mass, $\mathit{\Omega}$ is the angular velocity, and $\mathit{\theta}$ is the angle from the pole to the equator of the star. The first part of this equation is the gravitational potential and the second is the potential due to the \editfive{centrifugal acceleration}. The \editfive{effective gravity} is obtained by differentiating equation~\ref{eq:rotating_star_equipotential} with respect to $r$
\begin{equation}
	-g = \frac{\mathrm{d} \mathit{\Phi}(\theta)}{\mathrm{d} r} = \frac{\mathrm{G} M}{r^2} - \mathit{\Omega}^2 r \rm{sin^2} \theta .
	\label{eq:rotating_star_force}
\end{equation}
As the \editfive{centrifugal} acceleration is zero at the pole and $\rm{sin^2} \theta$ is unity at the equator, the ratio of the effective gravities between the equator ($g_{\mathrm{eq}}$) and pole ($g_{\mathrm{pol}}$) is
\begin{equation}
	\frac{g_{\mathrm{eq}}}{g_{\mathrm{pol}}} =  \frac { \mathrm{G} M / r_{\mathrm{eq}}^2 - \mathit{\Omega}^2 r_{\mathrm{eq} } } {\mathrm{G} M / r_{\mathrm{pol}}^2} ,
	\label{eq:initial_gravity_ratio}
\end{equation}
where $r_{\rm eq}$ and $r_{\rm pol}$ are the equatorial and polar radii of the star respectively.

At the breakup velocity, the acceleration due to gravity will precisely match the \editfive{centrifugal} acceleration at the equator
\begin{equation}
	\frac{\mathrm{G} M}{r_{\rm eq}^2} = \mathit{\Omega}_{\rm B}^2 r_{\rm eq} .
	\label{eq:rotating_star_breakup}
\end{equation}
We can use this result to substitute for $\mathrm{G} M$ in equation~\ref{eq:initial_gravity_ratio}
\begin{equation}
	\begin{split}
	\frac{g_{\mathrm{eq}}}{g_{\mathrm{pol}}} &= \frac { \mathit{\Omega}_{\rm B}^2 r_{\rm eq} - \mathit{\Omega}^2 r_{\mathrm{eq} } } { \mathit{\Omega}_{\rm B}^2 r_{\rm eq}^3 / r_{\mathrm{pol}}^2} \\
	&= \left( \frac{r_{\rm pol}}{r_{\mathrm{eq}}} \right)^2 \left( 1 - \left( \frac{\mathit{\Omega} }{\mathit{\Omega}_{\rm B}} \right)^2 \right) .
	\end{split}
	\label{eq:second_gravity_ratio}
\end{equation}
	
The potential at the pole is the same regardless of the rotation velocity
\begin{equation}
	\mathit{\Phi}_{\rm pol} = -\frac{\mathrm{G} M}{r_{\rm pol}} .
	\label{eq:rotating_star_equipotential_pole}
\end{equation}
As the surface follows an equipotential, even at breakup velocity the potential at the pole and equator will be the same
\begin{equation}
		-\frac{\mathrm{G} M}{r_{\rm pol}} = -\frac{\mathrm{G} M}{r_{\rm eq}} - \frac{1}{2} \mathit{\Omega}^2 r_{\rm eq}^2 .
	\label{eq:rotating_star_equipotential_breakup}
\end{equation}
Using equation~\ref{eq:rotating_star_breakup} to substitute for $\mathrm{G}M$ in equation~\ref{eq:rotating_star_equipotential_breakup} gives the standard result
\begin{equation}
	\frac{r_{\rm eq}}{r_{\rm pol}} = 1 + \frac{1}{2} \left( \frac{\mathit{\Omega}}{\mathit{\Omega}_{\rm B}} \right)^2 .
	\label{eq:rotating_star_equatorial_to_polar_radius}
\end{equation}
Finally we can substitute equation~\ref{eq:rotating_star_equatorial_to_polar_radius} into equation~\ref{eq:second_gravity_ratio}
\begin{equation}
	\frac{g_{\mathrm{eq}}}{g_{\mathrm{pol}}} = \frac { 1 - \left( \frac{\mathit{\Omega} }{\mathit{\Omega}_{\rm B}} \right)^2 } { \left( 1 + \frac{1}{2} \left( \frac{\mathit{\Omega}}{\mathit{\Omega}_{\rm B}} \right)^2 \right)^2 } .
	\label{eq:final_gravity_ratio}
\end{equation}

\section{$\sigma$ Orionis cluster spectra}
\label{app:sig_ori_spectra}

\begin{figure*}
	\centering
	\includegraphics[width=0.9\textwidth]{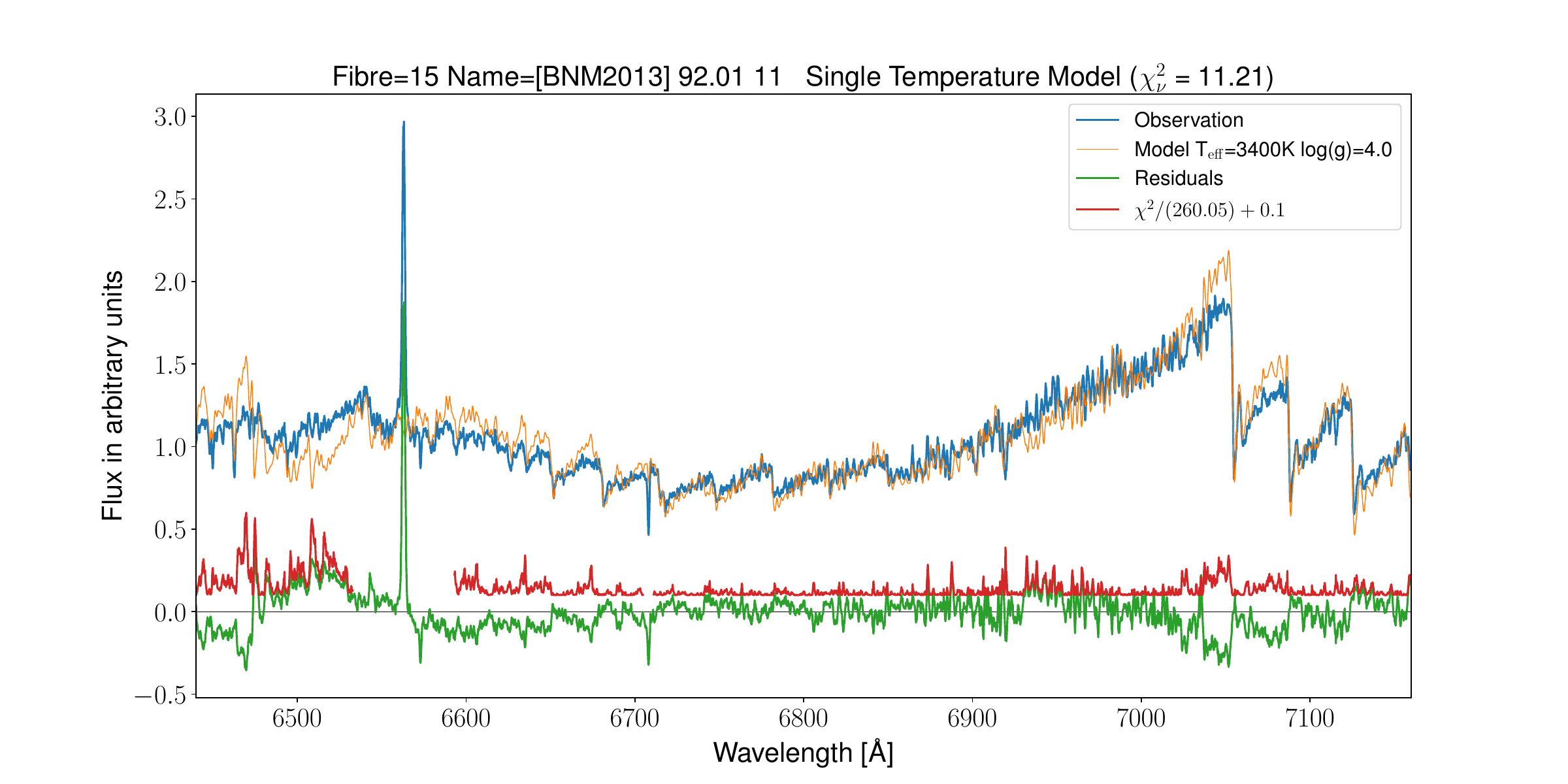}
	\includegraphics[width=0.9\textwidth]{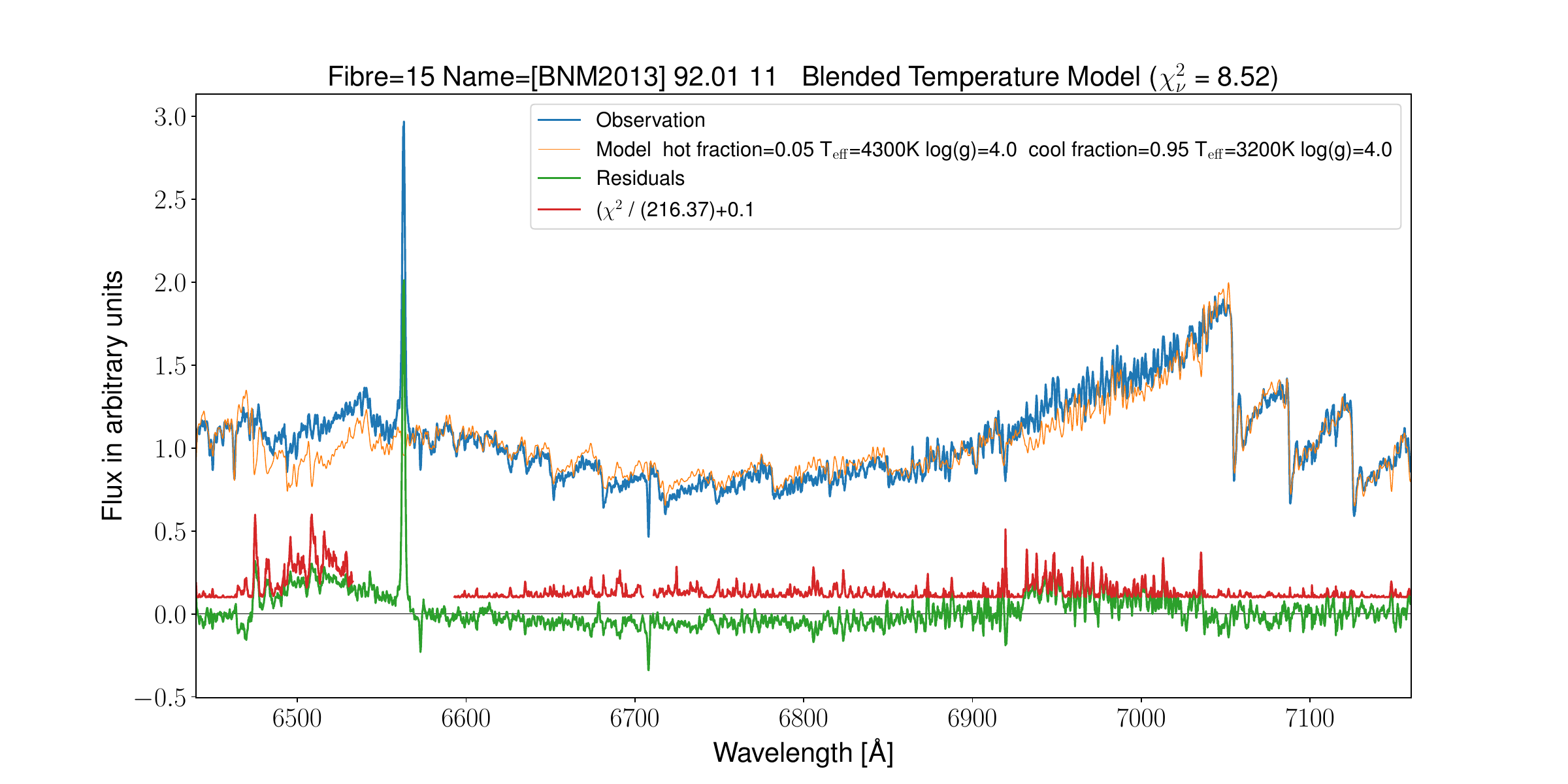}
	\includegraphics[width=0.9\textwidth]{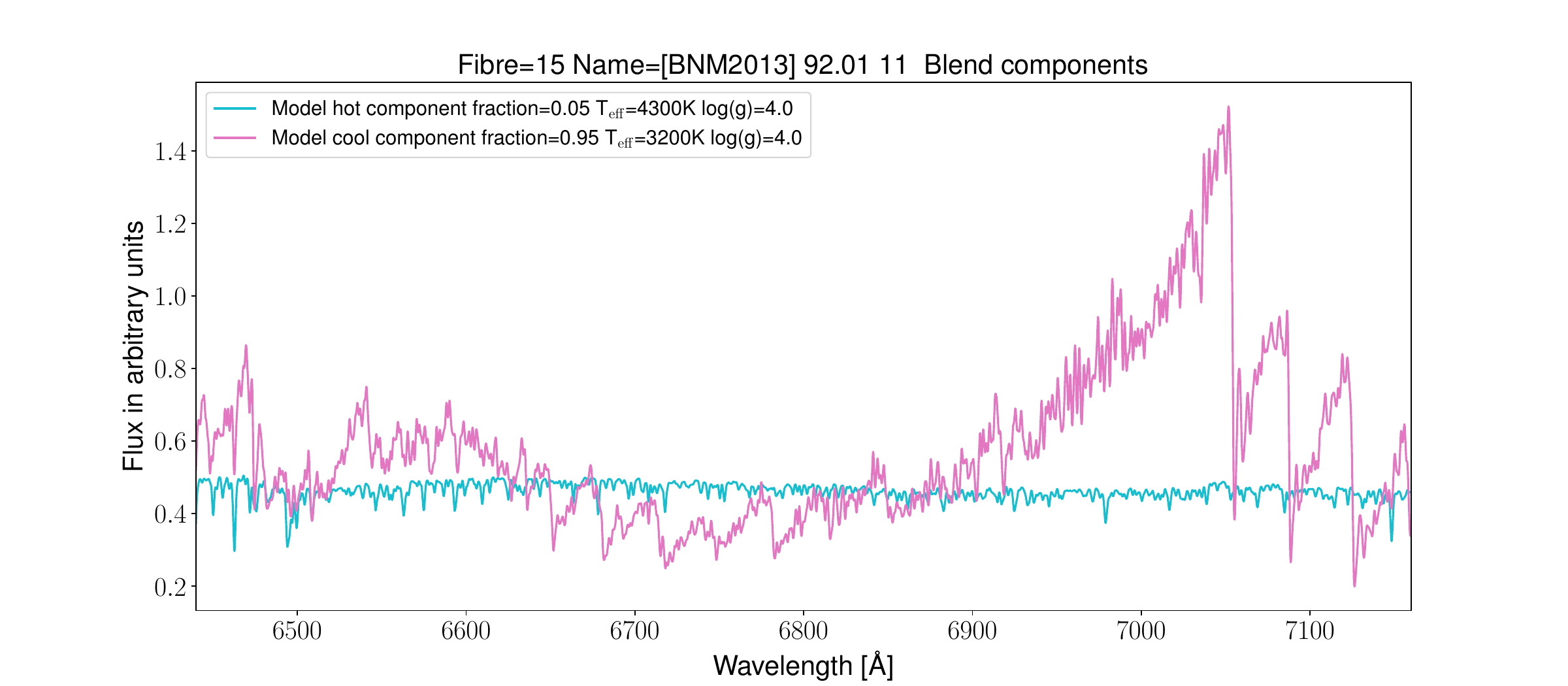}
	\caption{As Fig.~\ref{fig:fibre_27_spectra} for fibre 15.}
	\label{fig:fibre_15_spectra}
\end{figure*}

\begin{figure*}
	\centering
	\includegraphics[width=0.9\textwidth]{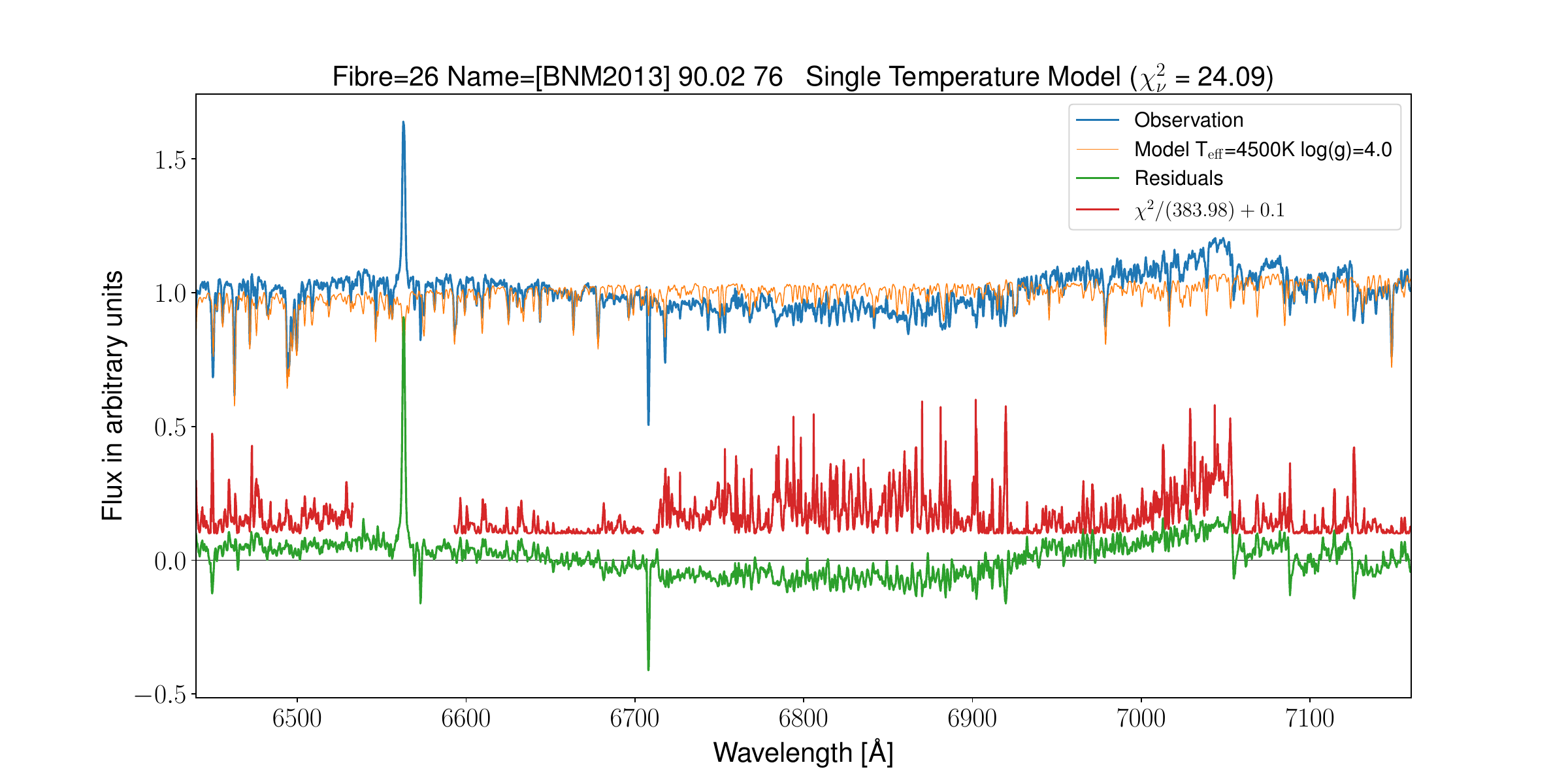}
	\includegraphics[width=0.9\textwidth]{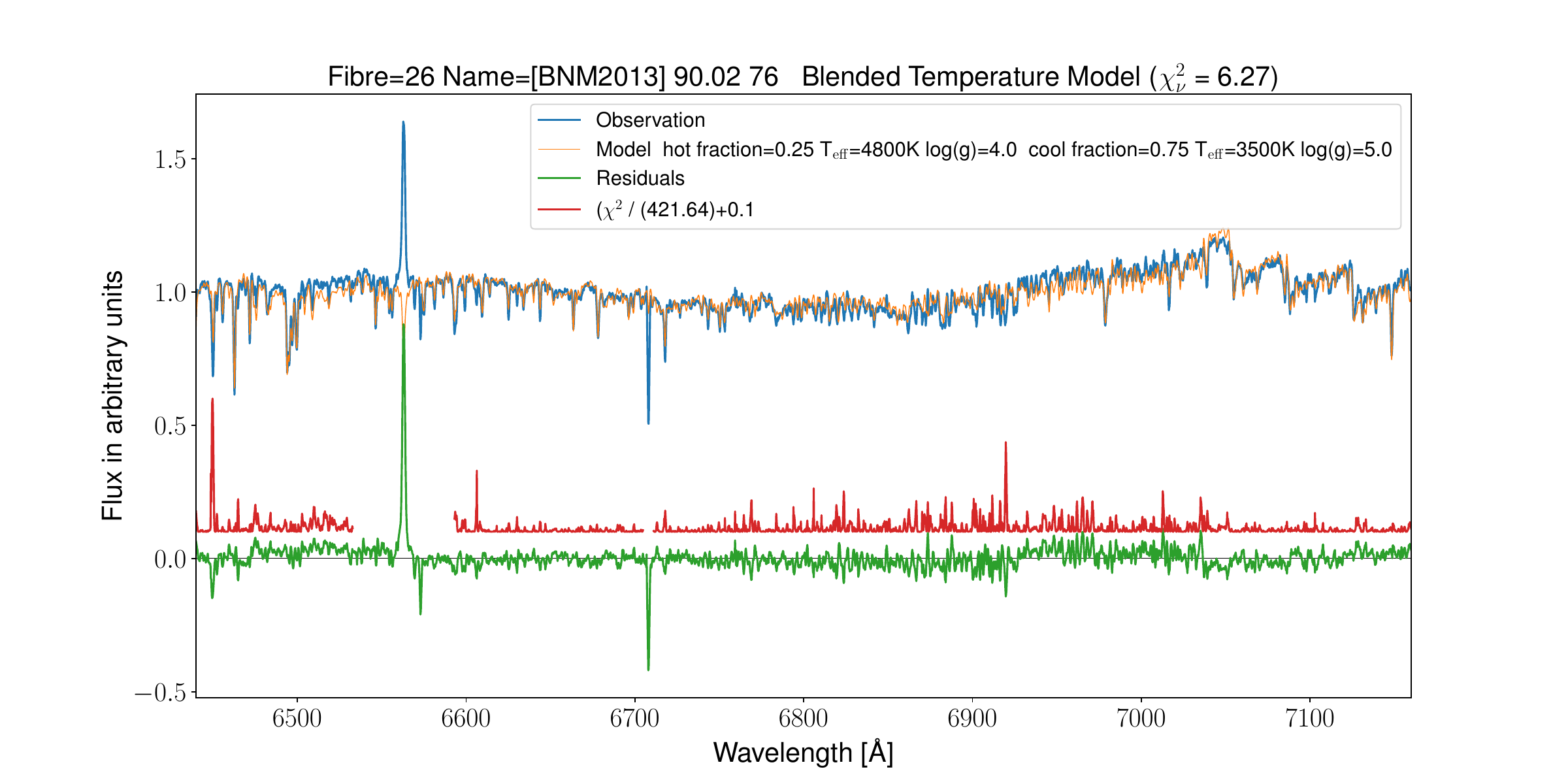}
	\includegraphics[width=0.9\textwidth]{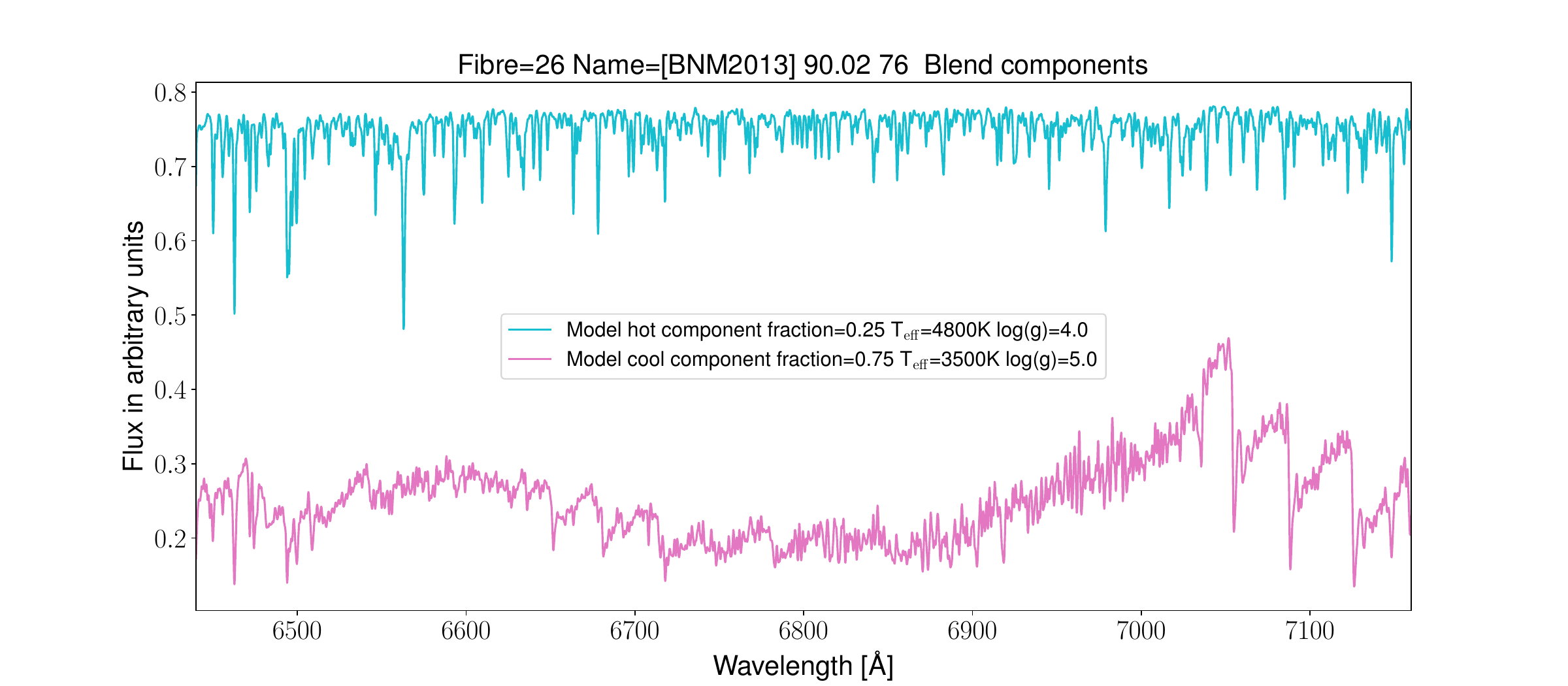}
	\caption{As Fig.~\ref{fig:fibre_27_spectra} for fibre 26.}
	\label{fig:fibre_26_spectra}
\end{figure*}

\begin{figure*}
	\centering
	\includegraphics[width=0.9\textwidth]{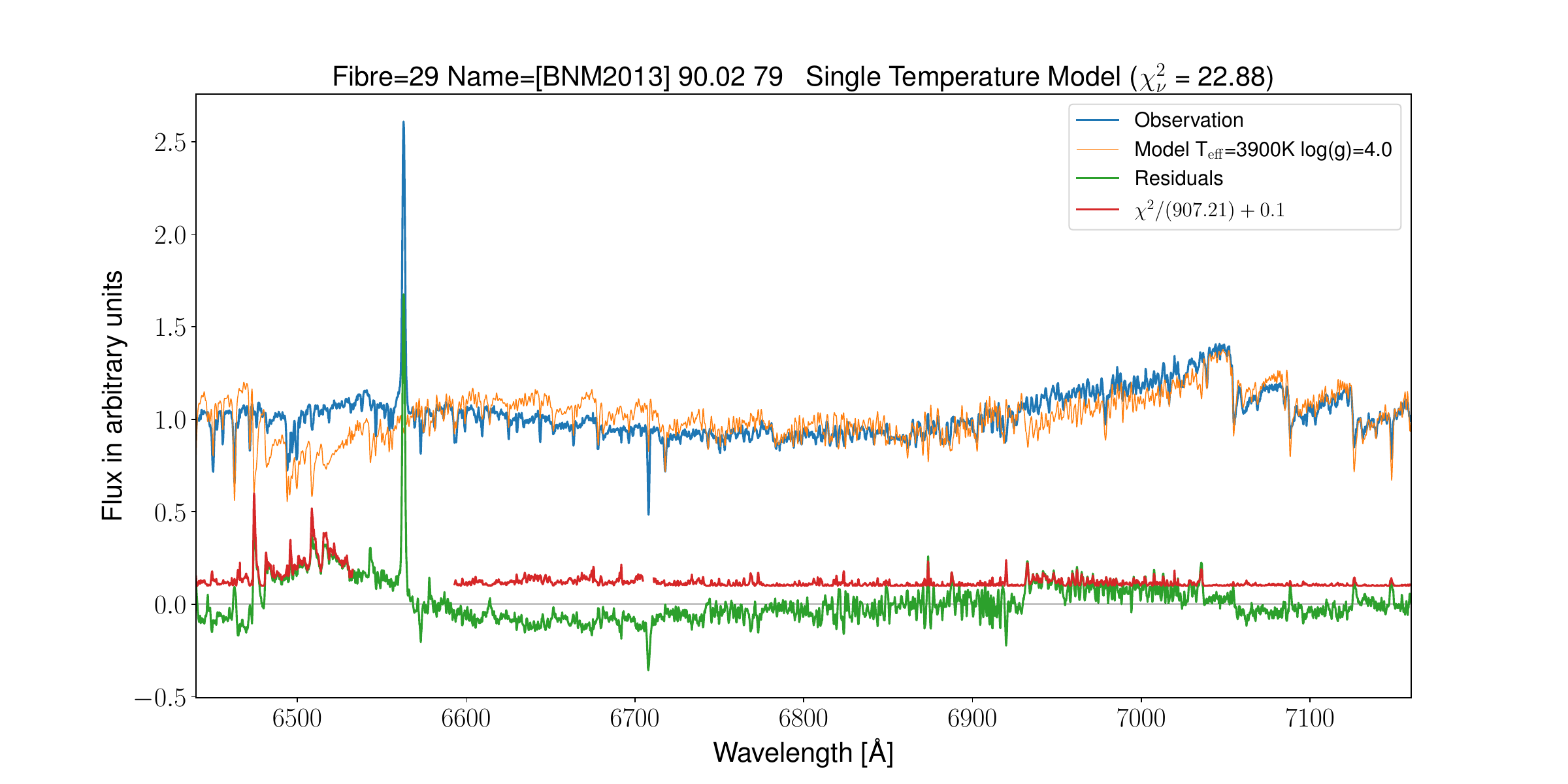}
	\includegraphics[width=0.9\textwidth]{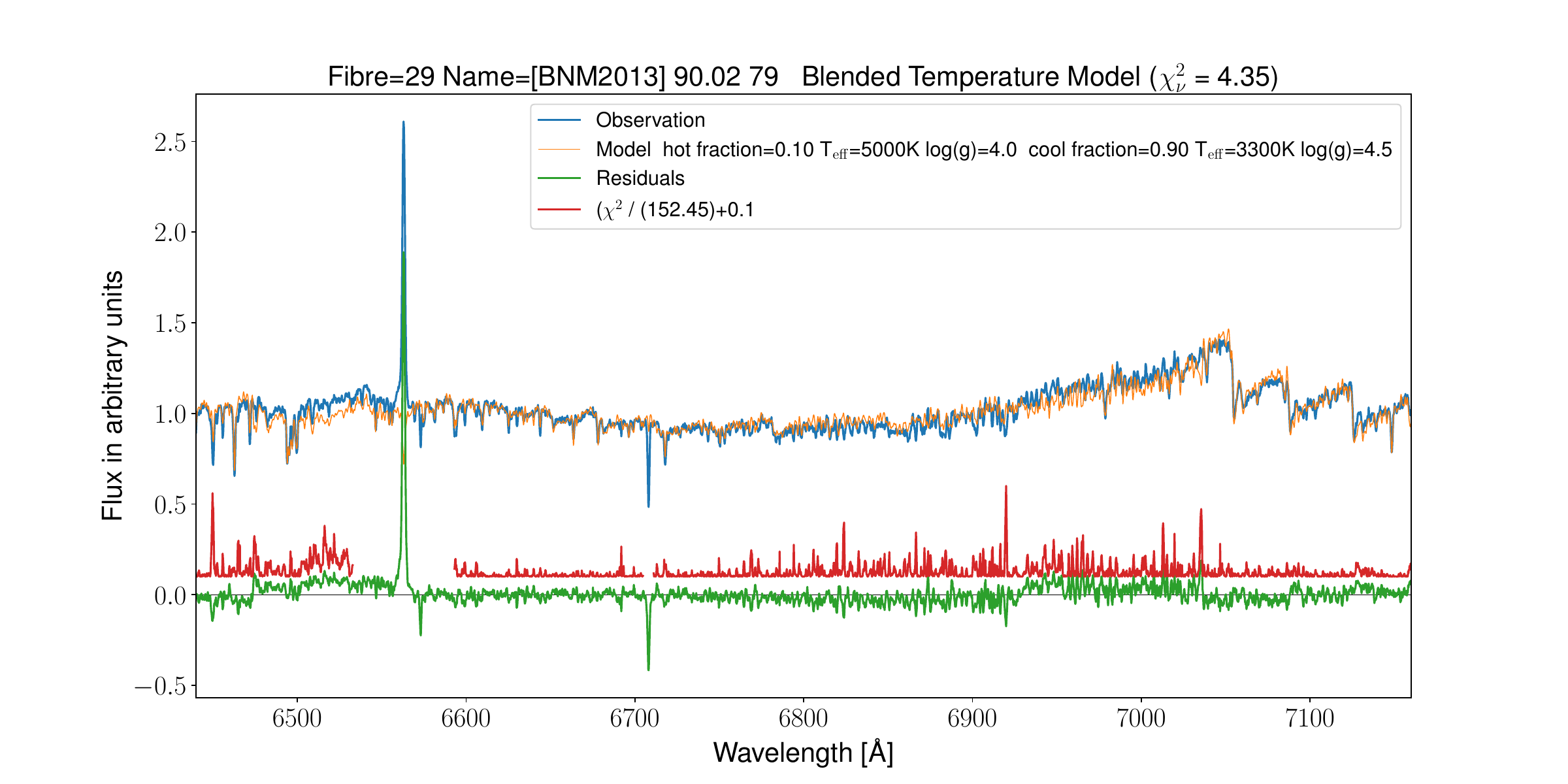}
	\includegraphics[width=0.9\textwidth]{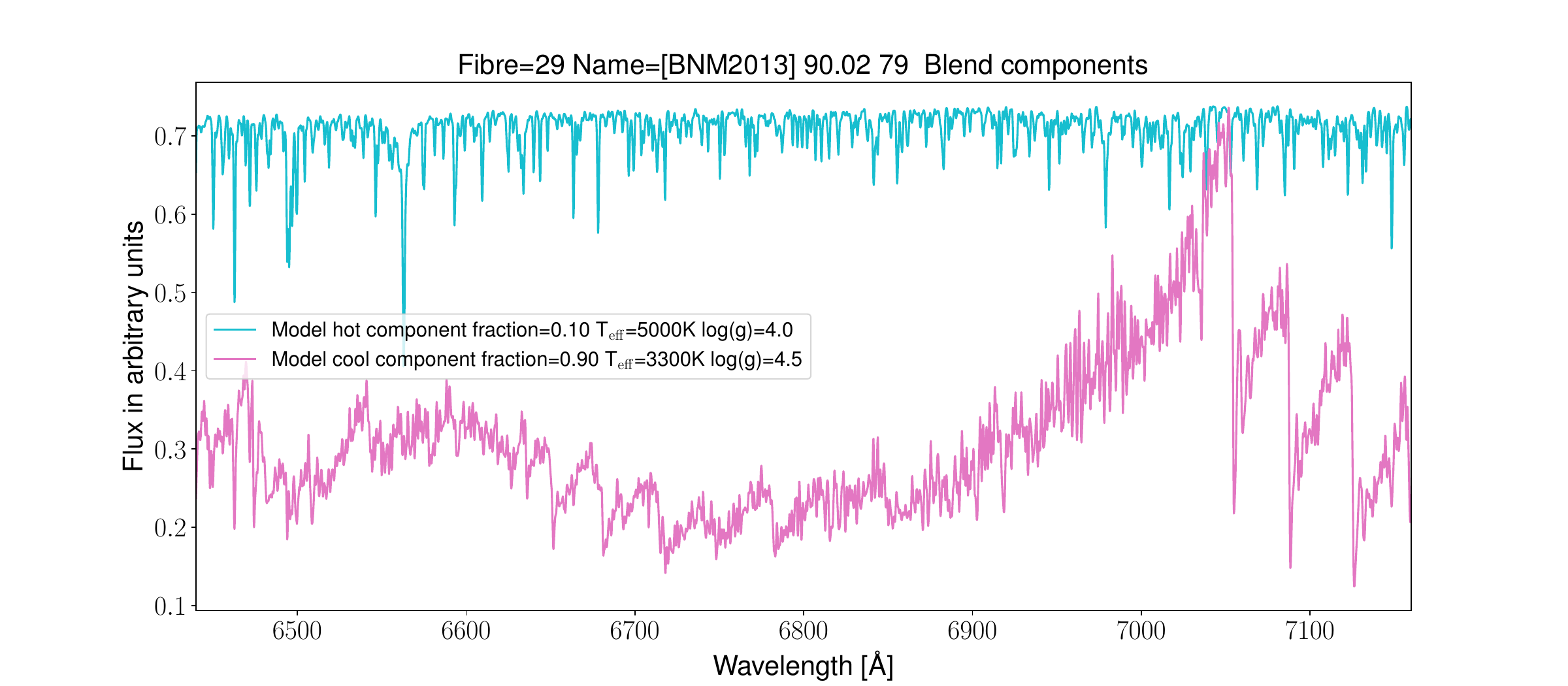}
	\caption{As Fig.~\ref{fig:fibre_27_spectra} for fibre 29.}
	\label{fig:fibre_29_spectra}
\end{figure*}

\begin{figure*}
	\centering
	\includegraphics[width=0.9\textwidth]{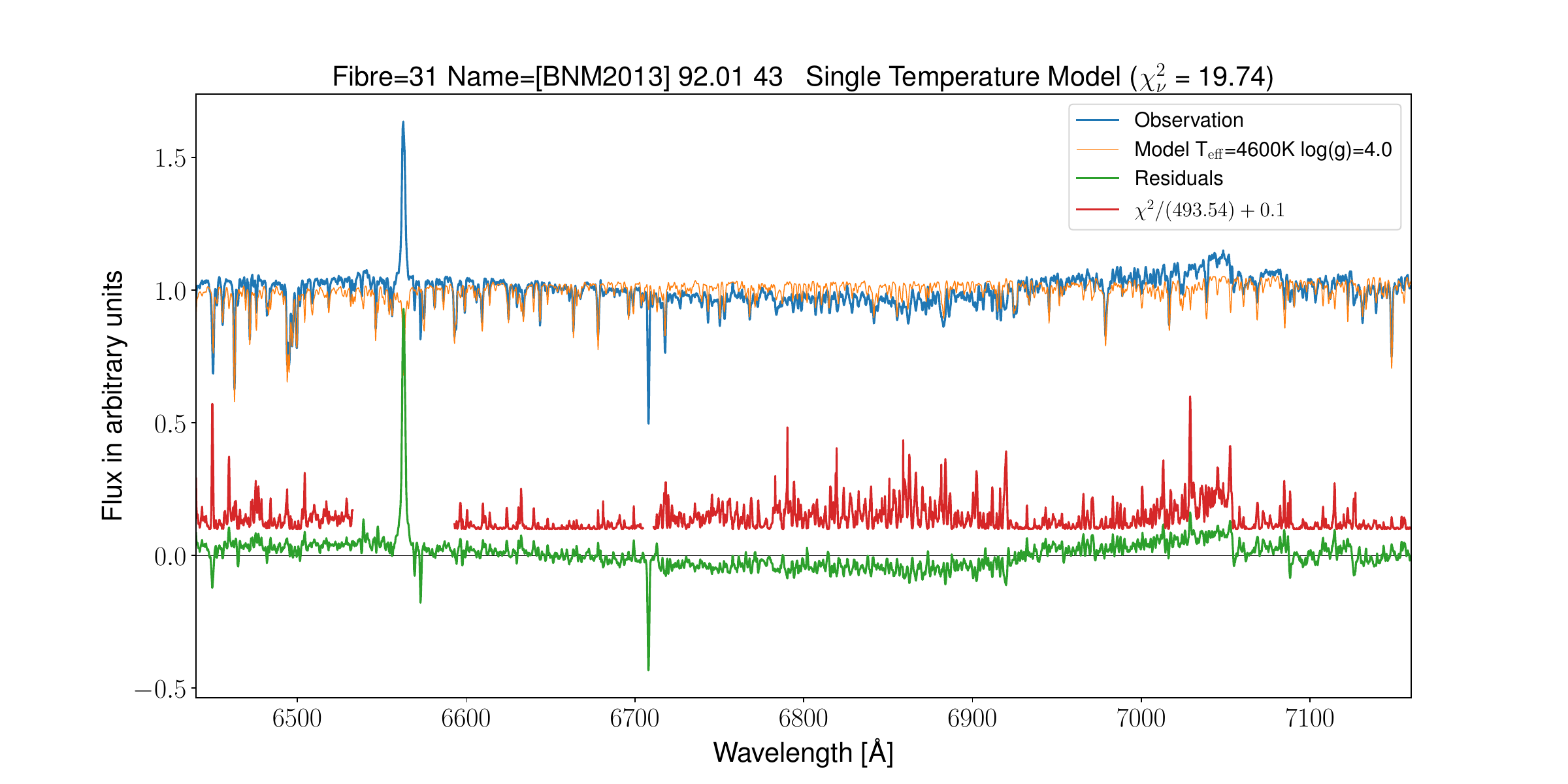}
	\includegraphics[width=0.9\textwidth]{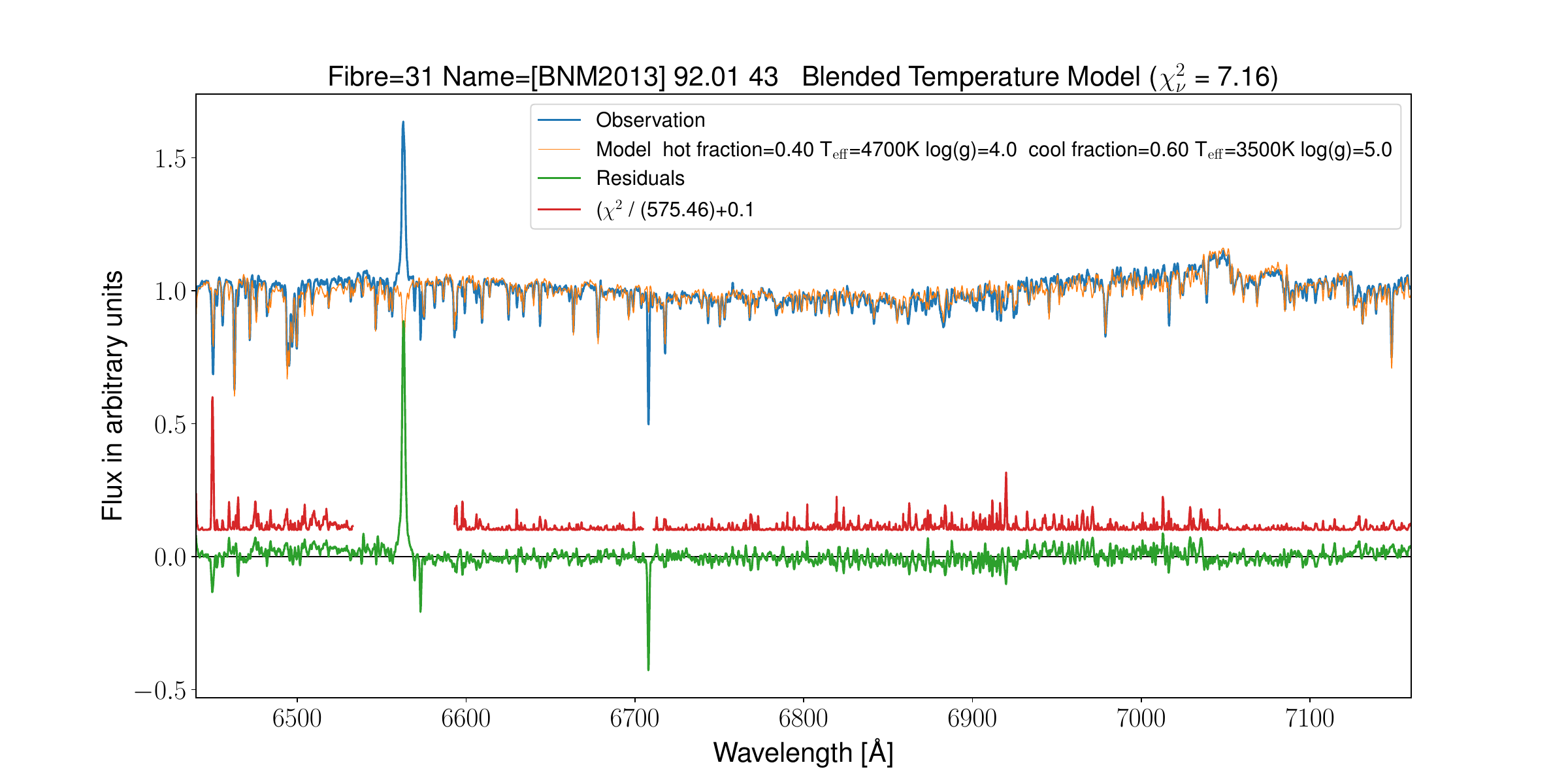}
	\includegraphics[width=0.9\textwidth]{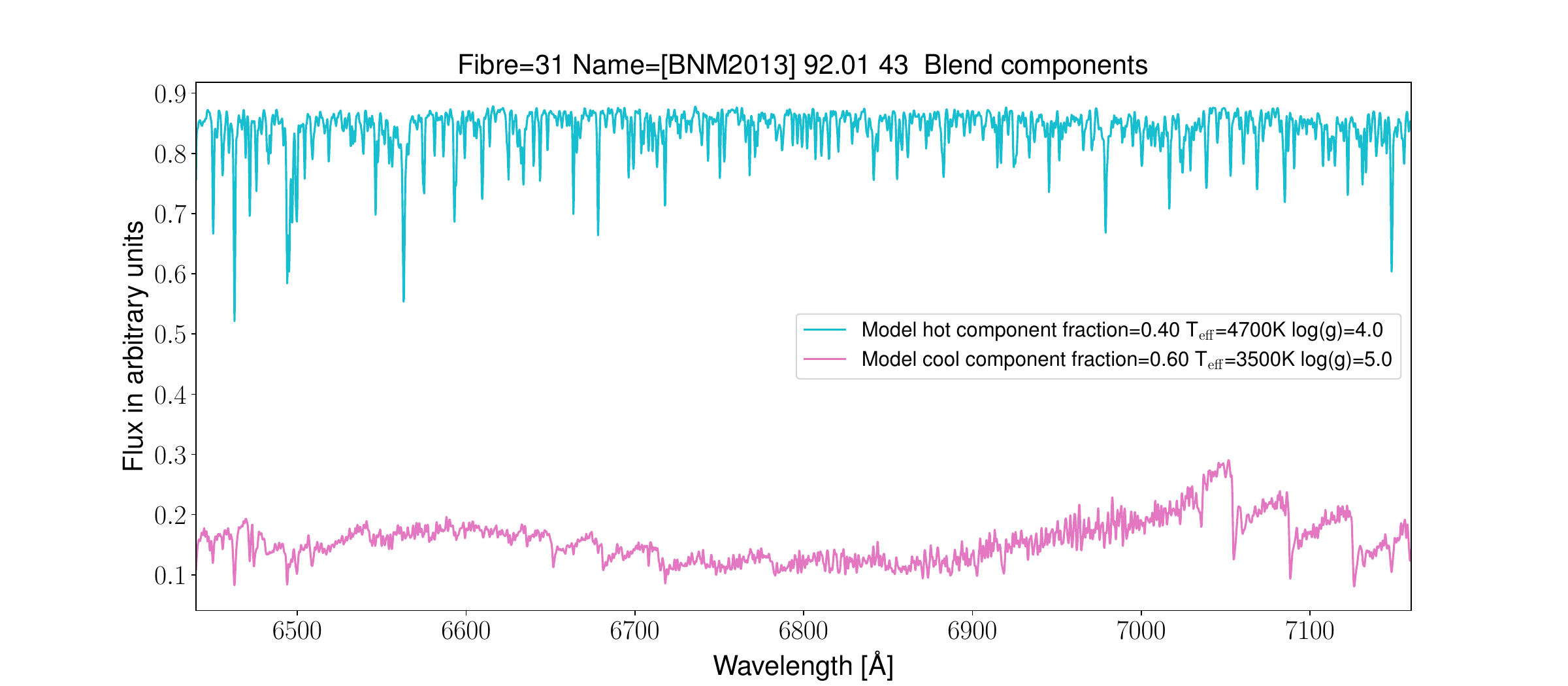}
	\caption{As Fig.~\ref{fig:fibre_27_spectra} for fibre 31.}
	\label{fig:fibre_31_spectra}
\end{figure*}

\begin{figure*}
	\centering
	\includegraphics[width=0.9\textwidth]{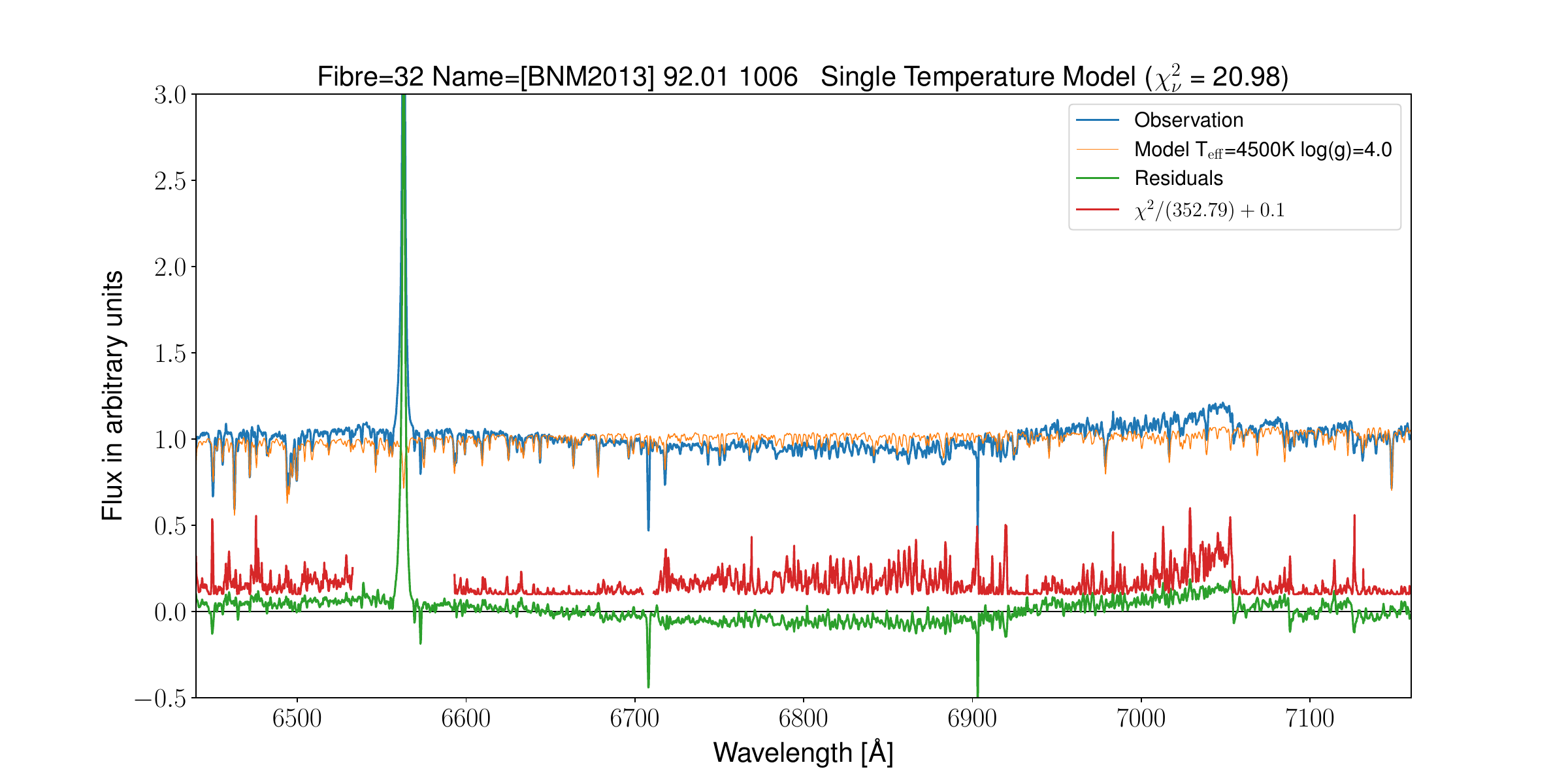}
	\includegraphics[width=0.9\textwidth]{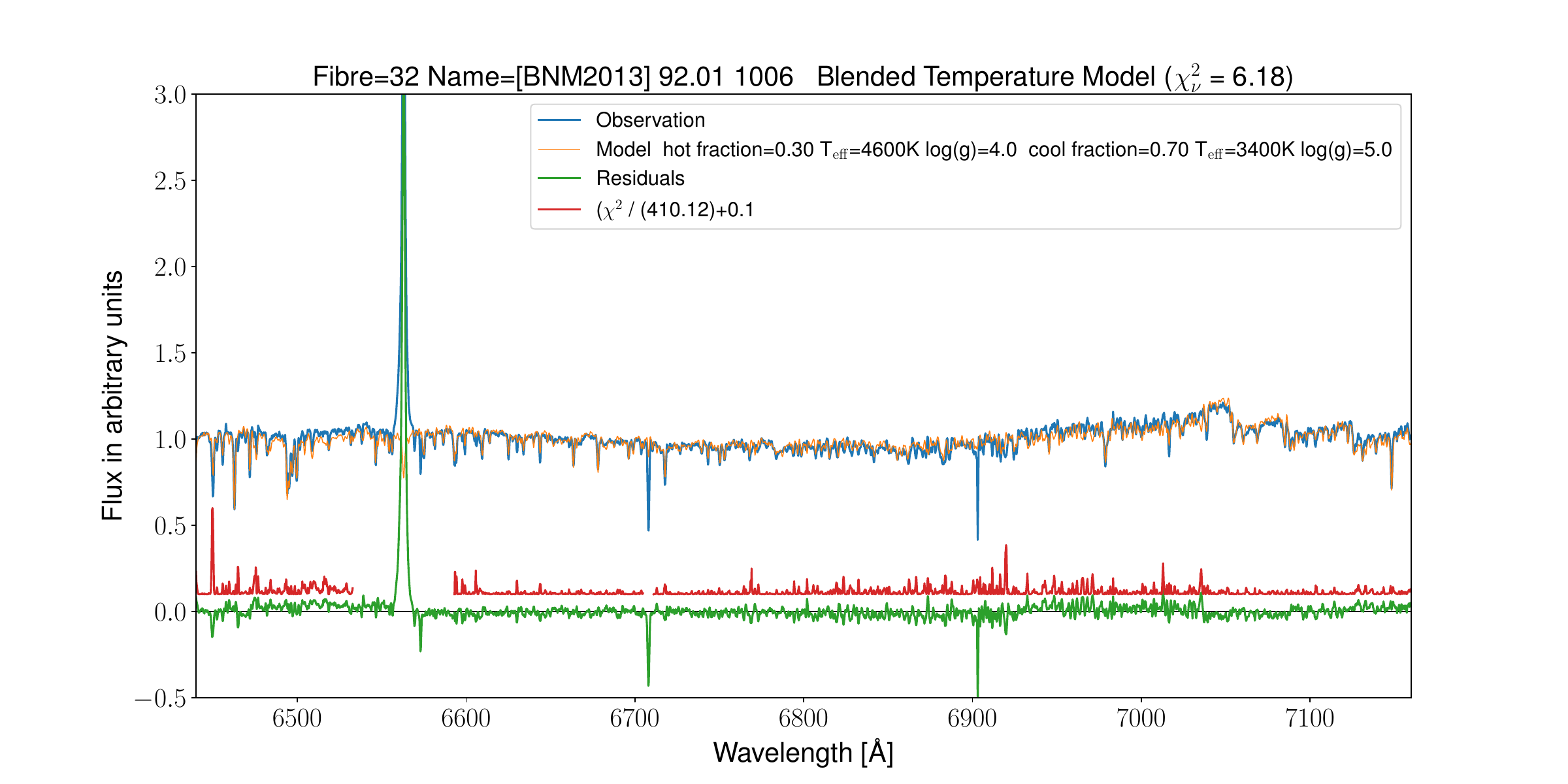}
	\includegraphics[width=0.9\textwidth]{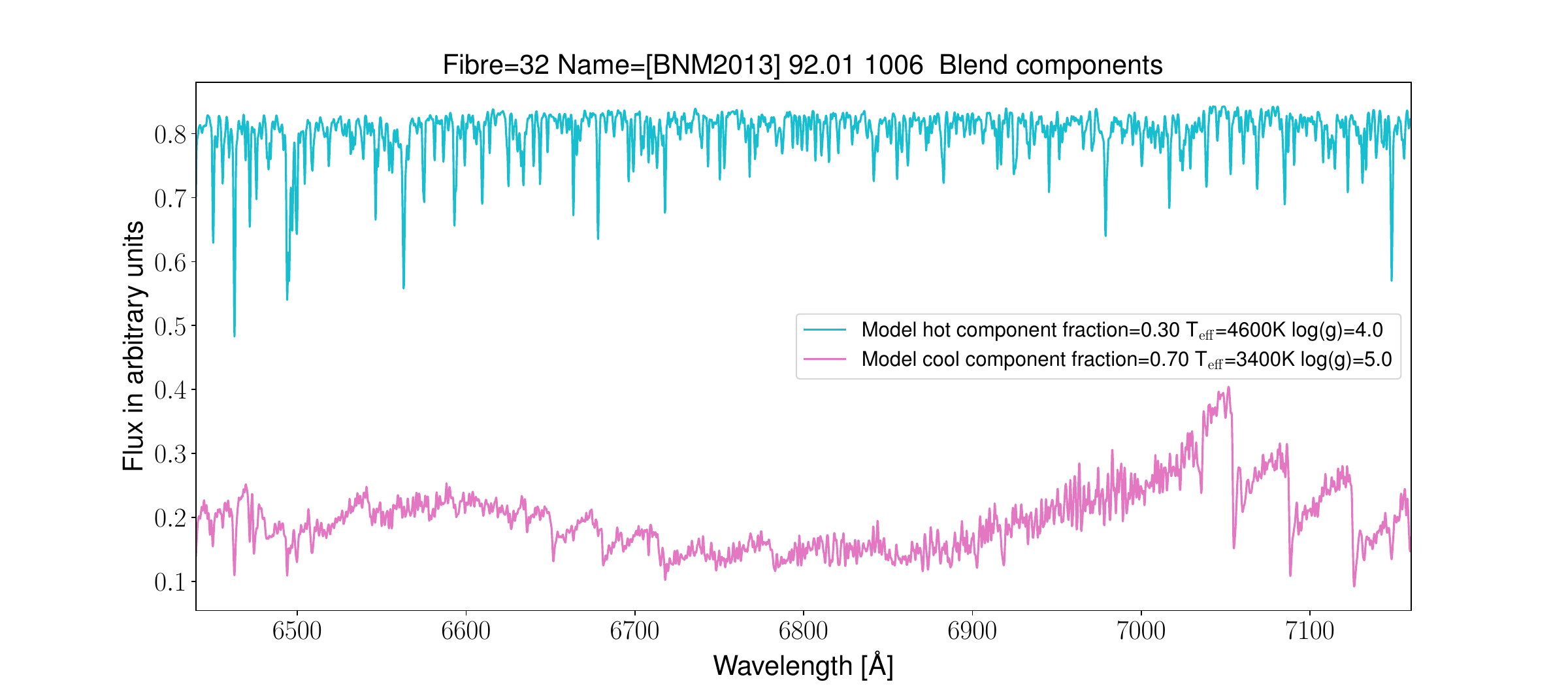}
	\caption{As Fig.~\ref{fig:fibre_27_spectra} for fibre 32.}
	\label{fig:fibre_32_spectra}
\end{figure*}

\begin{figure*}
	\centering
	\includegraphics[width=0.9\textwidth]{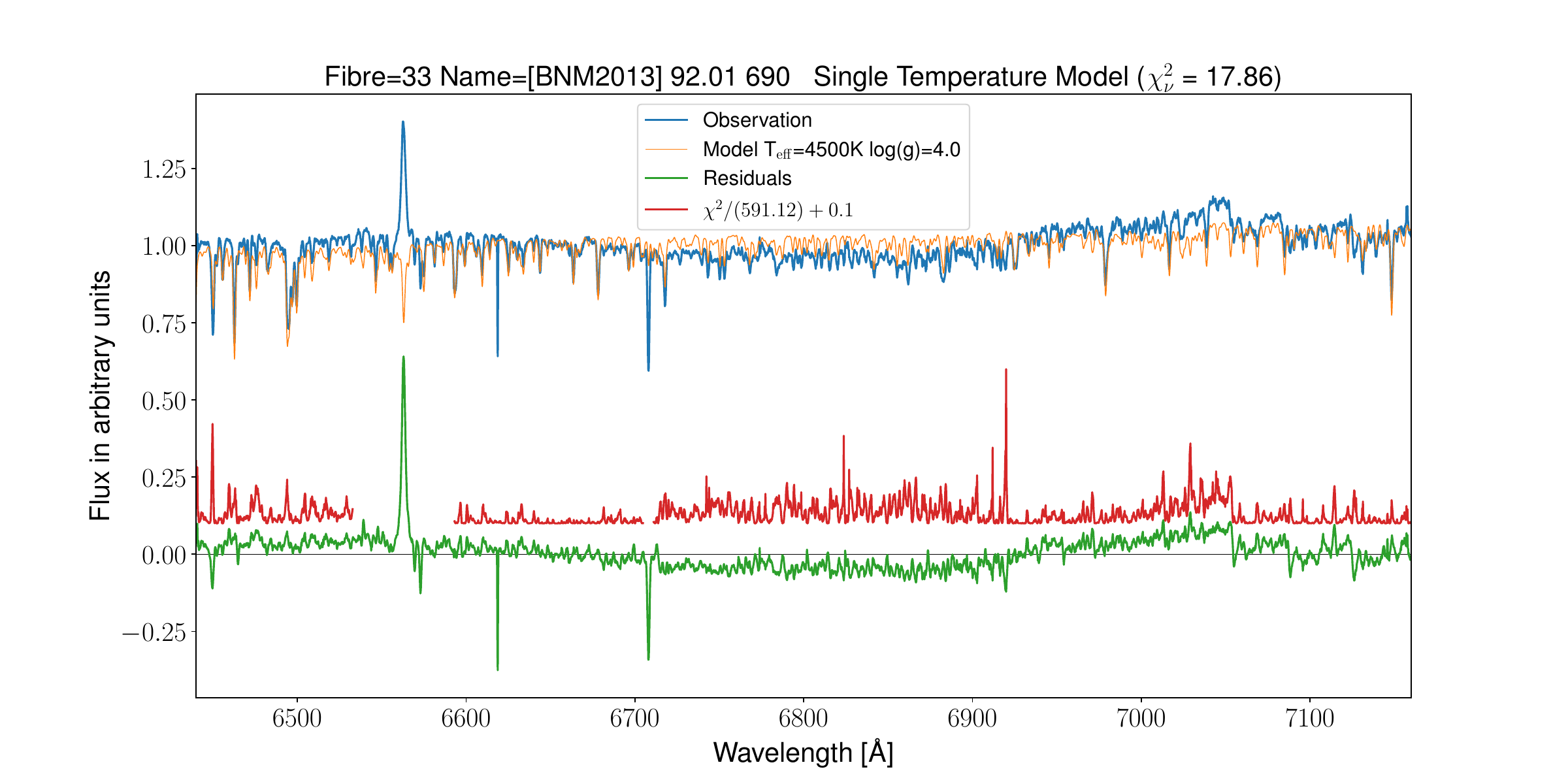}
	\includegraphics[width=0.9\textwidth]{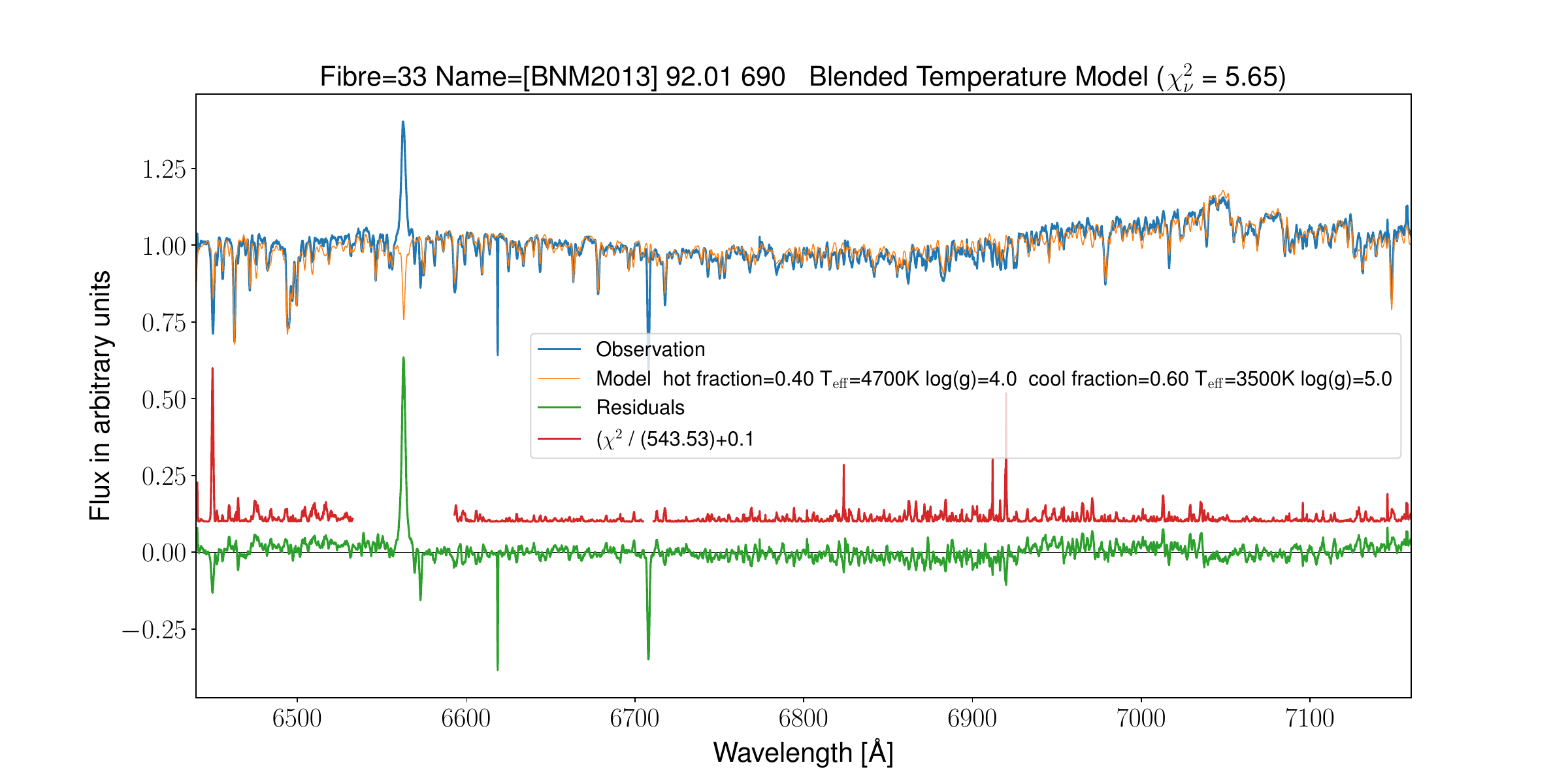}
	\includegraphics[width=0.9\textwidth]{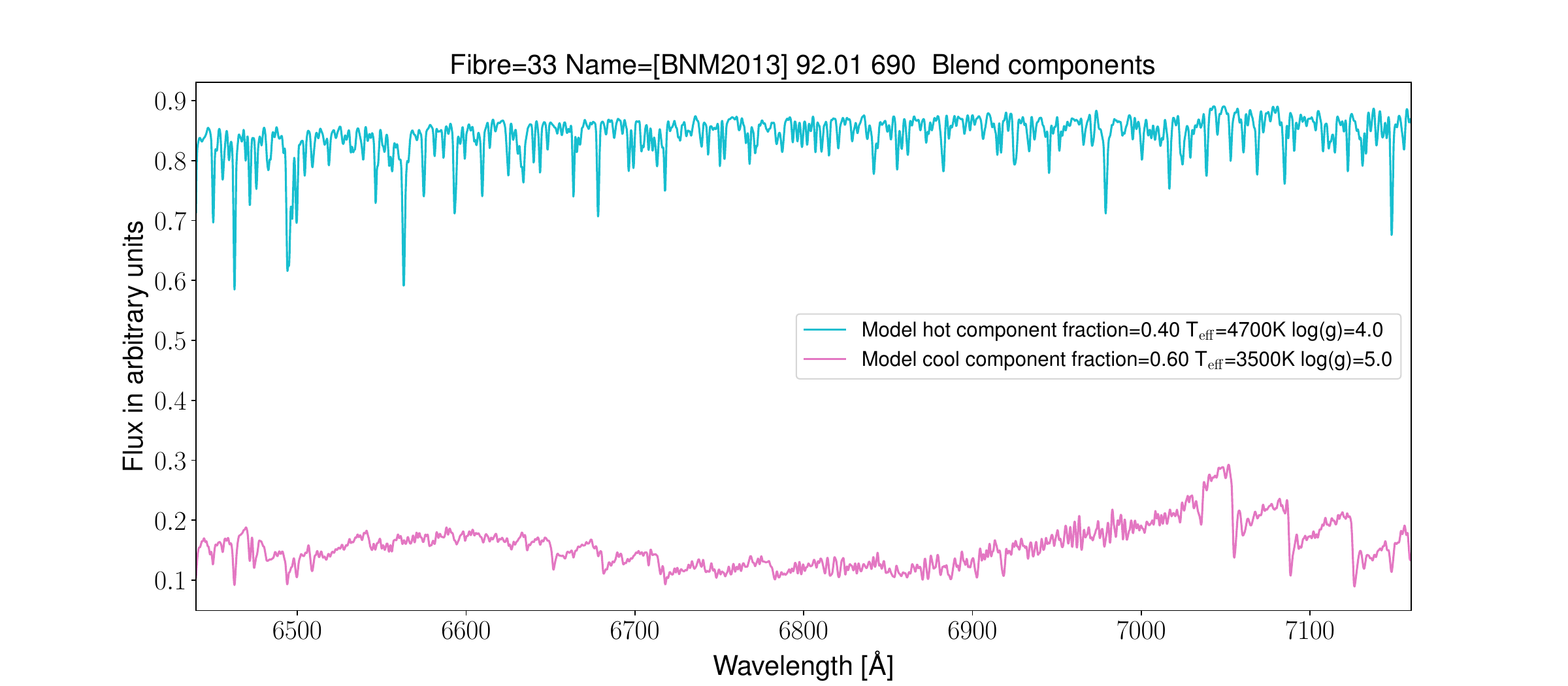}
	\caption{As Fig.~\ref{fig:fibre_27_spectra} for fibre 33.}
	\label{fig:fibre_33_spectra}
\end{figure*}

\begin{figure*}
	\centering
	\includegraphics[width=0.9\textwidth]{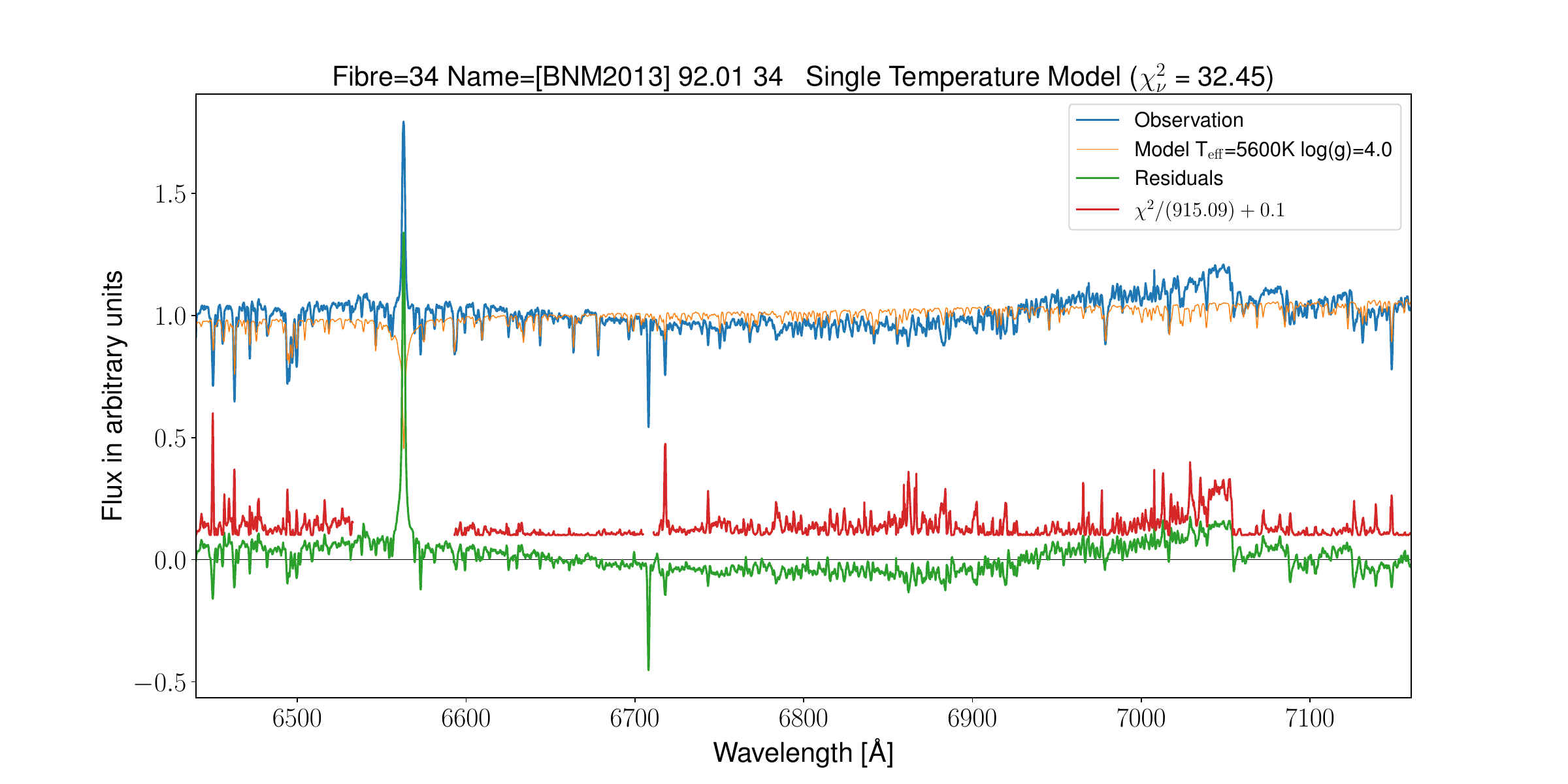}
	\includegraphics[width=0.9\textwidth]{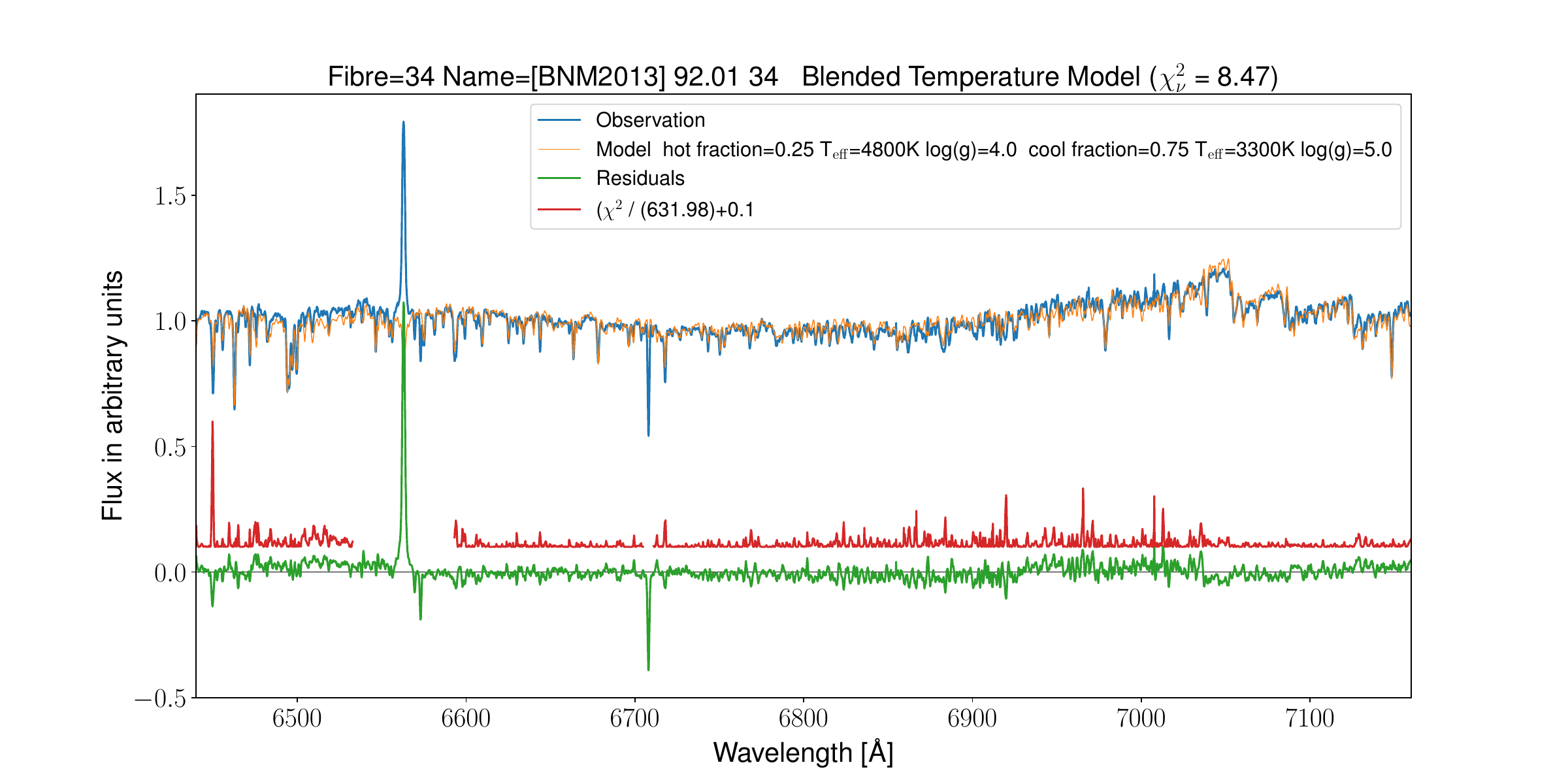}
	\includegraphics[width=0.9\textwidth]{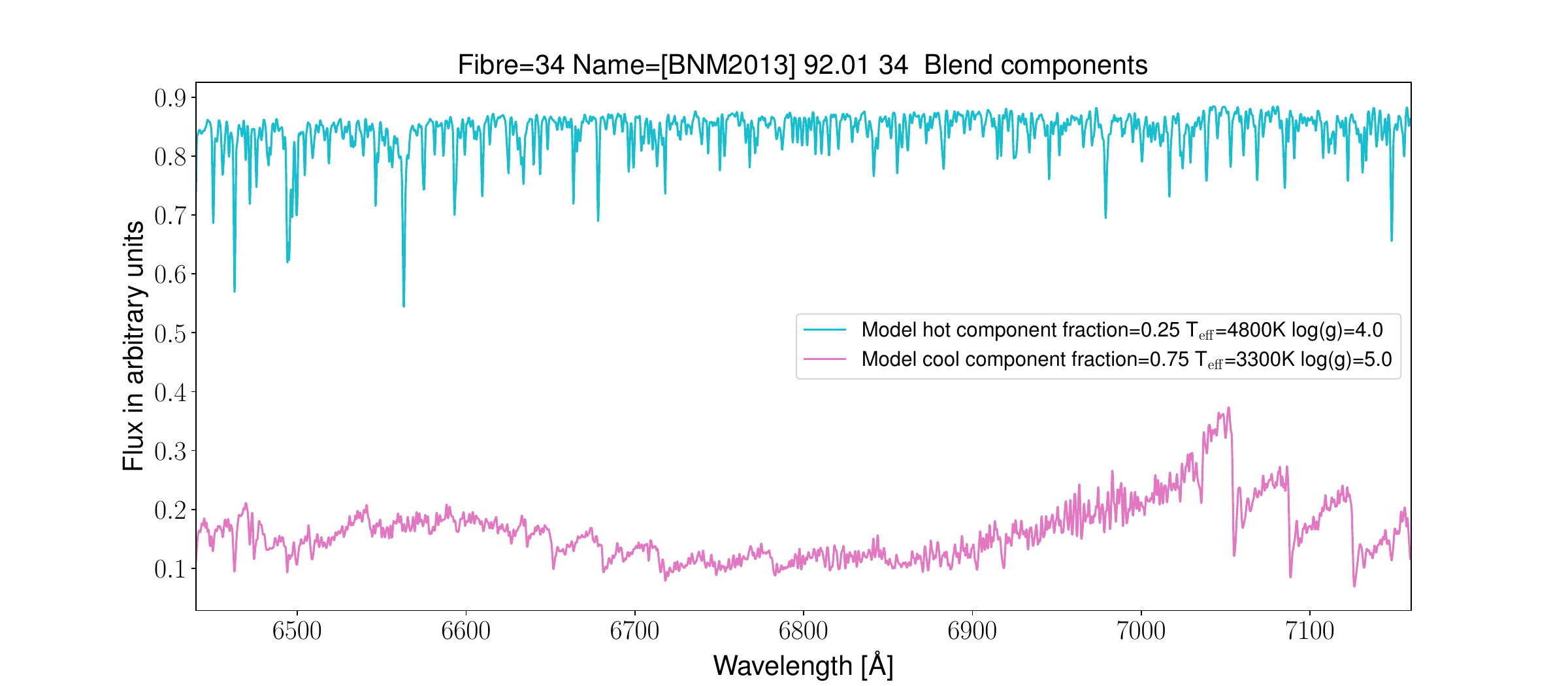}
	\caption{As Fig.~\ref{fig:fibre_27_spectra} for fibre 34.}
	\label{fig:fibre_34_spectra}
\end{figure*}

\begin{figure*}
	\centering
	\includegraphics[width=0.9\textwidth]{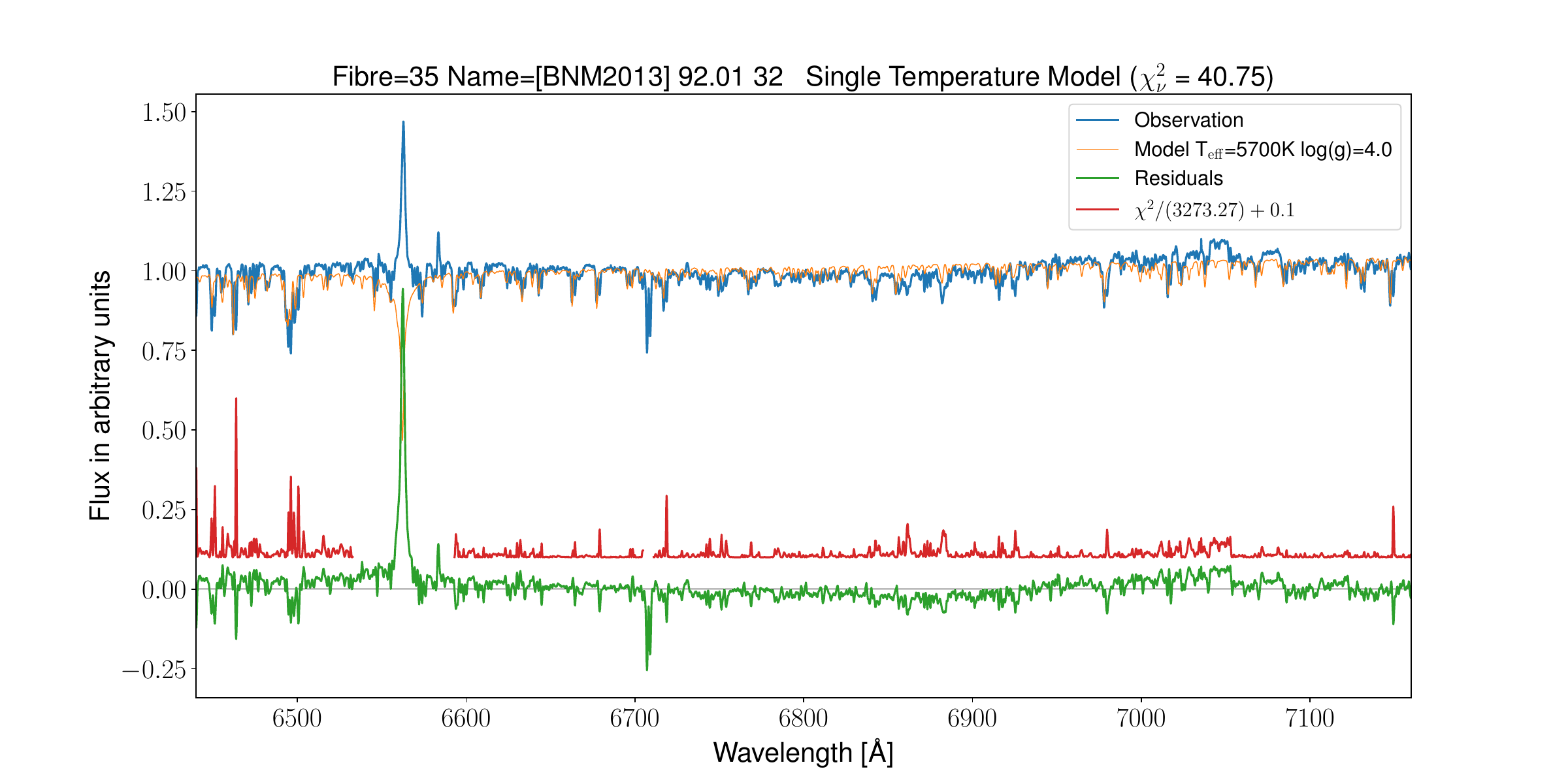}
	\includegraphics[width=0.9\textwidth]{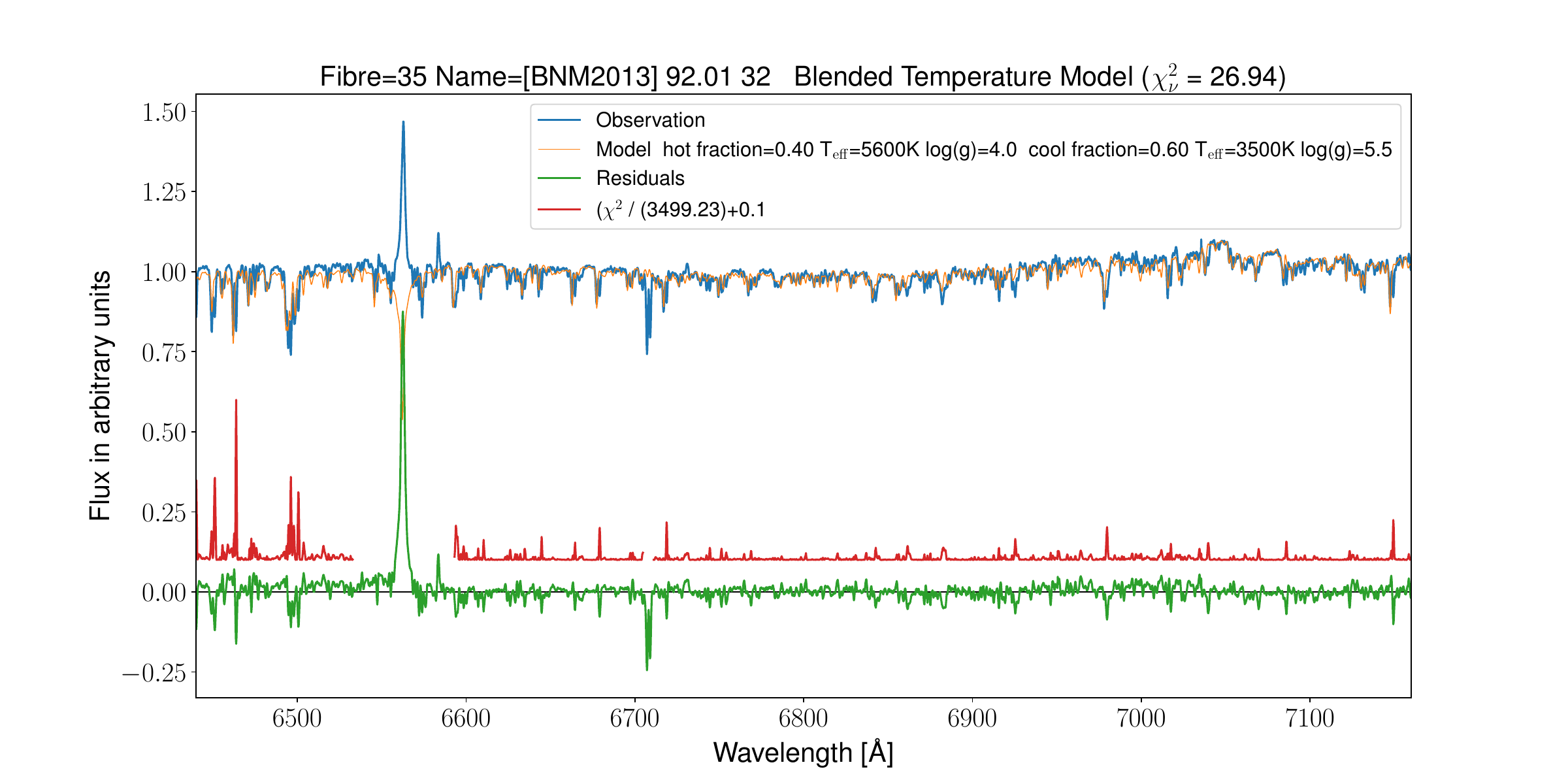}
	\includegraphics[width=0.9\textwidth]{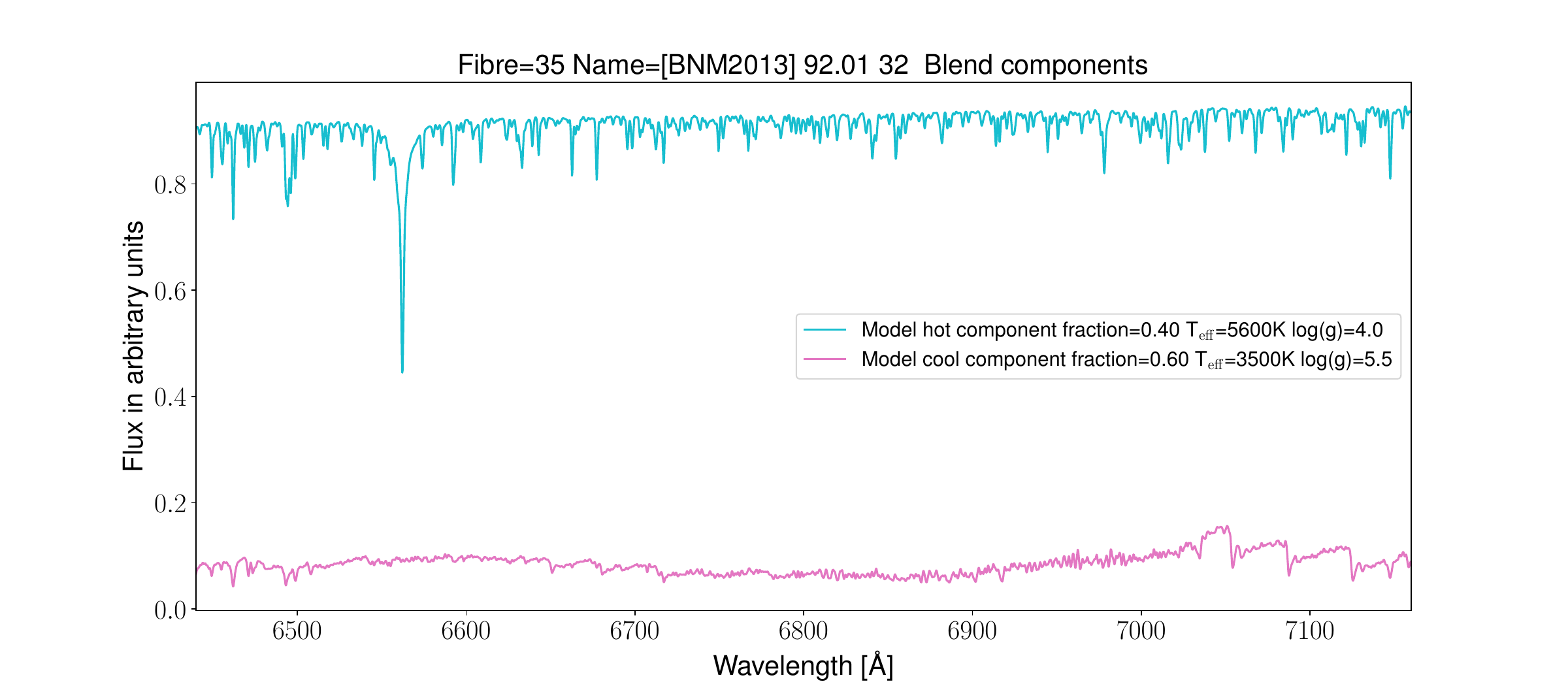}
	\caption{As Fig.~\ref{fig:fibre_27_spectra} for fibre 35.}
	\label{fig:fibre_35_spectra}
\end{figure*}

\begin{figure*}
	\centering
	\includegraphics[width=0.9\textwidth]{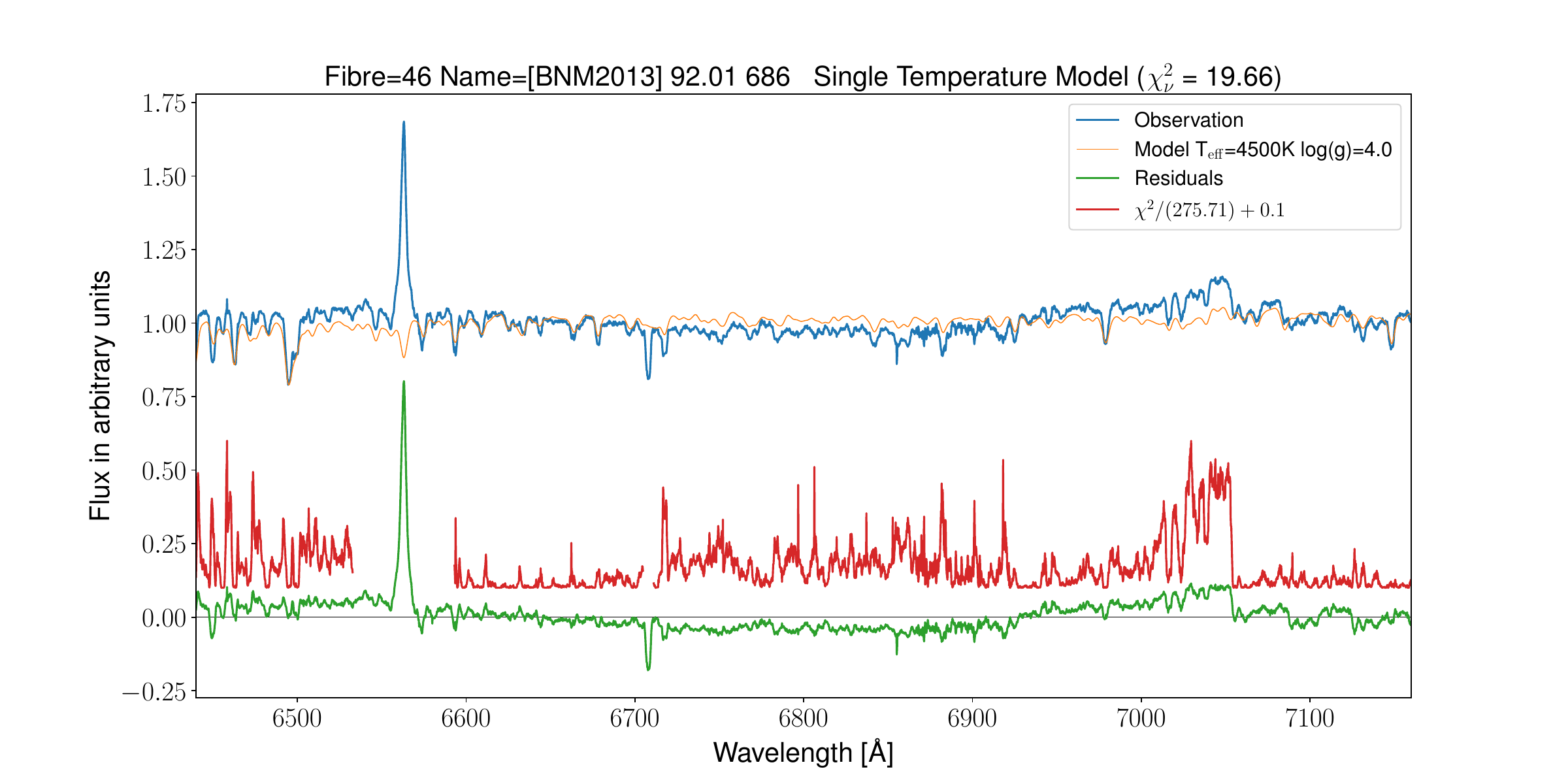}
	\includegraphics[width=0.9\textwidth]{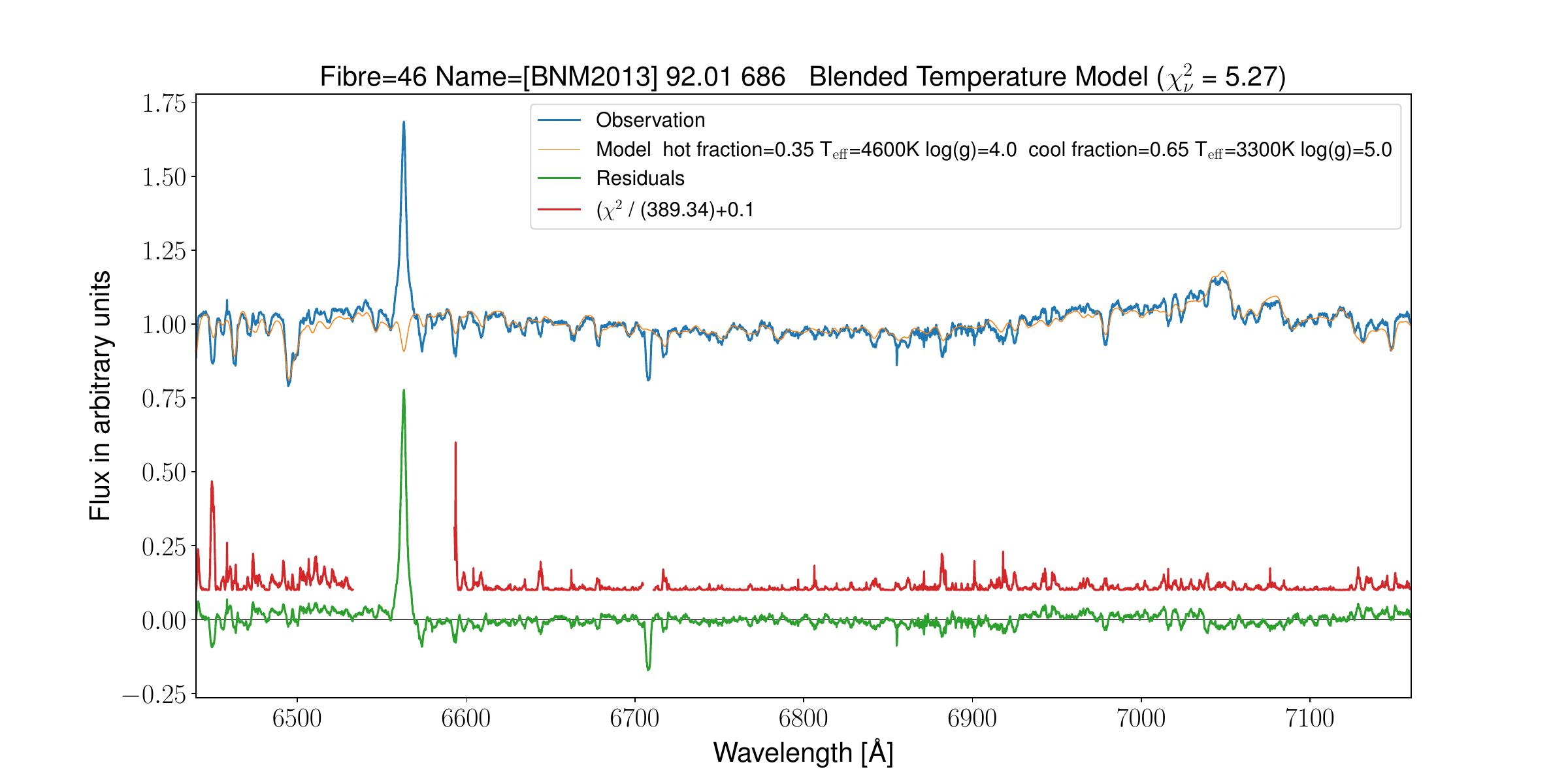}
	\includegraphics[width=0.9\textwidth]{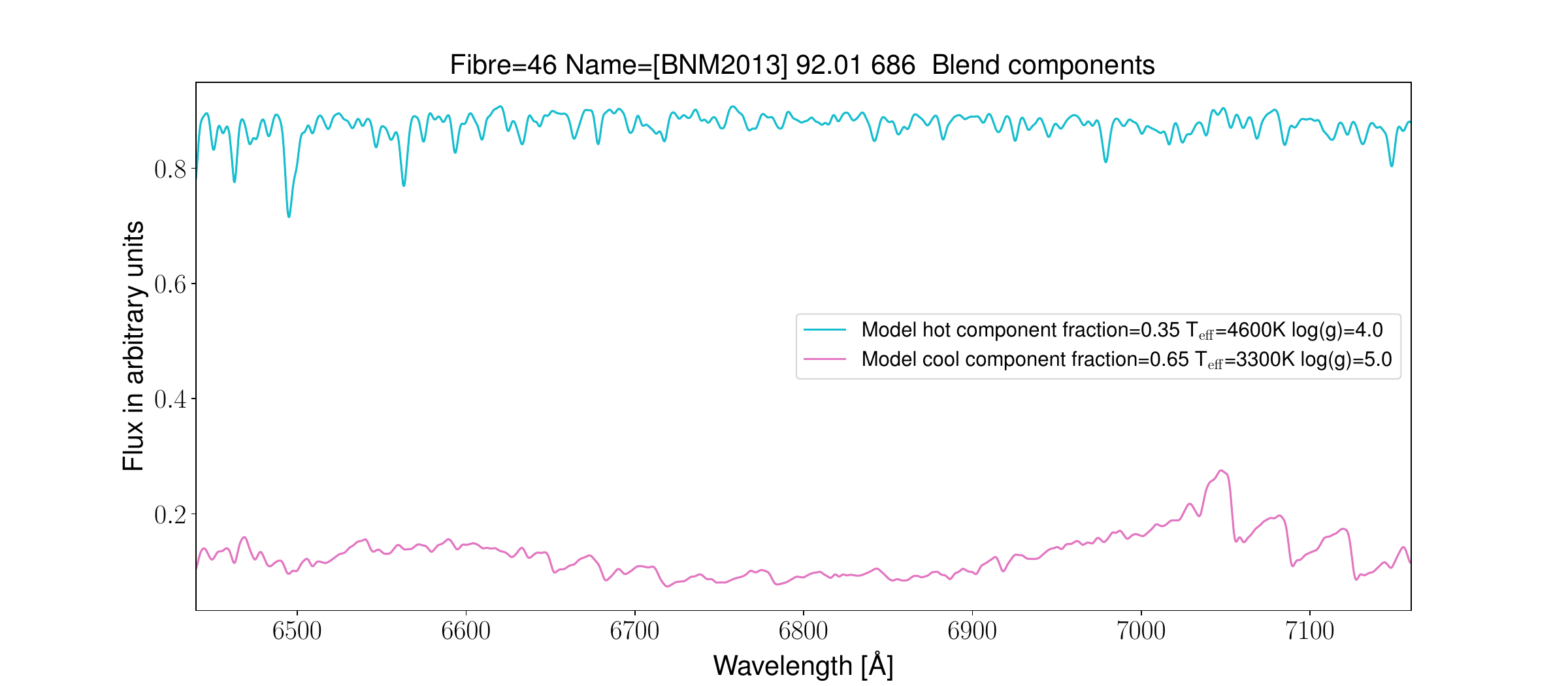}
	\caption{As Fig.~\ref{fig:fibre_27_spectra} for fibre 46.}
	\label{fig:fibre_46_spectra}
\end{figure*}

\begin{figure*}
	\centering
	\includegraphics[width=0.9\textwidth]{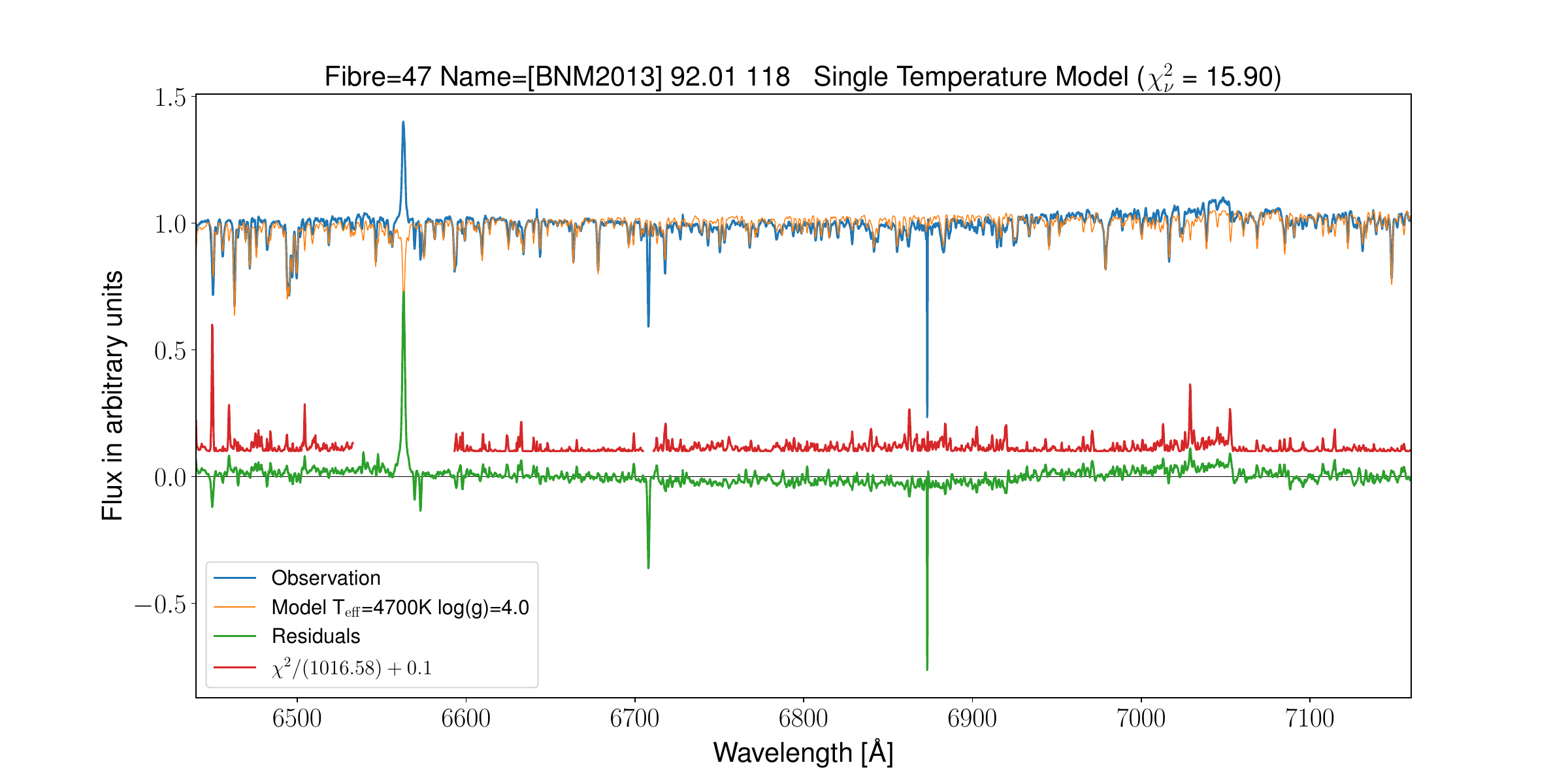}
	\includegraphics[width=0.9\textwidth]{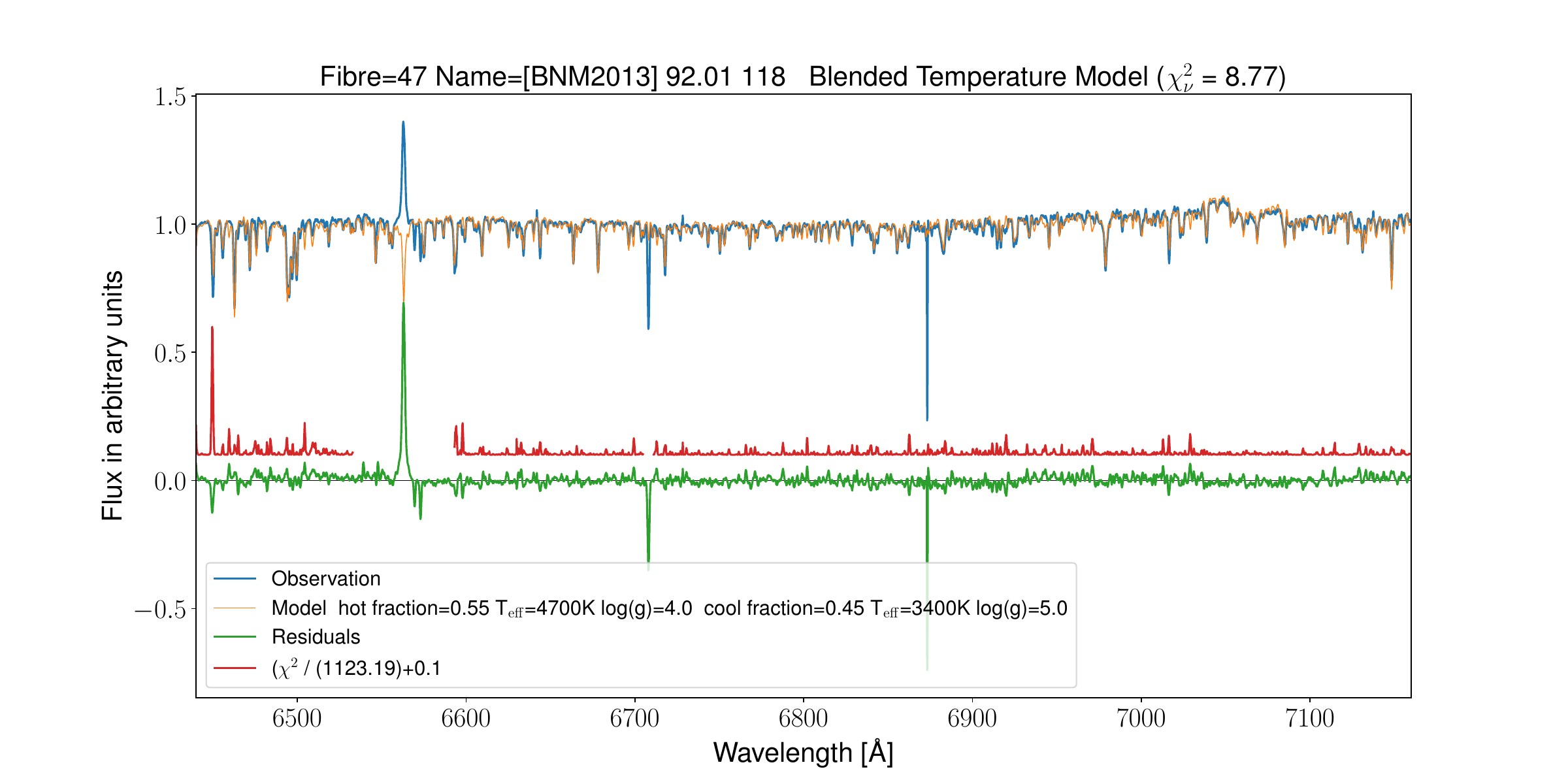}
	\includegraphics[width=0.9\textwidth]{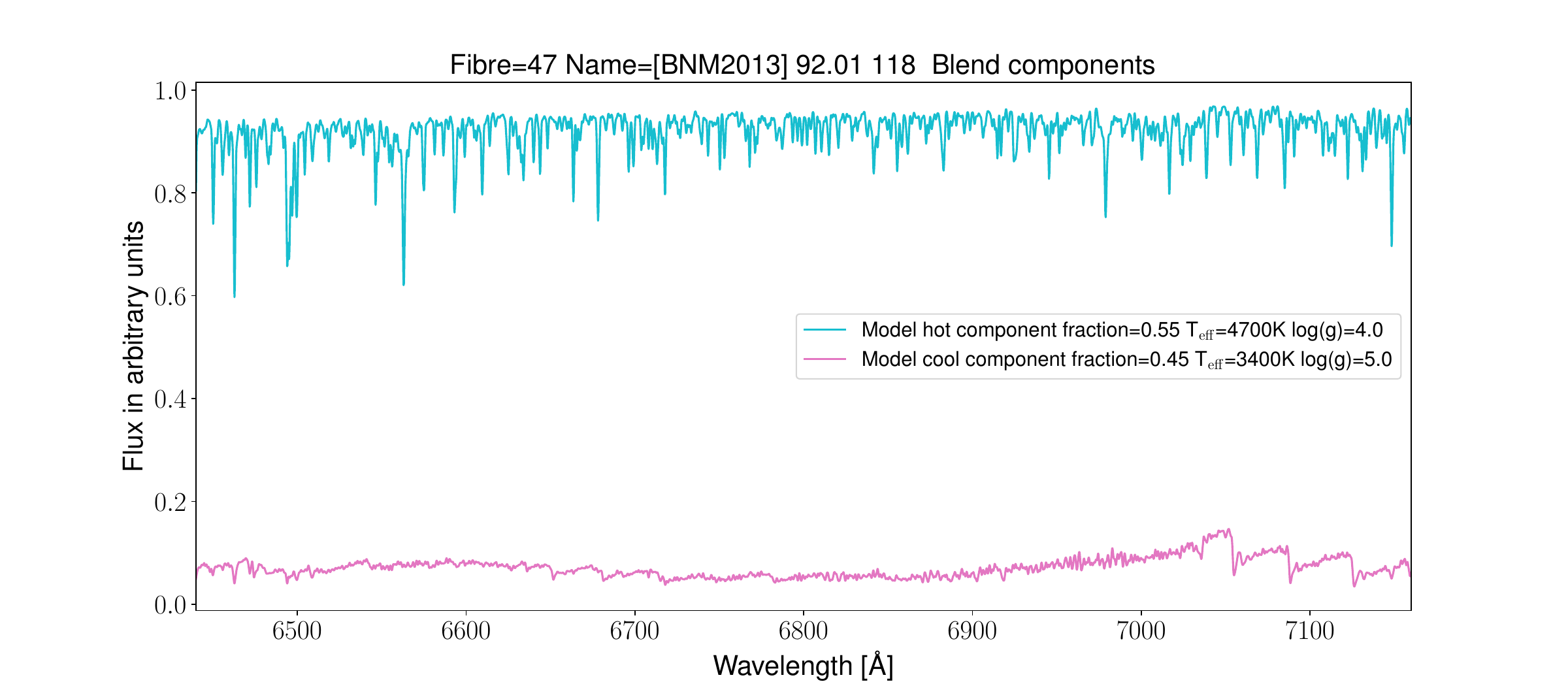}
	\caption{As Fig.~\ref{fig:fibre_27_spectra} for fibre 47.}
	\label{fig:fibre_47_spectra}
\end{figure*}

In Figs.\ref{fig:fibre_15_spectra} to \ref{fig:fibre_47_spectra}, we provide plots of the spectra with our single and blend of two temperature \editelev{models for the WTTS listed in Table~\ref{tab:sigori_review_members}} \editnine{that do not appear elsewhere within this paper}. Three plots are given for each WTTS. The top plot displays the fit to a single temperature photosphere, the middle plot is a blend of two photospheres, and the bottom plot is the two components of the blend scaled to the flux contributions from the fit. The vertical axis on the spectra are labelled 'flux in arbitrary units', an arbitrary scaling of energy per unit area per unit wavelength.

\section{$\sigma$ Orionis cluster photometry}
\label{app:sig_ori_photometry}

\begin{landscape}
\begin{figure}
	\centering
	\subfloat[]{\includegraphics[width=0.7\textwidth]{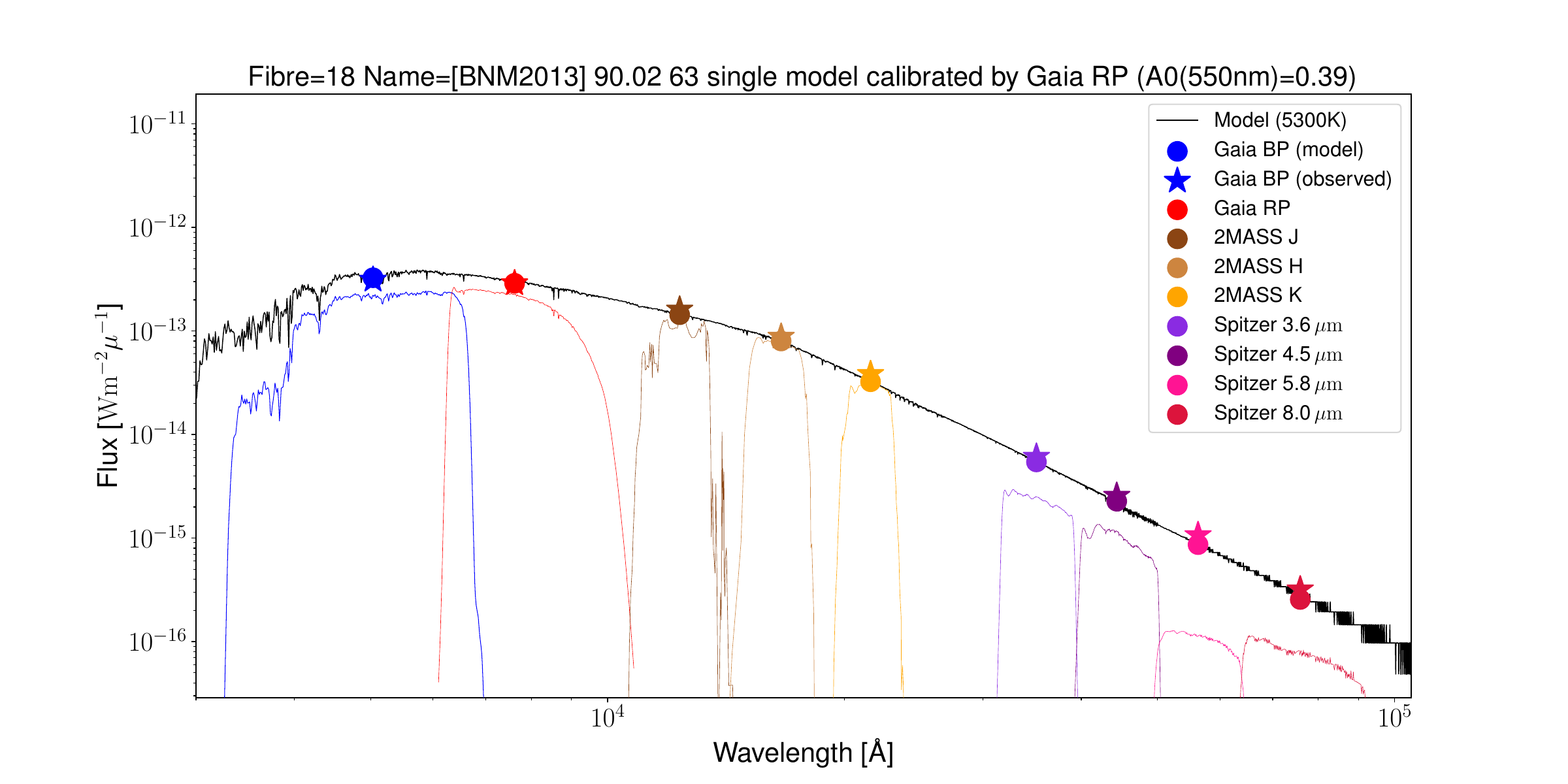}}
	\subfloat[]{\includegraphics[width=0.7\textwidth]{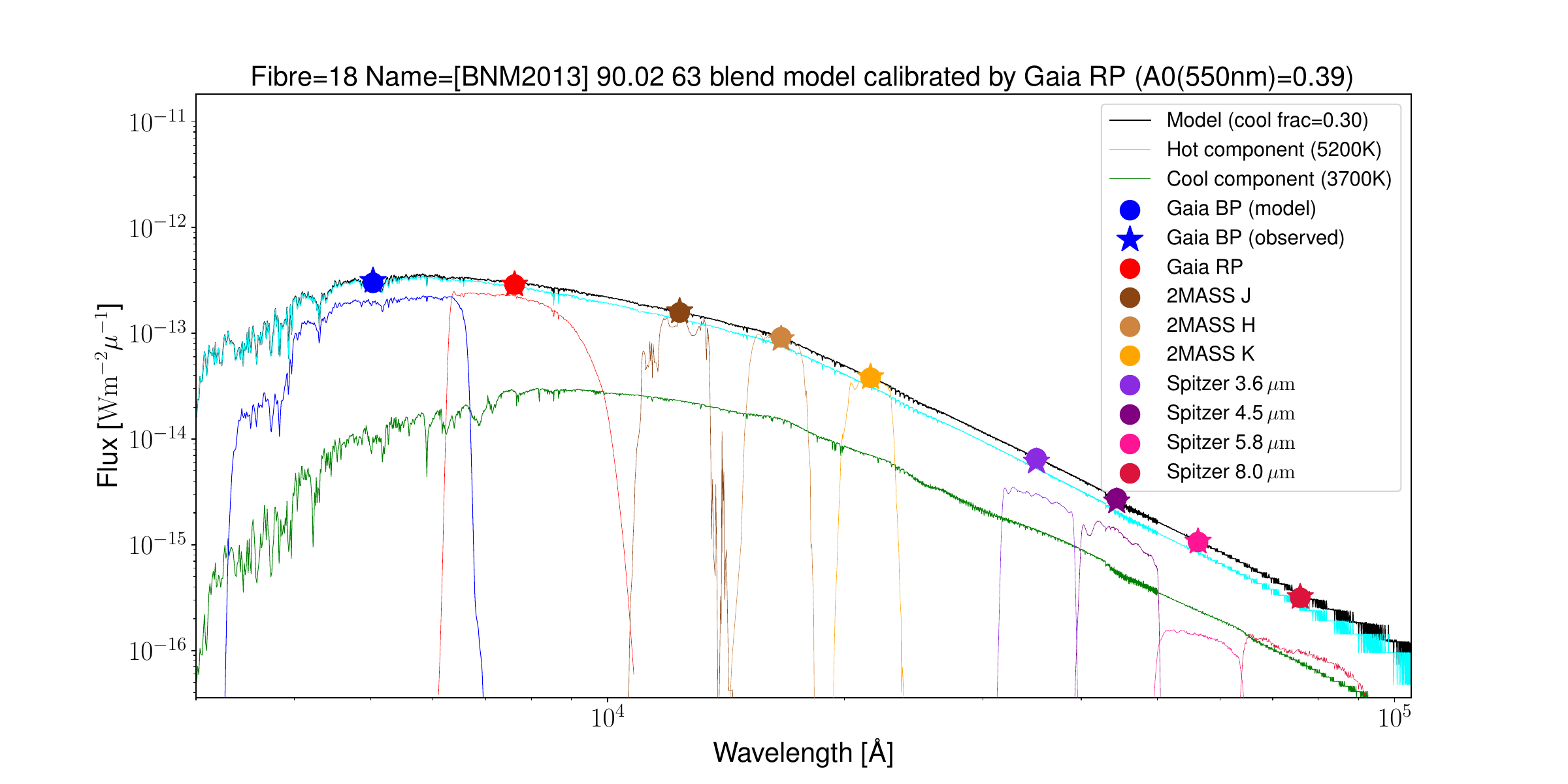}}\\
	\subfloat[]{\includegraphics[width=0.7\textwidth]{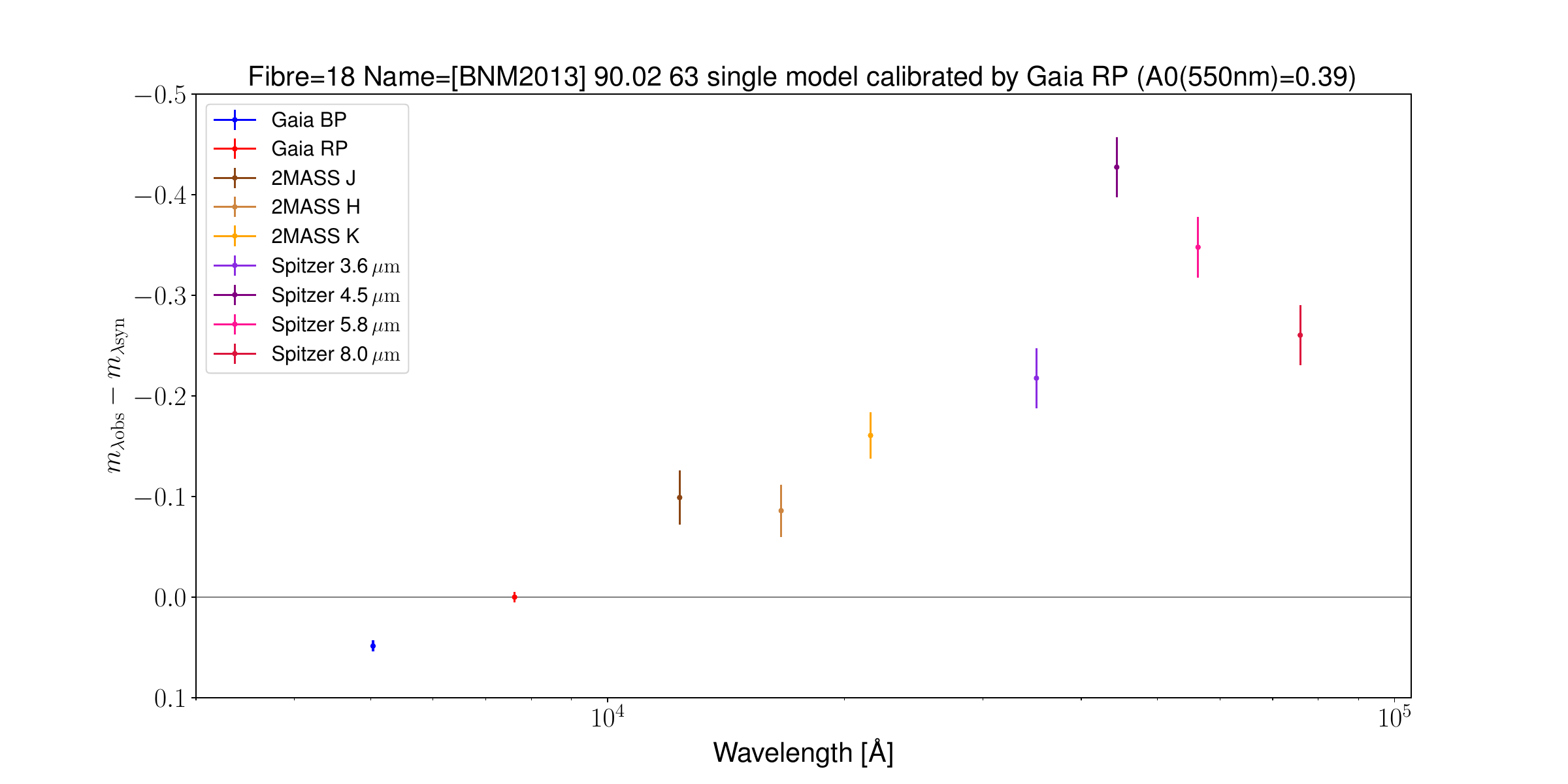}}
	\subfloat[]{\includegraphics[width=0.7\textwidth]{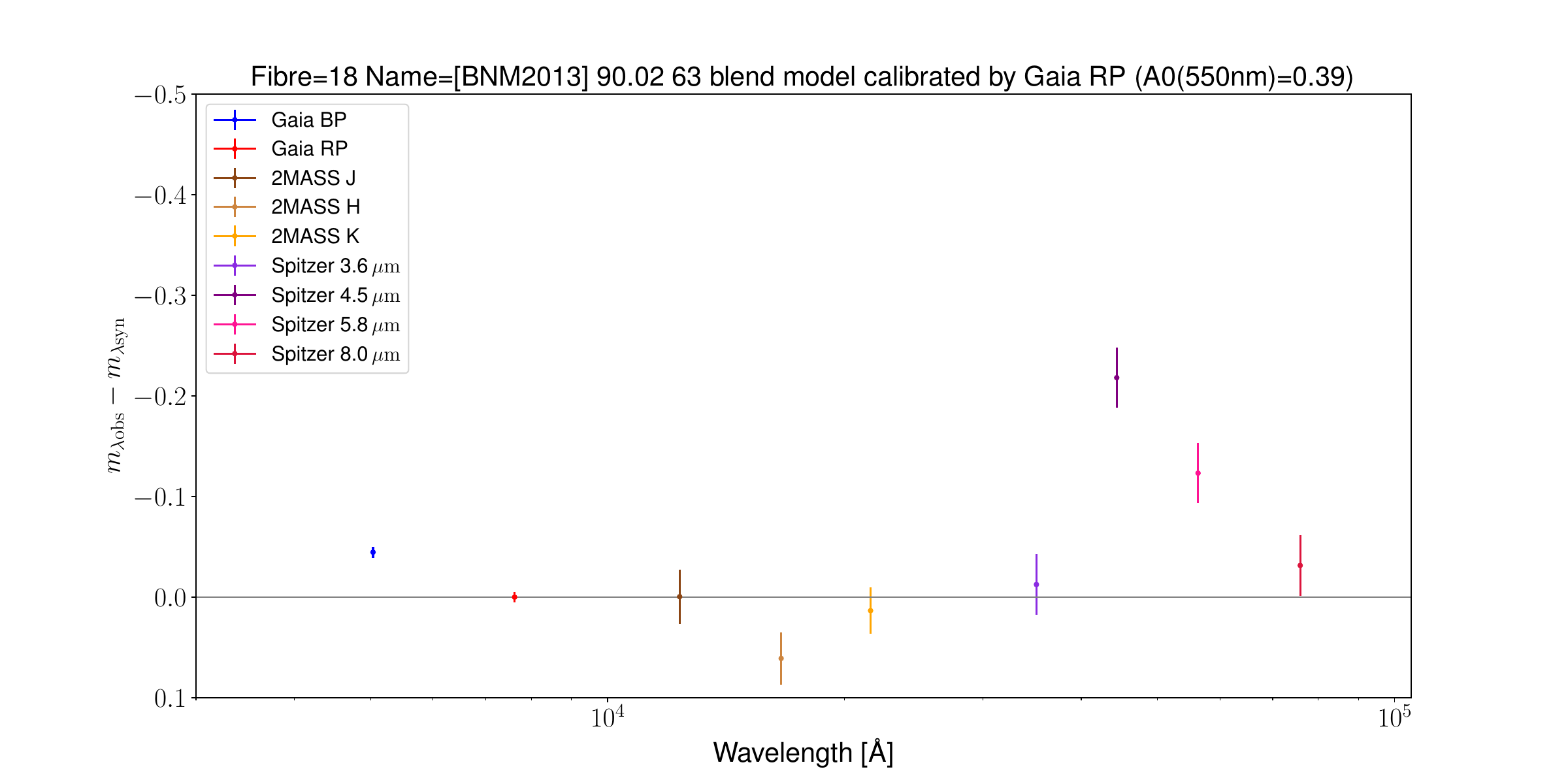}}
	\caption{As Fig.~\ref{fig:fibre_34_photometry} for fibre 18.}
	\label{fig:fibre_18_photometry}
\end{figure}
\end{landscape}

\begin{landscape}
	\begin{figure}
		\centering
		\subfloat[]{\includegraphics[width=0.7\textwidth]{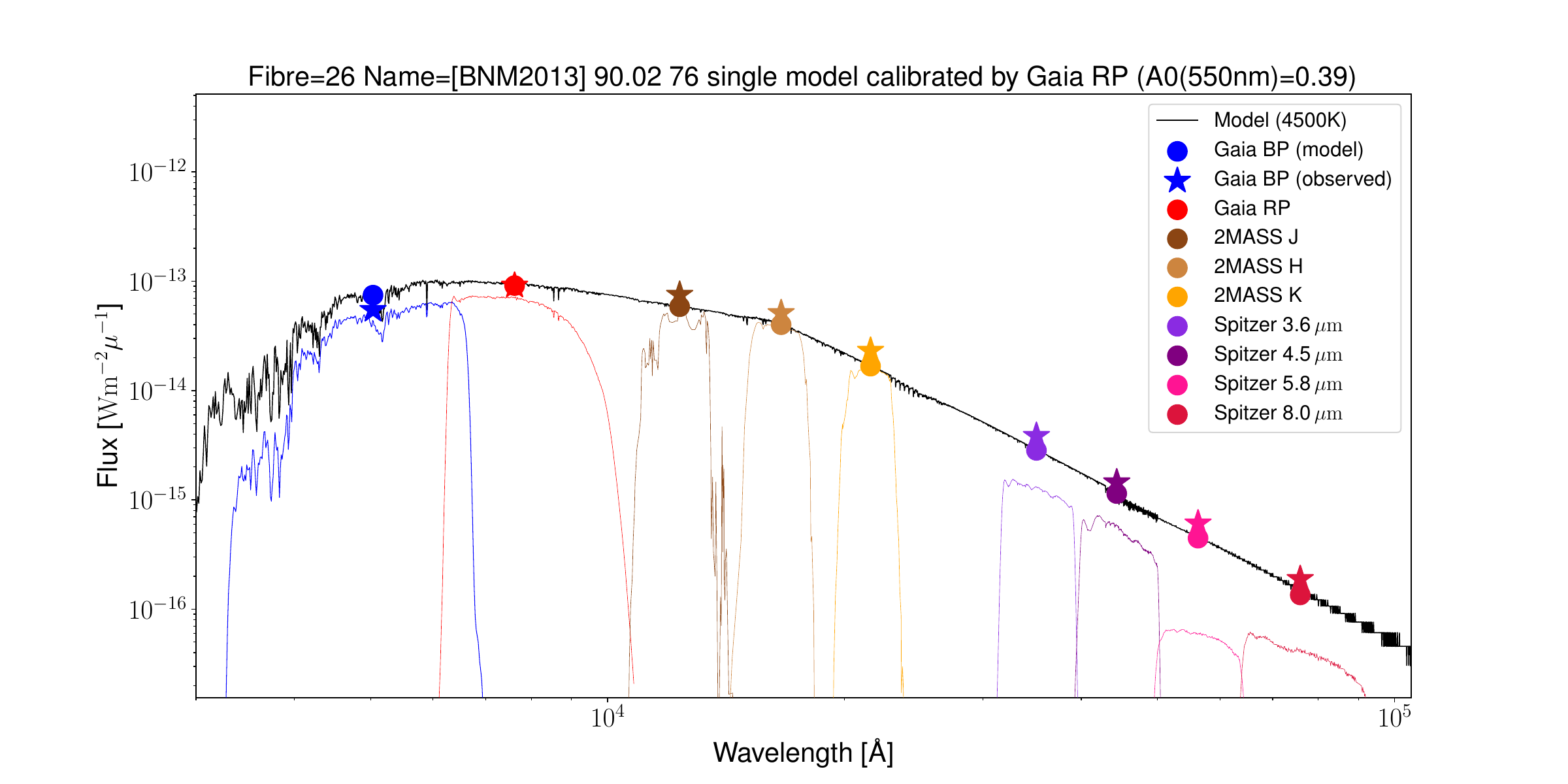}}
		\subfloat[]{\includegraphics[width=0.7\textwidth]{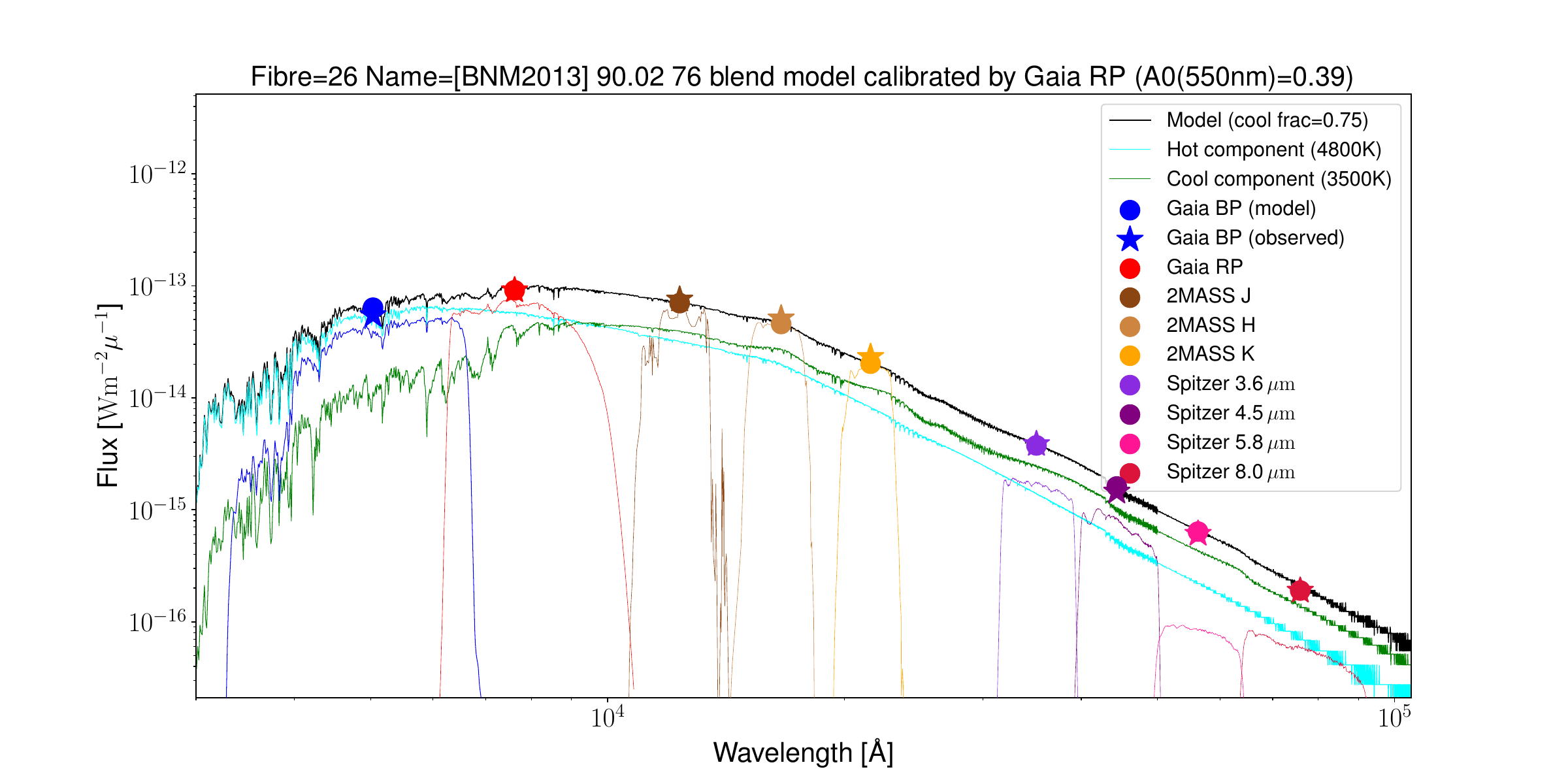}}\\
		\subfloat[]{\includegraphics[width=0.7\textwidth]{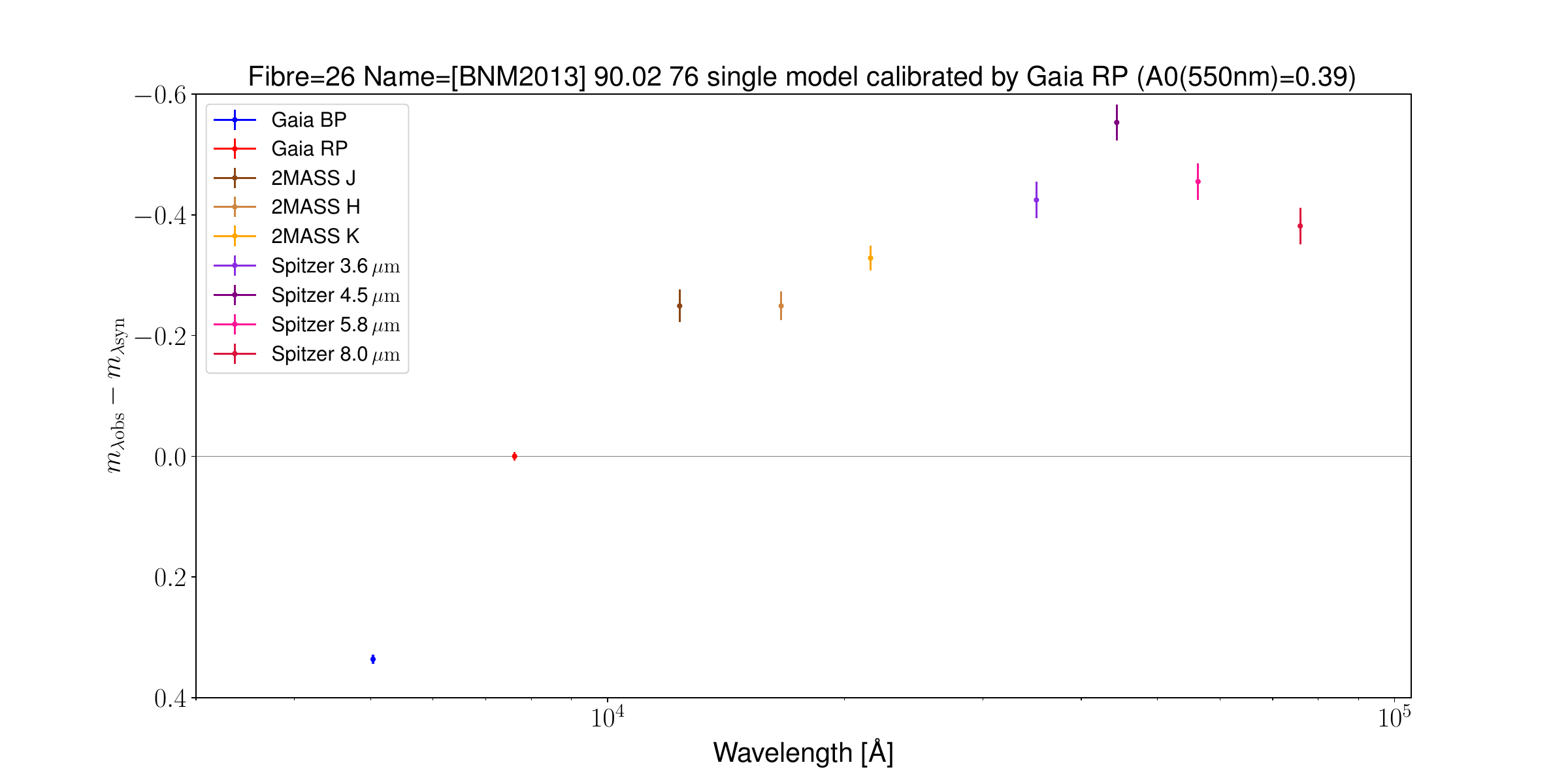}}
		\subfloat[]{\includegraphics[width=0.7\textwidth]{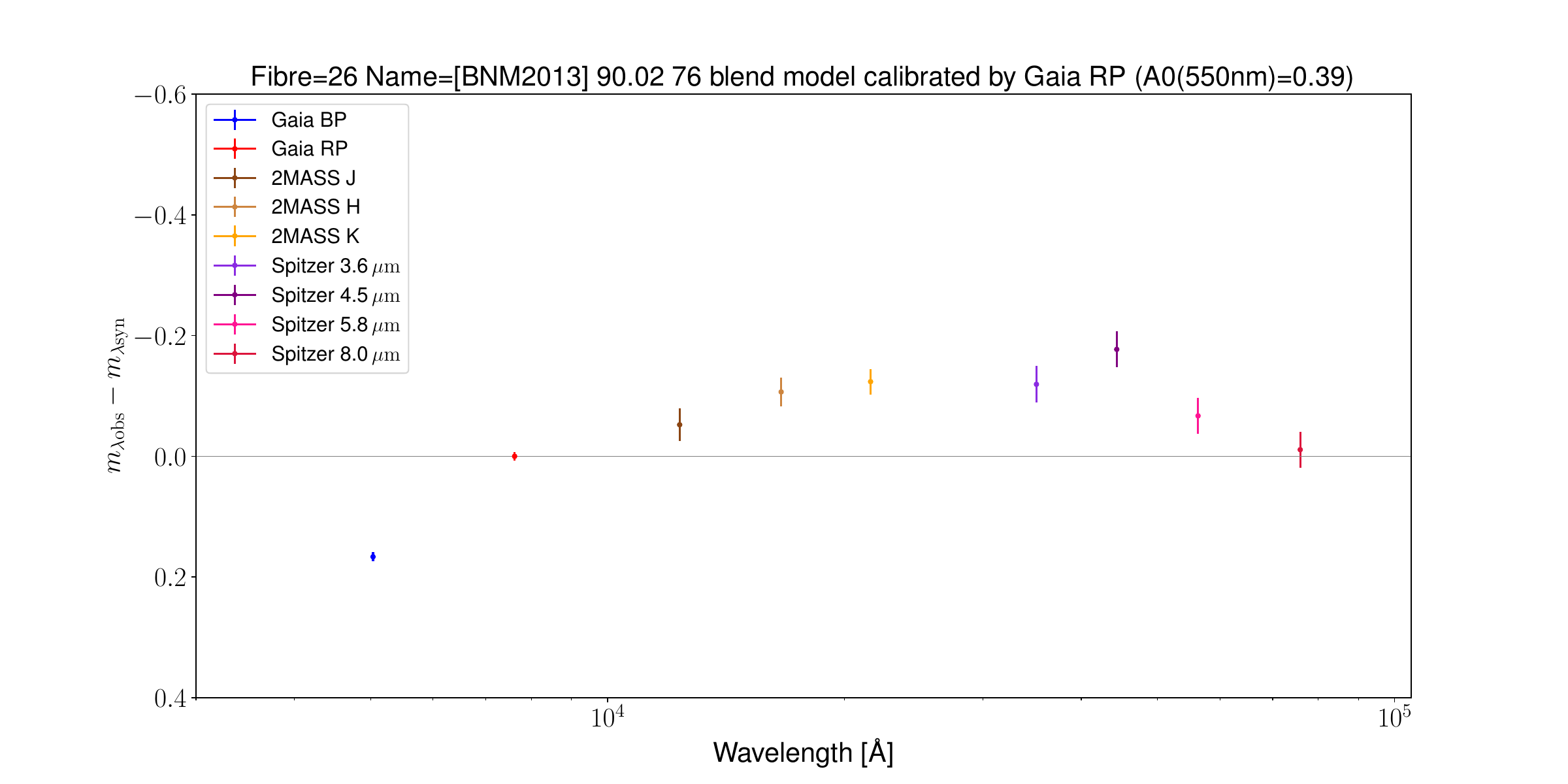}}
		\caption{As Fig.~\ref{fig:fibre_34_photometry} for fibre 26.}
		\label{fig:fibre_26_photometry}
	\end{figure}
\end{landscape}

\begin{landscape}
	\begin{figure}
		\centering
		\subfloat[]{\includegraphics[width=0.7\textwidth]{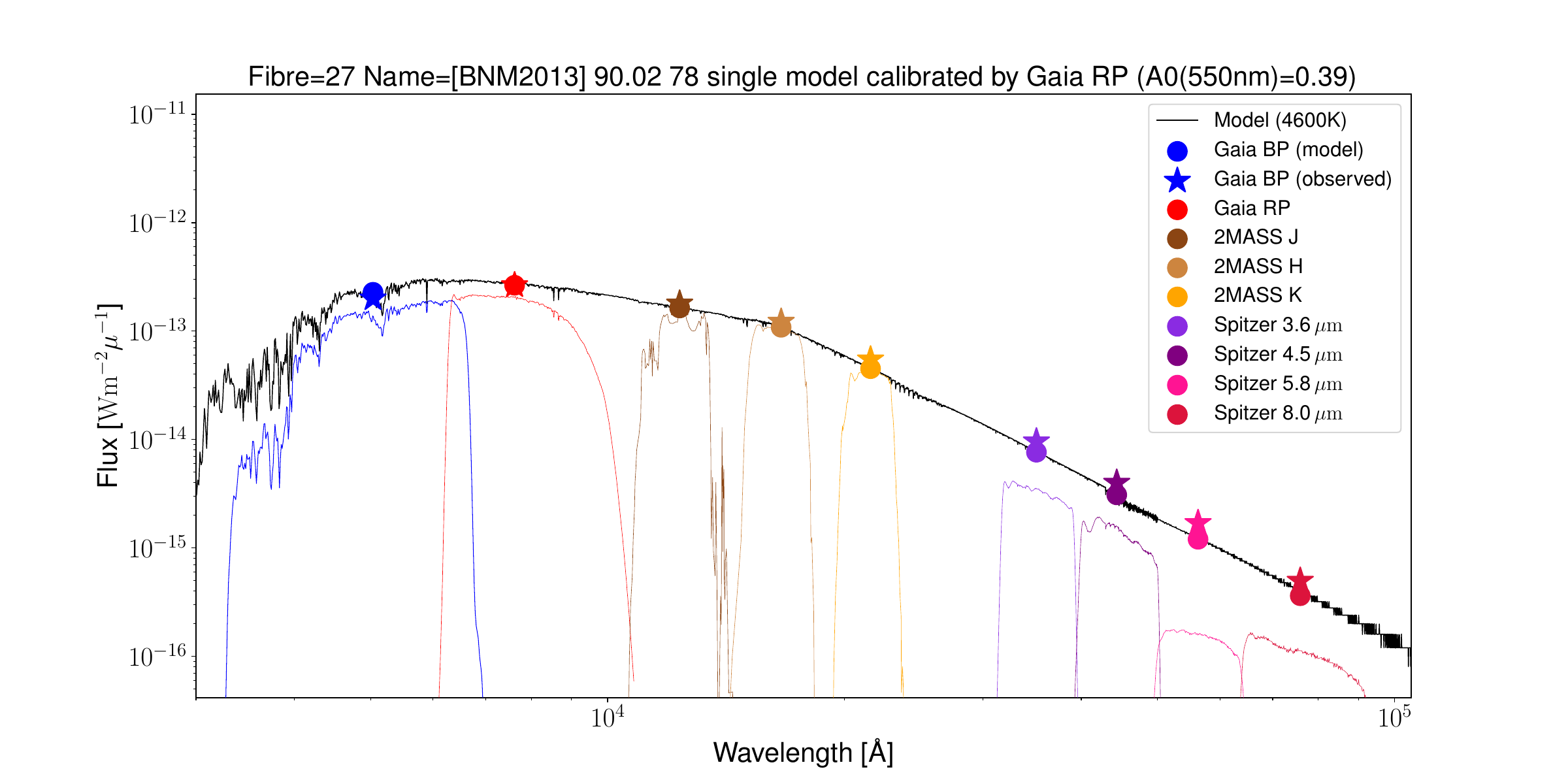}}
		\subfloat[]{\includegraphics[width=0.7\textwidth]{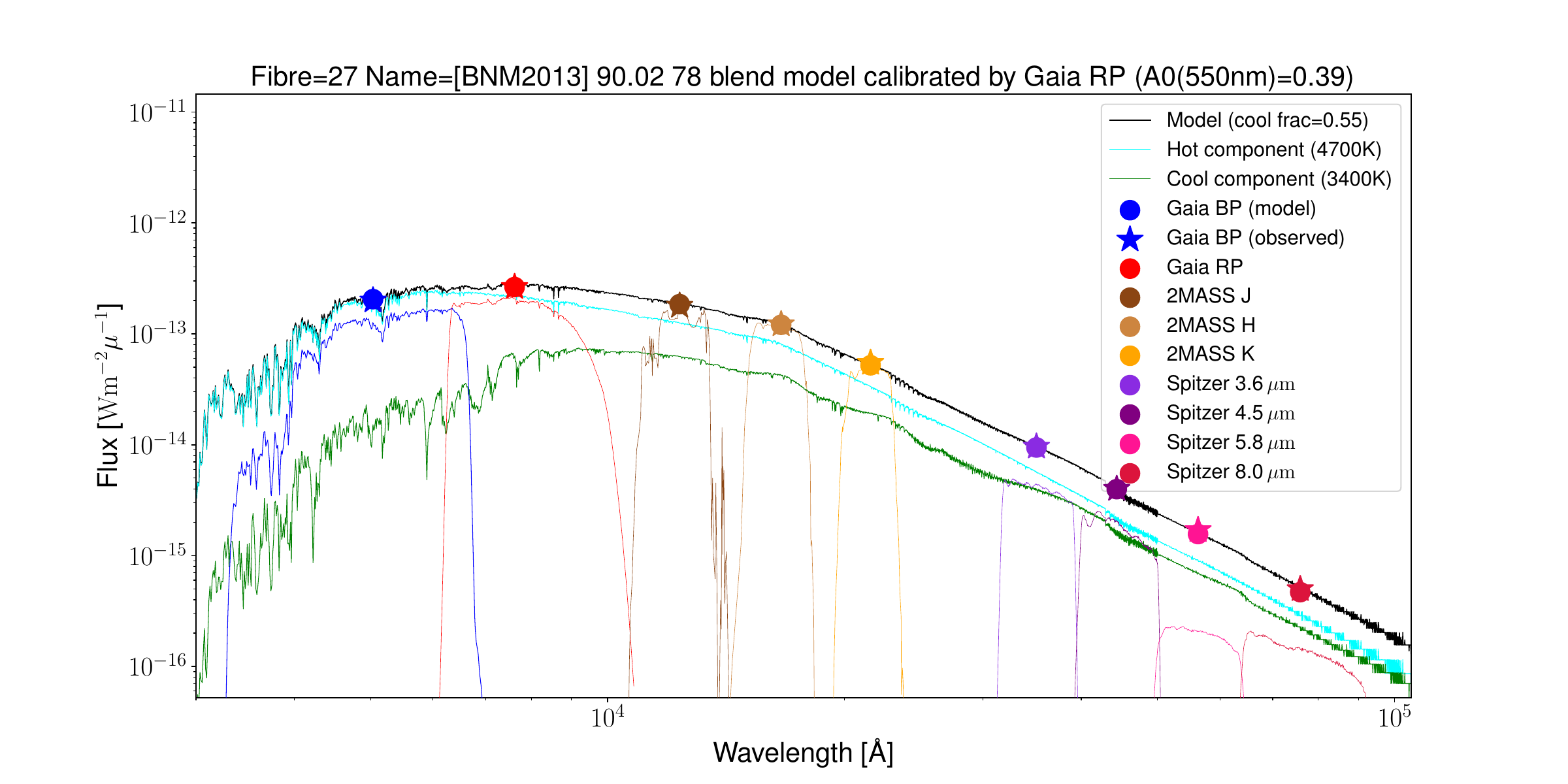}}\\
		\subfloat[]{\includegraphics[width=0.7\textwidth]{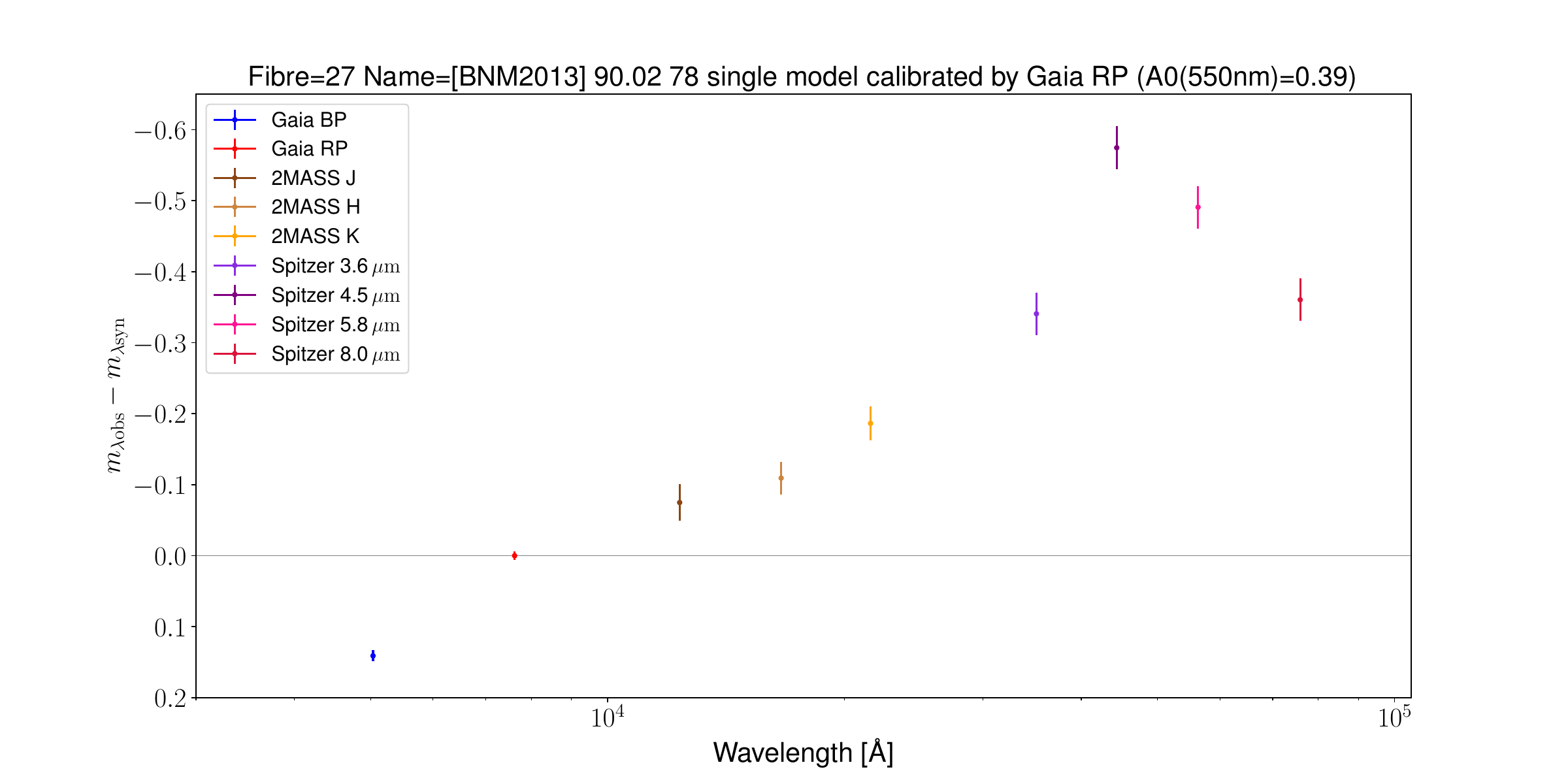}}
		\subfloat[]{\includegraphics[width=0.7\textwidth]{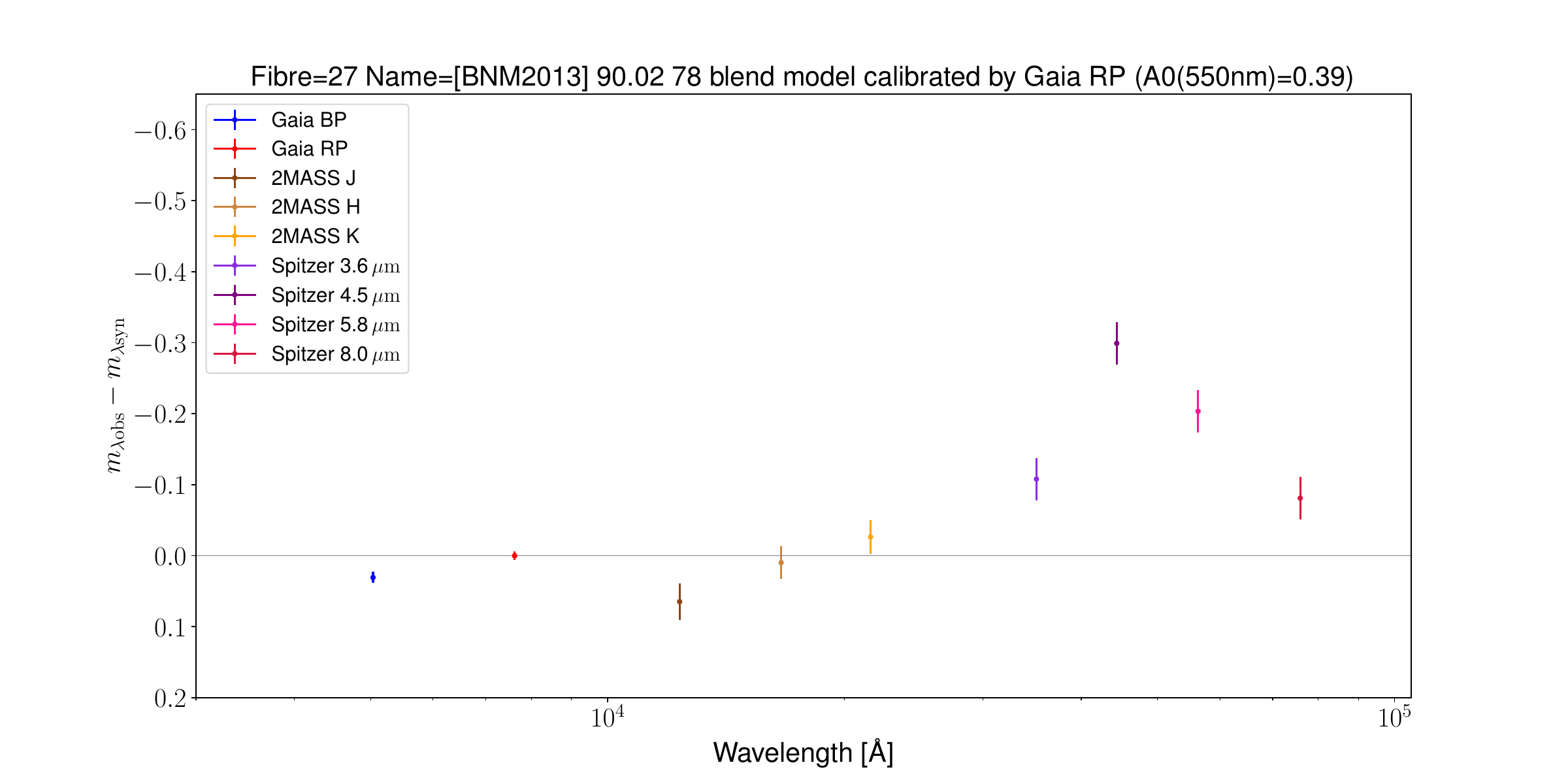}}
		\caption{As Fig.~\ref{fig:fibre_34_photometry} for fibre 27.}
		\label{fig:fibre_27_photometry}
	\end{figure}
\end{landscape}

\begin{landscape}
	\begin{figure}
		\centering
		\subfloat[]{\includegraphics[width=0.7\textwidth]{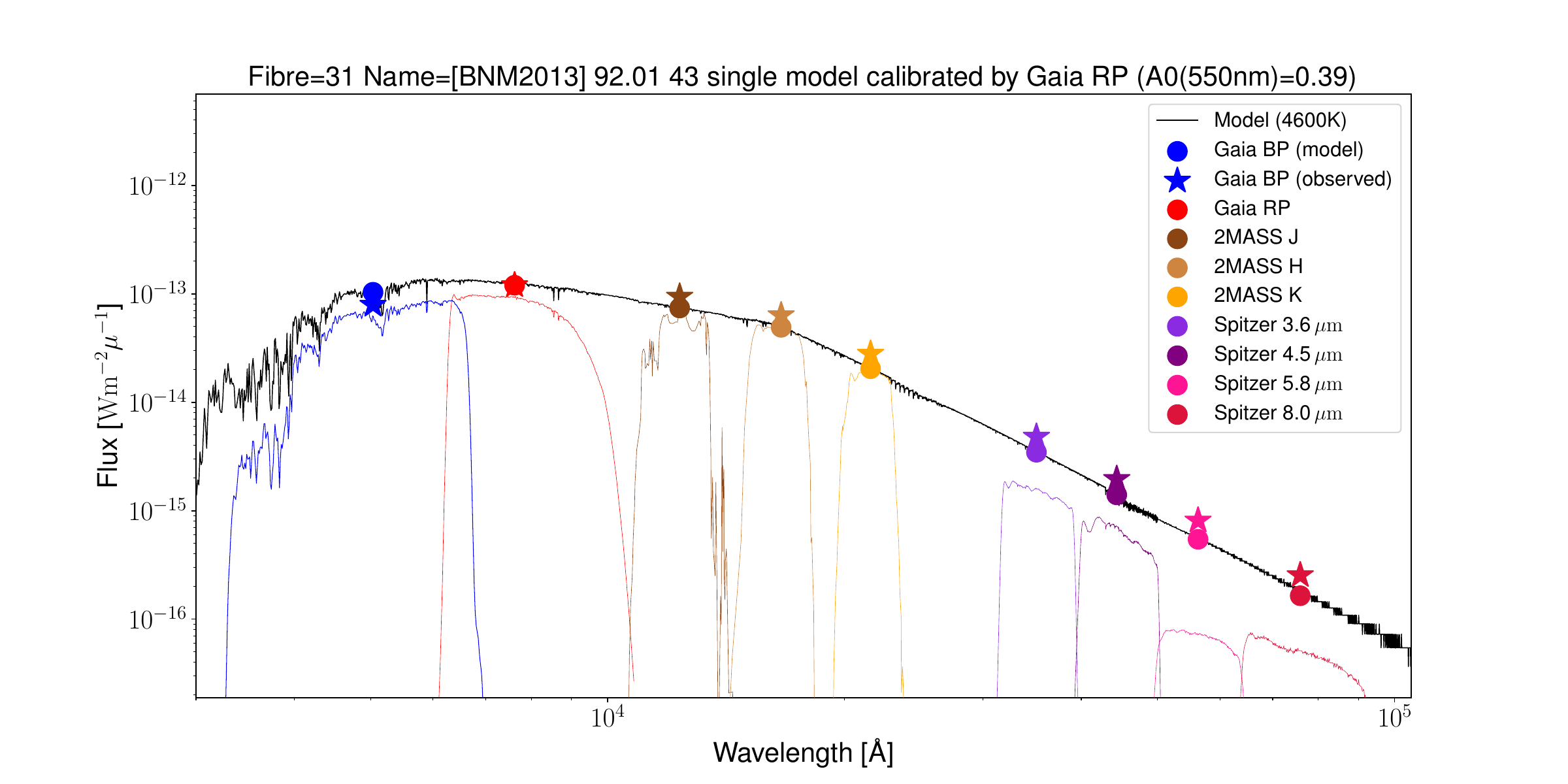}}
		\subfloat[]{\includegraphics[width=0.7\textwidth]{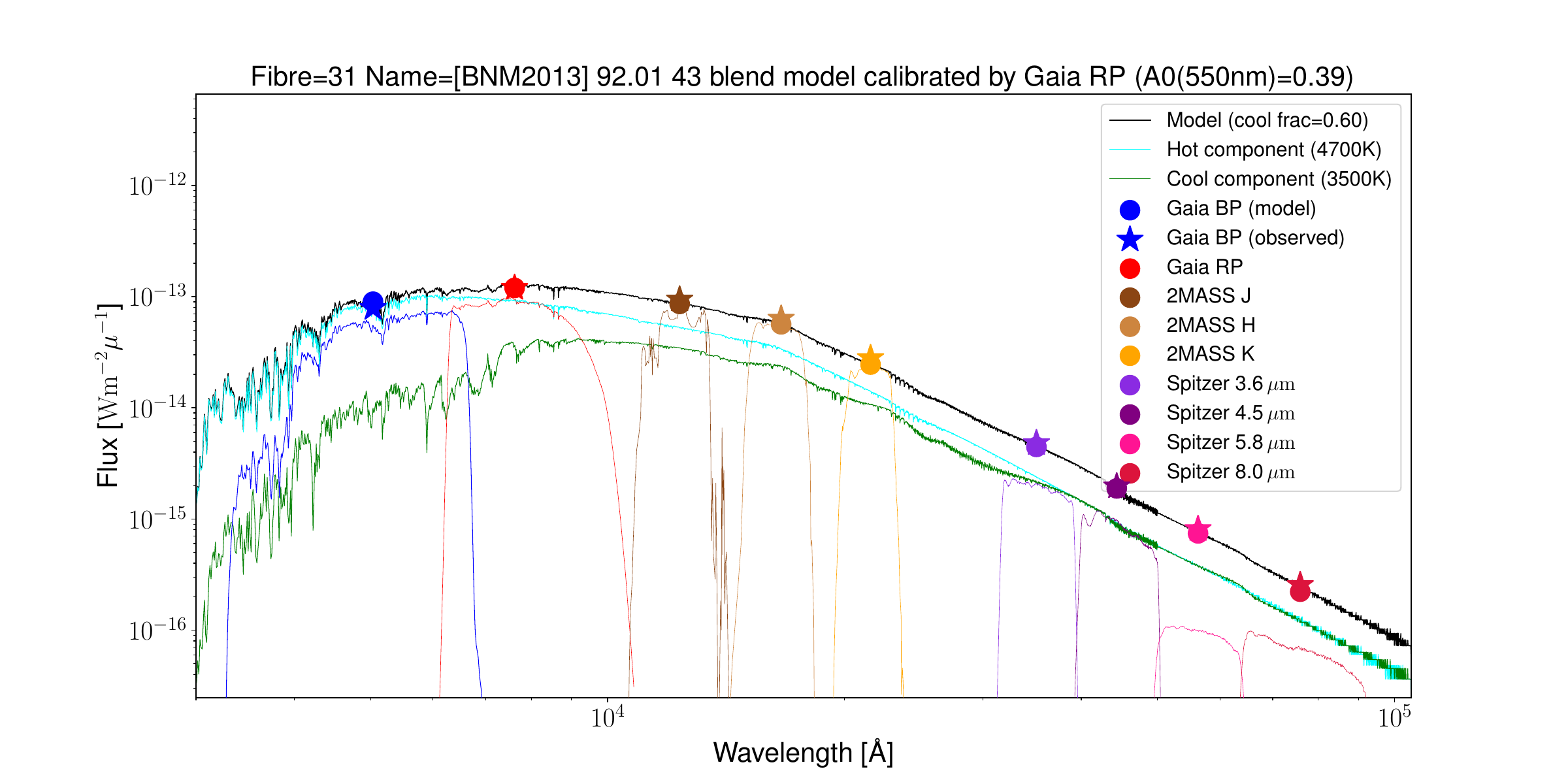}}\\
		\subfloat[]{\includegraphics[width=0.7\textwidth]{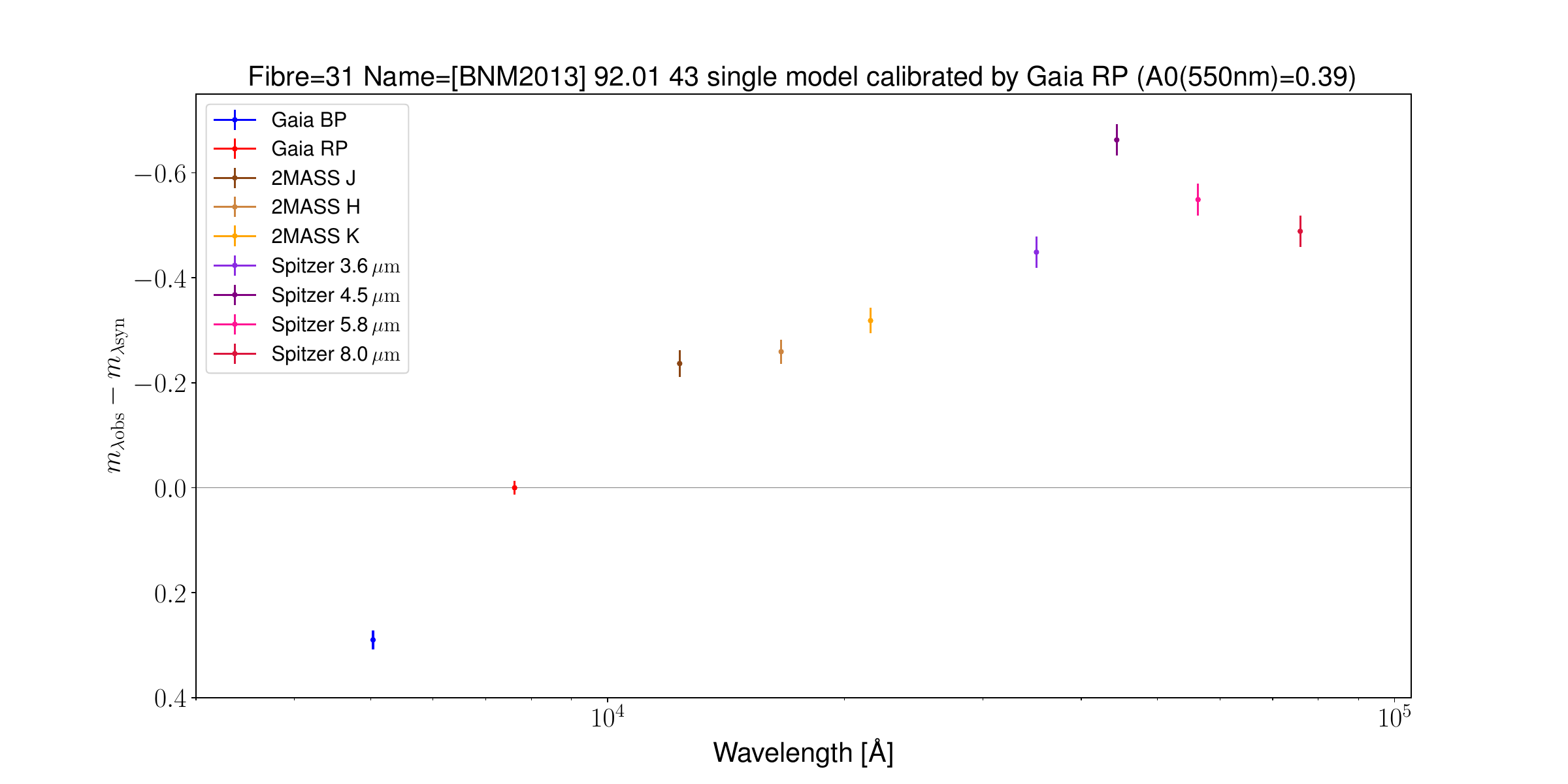}}
		\subfloat[]{\includegraphics[width=0.7\textwidth]{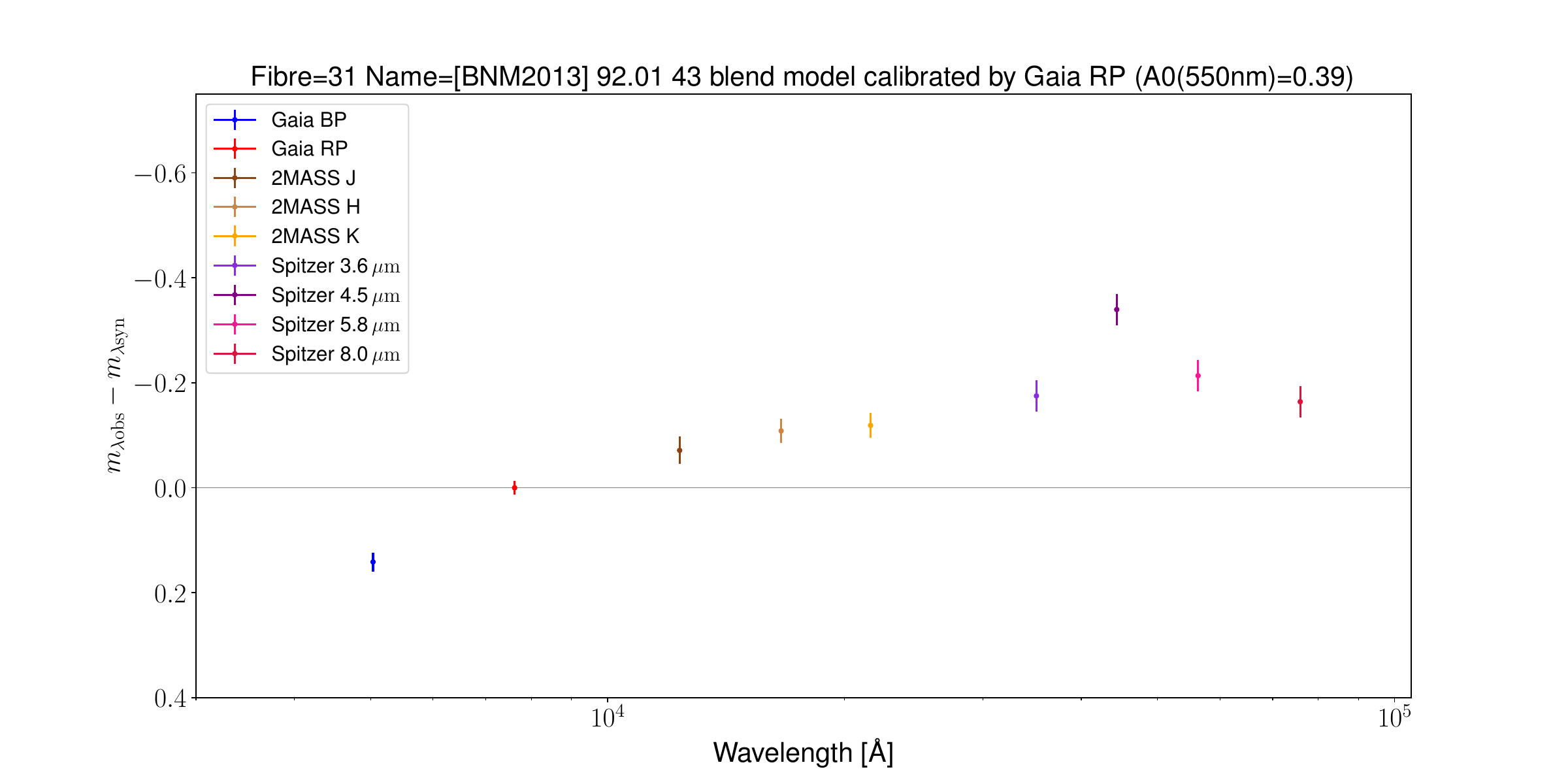}}
		\caption{As Fig.~\ref{fig:fibre_34_photometry} for fibre 31.}
		\label{fig:fibre_31_photometry}
	\end{figure}
\end{landscape}

\begin{landscape}
	\begin{figure}
		\centering
		\subfloat[]{\includegraphics[width=0.7\textwidth]{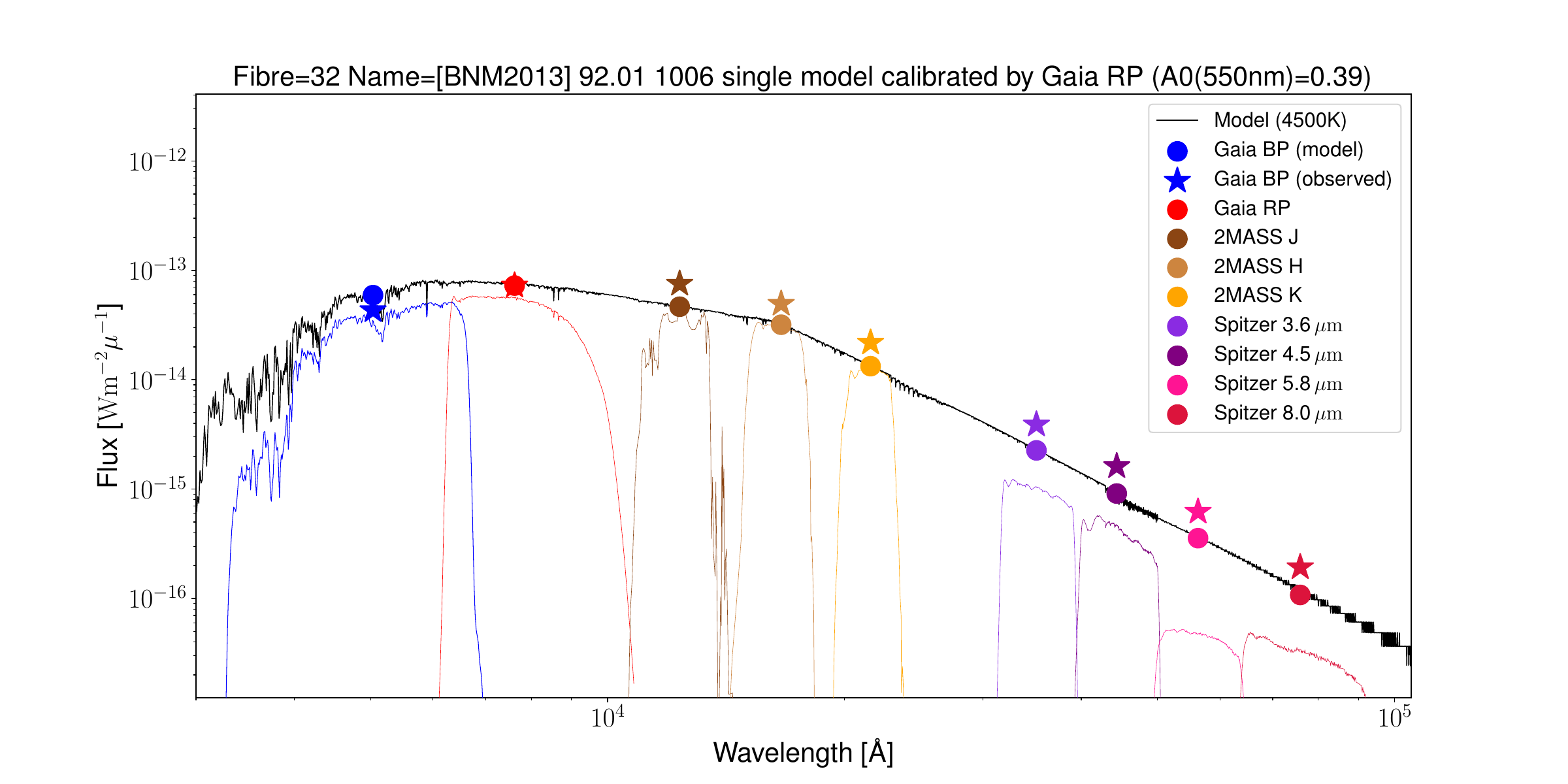}}
		\subfloat[]{\includegraphics[width=0.7\textwidth]{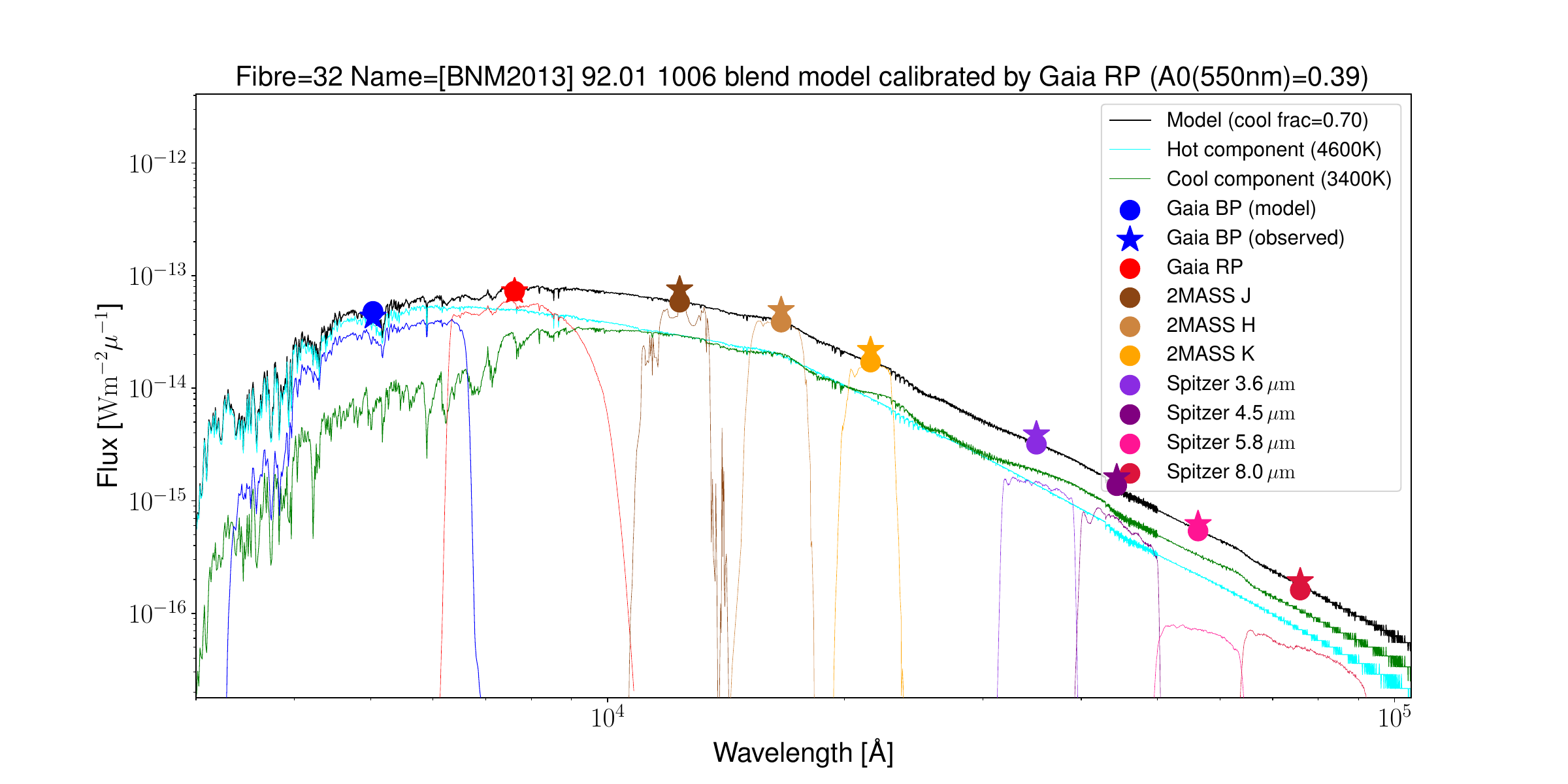}}\\
		\subfloat[]{\includegraphics[width=0.7\textwidth]{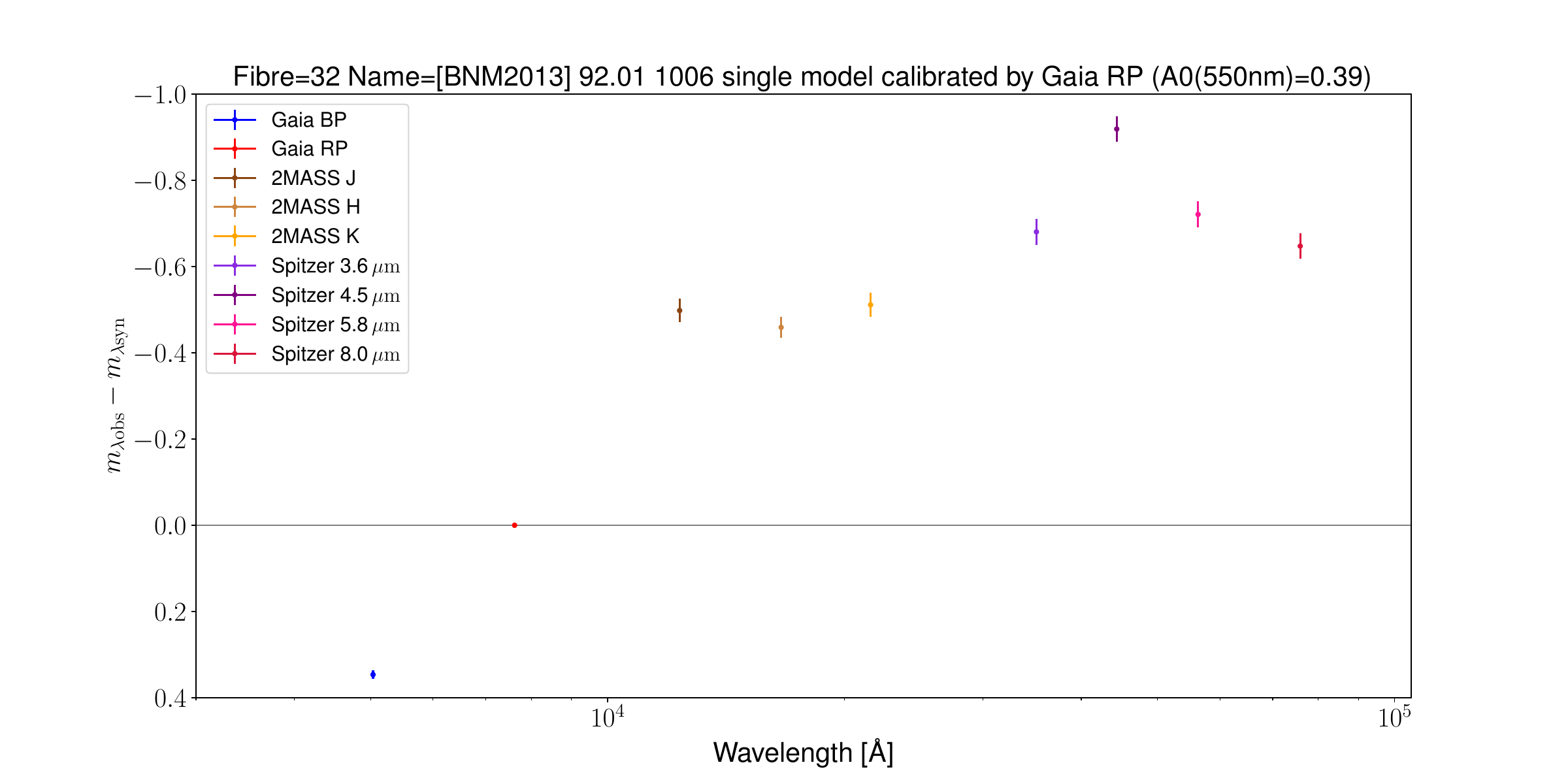}}
		\subfloat[]{\includegraphics[width=0.7\textwidth]{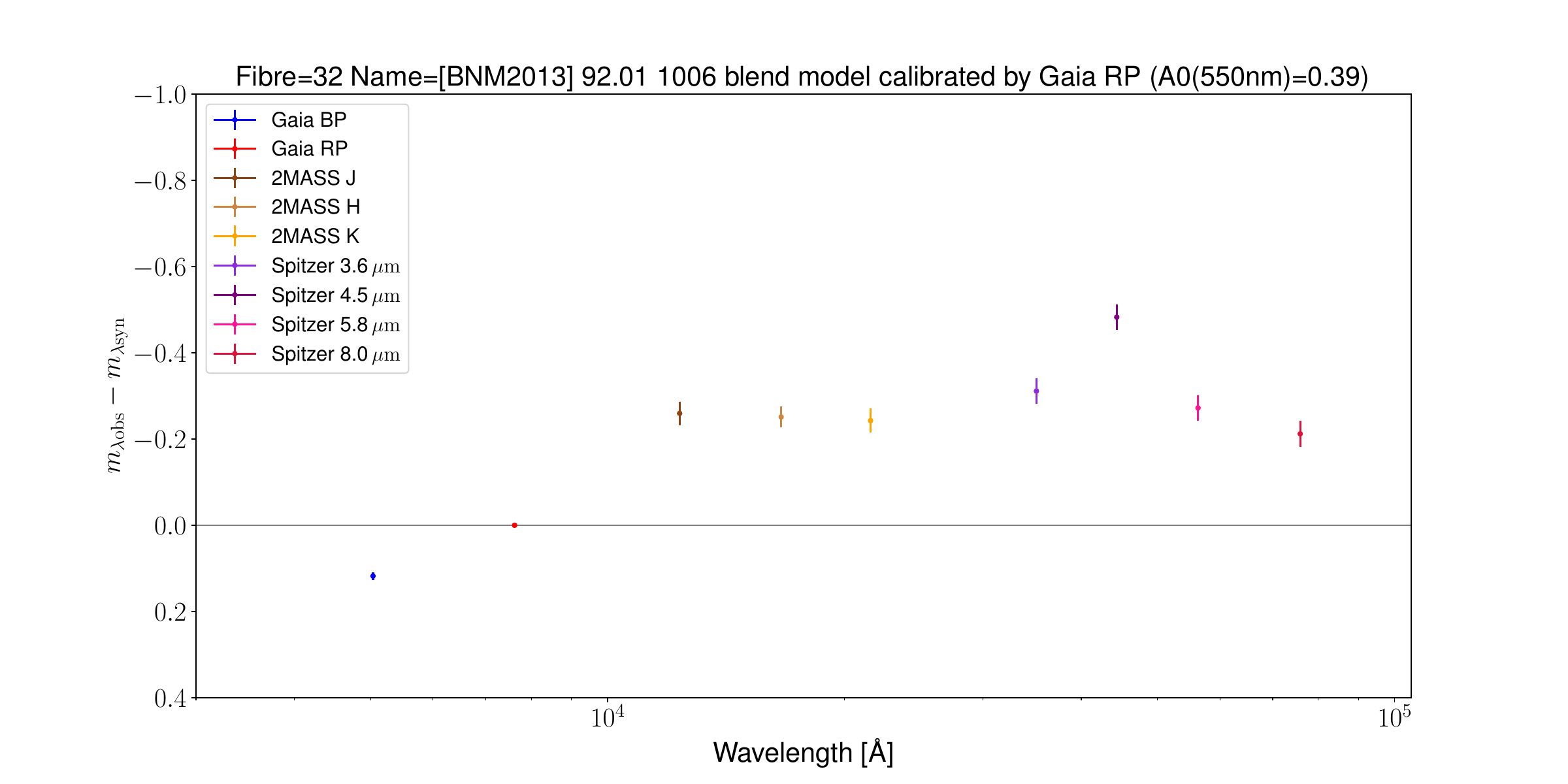}}
		\caption{As Fig.~\ref{fig:fibre_34_photometry} for fibre 32.}
		\label{fig:fibre_32_photometry}
	\end{figure}
\end{landscape}

\begin{landscape}
	\begin{figure}
		\centering
		\subfloat[]{\includegraphics[width=0.7\textwidth]{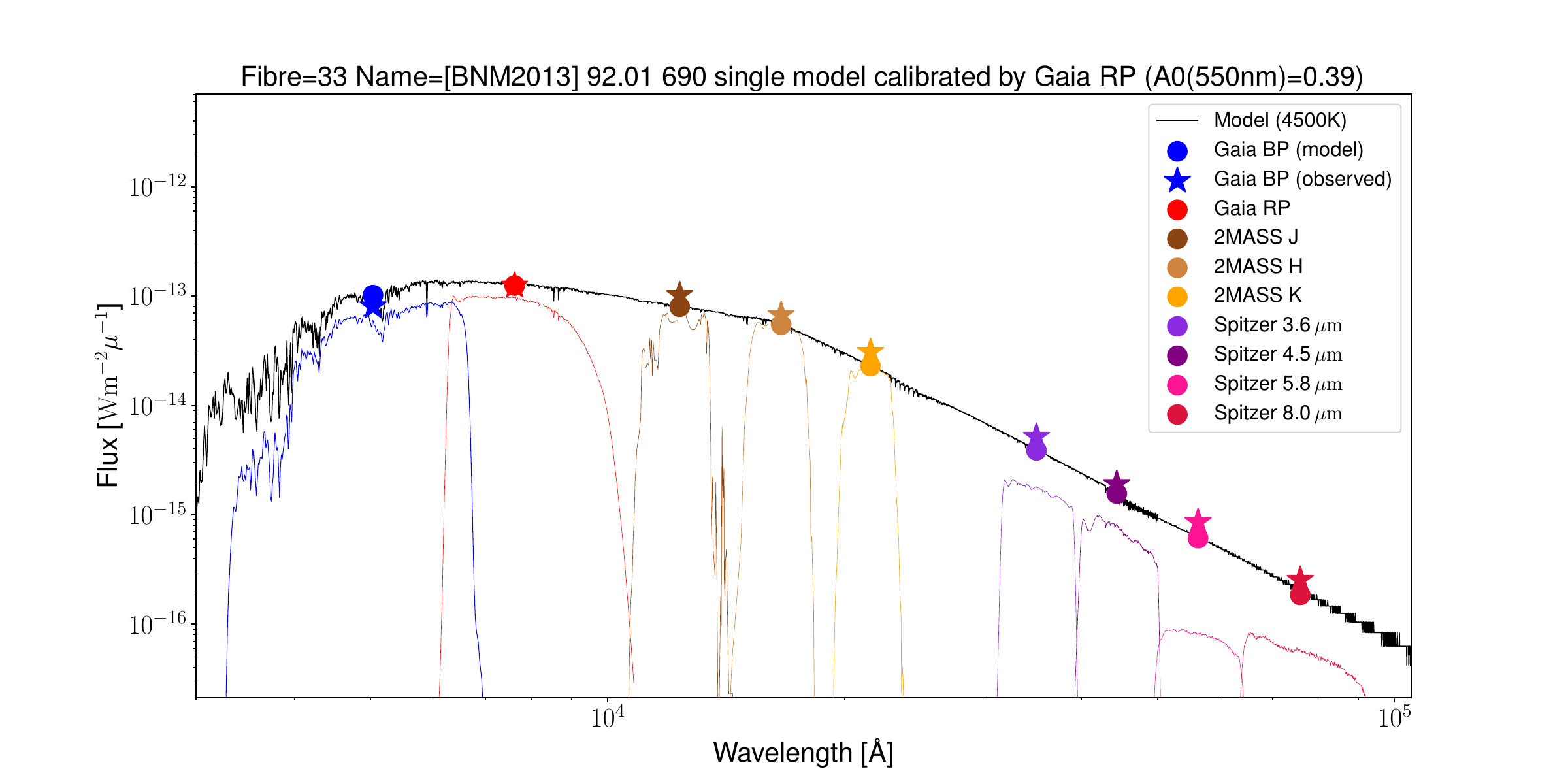}}
		\subfloat[]{\includegraphics[width=0.7\textwidth]{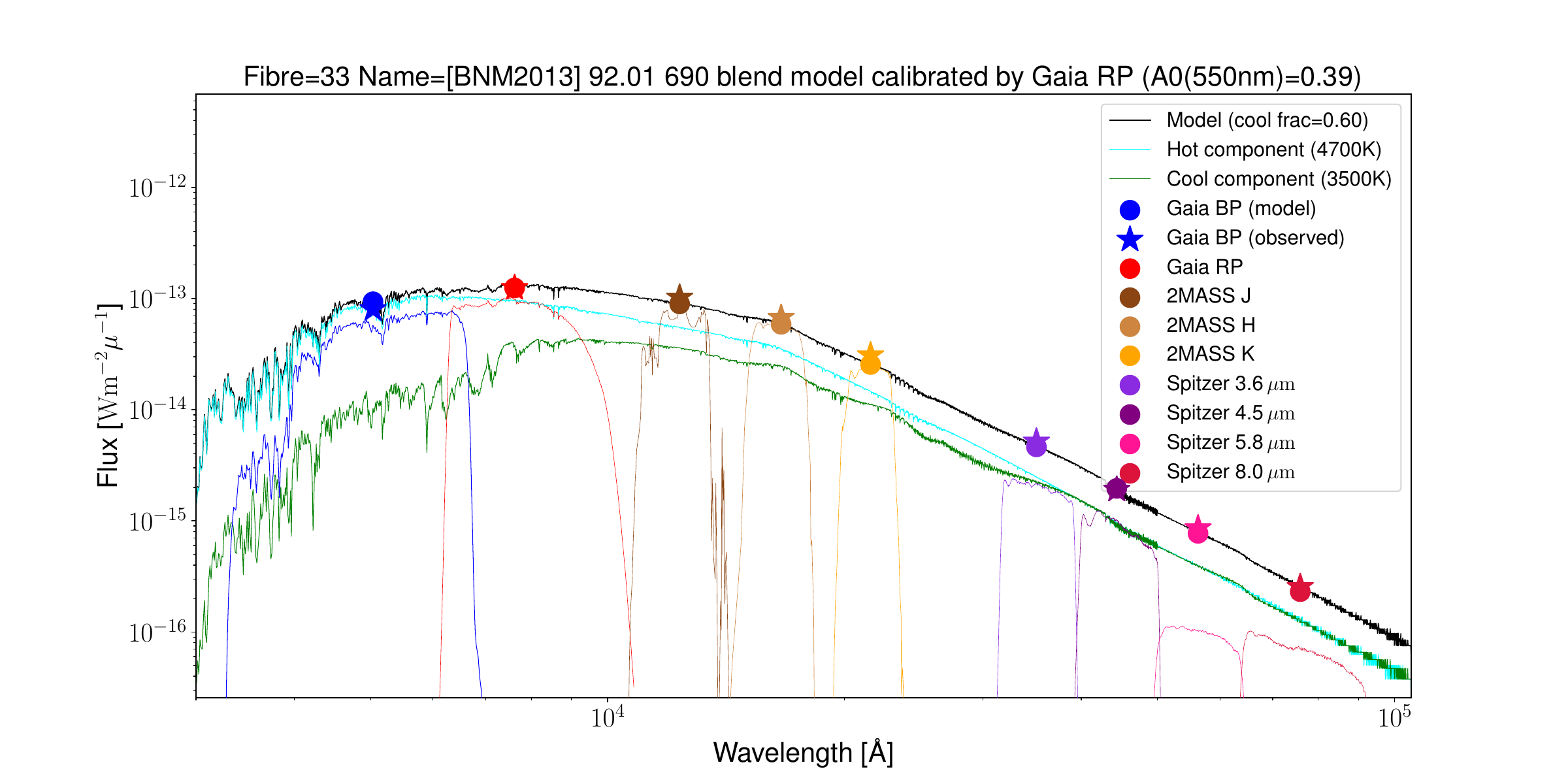}}\\
		\subfloat[]{\includegraphics[width=0.7\textwidth]{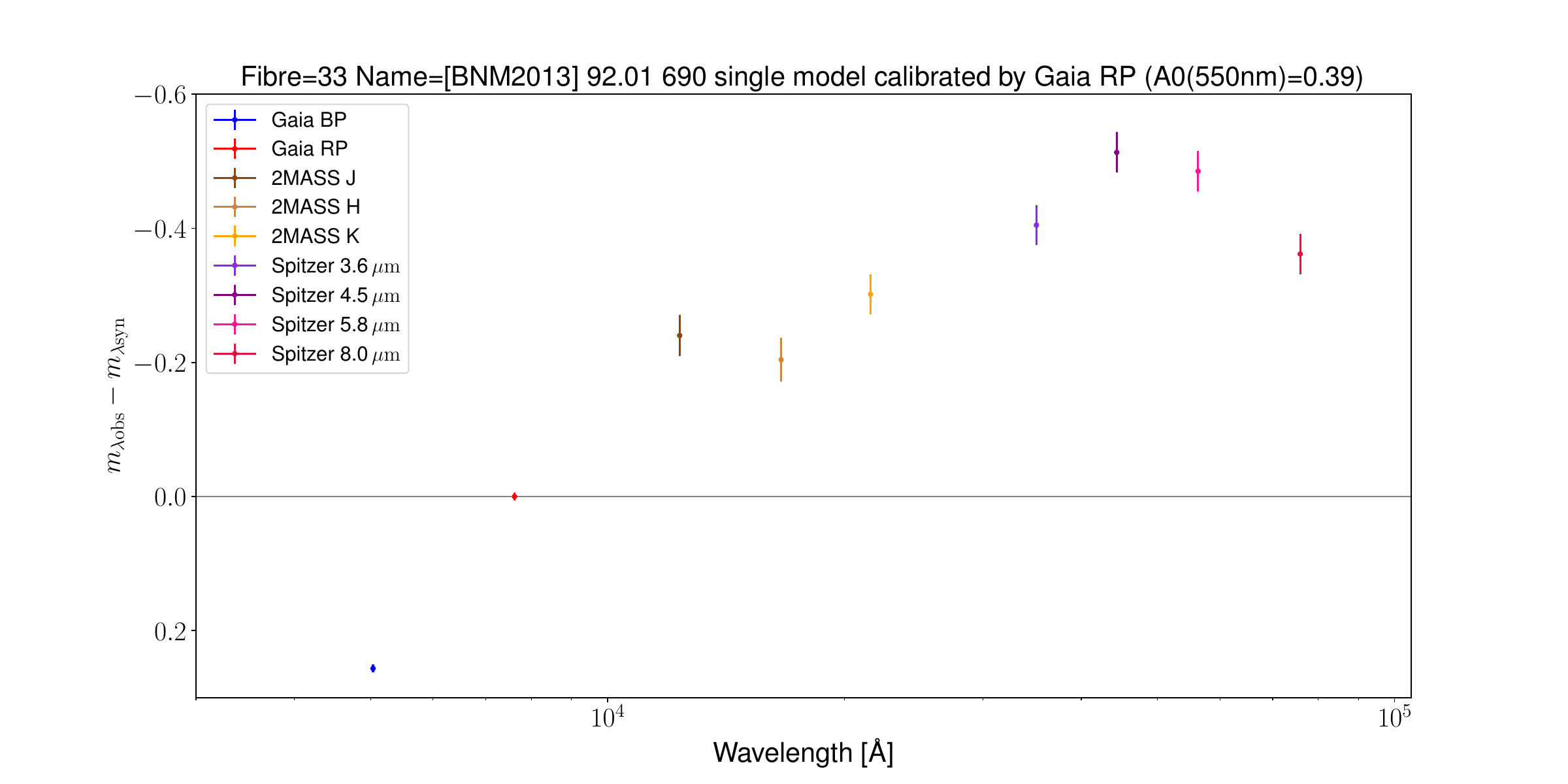}}
		\subfloat[]{\includegraphics[width=0.7\textwidth]{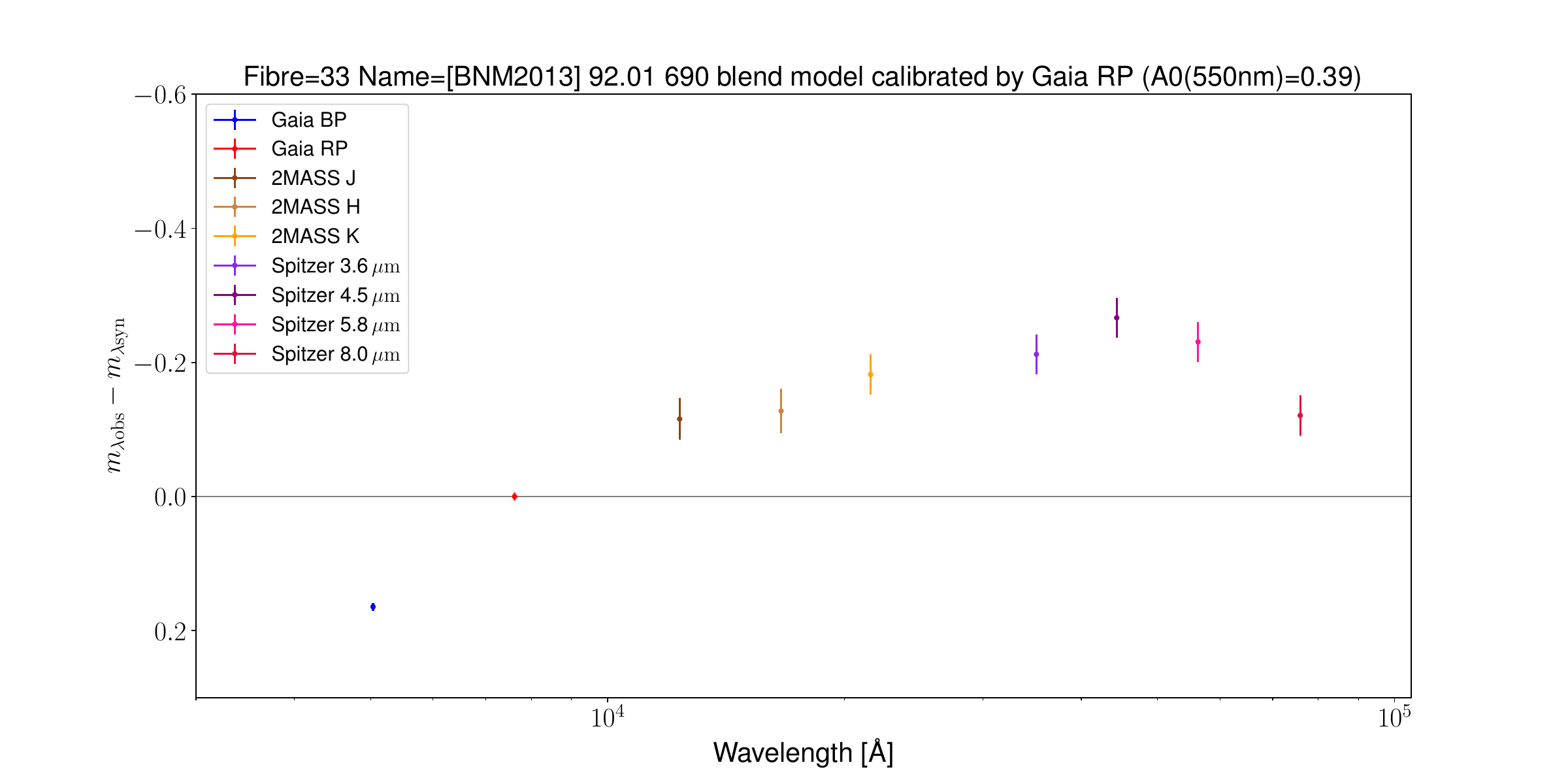}}
		\caption{As Fig.~\ref{fig:fibre_34_photometry} for fibre 33.}
		\label{fig:fibre_33_photometry}
	\end{figure}
\end{landscape}

\begin{landscape}
	\begin{figure}
		\centering
		\subfloat[]{\includegraphics[width=0.7\textwidth]{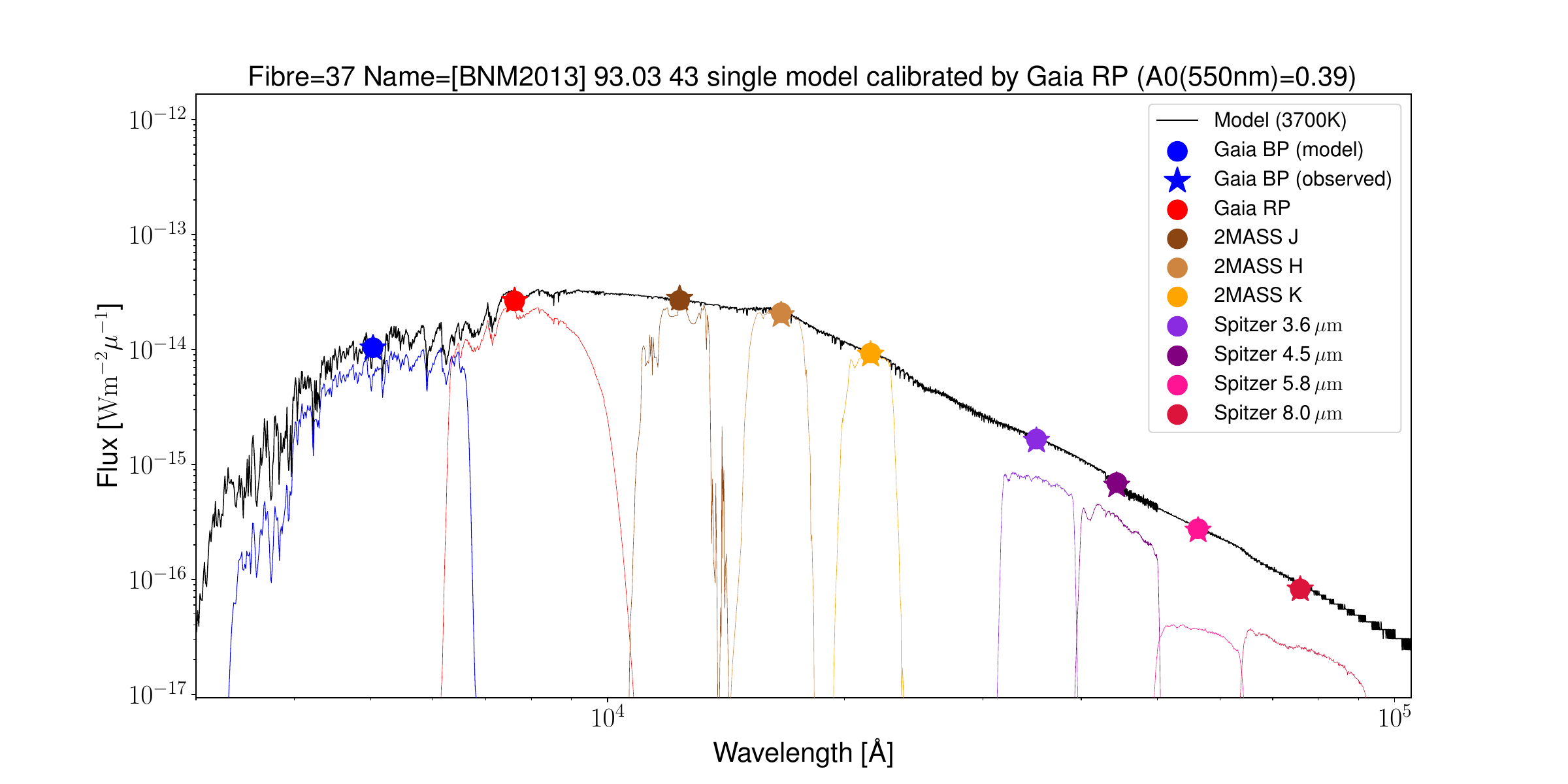}}
		\subfloat[]{\includegraphics[width=0.7\textwidth]{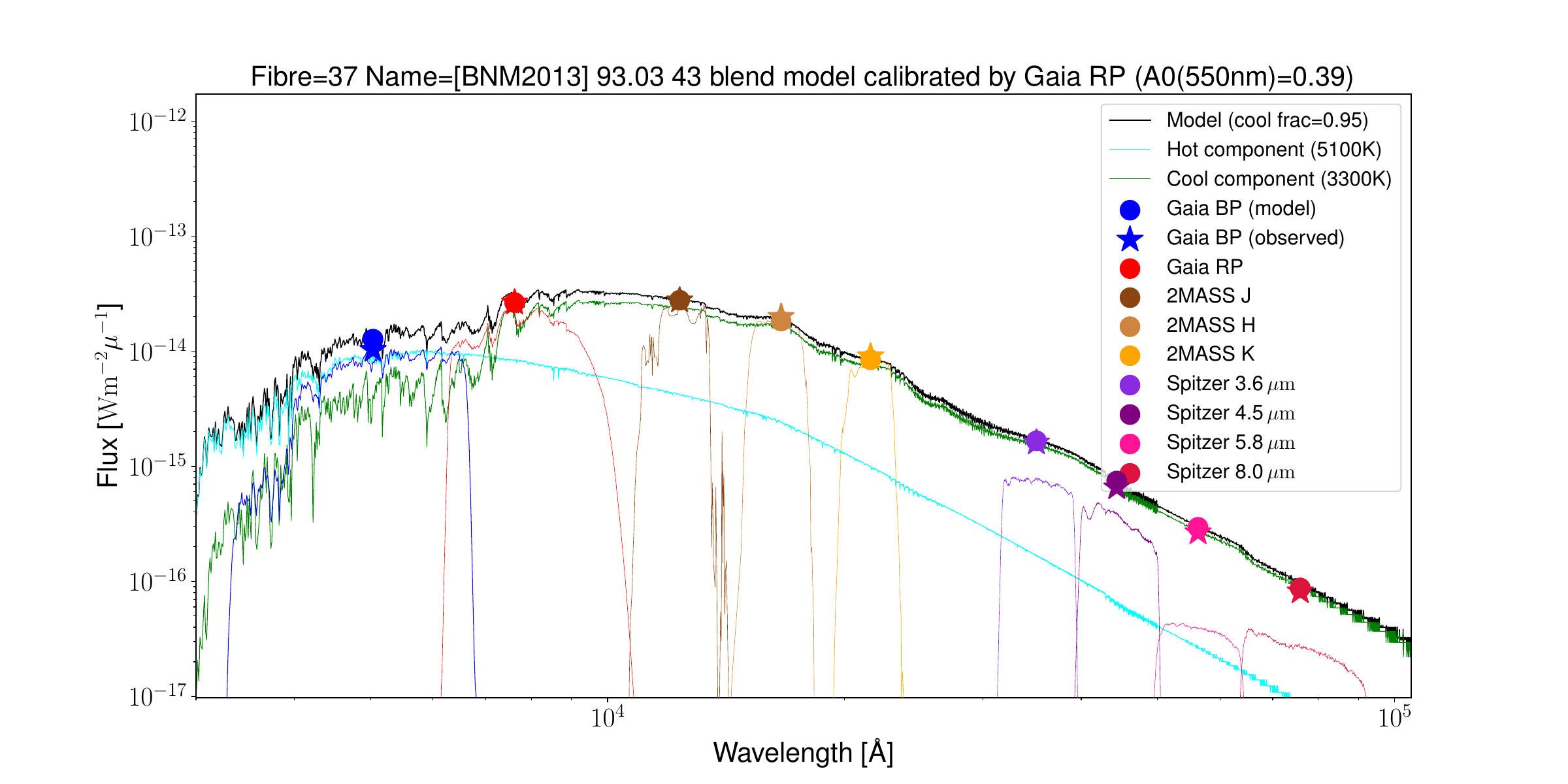}}\\
		\subfloat[]{\includegraphics[width=0.7\textwidth]{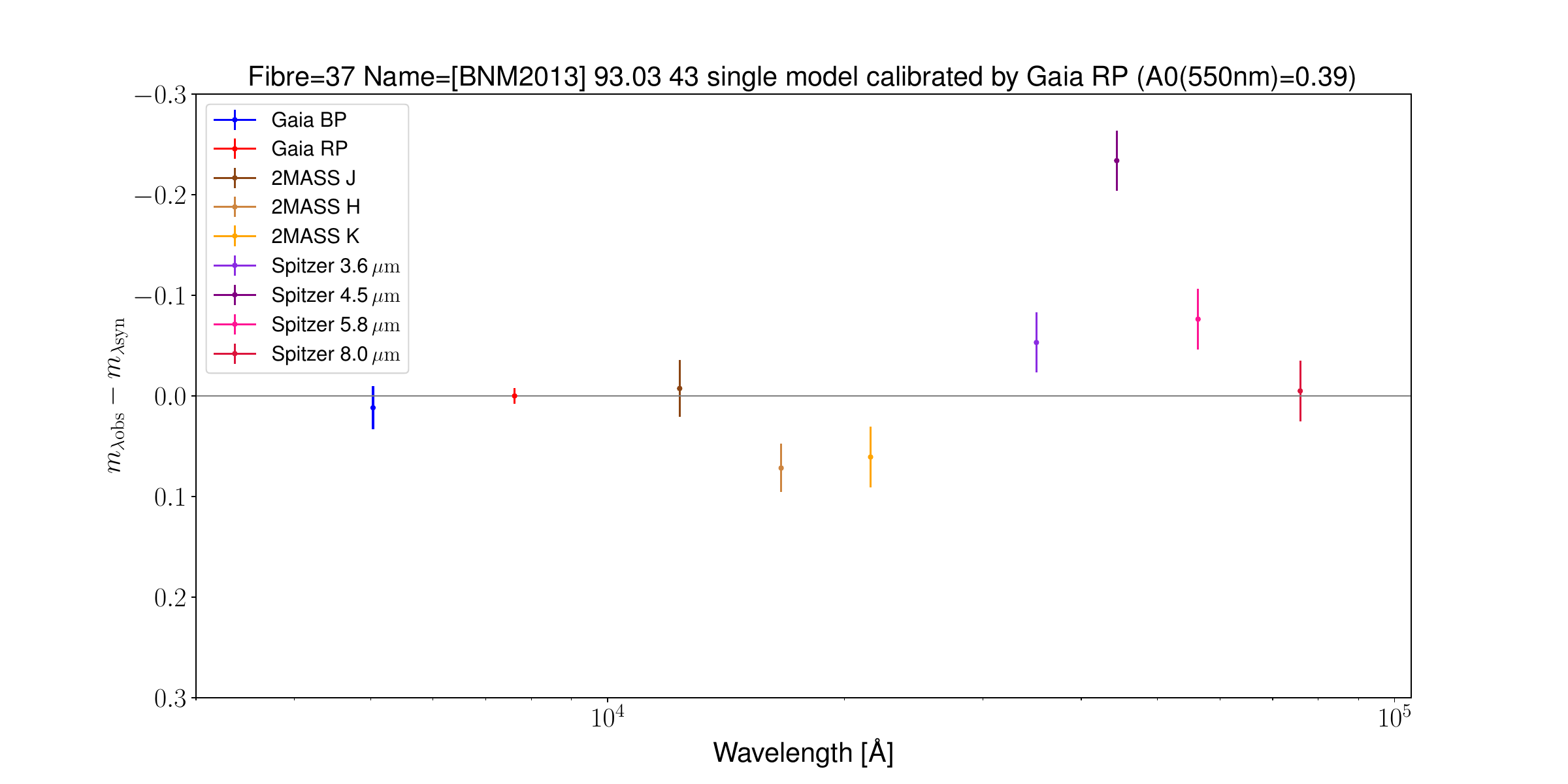}}
		\subfloat[]{\includegraphics[width=0.7\textwidth]{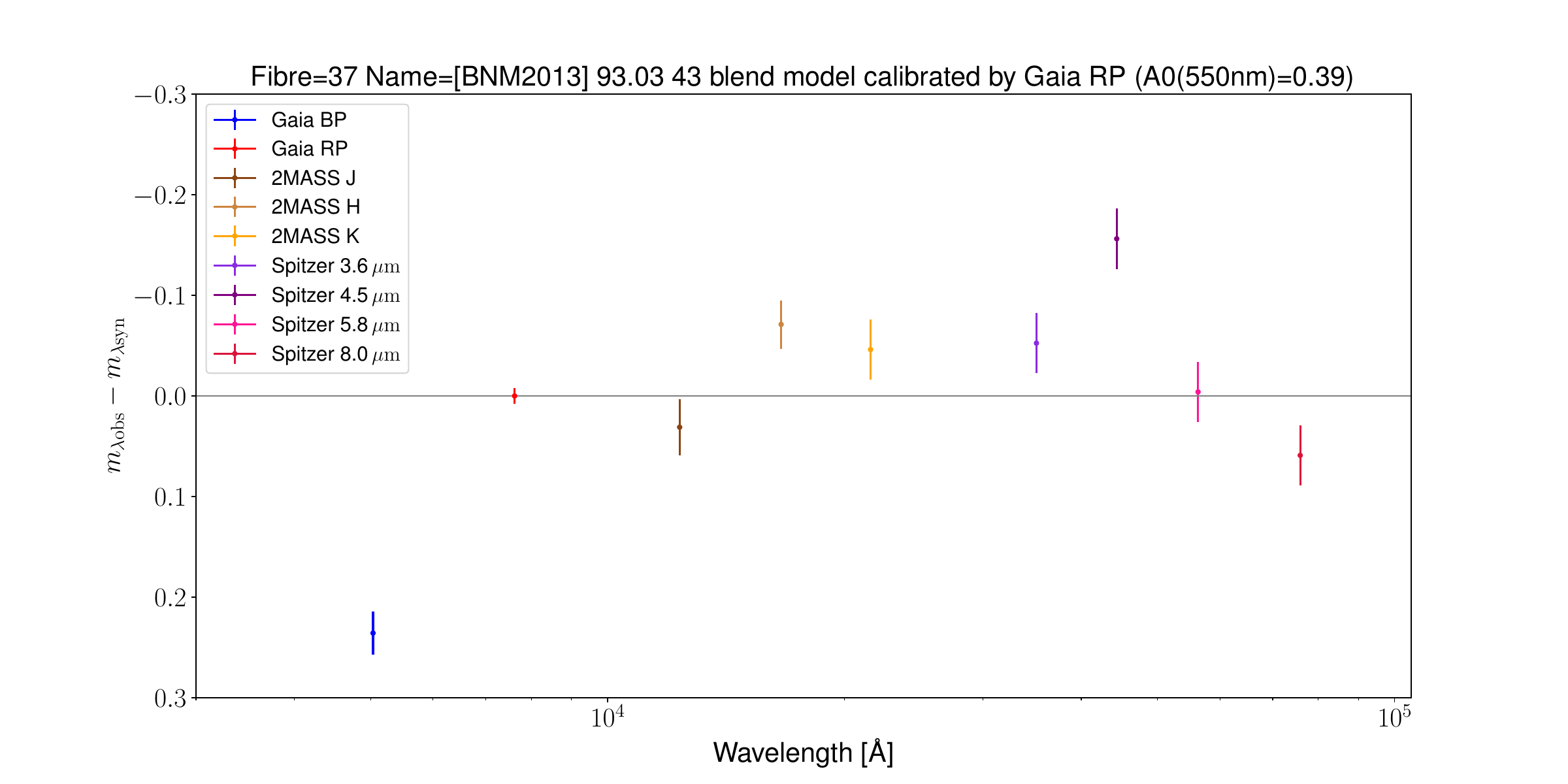}}
		\caption{As Fig.~\ref{fig:fibre_34_photometry} for fibre 37.}
		\label{fig:fibre_37_photometry}
	\end{figure}
\end{landscape}

\begin{landscape}
	\begin{figure}
		\centering
		\subfloat[]{\includegraphics[width=0.7\textwidth]{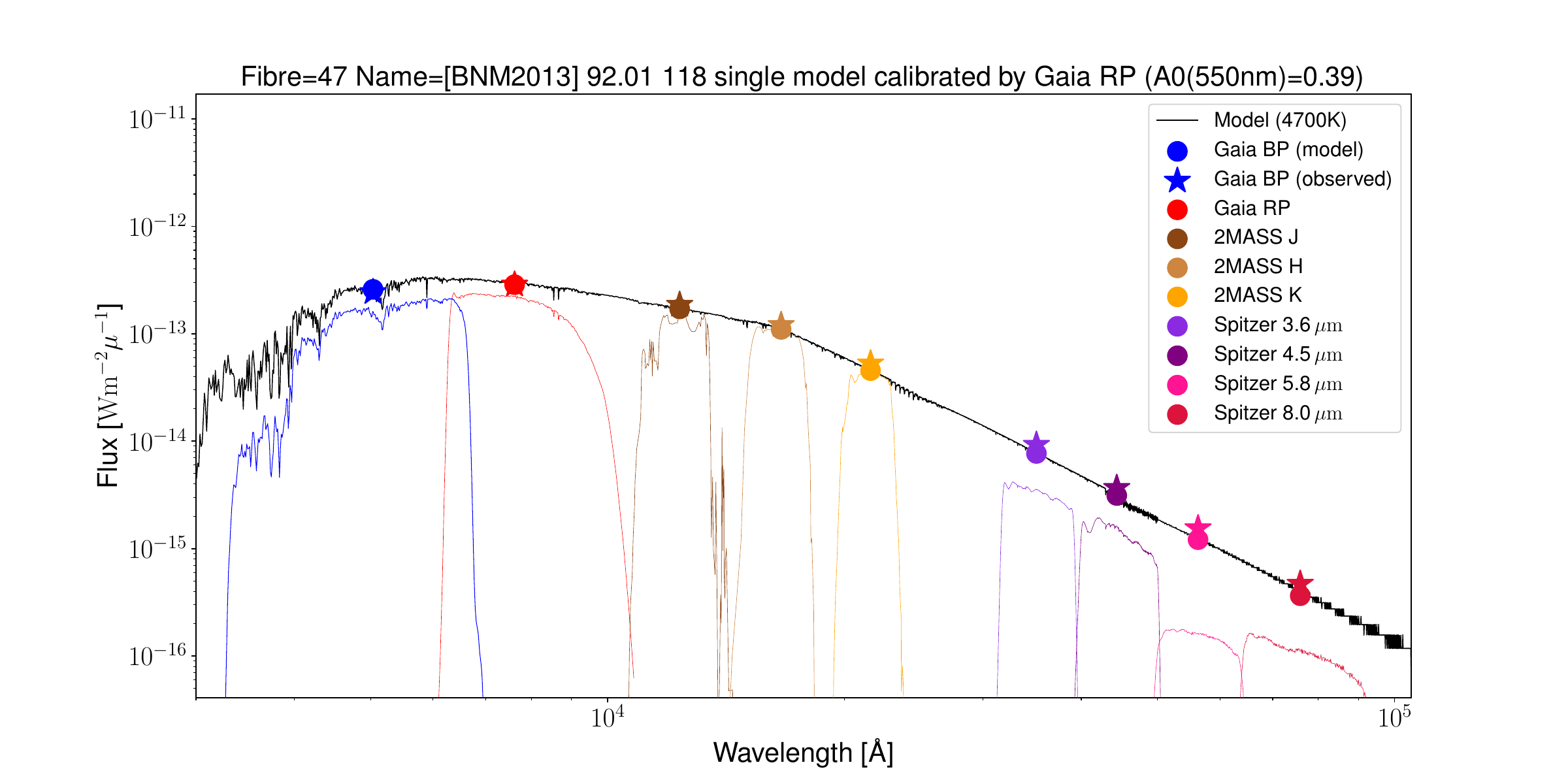}}
		\subfloat[]{\includegraphics[width=0.7\textwidth]{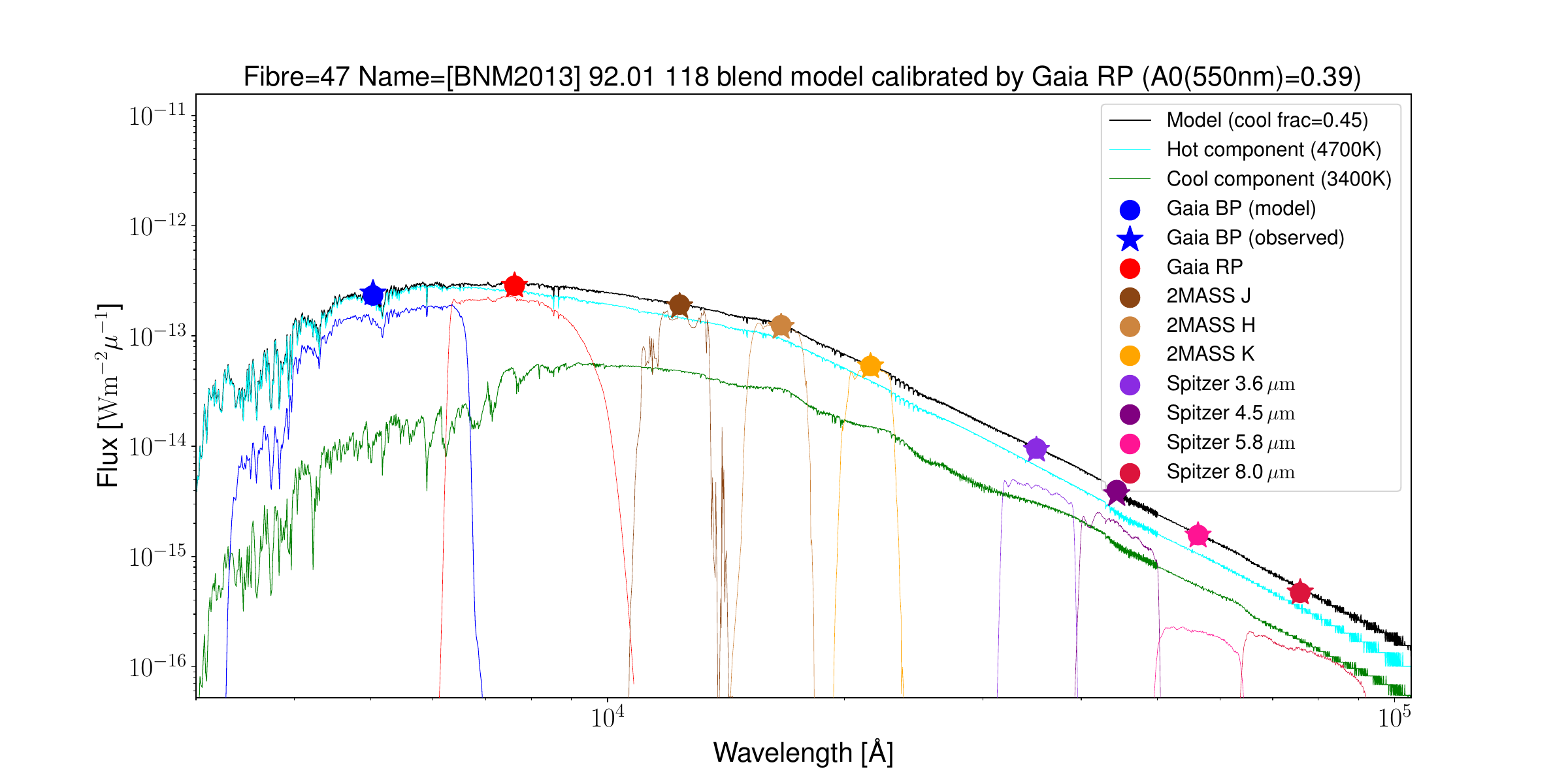}}\\
		\subfloat[]{\includegraphics[width=0.7\textwidth]{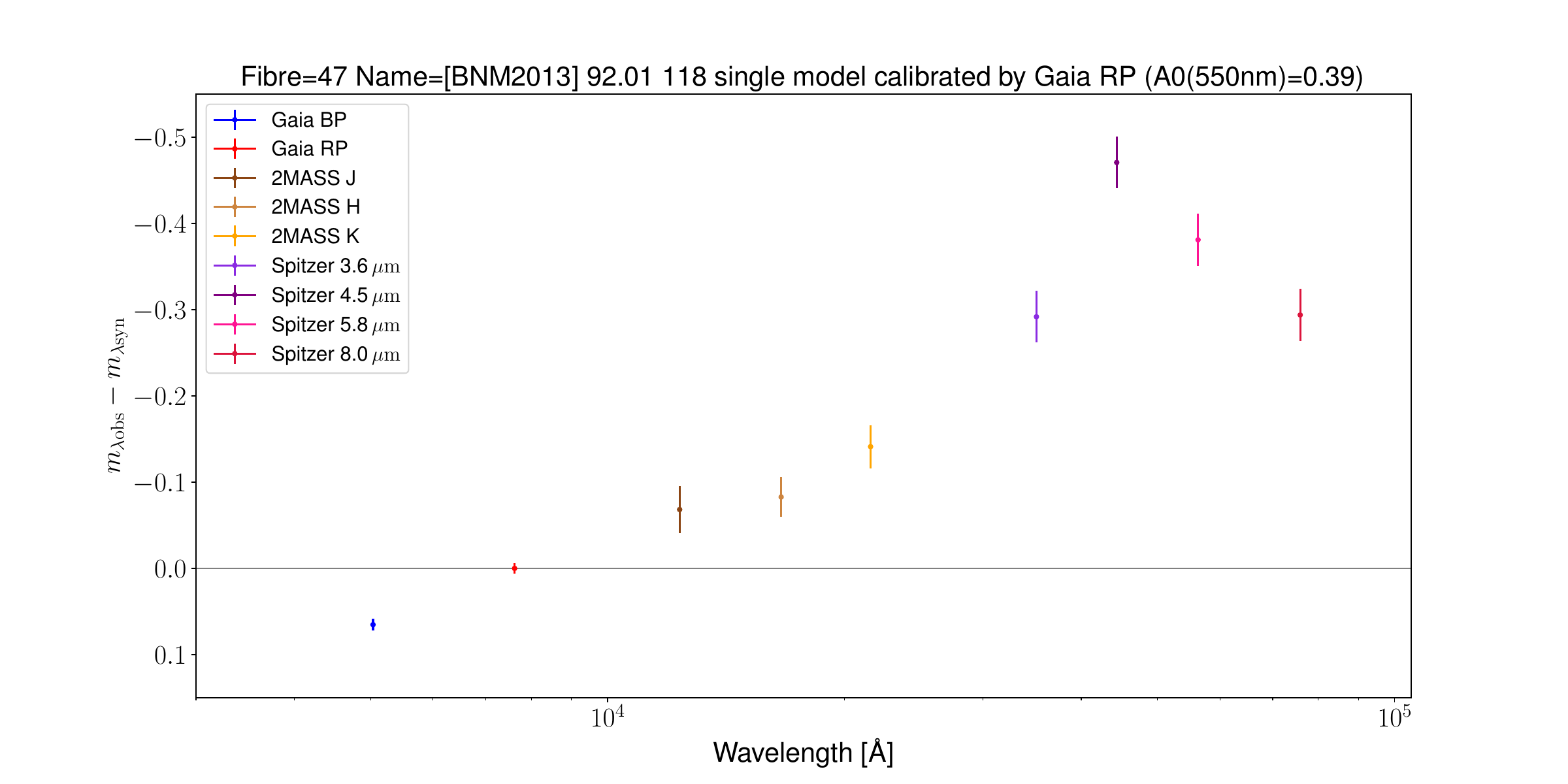}}
		\subfloat[]{\includegraphics[width=0.7\textwidth]{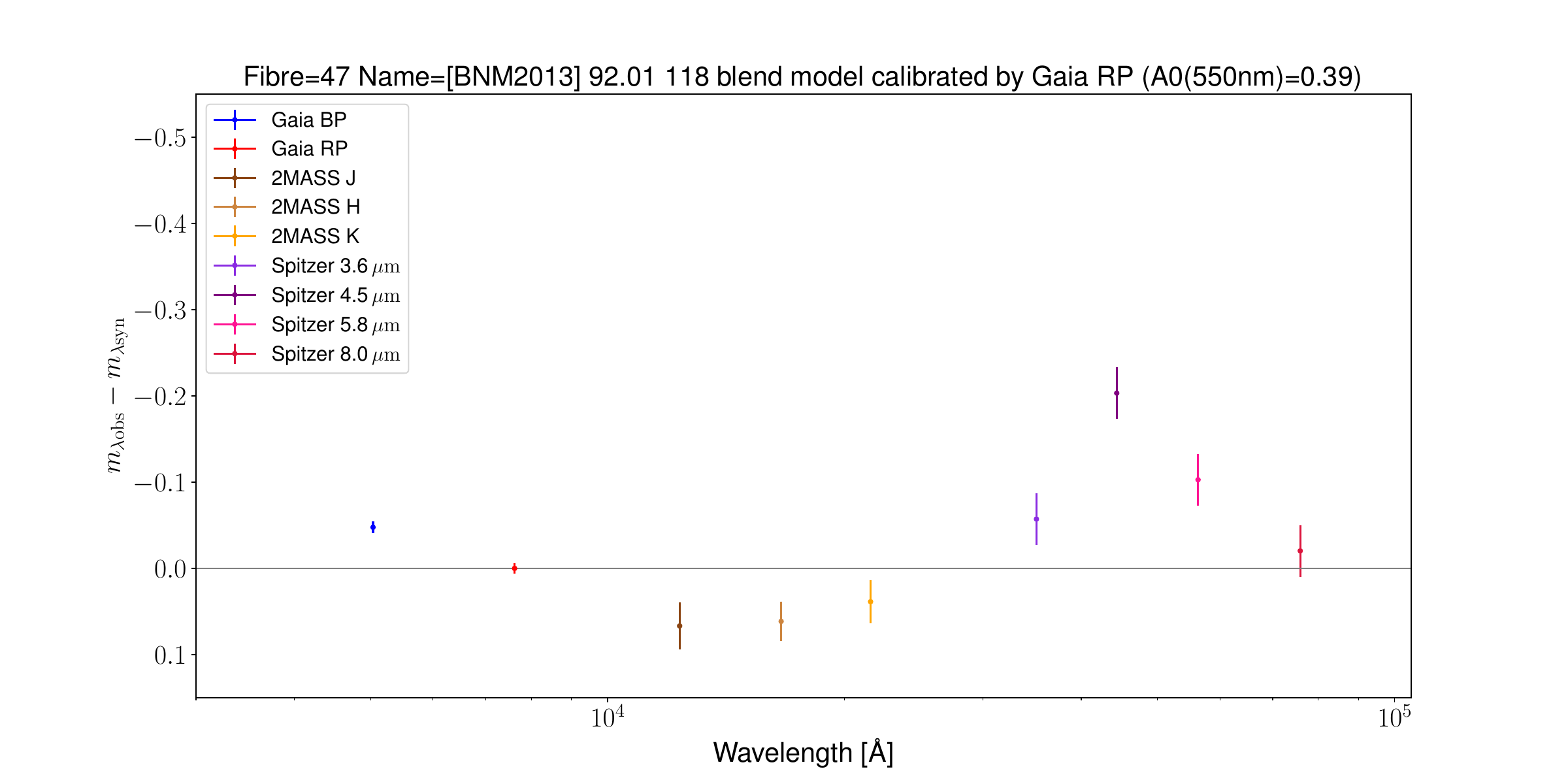}}
		\caption{As Fig.~\ref{fig:fibre_34_photometry} for fibre 47.}
		\label{fig:fibre_47_photometry}
	\end{figure}
\end{landscape}

We generated synthetic photometry from our single and blend of two temperature model spectra, and compared them with observed photometry from \textit{Gaia}, 2MASS and \textit{Spitzer}. Plots of the observed flux over the models for the \refone{WTTS K sample} listed in Table~\ref{tab:sigori_review_members} that do not appear as figures within the main body of the paper are displayed in Figs.\ref{fig:fibre_18_photometry} to \ref{fig:fibre_47_photometry}. The top plots in each figure display the model spectrum, with synthetic photometry from the spectrum and the observed photometry converted to flux using the zero points. The bottom plots show the excess in the observed flux over the models in magnitudes. \editten{To aid a comparison between the single and two-temperature models, the same magnitude scale is used for each pair of plots.}


\bsp	
\label{lastpage}
\end{document}